\documentclass{article}
\usepackage{authblk}

\usepackage[english]{babel}

\usepackage[letterpaper,top=2cm,bottom=2cm,left=3cm,right=3cm,marginparwidth=1.75cm]{geometry}

\usepackage{amsmath}
\usepackage{graphicx}
\usepackage{xcolor}
\usepackage{amssymb}
\usepackage{ulem}
\usepackage{threeparttable}
\usepackage{soul}

\usepackage{setspace,wrapfig}
\usepackage[leftcaption]{sidecap}

\usepackage{orcidlink}

\usepackage{hyperref}
\hypersetup{colorlinks,linkcolor={red!60!black},citecolor={blue!60!black},urlcolor={blue!60!black}}

\usepackage{natbib,twoopt}
\bibpunct{(}{)}{;}{a}{}{,} 
\makeatletter
\newcommandtwoopt{\citeads}[3][][]{\href{http://ui.adsabs.harvard.edu/abs/#3}%
{\def\hyper@linkstart##1##2{}%
\let\hyper@linkend\@empty\citealp[#1][#2]{#3}}}
\newcommandtwoopt{\citepads}[3][][]{\href{http://ui.adsabs.harvard.edu/abs/#3}%
{\def\hyper@linkstart##1##2{}%
\let\hyper@linkend\@empty\citep[#1][#2]{#3}}}
\newcommandtwoopt{\citetads}[3][][]{\href{http://ui.adsabs.harvard.edu/abs/#3}%
{\def\hyper@linkstart##1##2{}%
\let\hyper@linkend\@empty\citet[#1][#2]{#3}}}
\newcommandtwoopt{\citeyearads}[3][][]%
{\href{http://ui.adsabs.harvard.edu/\#abs/#3}
{\def\hyper@linkstart##1##2{}%
\let\hyper@linkend\@empty\citeyear[#1][#2]{#3}}}
\makeatother

\newcommand{\kms}{km\,s$^{-1}$}

\newcommand{\teff}{$T_\mathrm{eff}$}

\def\gapprox{\;\rlap{\lower 2.5pt            
    \hbox{$\sim$}}\raise 1.5pt\hbox{$>$}\;}       
\def\lapprox{\;\rlap{\lower 2.5pt           
    \hbox{$\sim$}}\raise 1.5pt\hbox{$<$}\;} 
\usepackage[section]{placeins}  

\usepackage{multicol}
\usepackage[font=small,labelfont=bf]{caption}
\usepackage{hyperref}

\title{Stellar Forcing of (exo)Planetary Environments}

\author[1]{Juli{\'a}n~D.~Alvarado-G{\'o}mez\footnote{Corresponding Author $|$ email: jullian.alvarado-gomez@aip.de}~\!\orcidlink{0000-0001-5052-3473}}
\author[2]{Erika~Palmerio~\!\orcidlink{0000-0001-6590-3479}}
\author[3,4]{Manuel~G{\"u}del~\!\orcidlink{0000-0001-9818-0588}}
\author[5]{Kosuke~Namekata~\!\orcidlink{0000-0002-1297-9485}}
\author[6]{Konstantin~Herbst~\!\orcidlink{0000-0001-5622-4829}}
\author[7,8]{N.~Eugene~Engelbrecht~\!\orcidlink{0000-0003-3659-7956}}
\author[9]{Antoine~Strugarek~\!\orcidlink{0000-0002-9630-6463}}
\author[9]{Judy~J.~Chebly~\!\orcidlink{0000-0003-0695-6487}}
\author[10,11]{Astrid~Veronig~\!\orcidlink{0000-0003-2073-002X}}
\author[12,13,14]{Maria~D.~Kazachenko~\!\orcidlink{0000-0001-8975-7605}}
\author[15]{Kriszti{\'a}n~Vida~\!\orcidlink{0000-0002-6471-8607}}
\author[3]{Kristina~G.~Kislyakova~\!\orcidlink{0000-0003-4680-6774}}
\author[1,16]{Katja~Poppenhaeger~\!\orcidlink{0000-0003-1231-2194}}
\author[9]{Antonio~Garc{\'i}a~Mu{\~n}oz~\!\orcidlink{0000-0003-1756-4825}}
\author[12,13,14]{Yuta~Notsu~\!\orcidlink{0000-0002-0412-0849}}
\author[4]{Harish~K.~Vedantham~\!\orcidlink{0000-0002-0872-181X}}

\affil[1]{Leibniz Institute for Astrophysics Potsdam, 14482 Potsdam, Germany}
\affil[2]{Predictive Science Inc., San Diego, CA 92121, USA}
\affil[3]{Department of Astrophysics, University of Vienna, A-1180 Vienna, Austria}
\affil[4]{ASTRON, Netherlands Institute for Radio Astronomy, 7991 PD Dwingeloo, Netherlands}
\affil[5]{Heliophysics Division, NASA Goddard Space Flight Center, Greenbelt, MD 20771, USA}
\affil[6]{Centre for Planetary Habitability (PHAB), Department of Geosciences, University of Oslo, 0371 Oslo, Norway}
\affil[7]{Centre for Space Research, North-West University, 2520 Potchefstroom, South Africa}
\affil[8]{National Institute for Theoretical and Computational Physics (NITheCS),  2520 Potchefstroom, South Africa}
\affil[9]{Universit\'e Paris-Saclay, Universit\'e Paris Cit\'e, CEA, CNRS, AIM, 91191 Gif-sur-Yvette, France}
\affil[10]{Institute of Physics, University of Graz, A-8010 Graz, Austria}
\affil[11]{Kanzelh{\"o}he Observatory for Solar and Environmental Research, University of Graz, A-9521 Treffen, Austria}
\affil[12]{Laboratory for Atmospheric and Space Physics, University of Colorado Boulder, Boulder, CO 80303, USA}
\affil[13]{National Solar Observatory, Boulder, CO 80303, USA}
\affil[14]{Dep. of Astrophysical and Planetary Sciences, University of Colorado Boulder, Boulder, CO 80305, USA}
\affil[15]{HUN-REN Research Centre for Astronomy and Earth Sciences, Konkoly Observatory, MTA Centre of Excellence, H-1121 Budapest, Hungary}
\affil[16]{Institute for Physics and Astronomy, University of Potsdam, 14476 Potsdam-Golm, Germany}

\begin{document}
\maketitle

\begin{abstract}
The environments of exoplanets are fundamentally shaped by the magnetic activity of their host stars through a combination of radiative, plasma, and particle-driven processes. This article presents a comprehensive overview of the four principal forms of stellar forcing that regulate atmospheric structure, chemistry, escape, and long-term planetary evolution: high-energy radiation, magnetized stellar winds, coronal mass ejections, and energetic particles. Using the Sun as a physically resolved benchmark, the discussion extends toward increasingly active cool stars to establish a broader picture for star--planet interactions across the main sequence.

The article first examines the production of stellar X-ray and extreme ultraviolet emission from chromospheres and coronae, together with the variability introduced by flares and, more generally, magnetic reconnection processes. Particular attention is given to spectroscopic diagnostics, activity scalings with stellar rotation and age, flare energetics, statistical flare behavior, and the observational connections between impulsive and gradual phases of magnetic energy release. The treatment then shifts to the continuous outflow of magnetized plasma, describing the thermodynamic and magnetic mechanisms that drive stellar winds and the role of multidimensional magnetohydrodynamic modeling in determining wind structure, angular momentum loss, and planetary interaction regimes. Solar and stellar coronal mass ejections are subsequently explored through their observational diagnostics, flare associations, propagation properties, and possible suppression by strong stellar magnetic fields. Finally, the transport and modulation of galactic and stellar energetic particles are discussed, together with current approaches for estimating particle environments in active planetary systems and their consequences for atmospheric chemistry and climate. The article concludes by outlining future observational and numerical developments required to connect these coupled stellar forcing processes within a unified exoplanetary framework.
\end{abstract}

\tableofcontents

\section{Introduction}



Magnetic activity is the unifying ingredient behind the diverse ways in which stars condition the physical and chemical state of their planetary companions. The outer atmosphere of a cool star is not a quiescent boundary layer but rather a highly structured, time-dependent plasma environment in which radiative output, particle acceleration, and magnetized outflows are all regulated by the stellar magnetic field. For planets, these processes are not merely external perturbations: they determine the ionization and chemical state of upper atmospheres, the deposition of energy in different atmospheric layers, the efficiency of escape, and, ultimately, the temporal evolution of habitability. The present article adopts a stellar perspective in which the Sun serves as the principal benchmark, while the broader aim is to connect solar phenomena to the range of conditions encountered around magnetically active stars across the main sequence. The discussion is organized into four major blocks (``Radiation'', ``Magnetized Stellar Winds'', ``Coronal Mass Ejections,'' and ``Energetic Particles''), preceded by this introduction and concluded with an outlook, all framed as the set of ``stellar forcings'' that shape exoplanetary environments.

The radiation section is centered on the magnetically-powered high-energy output of cool stars, with particular emphasis on the X-ray and extreme ultraviolet (hereafter XUV) domains that dominate the radiative coupling between stars and planetary atmospheres. It begins by characterizing the quiescent component of stellar emission, establishing its coronal and chromospheric origin and outlining how fundamental plasma properties such as temperature distributions and emission measures are inferred from spectroscopic diagnostics. The solar paradigm is used as the primary reference point, enabling a physically grounded extension to other late-type stars through empirical activity scalings involving rotation, magnetic flux, and age. Within this context, the section delineates how the spectral energy distribution and luminosity of the XUV output evolve across activity regimes, and how these variations define the baseline radiative environment experienced by orbiting planets.

Building on this quiescent picture, the radiation section moves onto transient radiative processes, focusing on flares as the dominant source of variability in stellar high-energy emission. The discussion follows the standard flare scenario from magnetic energy release and particle acceleration to chromospheric heating and evaporative plasma filling of coronal loops, culminating in enhanced soft X-ray emission. Particular attention is given to the Neupert effect as an observational constraint linking impulsive and gradual phases, as well as to statistical descriptions of flare populations through flare frequency distributions and their dependence on stellar activity level. The section further addresses the contribution of unresolved flaring to quiescent emission, the scaling relations between flare energetics, temperatures, and densities, and the multiwavelength characteristics of flares, including optical and near-ultraviolet continuum components. Collectively, these elements provide a comprehensive view of both steady and impulsive radiation processes, and their role in shaping the high-energy stellar radiation forcing.

The magnetized stellar winds section starts from the solar case to develop the theoretical and observational framework required to describe the continuous outflow of plasma and magnetic fields from cool stars, emphasizing both its thermodynamic driving mechanisms and magnetic structuring. It begins by introducing the basic physical picture of thermally driven winds, extending from the classical isothermal solutions to more realistic descriptions that incorporate energy deposition, wave pressure, and spatially varying heating. Particular attention is given to the role of magnetic fields in shaping wind acceleration and topology, highlighting how open and closed field regions regulate mass and angular momentum loss. The section then establishes the key scaling relations linking wind properties to stellar parameters such as mass, rotation rate, and magnetic activity, and discusses the challenges associated with observational constraints, including indirect diagnostics derived from astrospheric absorption and radio emission.

From these theoretical fundamentals and observational constraints, the magnetized winds section transitions to modern multi-dimensional modeling approaches that capture the inherently complex structure of stellar winds. It reviews the development of global three-dimensional magnetohydrodynamic (MHD) simulations, detailing how different numerical prescriptions treat coronal heating, magnetic field extrapolation, and boundary conditions based on spatial distributions of surface magnetic fields. The interplay between rotation and magnetic geometry is examined in the context of angular momentum loss and spin-down, while comparisons between models underscore the sensitivity of predicted mass-loss rates and wind structures to underlying assumptions. Together, these topics provide a cohesive description of the quasi-steady stellar wind forcing as a fundamental link for star--planet interactions, particularly in terms of magnetospheric compression and atmospheric erosion.

The third block introduces coronal mass ejections (CMEs) based on the knowledge gained from solar observations. It explains how CMEs are identified through both remote sensing and in-situ measurements, and how their association with solar flares became established. The main observational signatures used to detect and track stellar CMEs are presented. Promising diagnostics such as radio bursts, coronal dimmings and spectroscopic measurements are outlined, showing how they are complementary to one another. The diversity of CME properties and their dependence on the surrounding environment are emphasized, including their interactions with the large-scale magnetic field and the ambient stellar wind. This section also summarizes current observational efforts to put constraints on their occurrence rates beyond the Sun.

The CME section transitions to the extension of solar-based insights to other stars using empirical approaches. It examines how relationships between flares and CMEs have been used to estimate mass and energy losses, how these estimates depend on stellar activity levels, and how those estimates appear at odds with our current understanding of magnetized outflows and mass loss in cool stars. Different modeling efforts are presented to evaluate a possible explanation for these discrepancies, including the suppression of CMEs by strong overlying stellar magnetic fields. Results from state-of-the art stellar CME modeling efforts are discussed including the influence of magnetic suppression on classical observational signatures of CMEs in different wavelengths (e.g., coronal dimmings and Type II radio bursts) as well as the role of CMEs in angular momentum losses in cool stars. Finally, realistic 3D numerical experiments illustrate the challenges of detecting CMEs in unresolved stellar observations, describing how their signatures may appear in coronal spectroscopic data and outlining methods to infer their properties from spectral diagnostics.

The section focused on energetic particles introduces cosmic rays as charged, sometimes highly energetic particles whose origins are linked to star formation, stellar evolution, supernova explosions, and the interstellar medium. Within this framework, we examine galactic cosmic rays (GCRs) and solar energetic particles (SEPs), emphasizing their different energy domains and compositions, and briefly discuss the origins and composition of GCRs that propagate through the interstellar medium before entering a stellar system through its astrosphere. The discussion then turns to their transport through matter, magnetic fields, and radiation fields, highlighting diffusion, drift, convection, and adiabatic energy changes. These processes modulate GCRs, meaning that the intensity observed at Earth or on an exoplanet is not necessarily equal to the local interstellar intensity. This modulation depends on the plasma properties of the host environment and produces a time-dependent radiation environment that reflects both the star's position in the galaxy and the changing state of its magnetic and plasma conditions, including the stellar activity and magnetic cycles.

The section then moves from GCR transport to SEP production and inference. We revisit empirical and semi-empirical relations that connect flare and CME observables to SEP outputs, including peak size distributions, the correlation between soft X-ray flux and flare energy, and scaling relations for CME kinetic energy and mass. These relations are then used to motivate attempts to estimate stellar energetic particle environments beyond the solar system, where direct measurements are unavailable. We show how solar scaling laws and extreme-event methodologies have been applied to infer worst-case stellar particle environments and why such simple extrapolation can become unreliable at large flare classes. We conclude this section by summarizing recent numerical efforts to model stellar energetic particle transport in systems such as TRAPPIST-1, GJ 436, and Proxima Centauri, to name a few, showing how CME-driven shocks, magnetic turbulence, and flare-constrained injection profiles can substantially modify stellar energetic particle fluxes (compared to solar expectations) and alter the associated atmospheric chemistry and climate of orbiting planets.

Having established the physical basis and current understanding of radiation, magnetized stellar winds, eruptions in the form of CMEs, and energetic particles as the four principal channels of stellar forcing, we conclude this article by turning to a forward-looking synthesis of these processes. In the final section, we outline key directions for future progress, including the development of next-generation observational diagnostics capable of probing stellar coronae, winds, and eruptive phenomena, as well as advances in multidimensional numerical modeling that can capture the interplay between these distinct yet interconnected forcings.

\section{Radiation Forcings}\label{radiation}


%

\begin{figure}[th!]
    \includegraphics[width=0.95\textwidth]{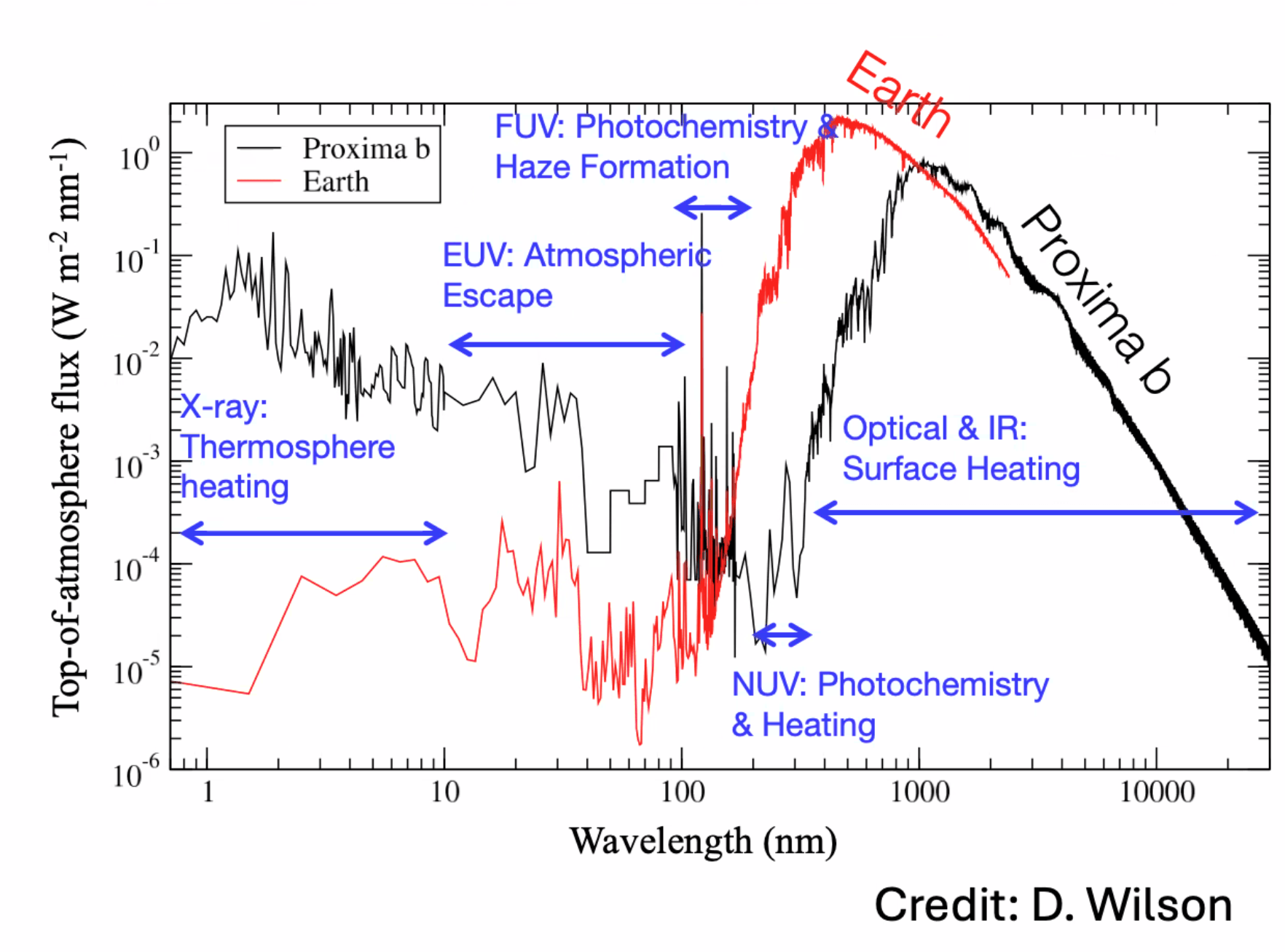}
    \caption{
    Comparison of the spectral energy distributions at the top of the atmospheres of Proxima b (black, the Earth-like planet orbiting around the M5.5 dwarf Proxima Centauri) and Earth (red), with the radiation effects on planetary atmosphere in each wavelength. Flux units are W m$^{-2}$ nm$^{-1}$ at 1~au for Earth and at 0.0485~au for Proxima b. The original figure is from \citet{ribas2017} and was modified by D.~Wilson.}\label{fig:AY_Wilson_Spectrum}
\end{figure}

Cool stars exhibiting magnetic activity are panchromatic astrophysical objects. They emit radiation across the breadth of the electromagnetic spectrum, from radio waves to gamma rays. While the predominant irradiance is accounted for by a blackbody spectrum for the star's spectral type, other emission processes (including bound-bound, bound-free, and free-free transitions) lead to significant deviations from a Planck function ~\citep{Rutten:RadiativeTransfer}. As illustrated in Figure~\ref{fig:AY_Wilson_Spectrum}, this broad-spectrum emission encompasses various regimes that interact with different layers of a planetary environment.

Although the bulk of a star's radiant energy may fall at other wavelengths, this section deliberately focuses on the high-energy radiation regime---such as X-ray and extreme ultraviolet (EUV), collectively XUV (cf.\  Figure~\ref{fig:AY_Wilson_Spectrum}). 
As the figure demonstrates, it is these highly energetic photons that dictate the physical and chemical evolution of the uppermost atmospheric layers. High-energy radiation acts as the primary driver for extreme planetary forcing, triggering critical processes like atmospheric heating, hydrodynamic escape, and complex photochemical reactions that ultimately govern long-term atmospheric retention and planetary habitability.

To thoroughly examine this radiative forcing, the following subsections decouple the quiescent stellar output from highly variable transient events: Section~\ref{subsec:quiescent} explores the steady-state, or quiescent, high-energy radiation, detailing the continuous emission mechanisms originating from the stellar corona and chromosphere. Section~\ref{sec:flares} shifts focus to magnetically-powered transients, specifically flares. Using the solar magnetic environment as a benchmark, this subsection breaks down the physical processes and observational signatures of flares across various wavebands, from the optical to the X-ray regimes. Section~\ref{sec:Rad_Stellar_Flares} synthesizes these concepts by addressing the role of flares in high-energy radiation, evaluating how these explosive and impulsive events dominate the overall high-energy budget and shape the long-term radiative environment experienced by orbiting planets.


\subsection{High-Energy Radiation: Quiescent Emission}\label{subsec:quiescent}

\subsubsection{The Sun as a Benchmark for High-Energy Emission}

High-energy radiation from the Sun spans many orders of magnitude in photon energy, conventionally separated into gamma rays (photons of order 100~keV and higher, corresponding to wavelengths shorter than $\sim$0.1~\AA\ = 0.01~nm), X-rays ($\sim 0.1$--100~keV, corresponding to wavelengths of $\sim$0.1--100~\AA\ or 0.01--10~nm), and the EUV range (0.014--0.1~keV, corresponding to $\sim$100--912~\AA\ or 10--91~nm). All high-energy radiation is generally ionizing, and the limit for EUV at $912$~\AA\ corresponds to the ionization energy of hydrogen (13.6~eV) and, approximately, of molecular hydrogen as well. Longer-wavelength radiation can ionize atoms with lower ionization potential such as Fe or Mg, which are observable in exoplanet atmospheres and can be employed as diagnostics of overall atmospheric metallicity.

High-energy radiation from the Sun originates in the magnetized outer layers of the solar atmosphere. X-rays are emitted by the hot (${>}10^6$~K) corona, while EUV radiation receives contributions from a wide range of temperatures associated with the corona, transition region, and chromosphere.
Most of the hard X-rays ($\sim$10--100~keV) are emitted in the chromosphere, where during flare events high-energy particles collide with the dense plasma resulting in emission of non-thermal bremsstrahlung. 
In the following, we will not discuss further non-thermal hard X-rays, which have remained somewhat elusive from other stars---given their low fluxes and the overwhelming thermal soft X-ray spectrum. 

\begin{figure}[!th]
\centering
\includegraphics[width=0.85\textwidth]{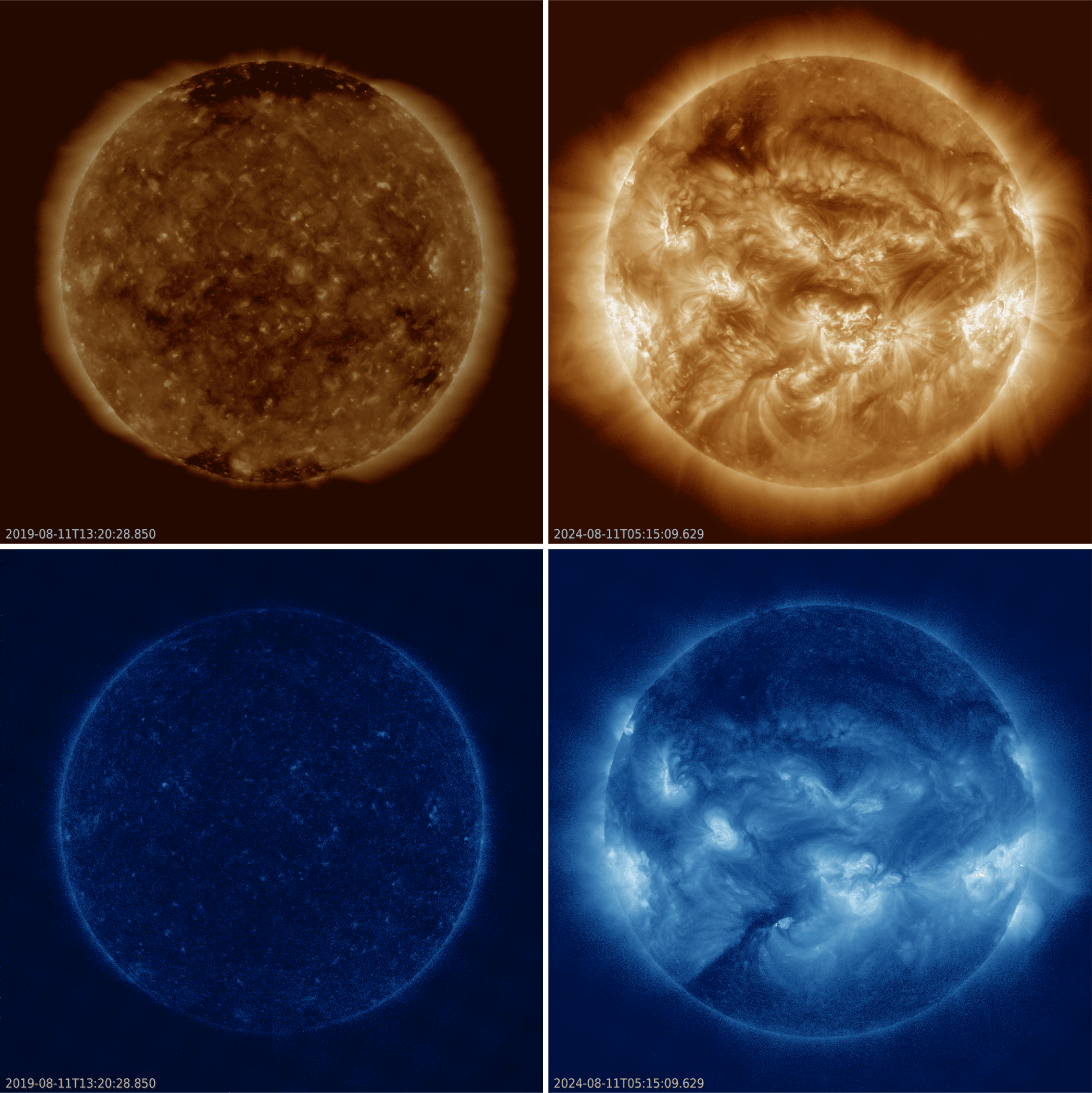}
\caption{Observations of the solar corona in two narrow-band EUV filters by SDO/AIA; top: 195~{\AA} ($T \sim 1.5$~MK), bottom: 335~{\AA} ($T \sim 2.5$~MK). The left panels are recorded during solar minimum, showing quiet-Sun emission and polar coronal holes. The right panels show the solar corona during the maximum of Solar Cycle 25, where the emission is dominated by active regions.}
\label{fig:euv-sun}
\end{figure}

\begin{figure}[!th]
\centering
\includegraphics[width=0.65\textwidth]{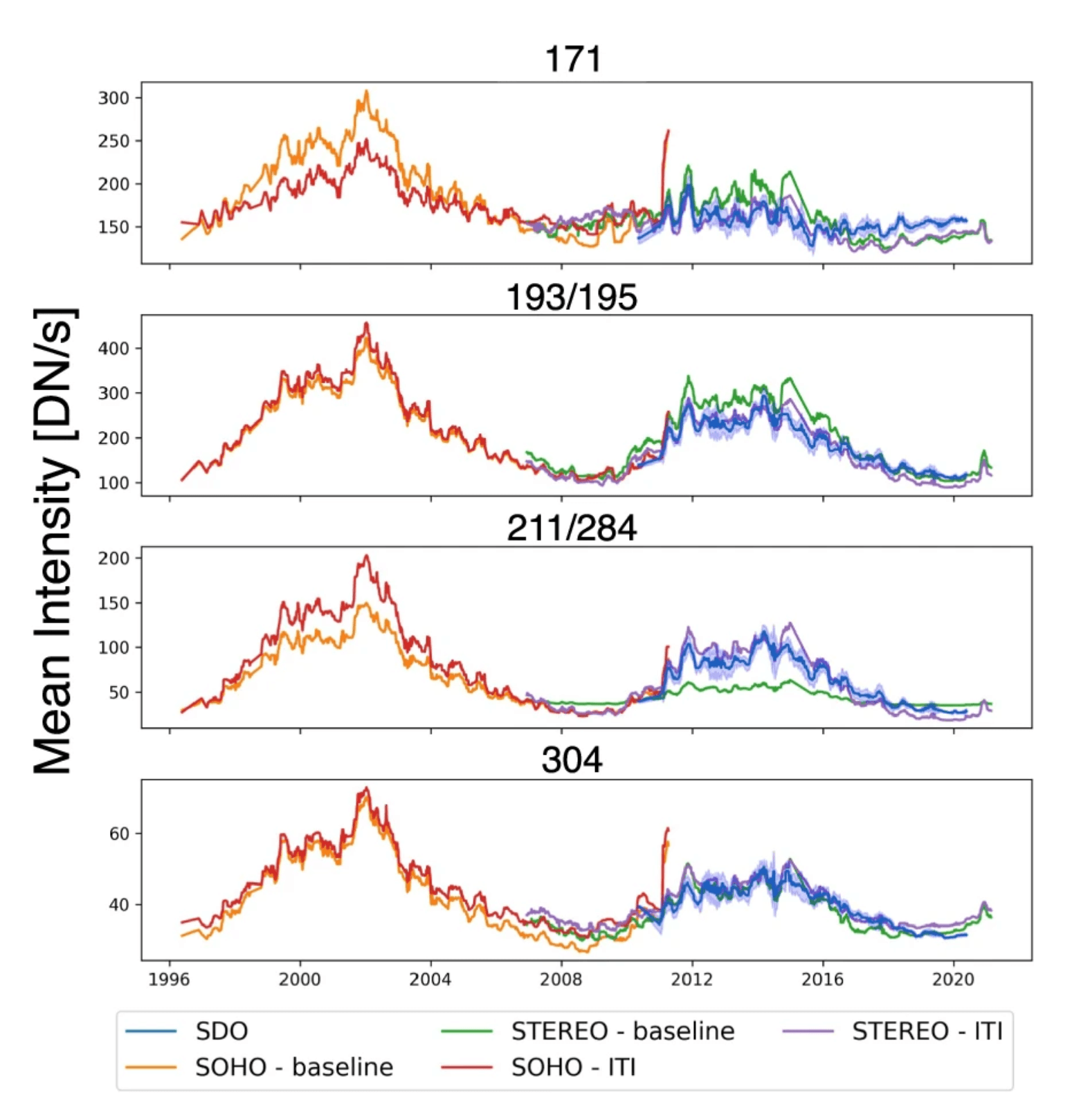}
\caption{Evolution of the solar EUV emission in four different wavelength bands over two solar cycles (nos.~23 and 24). The plots are composites from different instruments (SOHO/EIT, STEREO/EUVI, SDO/AIA) observing the Sun over different periods, and cross-calibrated curves derived by the instrument-to-instrument (ITI) method. Adapted from \cite{Jarolim_2025}. }
\label{fig:euv-cycle}
\end{figure}

Observations of solar coronal emission have a long history that was boosted by many space missions, initially perhaps most dramatically through soft X-ray imaging by  Skylab that revealed complex magnetic structures in the solar corona, with closed magnetic loops and open-field regions observed as ``coronal holes'' \citep{Vaiana1973,Timothy1975}. 
Figure~\ref{fig:euv-sun} shows the Sun in two EUV filters of the Atmospheric Imaging Assembly (AIA) onboard SDO during solar minimum (left) and solar maximum (right). The solar EUV and X-ray emissions during quiescent phases (i.e., outside transient events like flares) are dominated by active regions, where plasma is trapped and heated by the concentrated magnetic fields. These emissions therefore vary strongly with the presence of sunspots, i.e.\ on time scales associated with the solar rotation period (${\sim}$27 days) and the solar cycle (${\sim}$11 years). Depending on the specific wavelength range, the solar XUV emission can vary up to an order of magnitude over the solar cycle. 
A difficulty in studying the solar cycle evolution of the solar EUV and X-ray fluxes is the cross-calibration of different instruments (see Figure~\ref{fig:euv-cycle}).

\subsubsection{High-Energy Emission in Other Stars}

It is through the nearby example of the Sun that much of the origin of stellar XUV emission has been understood. The main questions underlying the ``solar--stellar connection'' are the following: i) How does the Sun’s XUV activity compare with that of other stars?; ii) Can the Sun be used as a template for understanding the XUV activity of stars with different properties, such as magnetic flux, mass, rotation period, and age?; and iii) Are the mechanisms responsible for the diversity of XUV phenomena observed in stars fundamentally the same as those observed on the Sun, differing mainly in their energy output?

To address these questions, several complementary approaches have proven indispensable. First, large samples of stars---including coeval populations in stellar clusters---have provided insight into the operation of the magnetic dynamo, its relation to rotation, and the evolution of magnetic activity as stars spin down due to angular momentum loss via magnetized winds. Second, XUV spectroscopy has enabled access to key coronal plasma parameters such as temperature, density, pressure, elemental abundances, and velocity fields. Third, sufficiently long or repeated observations have revealed various mechanisms of variability, ranging from flares to magnetic activity cycles. Space missions have been essential for advancing this field of research. Among others, major breakthroughs in stellar high-energy astronomy have come from the Einstein Observatory, which provided the first large stellar X-ray samples; ROSAT, which conducted an all-sky survey and vastly expanded the sample of X-ray-emitting stars; and XMM-Newton together with the Chandra X-ray Observatory, which enabled systematic, sensitive CCD imaging and high-resolution grating X-ray spectroscopy. The Extreme Ultraviolet Explorer (EUVE) provided complementary and still unique observations in the EUV range.

It is now known that essentially every star of every late-type main-sequence spectral class (F, G, K, M) is a coronal X-ray emitter \citep{schmitt1997}. This also applies to pre-main sequence stars such as T Tauri stars and younger protostars \citep{preibisch2005, guedel2007}, to substellar objects of the Brown Dwarf class \citep{2006A&A...448..293S, berger2008}, and to subgiants in close binary systems \citep{dempsey1993, singh1996}. Additional regions of the Hertzsprung--Russell Diagram (HRD) may host magnetically active X-ray emitters, including rapidly rotating giants of the FK Comae Berenices stars class \citep{gondoin1999}.

\begin{figure}[t]
\begin{center}
\includegraphics[width=0.53\textwidth]{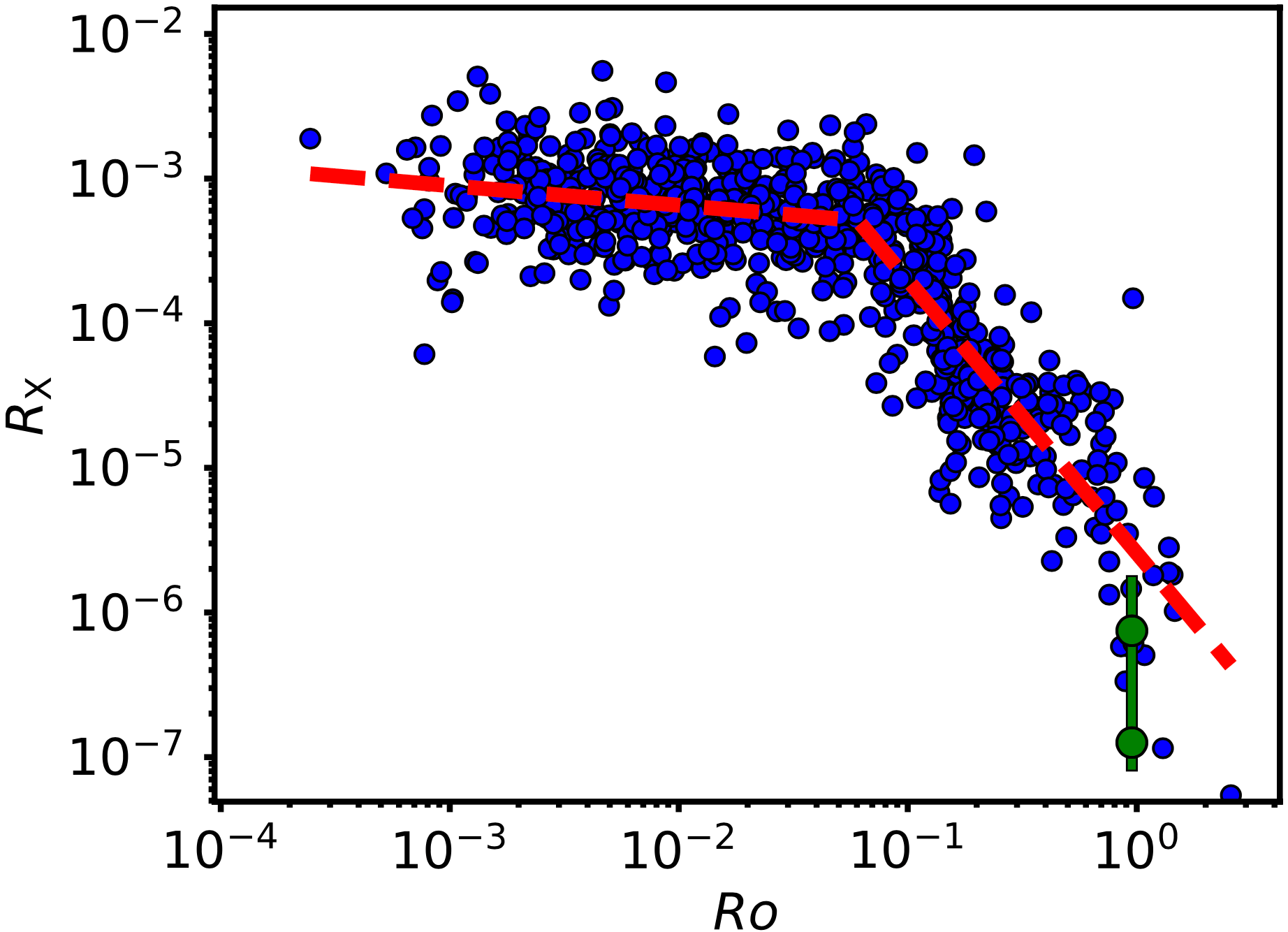} 
\caption{Relation between $R_{\rm X} \equiv L_{\rm X}/L_{\rm bol}$ and the Rossby number $Ro = P/\tau$ where $P$ is the rotation period and $\tau$ the convective turnover time.  The dashed red line is an approximate fit. The range of solar values is given by the green line, the two green bullets marking the 10th and 90th percentiles.
(From \citealt{johnstone2021a} using the stellar sample of \citealt{wright2011}.)}
\label{fig:RoRx}
\end{center}
\end{figure}
One of the first systematic trends found for X-ray emission is that the stellar X-ray luminosity $L_{\rm X}$ is closely correlated with the stellar rotation period or, somewhat equivalently, equatorial rotation velocity, very roughly declining with increasing rotation period, $P$, as
$L_{\rm X} \propto P^{-2}$ (\citealt{pallavicini1981}; see also \citealt{wright2011, 2014ApJ...794..144R}).
This observation links to an internal magnetic dynamo for which (differential) rotation is crucial.
However, for the most rapidly rotating stars, $L_{\rm X}$ ``saturates'' at a maximum level of $\sim$10$^{-3}L_{\rm bol}$, where $L_{\rm bol}$ is the stellar bolometric luminosity (Figure~\ref{fig:RoRx}). The cause of this saturation is not clear; it could be related to 100\% surface coverage with strong, active region-like magnetic fields \citep{vilhu1984, fleming1989} although higher pressures or a larger coronal extent than common on the Sun would be required to explain the observed extreme $L_{\rm X}$ \citep{vaiana1978, wood1994, vilhu1984}; or it could be the result of a maximum efficiency of the internal dynamo \citep{vilhu1987}. Close binaries of the RS CVn or Algol type are tidally locked and therefore forced into rapid rotation; these stars indeed reach maximum $L_{\rm X}$ values (e.g., \citealt{dempsey1993, singh1996}). The relation between rotation and $L_{\rm X}$ reads
\begin{equation} 
\left. 
   \begin{array}{ll}
     {\displaystyle{\frac{L_{\rm X}}{L_{\rm bol}}}} \propto Ro^{\prime-2}  \quad {\rm and}\quad  L_{\rm X} \propto P^{-2}  &\quad\quad {\rm for}\quad\  P \gapprox  \\
     {\ } \\
     {\displaystyle{\frac{L_{\rm X}}{L_{\rm bol}}}} \approx 10^{-3}                                                 &\quad\quad {\rm for}\quad\ P \lapprox        
   \end{array} 
   \right\} 
   P_{\rm sat} \approx 1.2\left({\frac{L_{\rm bol}}{L_{\odot}}}\right)^{-1/2},
\end{equation}
where $Ro^{\prime}$ is the Rossby number $P/\tau^{\prime}$, $\tau^{\prime}$ being an empirically modified convective turnover time, and $L_{\odot} = 3.85\times 10^{33}$~erg~s$^{-1}$ is the bolometric luminosity of the Sun \citep{pizzolato2003}. 
We provide a closer look at the rotation--X-ray relation for cool main-sequence stars and the consequences for stellar X-ray evolution in \citet{Chapter4}.  

\begin{figure}[!h]
\begin{center}
\includegraphics[width=0.59\textwidth]{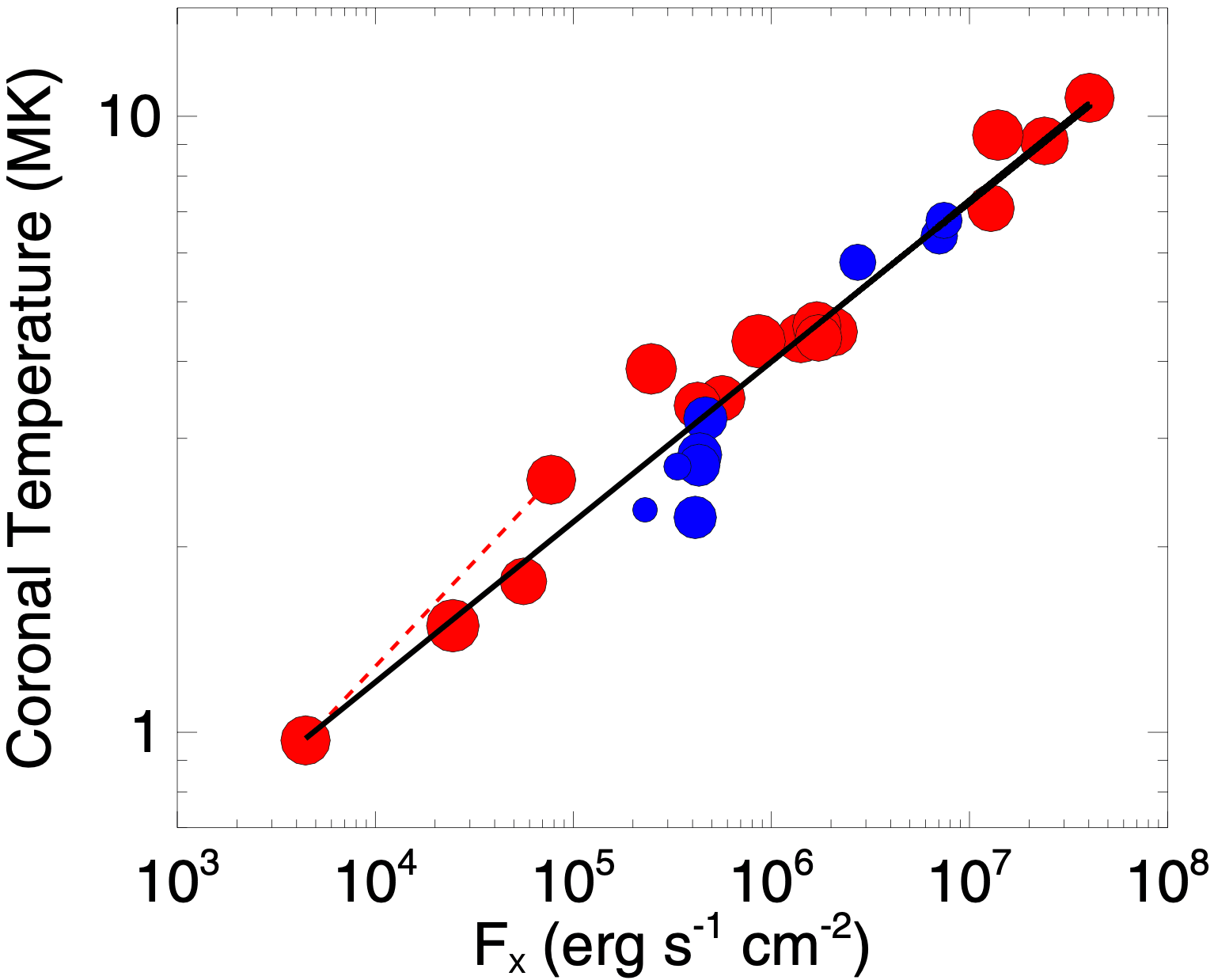} 
\caption{Correlation between the average coronal temperature and the average surface X-ray flux of stars; blue and red bullets  refer to stellar masses below and above $0.65\,M_{\odot}$, respectively; the black line is a fit given by Equation~\ref{eq:TFX}; the two red bullets connected by a dashed line show the range of the Sun during its activity cycle.
(From \citealt{johnstone2015b}.)}
\label{fig:FXT}
\end{center}
\end{figure}

Coronae characteristically reach temperatures in excess of 1~MK, as determined by X-ray spectroscopy. Stellar samples revealed that coronae of stars at a higher activity level (in the sense of the $L_{\rm X}/L_{\rm bol}$ ratio) are on average hotter (see, for example, early reports by \citealt{vaiana1983, schrijver1984, stern1986, schmitt1990, johnstone2015b}); the hottest coronae reach, outside some obvious, strong flares, temperatures of up to 40--50~MK \citep{audard2004,Notsu+2025_ApJ}, to be compared with a typical temperature in solar active regions of only a few MK \citep{peres2000}. The narrowest correlation across all late-type main-sequence stars is found between
the average coronal temperature $\bar{T}$ (in MK) and the average X-ray surface flux, $F_{\rm X}$ (in erg~cm$^{-2}$~s$^{-1}$)
\begin{equation}\label{eq:TFX}
    \bar{T} = 0.11F_{\rm X}^{0.26}
\end{equation}
\citep{johnstone2015b}.
However, stellar coronae are not isothermal; medium and high-resolution X-ray spectroscopy of stars, but also spatially resolved analysis on the Sun \citep{peres2000, orlando2000}, reveal that the coronal emission measure, i.e., the quantity $n_en_i {\rm d}V$, where $n_i$ is the ion density, $n_e$ the electron density, and ${\rm d}V$ a coronal volume element, is distributed across a wide range of temperatures. Characteristically, the distribution of emission measure in $\ln T$ shows an increasing power law on the cooler side up to a peak, continued by a decreasing power law beyond the peak (see Figure~\ref{fig:DEM} below). The temperature of the peak is close to the emission measure weighted averaged temperature \citep{telleschi2005}. Various theories have been presented to explain the slopes of the emission measure, from the structure of static coronal loops and their internal temperature profiles \citep{rosner1978, stern1986, vesecky1979, antiochos1986}, to flaring structures \citep{antiochos1980}, episodic heating due to stochastic flares \citep{sturrock1990, cargill1994},  and the effect of a distribution of stochastic flares with peak temperatures in relation to their total released energies (\citealt{guedel1997b, guedel2003}; see summary in \citealt{guedel2004} and also Section~\ref{sec:Rad_Stellar_Flares} below). Emission measure distributions have also been derived for stars other than the Sun from high-resolution X-ray and UV spectra \citep{Sanz-Forcada2003A&A...408.1087S, Wood2018ApJ...862...66W, Pillitteri2022A&A...666A.198P}; however, those studies are typically limited to magnetically at least moderately active stars, since a high X-ray flux is necessary to obtain spectra with sufficient signal-to-noise ratio.

Coronal densities have become accessible through high-resolution X-ray spectroscopy revealing density-sensitive line ratios, for example ratios of Fe lines or ratios in line triplets of He-like ions.  The principal result of such studies (e.g., \citealt{bowyer2000, ness2004, testa2004,Osten+2006_ApJ}) is that coronae at higher activity levels show higher electron densities of several times $10^{10}$~cm$^{-3}$, and potentially up to $10^{13}$~cm$^{-3}$, than low-activity stars with $n_{\rm e} < 10^{10}$~cm$^{-3}$ including the Sun outside individual flares. This also indicates that cooler coronal plasma is less dense than hotter plasma, a finding that also reflects in (solar or stellar) flares being hotter and denser than the average ``quiescent'' values.
Spectral density measurements are, however, notoriously plagued by measurement bias and limited ranges of density sensitivity of line ratios (for a broader discussion, see \citealt{guedel2004, ness2004, testa2004}).

Understanding the XUV properties of host stars is crucial for (exo-)planetary science. First, as has been studied in much detail in the past, XUV radiation ionizes and chemically processes planetary upper atmospheres and therefore modifies their radiative (heating/cooling) properties \citep{garciamunoz2007}, depending on the X-ray flux. The presence of cooling molecules such as CO$_2$, H$_2$O, or NO depends on the photodissociation by high-energy photons \citep{tian2008, johnstone2018,garciamunoz2023}---a steady-state model solution requires accurate inputs of atmospheric composition and XUV flux. 
Second, the penetration depth of radiation into an atmosphere is a function of photon energy (for given atmospheric composition and density); X-rays normally penetrate deeper into an atmosphere than EUV. The way in which the stellar photons at each wavelength range are progressively deposited in the atmosphere produces a variety of peaks in the ionization fractions and temperature. The spectral hardness owing to the coronal temperature structure is therefore important to determine  heating and chemical reactions in an atmosphere.
Third, atmospheric mass loss crucially depends on XUV irradiation that heats upper atmospheres as it ionizes the  gas and induces chemical reactions that themselves will contribute to heating \citep{johnstone2018, garciamunozetal2021}. The longer wavelengths also contribute to heating, which proceeds by photodissociation and subsequent thermal relaxation of the heavy fragments. XUV heating leads to atmospheric escape via Jeans mass loss or hydrodynamic loss both for primordial H/He atmospheres \citep{Yelle2004Icar..170..167Y,murray2009, johnstone2015c, owen2016, Kubyshkina2018ApJ...866L..18K} and for secondary atmospheres  \citep{tian2008, johnstone2018,garciamunozetal2021}. These mechanisms are discussed in \citet{Chapter3}. 
Fourth, XUV evolves throughout the main-sequence life of a star, generally decreasing with time due to stellar spin-down; the way a planetary atmosphere reacts is therefore time-dependent, reactions such as chemical alteration and mass loss being most effective in the earlier periods of evolution \citep{tu2015, johnstone2021b, Ketzer2023MNRAS.518.1683K, France+2025_AJ} 
The importance of  XUV evolution and its role in habitability are discussed in \citet{Chapter4}. 
Fifth, that XUV radiation significantly affects the strength of atmospheric tracers such as the He triplet line at 1.08~$\mu$m with which atmospheric escape is currently being probed \citep{Poppenhaeger2022MNRAS.512.1751P, garciamunozetal2025,sanz-forcadaetal2025}. In such efforts, the ideal situation is to perform atmospheric measurements (typically through transmission spectroscopy) contemporaneously with X-ray measurements from which the XUV emission can be estimated \citep{damonteetal2026}.

\subsection{Magnetically-Powered Transients: Flares}\label{sec:flares}

\subsubsection{The Magnetic Environment of Solar Flares} \label{sec:Rad_solarflare_mag}


Unlike other stars, solar flares provide the only environment where photospheric magnetic fields and radiative signatures of flares can be spatially resolved. High-resolution solar instruments are now capable of resolving vector magnetic fields in the photosphere down to scales of $50$ km and tracking emission signatures down to scales of $20$~km. High spatial resolution combined with detailed spectra allow one to perform quantitative studies of flare energy storage, release, and transport that are inaccessible for other stars. Consequently, solar flare studies provide the physical foundation for interpreting stellar flares. The following discussion summarizes how solar observations constrain the magnetic energy buildup, reconnection, and eruptive consequences of flares.

Solar flares are powered by the rapid release of magnetic free energy stored in the solar corona above active regions. This energy accumulates gradually over days of photospheric evolution through shearing, twisting, flux emergence, and flux cancellation. Observations and models indicate that these pre-eruptive configurations often take the form of sheared arcades \citep[e.g.,][]{Moore1992, 1999ApJ...510..485A} or magnetic flux ropes \citep[a twisted flux tube where the majority of the interior magnetic field lines wind about a common tube axis; e.g.,][]{Amari.etal2000_rcc, torok2005}. These current-carrying structures store magnetic energy that can later power eruptive phenomena such as flares and CMEs. For details of the pre-eruptive magnetic field configurations, we refer the reader to the review by \citet{patsourakos2020}.

The Standard Solar Flare Model describes the standard conceptual picture of magnetic field structure and evolution during solar flares. The original 2D formulation, known as the CSHKP model \citep[after][]{Carmichael1964, Sturrock1968, Hirayama1974, Kopp1976}, together with its subsequent 3D extensions \citep[e.g.,][]{Longcope2007,Aulanier2012,Savcheva2016}, outlines the following general scenario.
Magnetic reconnection occurs between pairs of magnetic field lines under a pre-existing rising magnetic flux rope or a sheared arcade \citep{1989ApJ...343..971V, Longcope2007, Priest2017, Pontin2022, Kazachenko2022r}.
As a result, newly reconnected magnetic loops form below the reconnection region, while an expanding flux rope forms above it. This process simultaneously transfers magnetic flux into the rising CME and contributes to acceleration of non–thermal particles propagating along the closed loops (e.g. \citealt{Fletcher2008}). When these energetic particles encounter the denser upper chromosphere, they emit hard X-rays through bremsstrahlung and deposit energy that heats the plasma and drives chromospheric evaporation \citep{Fisher1985}. Both bremsstrahlung emission and chromospheric heating can be observed as hard X-ray emission and intense chromospheric brightenings or flare ribbons in EUV, respectively. The chromospheric plasma heated and evaporated by these particles fills the newly formed magnetic loops, emitting light in EUV and soft X-ray.

The CSHKP model explains a wide range of observations of solar flares: an expanding flux rope that forms the core of a CME, cusp-shaped post-reconnection loops, coronal dimmings, flare ribbons, and the growing arcade of post-reconnection loops.
As the flux rope above the reconnection region rises, the reconnection site below also rises, leading to the separation of flare ribbons away from the underlying polarity inversion line. The magnetic flux swept by flare ribbons measures the amount of magnetic field participating in reconnection or reconnection flux \citep{Forbes.Priest1984_rcc}. Furthermore, beyond the standard model, ribbon motion could constrain the electric fields and the geometry of the guide-field in the reconnection current sheet above \citep{Qiu2002, Tamburri2024}.
The expansion of the erupting flux rope stretches the surrounding overlying magnetic field lines. This stretching reduces plasma density in the coronal volume rooted at the footpoints of the flux rope and the surrounding overlying field lines. The resulting density depletion produces transient dark regions known as coronal dimmings \citep{Hudson1996,veronig2025}.
Thus, coronal dimmings trace the footpoints of expanding magnetic structures and could help in quantifying the magnetic flux carried away by the CME \citep[e.g.,][]{Qiu2007,Dissauer2018,Dissauer2019}. In terms of flare energy partitioning, magnetic reconnection converts stored magnetic free energy into multiple forms: non-thermal particle acceleration, direct  plasma heating , radiation across the electromagnetic spectrum, and, in eruptive events, CME kinetic and gravitational energy \citep[e.g.,][]{Emslie2012,Warmuth2016,Aschwanden2014}. Observationally, the observed magnetic reconnection flux strongly correlates with the peak soft X-ray flux and fluence, indicating that the total magnetic flux processed by reconnection controls the thermal flare energy output \citep{kazachenko2017}.

Solar flares provide a natural laboratory for studying magnetic reconnection. While the flare reconnection region is not directly observable, its properties can be inferred from indirect signatures such as the footpoints of reconnected magnetic field lines and flare ribbons. Recent observations from high-resolution instruments (such as IRIS, the Swedish Solar Telescope, Solar Orbiter, and NSF's D.\ K.\ Inouye Solar Telescope) have revealed a variety of fine structures within flare ribbons, including zigzags, spirals, and waves (e.g. \citep[e.g.,][]{Brannon2015, ThoenFaber2025, Yadav2025, French2025b, CorchadoAlbelo2026}. Analytical and modeling works suggest that these structures are consistent with patchy and intermittent reconnection, possibly driven by tearing instabilities and turbulence within the current sheet \citep{Wyper2021,Dahlin2025}. Thus, while the CSHKP framework provides the global geometry, the energy release process itself is fragmented and dynamic.
Nevertheless, the realistic evolution of 3D coronal magnetic fields during flares and CMEs is often more complex than described by the standard flare model. Because direct measurements of coronal magnetic fields remain highly limited, numerical simulations are required to reconstruct their evolution. With the launch of the Solar Dynamics Observatory, which started to observe vector magnetic field maps in the photosphere routinely, great progress has been made in developing data-driven simulations that use these maps as a lower boundary condition. These models now enable tracking of the 3D magnetic connectivity changes throughout the flare, linking photospheric driving to coronal restructuring both before and during the flare (e.g., \citealt{2024ApJ...975..206F,Kazachenko2026} and reviews by \citealt{Schmieder2024,Jiang2022}).

Solar flares can be broadly classified as either confined or eruptive, depending on whether the energy release leads to a CME. Confined flares are particularly important for stellar studies because stellar flares appear to be relatively CME-poor, suggesting that many stellar events may resemble confined solar flares. Within the framework of the standard CSHKP flare model, magnetic reconnection beneath a rising magnetic structure produces both flare radiation and restructuring of the coronal magnetic field. In eruptive events, reconnection injects poloidal flux into the rising flux rope and drives CME acceleration, while newly reconnected loops form the growing post-flare arcade. Observations show strong correlations between ribbon reconnection flux, coronal dimming flux, and CME properties, supporting this coupled flare–CME magnetic feedback scenario \citep[see the review by][]{Kazachenko2022r}.
However, the CSHKP reconnection geometry can occur both with and without a successful eruption. In confined flares, reconnection proceeds but the rising magnetic structure is prevented from escaping by strong overlying (``strapping'') magnetic fields. The torus instability criterion, quantified by the decay index of the overlying field $n=-\partial \ln B_H / \partial \ln z$, provides a physical framework that could distinguish these regimes: eruptions would occur when the flux rope reaches heights where $n$ exceeds the critical value $n_\mathrm{cr}\approx1.5$. Large statistical studies are consistent with this framework, showing that the key parameter controlling eruptivity is the ratio of magnetic flux participating in the flare to the total active region flux \citep{Li2020}. For flares of the same class and reconnection flux, events occurring in active regions with larger unsigned flux $\Phi_\mathrm{AR}$ are more likely to be confined. A strong positive correlation ($0.86$) between the critical decay-index height and $\Phi_\mathrm{AR}$ suggests that large-flux active regions form strong magnetic ``cages'' that inhibit eruptions \citep{Amari2018}. Thermodynamically, statistical studies have found that while confined and eruptive flares of the same peak X-ray class process comparable amounts of reconnection flux, confined flares typically exhibit higher peak reconnection rates \citep{Tschernitz2018, Kazachenko2023}. These findings suggest that confined flares result from reconnection between stronger, lower-lying magnetic fields that reorganize a smaller fraction of the active-region flux. Reconnection in such configurations proceeds more rapidly and terminates earlier, which may also favor more efficient particle acceleration in confined flares.

To summarize, solar flares are defined by gradual free-energy buildup and 3D reconnection that restructures magnetic field connectivity. Whether the outcome is eruptive or confined depends on the interplay between magnetic non-potentiality and the strength and geometry of the overlying field. The Sun's uniquely resolved magnetic and radiative diagnostics make it the cornerstone for understanding magnetic reconnection and flare energy release not only in our heliosphere but throughout magnetically active stars.


\vspace{1mm}
\subsubsection{Solar and Stellar Flare Frequency Distributions}
\label{sec:flares-FFD}

Solar observations have shown that the number of flares  per unit time $N({>}E)$ exceeding an energy threshold $E$ is a function of the threshold energy itself; the flare rate follows a power-law in energy, implying that the flare occurrence is scale-free over a wide range of energies. This behavior is commonly expressed in terms of flare frequency distributions (FFDs), 
\begin{equation}\label{flareenergies_int}
    N({>}E) \propto E^{-\beta}\, ,
\end{equation}
where the power-law index $\beta$ is an empirical value. In differential form, the distribution reads
\begin{equation}\label{flareenergies_diff}
    \frac{{\rm d}N}{{\rm d}E} = aE^{-(\beta+1)} \equiv aE^{-\alpha} \, ,
\end{equation}
where $a$ is a constant to be determined from observations. The index $\alpha \equiv \beta + 1$ is often found to be in the range of 1.35--2.9 for the Sun's XUV emission \citep{crosby1993, krucker1998, parnell2000, jess2019}. It is little surprising that stellar flares follow similar distributions, and they do so both in the optical and in the XUV regimes, as will be discussed separately below (Sections~\ref{sec:flares-optical} and \ref{subsubsec:X-Ray_regime}). We note that $\alpha$ values measured in the optical or in the XUV, even in different XUV energy bands, do not need to be identical; in particular, optical and X-ray flare energies are not necessarily proportional.

The value of $\alpha$ is critical. If $\alpha < 2.0$, then energetically the large flares dominate the released coronal energy unless there is a physical upper cutoff (which must be lower than the total magnetic energy of a large active region). On the other hand, for $\alpha \ge 2.0$, the integrated energy of all small flares diverges to infinity, making flares plausible contributors to ``quasi-steady'' coronal heating \citep{hannah2011}. In this case, the presence of a physical lower limit for flare energies is required, or a change of $\alpha$ toward smaller values at low energies. The empirically supported concept of power-law flare energy distributions immediately implies that ``flare rates'' on a star have no meaning unless a lower energy threshold is provided for such statistics. Comparing flare rates between two stars requires the same energy threshold to be adopted for both stars.


\vspace{1mm}
\subsubsection{Stellar Flares in the Optical Regime}
\label{sec:flares-optical}

Flares in the optical regime are widely observed across a diverse range of stellar types, most notably among G, K, and M spectral types. The Kepler and TESS space observatories brought a new era in studying flares on a statistical basis---the data from these instruments, for example, showed the existence of superflares on G-type dwarf stars, revealing that their energy distributions are consistent with those of solar flares, and that superflares occur more frequently on faster-rotating, and thus generally younger, stars (see also the reviews by \citealt{Kowalsk+2024_LRSP} and \citealt{2024Univ...10..313V}). 

Studies of solar and solar-like stellar flares show that, while the Sun's largest observed events---such as the 1859 Carrington event with a bolometric energy of $10^{32}$ erg \citep{cliver1859SpaceWeather2013, Hayakawa+2023_ApJ}---are highly significant for space weather at Earth, they are nevertheless orders of magnitude weaker than the most energetic flares observed on other stars \citep{Shibayama2013}. Historical records of aurorae, sunspots, and cosmogenic radionuclides such as $^{14}$C and $^{10}$Be indicate that the Sun may have experienced stronger outbursts in the past as well---if solar flares were the cause of these, rather than other high-energy sources, their estimated energy  could be at least five times stronger than any instrumentally recorded solar event \citep{miyakeLengthsSchwabeCycles2013, 2012Natur.486..240M, usoskinHistorySolarActivity2013,Usoskin+2023_SSRv}. Further details are discussed in Section~\ref{particles}.
To assess the full range of solar flare potential, it is useful to examine observations of solar analogs \citep{maehara2012, Shibayama2013, Notsu+2019_ApJ, okamotoStatisticalPropertiesSuperflares2021}. Kepler's four-year dataset of roughly 80,000 solar-like stars identified over 2,000 superflares on G-type dwarfs, revealing a power-law energy distribution similar to solar flares. These studies suggest that solar superflares of $7\times10^{33}$ erg of bolometric energy could occur roughly once every 3,000 years, with even stronger events possible on longer timescales, but there are estimates that suggest that solar-like stars could produce a superflare once per century \citep{Vasilyev+2024_Science}.

\begin{figure}[!t]
\includegraphics[width=1.0\textwidth]{"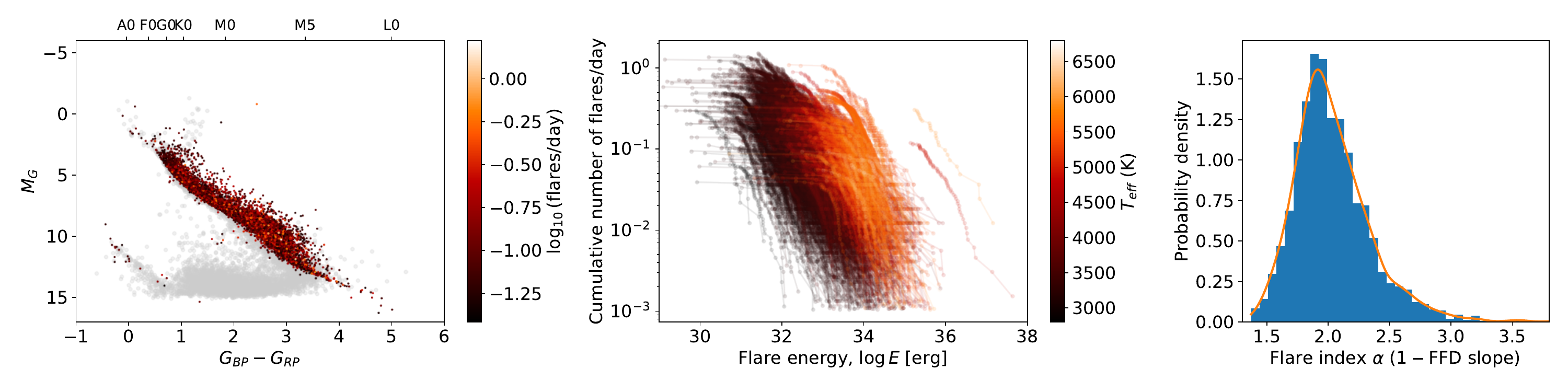"}
\caption{Flaring stars from TESS light curves, based on the data of \cite{seliStellarFlareMorphology2024}. \textit{Left:} Hertzsprung--Russell diagram of stars with detected flares, colored by their flare ratio. As comparison, a sample of nearby Gaia target stars (${<}150$~pc) was added with gray points. \textit{Middle:} flare frequency distribution of flaring stars with at least ten flares, colored by the $T_\mathrm{eff}$ of the star. \textit{Right:} Distribution of the FFD slopes. Continuous line shows the kernel density estimation (KDE) of the data. }
\label{fig:tess-ffd}
\end{figure}

Lower-mass stars, especially M dwarfs, also display considerable flaring activity. 
For the high-energy flares on cool dwarfs, strong correlations exist between their amplitude, duration, and energy:
high-energy flares typically have both long duration and high amplitude, and complex flares (having more than one peak) have longer duration and higher energy at the same amplitude \citep{hawleyKeplerFlaresActive2014}. Even ultracool dwarfs---classified as M7 or later--- that are often scrutinized for planetary habitability are known to produce (super)flares with bolometric
energies up to $10^{33}$ erg, although less frequently than earlier M dwarfs. 
A notable example is TRAPPIST-1---an object in the focus of researchers for its planetary system's habitability interest---shows flaring activity typical for its type. Its flares were detected not only in the optical passband \citep{vidaFrequentFlaringTRAPPIST12017}, but in infrared as well using the James Webb Space Telecope \citep{howardCharacterizingNearinfraredSpectra2023a}. 

Flares are known to follow a power-law energy distribution (\citealt{Lacy+1976_ApJS}; see Section~\ref{sec:flares-FFD}), with FFDs having been significantly refined by space-based missions such as Kepler and TESS (see, e.g., \citealt{guntherStellarFlaresFirst2020}). Analyses based on large flare catalogs, including the compilation of \citet{seliStellarFlareMorphology2024} from the first five years of the TESS mission, now provide FFDs for thousands of stars and enable robust statistical characterization of flare populations.
Observationally, stellar FFDs are well described by such power laws with indices typically around $\alpha \approx 2$ (see Figure~\ref{fig:tess-ffd} and Section~\ref{sec:Rad_Stellar_Flares}), consistent with the idea of scale invariance and the absence of a preferred energy scale in the underlying process.
Observations show that flare occurrence peaks in mid-M dwarfs and decreases toward ultracool dwarfs (see Figure~\ref{fig:tess-ffd}), although even the coolest stars can produce (super)flares with energies up to ${\sim}10^{33}$~erg. The flare energy distributions in these stars are broadly consistent with those of solar-type stars, following a power law with a slope of $\alpha \approx 1.8$ \citep{Gizis2017ApJ...838...22G, 2024MNRAS.527.8290P}. 
The analysis of TESS flares presented in Figure~\ref{fig:tess-ffd} shows similar results, with no strong correlation to be found between \teff{} and FFD slopes. It may be interesting to note that in the optical range, stars with higher rotation rates tend to have larger $\alpha$ values \citep{aschwanden2021}.

However, it is important to note that comparing flare occurrence rates across different spectral types requires adopting a common low‑energy limit for all stars (Sect.~\ref{sec:flares-FFD}); otherwise, arbitrary results can be obtained. When such a common energy threshold is used (e.g., above $10^{33}$~erg), the highest flare rates are actually found on active, young G dwarfs (e.g., EK Dra) and early K dwarfs (e.g., AB Dor)---not on M dwarfs \citep{yang2019}. M dwarf flares rarely reach the highest energies, hence their rates above a high threshold are lower than those of active G dwarfs. Nevertheless, M dwarfs remain the most prolific sample for flare studies for two kay reasons: (1) their magnetic activity persists over much longer timescales than in G/K dwarfs, resulting in a larger fraction of active stars at any given age, and (2) the large temperature contrast between flare footpoints and the cool photosphere makes flares easier to detect. Thus, the statement that flare occurrence ``peaks in mid‑M dwarfs'' holds true for low‑energy thresholds (or for detected events in magnitude‑limited surveys), but it no longer applies for absolute comparisons above a fixed high‑energy threshold--where G dwarfs dominate. 
Similar findings about very strong superflares on G dwarfs in relation to the FFDs for G dwarfs were already reported by \citet{maehara2012}  based on Kepler data.
This nuance is essential when interpreting FFDs and stellar activity trends (see also Section~\ref{sec:Rad_Stellar_Flares}).

Flares are also known to occur on red giant stars, where their energies can range from roughly $10^{32}$ to $10^{38}$ ergs in the Kepler bandpass, often surpassing the most energetic solar flares. Detecting these events in the visible spectrum on giants is more challenging due to their inherently higher background brightness.
In the Kepler field, only about 0.3\% of giant stars ($\log g \leq 3.5$) exhibit flaring activity, corresponding to 61 identified objects, while the number of oscillating red giants is roughly ten times higher. Some stars display both phenomena, with eight such cases reported by \cite{Olah2021A&A...647A..62O}. Magnetic activity may also suppress oscillations in close binary systems \citep{Gaulme2014ApJ...785....5G}. The flare morphologies on giants resemble those on dwarf stars. Furthermore, the flare energy--duration relation for giants departs from the near-linear trend seen in main-sequence stars, instead suggesting a dependence on surface gravity \citep{Olah2022A&A...668A.101O}.

\begin{figure}
\centering
\includegraphics[width=0.95\textwidth]{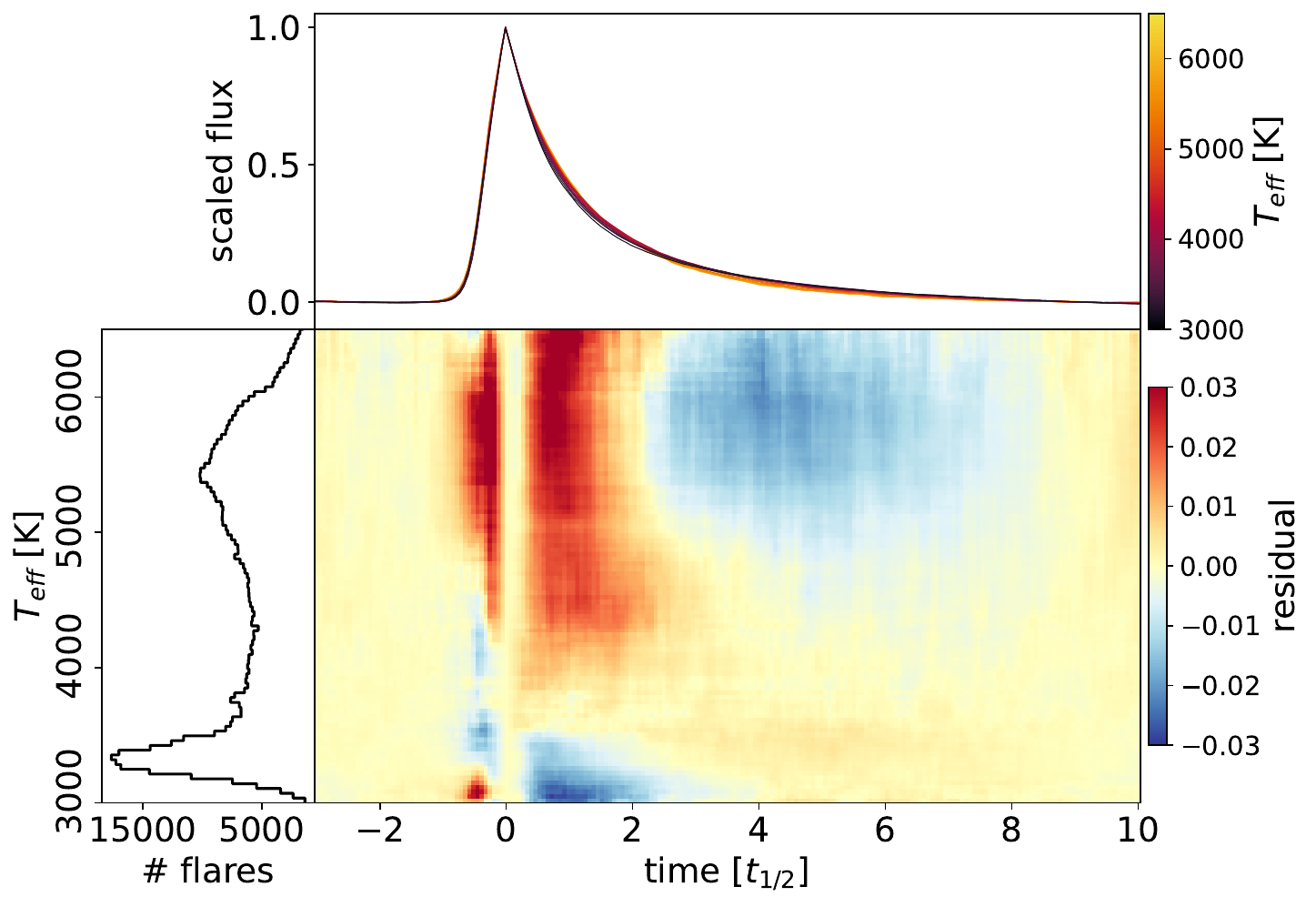}
\caption{Flare shapes along the main sequence. \textit{Top:} median flare shapes colored with \teff{}. \textit{Left:} number of flares (binned) along the main sequence. \textit{Middle:} Residual flare shapes after removing the sample median from the median shape in each bin (from \citealt{seliStellarFlareMorphology2024}).}
\label{fig:FlareMorphology}
\end{figure}

Flare properties and average morphologies are known to vary systematically with stellar spectral type: equivalent duration, amplitude, and $t_{1/2}$ (the flare time scale) change with stellar effective temperature, with hotter stars showing lower amplitudes, shorter equivalent durations, and longer $t_{1/2}$. Flare morphology also varies, with hotter stars producing broader, more complex flares that decay more slowly in the conductive cooling phase, suggesting lower plasma densities and higher atmospheric origins compared to M-dwarf flares (see Figure~\ref{fig:FlareMorphology} and \citealt{seliStellarFlareMorphology2024}).
TESS data with 20-s sampling indicates that a large fraction (46\%) of the flares show a complex behavior during the rise phase, explaining sharply-peaked/weakly-peaked flares in longer-cadence data \citep{Howard+2022_ApJ}.
Ultrafast-cadence (0.3\,s) ground-based data suggest that the optimal sampling of flares is around 4--5 seconds: information content does not increase significantly with higher cadence \citep{Schmercz2025_ADLeoFlares}. While flaring is linked to stellar rotation and magnetic complexity, many open questions remain regarding the physical mechanisms behind FFD breaks, flare periodicity, and cycle dependence, making stellar flares a rich laboratory for magnetic activity studies across the HRD.

During both solar and stellar flares, the majority of the radiative energy release occurs in the optical and near-ultraviolet (NUV) continuum, commonly referred to as the white-light continuum emission \citep{2015ApJ...809...79O, Kowalsk+2024_LRSP}.
The largest uncertainty in the Kepler/TESS studies arises from the widely-held assumption that the white-light continuum follows an optically-thick ``9,000--10,000~K blackbody spectrum'' \citep{Hawley+1992, Kowalski+2013}, when estimating the bolometric energy \citep{maehara2012,Shibayama2013,okamotoStatisticalPropertiesSuperflares2021}. 
Much higher temperature values of $\sim$40\,000~K \citep{Howard+2020_ApJ}
and lower temperature of $\sim$5,000--6,000~K \citep{Bicz2025A&A...699A..90B,2020PASJ..tmp..253M,Osten+2016} 
have been also recently suggested for some early/mid M-dwarf flares, on the basis of broad-band multi-color photometry observations. 
Lower temperatures of 3,000--4,000~K were recently reported and interpreted for flares on the late cooler M-dwarf TRAPPIST-1 \citep{Howard_Ward+2023, Shapiro2026}.

\begin{figure}[ht!]
\begin{center}
\includegraphics[width=0.75\textwidth]{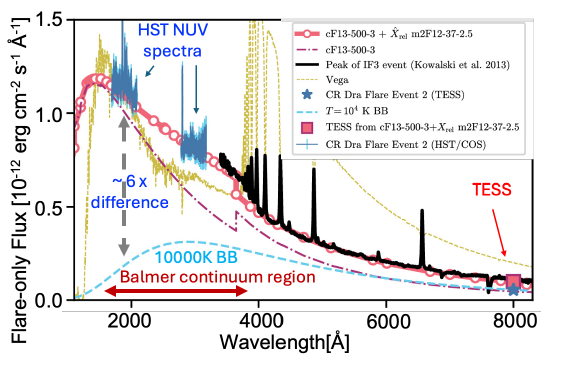} 
\caption{HST/COS NUV spectra and TESS data from a superflare 
observed in \citet{Kowalski+2025_ApJ} with the largest NUV flare luminosity observed to date from an M star. RHD modeling results and the scaled previous optical flare spectrum are overplotted for comparison. Most importantly, the HST NUV spectra are much (${\sim}6\times$) brighter than the extension of 10\,000~K blackbody emission scaled into the TESS optical flux.}
\label{fig:NUV_HST_flare}
\end{center}
\end{figure}

Moreover, recent simultaneous NUV spectroscopy and TESS photometry observations of M-dwarf superflares (\citealt{Kowalski+2025_ApJ}) have showed that 
the composite NUV flare spectra are not well represented by the single blackbody 
that is commonly assumed in the literature.
Rather, they showed that 
continuum flux rises toward shorter wavelengths into the FUV and 
an optical 10\,000~K blackbody can underestimate 
the short-wavelength NUV flux by a factor of up to $\sim$6 (Figure~\ref{fig:NUV_HST_flare}), 
which would suggest much stronger particle acceleration 
processes compared with solar flares.
Radiative hydrodynamic (RHD) modeling studies of M-dwarf flares (e.g., \citealt{Kowalski+2024_ApJ}) have also revealed that the shape and amount of UV through optical radiation (e.g., Balmer continuum in Figure~\ref{fig:NUV_HST_flare}) 
is a key parameter to understand the particle acceleration processes 
in the stellar atmosphere generating stellar flare radiations.
These results may suggest larger contribution of NUV wavelength range
into the total flare energy release in case of superflares.
However, simultaneous NUV and optical spectral observations are still limited 
(especially for stars other than M-dwarfs, especially G-dwarf superflares).
It is still uncertain how we can physically explain the white-light continuum emission and whether knowledge from recently investigated M-dwarf superflares (e.g., larger NUV emission than the 10\,000~K blackbody assumption) could be applied
especially into G/K dwarf flares. Relations between NUV/optical white-light and X-ray thermal emissions can be also an important aspect of further studies \citep[][see also Section~\ref{subsubsec:X-Ray_regime}]{Tristan+2023_ApJ,Notsu+2025_ApJ}.

\subsubsection{Stellar Flares in the XUV Regime}\label{subsubsec:X-Ray_regime}

Flares observed in the X-ray and EUV regimes are the result of coronal plasma heating to temperatures up to tens of MK. The sequence of key processes is relatively well understood from solar observations, although there may be deviations from the standard flare model (see Section~\ref{sec:Rad_solarflare_mag}). Initially, electrons are accelerated in the corona to high energies, generally as a consequence of magnetic reconnection involving tangled magnetic fields and motion. Trapped electrons precipitate along the closed magnetic field lines into the chromosphere where they collisionally heat the gas. Electrons with non-zero pitch angles promptly radiate gyrosynchrotron radiation usually detected as centimeter to millimeter radio emission (``microwaves''). The collisions in the chromosphere immediately release energy that can be observed as non-thermal hard X-ray bremsstrahlung, but also as optical line+continuum radiation from the heated, dense gas with temperatures  of order $10^4$~K. Radio, optical, and hard X-ray emissions are very closely correlated in time \citep{kosugi1988, hudson1992}, on time scales of only seconds; collectively, these emissions are called ``impulsive emissions'' as they are related to the impulsive phase of a flare, i.e.\ the phase where the bulk of the flare energy is released. As a consequence of chromospheric heating, the gas pressure increases, and the plasma flows upward along the field lines, filling the entire coronal loops (``chromospheric evaporation''). As the coronal plasma becomes denser and reaches millions of K, it brightens in X-rays, reaching peak luminosities only significantly after the impulsive phase. However, at the time of the X-ray peak, the temperature is already declining as a consequence of efficient cooling, both by  conduction back to the surface and by radiation. The highest temperatures are reached at the beginning of the heating phase, i.e., during the impulsive phase when the flare X-ray luminosity $L_{\rm X}$ is still low. Cooling may take minutes to tens of minutes, defining the gradual phase of a flare.

The most direct support for this model comes from the so-called ``Neupert Effect'' \citep{neupert1968}. While prompt, impulsive flare emission traces the input power related to the accelerated particles, the gradual flare X-ray emission roughly scales with the slowly accumulating amount of hot plasma---more precisely, the emission measure that is proportional to the square of the electron density, apart from a dependency of the emissivity on temperature \citep[e.g.,][]{Veronig2005}. To first order, one would therefore expect that the flare luminosities $L_{\rm radio,~ optical,~ HXR}$ (after subtraction of the non-flare ``background'' emission) obey two equivalent relations in time of the form
\begin{equation}\label{neupert}
    L_{\rm X}(t) \propto \int_{t_0}^t L_{\rm radio,~ optical,~ HXR}(u) {\rm d}u, \quad\quad\quad\quad
    L_{\rm radio,~ optical,~ HXR}(t) \propto \frac{{\rm d}}{{\rm d}t} L_{\rm X}(t),
\end{equation}
where $t_0$ is a fixed time before the start of the flare. These relations express that the impulsive radiation roughly corresponds to the time derivative of the X-ray luminosity. 
Because $L_{\rm radio,~ optical,~ HXR}$ must be positive, this relation holds only as long as the X-ray emission is increasing before reaching its peak, or equivalently, as long as there is positive power in the impulsive phase. The relation is only qualitatively correct because the bandpass of observations matters as the luminosity in an energy range is related to the continuously developing plasma temperature. Furthermore, heating may continue after the impulsive flare phase, and cooling acts continuously also during the heating process; also, the cooling rate itself is a function of the momentary temperature and density. A better approximation can be achieved if Equation~\ref{neupert} is formulated for the precipitating electron kinetic energy flux and the accumulated thermal energy in the loop plasma taking into account continuous energy loss via radiation (and/or conduction). 
A generalized Neupert Effect for the X-ray luminosity $L_{\rm X}(t)$ based on electron energy flux $F(t)$ and thermal energy content in the X-ray emitting plasma can be formulated as follows,
\begin{equation}\label{neupert2}
    L_{\rm X}(t) \propto \frac{1}{\tau(t)} 
    \int_{t_0}^t F(u)e^{-(t-u)/\bar{\tau}(t, u)}{\rm d}u
\end{equation}
where $\tau(t)$ is the $e$-folding decay time of the thermal energy (e.g., via radiative cooling) as a function of electron temperature and density at time $t$,  and $\bar{\tau}(t, u)$ is the harmonic mean of $\tau$ over the interval $[u, t]$ (see \citealt{guedel1996} for details). Observed radio, optical, or HXR emission would reasonably be assumed to be proportional to $F(u)$.

\begin{figure}[t!]
\begin{center}
\includegraphics[width=0.75\textwidth]{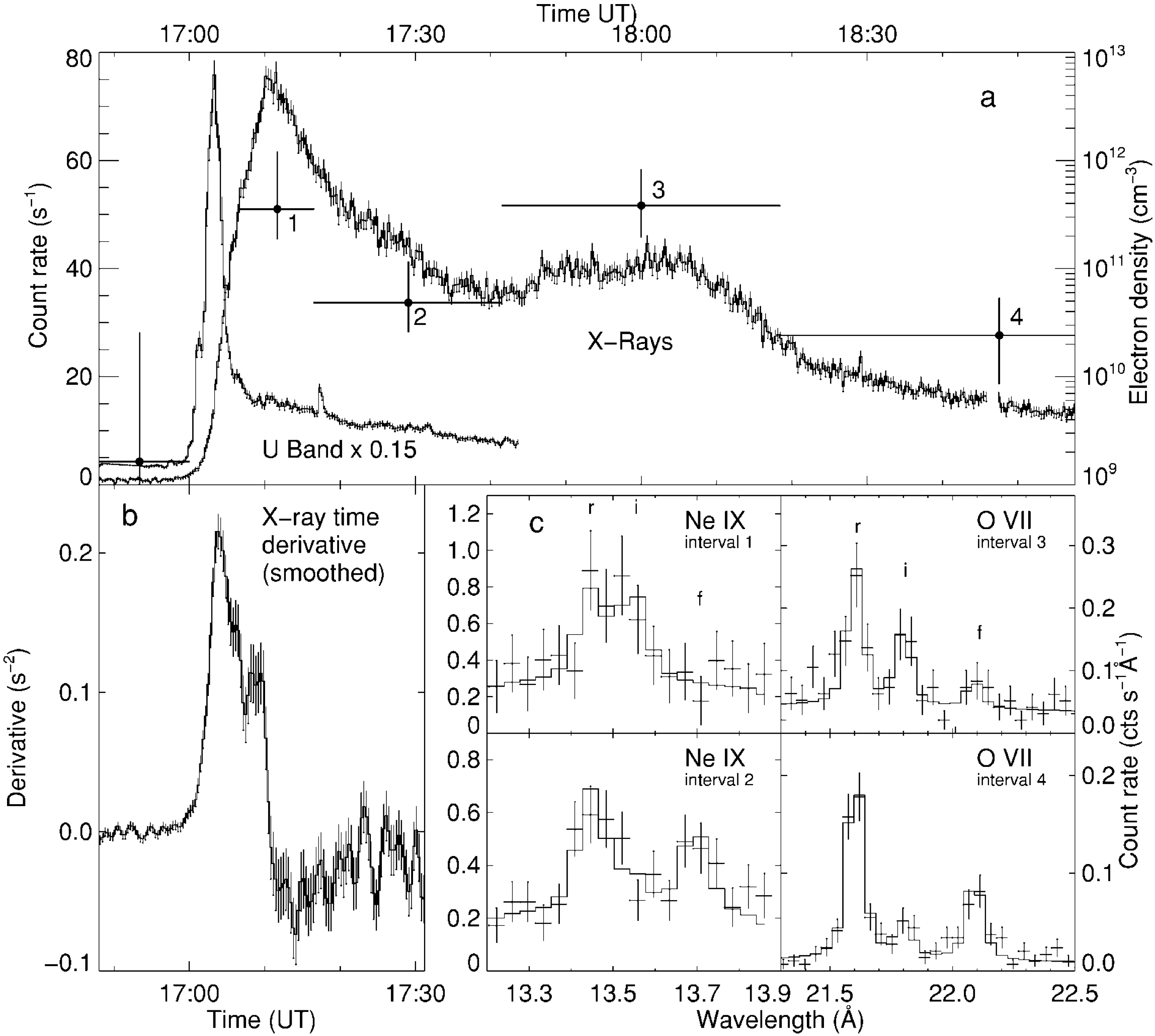} 
\caption{The X-ray vs. $U$-band ``Neupert Effect'' for a flare on Proxima Centauri. In the upper panel, the impulsive and the gradual phases are observed in the $U$ band and in X-rays, respectively. The lower left panel shows the time derivative of the X-ray light curve, closely resembling the $U$-band light curve. The four small panels in the lower right show the density-sensitive Ne~{\sc ix} and O~{\sc vii} line triplets during the four time intervals `1', `2', `3', and `4' marked by horizontal bars in the upper panel. A large flux ratio between the intercombination and the forbidden lines (middle and right lines of the triplets, labeled as `i' and `f', respectively) indicates higher densities than a small ratio, as shown by the density values of the bars in the upper plot (referring to the $y$ axis on the right). The densities are thus increased during the two X-ray peaks `1' and `3' to which the two upper spectral line plots refer. (From \citealt{guedel2002}.)}
\label{fig:neupert}
\end{center}
\end{figure}

The Neupert effect is commonly observed on the Sun, especially for thermal soft and non-thermal hard X-ray emission (e.g., \citealt{dennis1993,veronig2002}). It has also often been seen in stellar flares, where the impulsive phase is observed either in the optical or in radio waves, and instead of X-rays the EUV range has also been used \citep{hawley1995, guedel1996, osten2004,Tristan+2023_ApJ}. Figure~\ref{fig:neupert} shows an example of a Neupert Effect on Proxima Centauri, recorded in observations in the $U$ band (impulsive phase) and the soft X-rays (gradual phase; \citealt{guedel2002}). During the X-ray peaks, the electron densities are enhanced, as measured by line ratios in the density-sensitive line triplets of He-like Ne~{\sc ix} and O~{\sc vii}. It should be noted that a fraction of flares, both solar and stellar, do not follow  this behavior (see \citealt{Tristan+2023_ApJ} for classification and statistics of deviating behavior).

\subsection{The Role of Flares in High-Energy Radiation} \label{sec:Rad_Stellar_Flares}



\begin{figure}[t!]
\begin{center}
\includegraphics[width=0.74\textwidth]{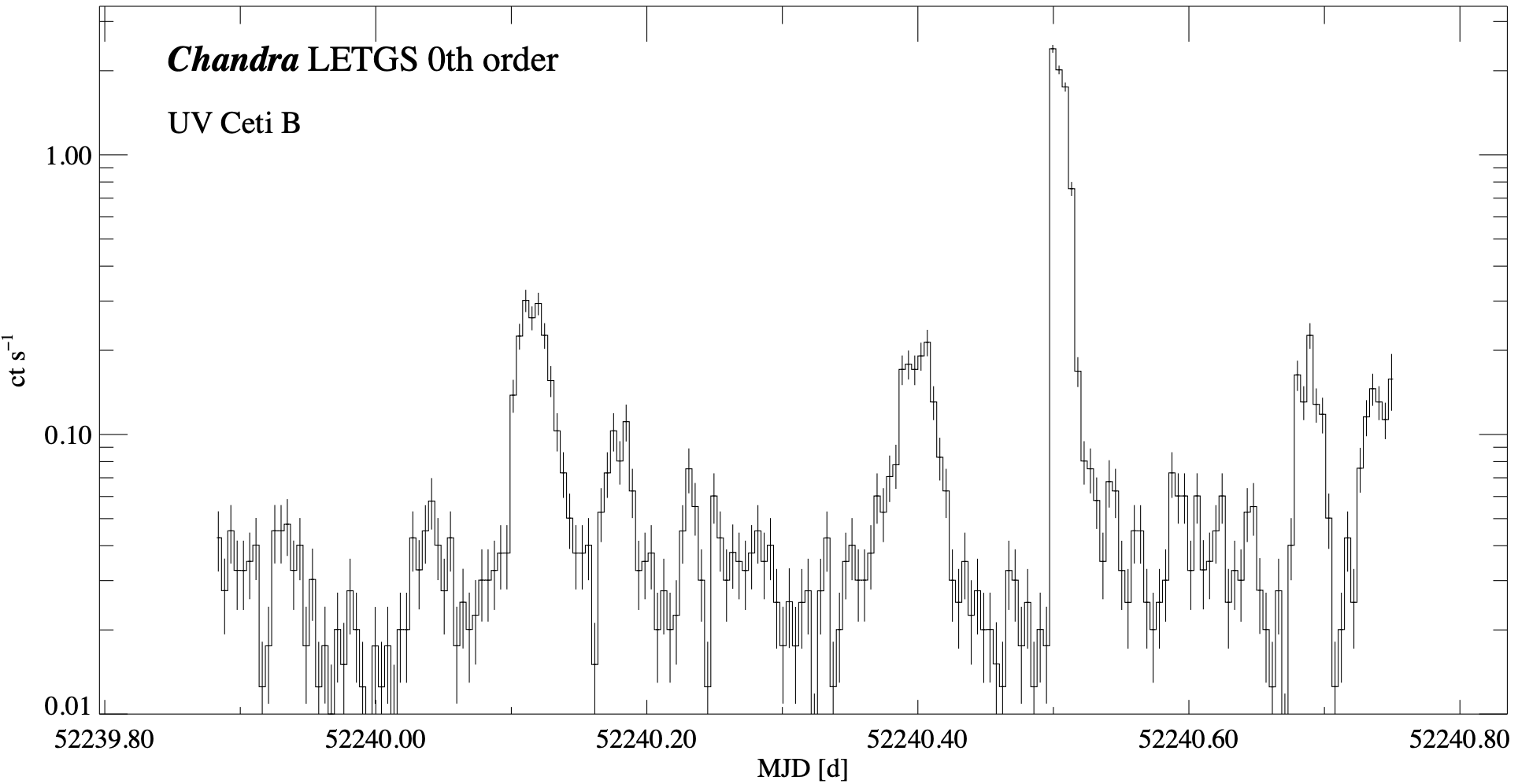} 
\caption{X-ray Light curve of UV~Cet over nearly one day, on a logarithmic flux axis. Note that a ``low'' level cannot easily be defined in this observation. (From \citealt{guedel2004}, courtesy of M.~Audard, after \citealt{audard2003}.)}
\label{fig:lightcurve}
\end{center}
\end{figure}

Observations of the Sun reveal flare events at almost any time, from very frequent small events to rare but giant flares \citep[e.g.,][]{2016SoPh..291.1761H, 2017LRSP...14....2B}. This is no different for other cool main-sequence stars (see \citealt{damonteetal2026} for Proxima Centauri). Flares appear to occur stochastically, and flare radiative energies or flare amplitudes are seemingly uncorrelated with each other. X-ray or EUV light curves of cool stars therefore show, on top of a quasi-steady level, a ``flickering'' contribution from individual flares (see Figure~\ref{fig:lightcurve} and further examples described in \citealt{ambruster1987, marino2000, audard2000, stelzer2002, audard2003, guedel2003}, and others). In extreme cases, flares are so frequent and dominant in light curves that the concept of a ``background'' quasi-steady or ``quiescent'' XUV emission becomes questionable \citep{audard2003, damonteetal2026}. The stars are continuously flaring, and such episodes can easily be seen in X-rays on the (whole) Sun as well, for days in a row \citep{guedel2004}.

The range of flare energies is huge. The smallest solar events were coined ``nanoflares'' and ``microflares'',
corresponding to a total XUV radiative energy of order $10^{24}$~ergs and $10^{27}$~ergs, respectively \citep{lin1984, parker1988, hannah2011}, in relation to the largest observed solar flares with bolometric energies of order $10^{33}$~ergs \citep{hannah2011}. 
Such events cannot be detected on other stars, both because the observational sensitivity is too low and because the stellar surfaces are not resolved in observations.
On the other hand, stars with higher activity levels may reveal gigantic flares, also called ``superflares'' with radiative energies up to $10^4$ times the strongest solar flares or about $10^{37}$~erg (including optical observations; see \citealt{schaefer2000, maehara2012, shibata2013} and Section~\ref{sec:flares-optical}), which seem to have no analog on the present-day Sun. Extremely large flares are also observed beyond the main-sequence, e.g., in subgiant RS CVn-type and Algol binaries \citep{osten2007, schmitt1999}, in giants \citep{ayres1999, ayres2001}, rapidly rotating giants of the FK Com type \citep{welty1994}, and pre-main sequence T Tauri stars \citep{wolk2005, franciosini2007}.

Before the role of flares in the overall XUV luminosity can be addressed, it is necessary to quantify the statistics of flares. As already discussed in Sections~\ref{sec:flares-FFD} and \ref{sec:flares-optical}, flares follow power-law FFDs, and this is no different in high-energy emission (Figure~\ref{fig:energydistrib}). In the XUV range, various observations reveal $\alpha$ values in the range of 1.5--2.5 with a dominance of values at or above a value of 2.0, typically in the range of 2.0--2.4 \citep{collura1988, pallavicini1990, audard1999, audard2000, kashyap2002, guedel2003, arzner2004, wolk2005, stelzer2007, 2007A&A...471..645C, 
albacetecolombo2007}. This suggests that a large number of small flares contribute significantly to the overall coronal XUV emission, as we will discuss in more detail below.

\begin{figure}
\begin{center}
\includegraphics[width=0.55\textwidth]{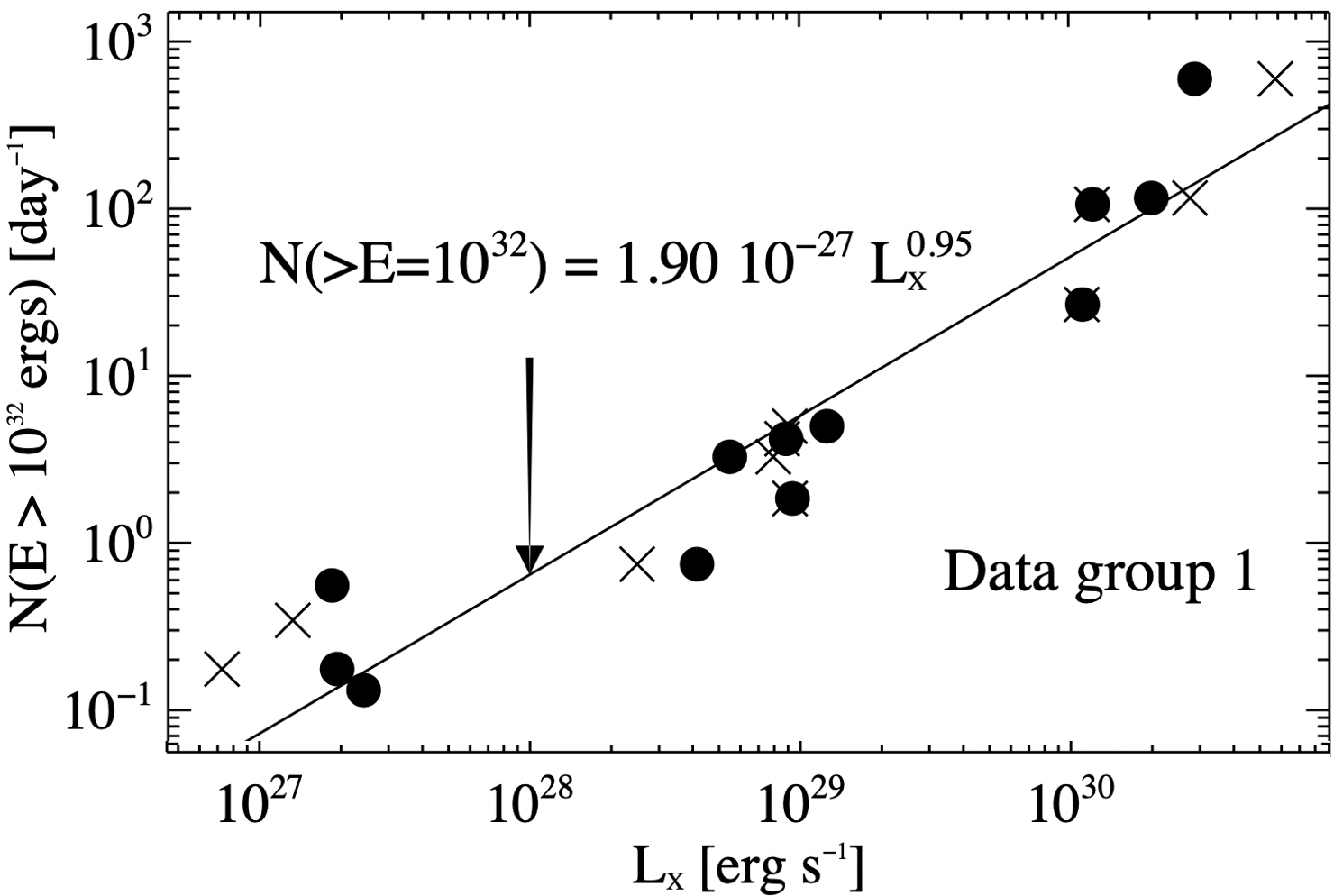} 
\caption{
Correlation between cumulative EUV flare rate above a threshold of $10^{32}$~erg in relation to the average stellar $L_{\rm X}$. 
(From \citealt{audard2000}.)}
\label{fig:flareratevsLX}
\end{center}
\end{figure}

\begin{figure}[t!]
\begin{center}
\includegraphics[width=0.97\textwidth]{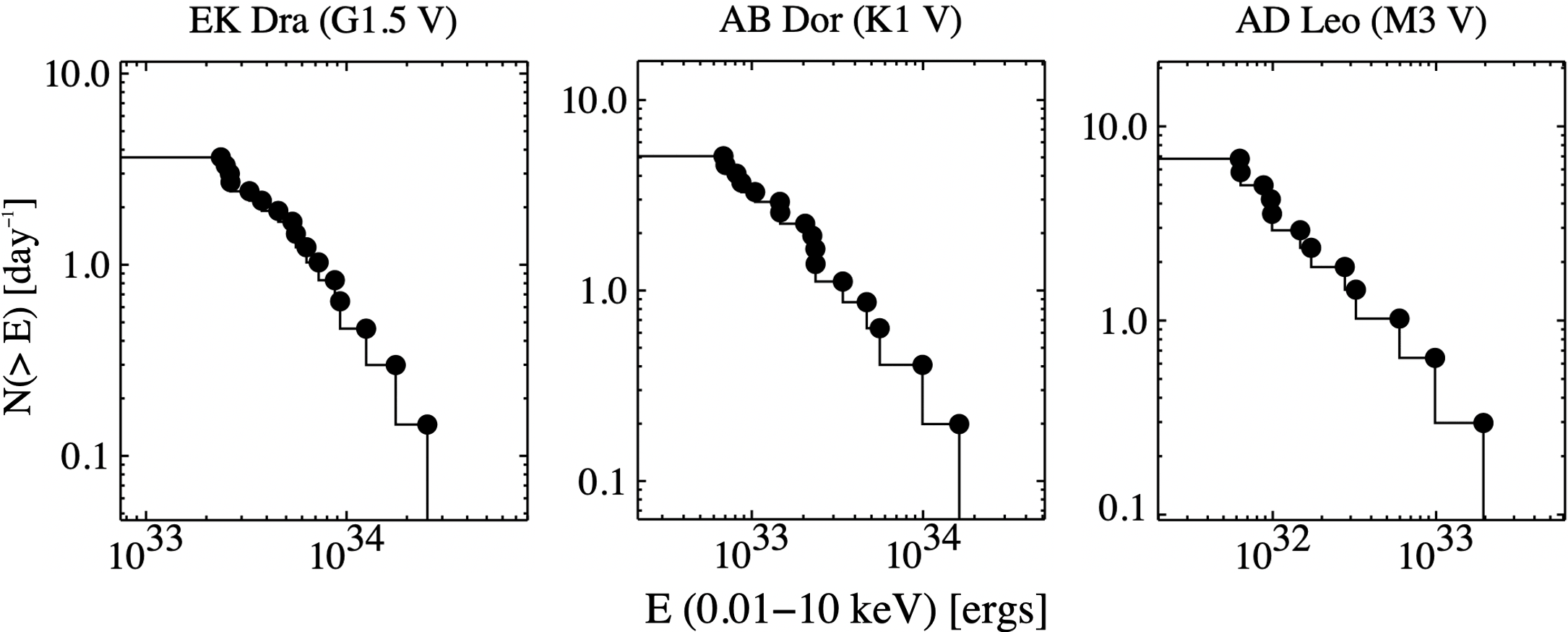} 
\caption{Cumulative flare energy distributions (Equation~\ref{flareenergies_int}) for extremely active stars; from left to right, EK Dra (G1.5V), AB Dor (K1V), and AD Leo (M3V). Note the decreasing rate with progressively later spectral type at a given energy threshold, e.g. $10^{33}$~erg, in conjunction with a lower energy range for measured flares. 
(From \citealt{audard2000}.)}
\label{fig:energydistrib}
\end{center}
\end{figure}

As far as ``flare rates'' (above a fixed lower energy threshold, see Section~\ref{sec:flares-FFD}) for  different activity levels and different spectral classes are concerned, the coefficient $a$ in Equation~(\ref{flareenergies_diff}) increases linearly with the activity level of a star (as measured by the X-ray luminosity; see Figure~\ref{fig:flareratevsLX} and \citealt{audard2000}); because the X-ray luminosity  of stars at a given relative $L_{\rm X}/L_{\rm bol}$ activity level increases from M to G spectral classes (following the increase of $L_{\rm bol}$), the more
massive stars produce more flares above a given energy threshold and therefore stronger flares.
This is explicitly illustrated in  Figure~\ref{fig:energydistrib}: the value of $a$ for a comparable relative activity level---here for a sample of (near-)saturated stars with $L_{\rm X}/L_{\rm bol} \approx 10^{-3}$ and with different spectral types---increases toward earlier spectral types or, equivalently, with increasing stellar mass up to about solar mass. These observations show that very active G dwarfs produce the highest flare rates (above a given energy threshold) and therefore, within a given time period, statistically the most energetic flares. This parallels the finding for optical flares as discussed in Section~\ref{sec:flares-optical}.
Even considering that magnetic activity of M dwarfs decays on much longer timescales (up to Gyr) than that of G dwarfs (hundreds of Myrs; \citealt{johnstone2021a}; see also \citealt{Chapter4}), G dwarfs on average produce a higher frequency of flares above a given threshold and therefore larger flares at any given age on the main sequence (see Figure~19 of \citealt{johnstone2021a} and Figure~7 of \citealt{Chapter4})---at the limit, comparable levels may be reached for early-to--mid M dwarfs and early G dwarfs.

This contrasts with the commonly held opinion that the ubiquitous and notoriously active M dwarfs flare the most energetically (hence the often used expression ``flare stars''). Although apparently supported by observations of flares with high amplitudes relative to the average low-level emission (e.g., \citealt{audard2003, guedel2003}, see also Figure~\ref{fig:lightcurve}), it contradicts the statistics of flares; the physical reason for the conspicuous flare activity in M dwarfs is the result of contrast between flares and low-level (``quiescent'') emission. A single flare in an active region on an M dwarf constitutes a large perturbation of its corona while on a much larger surface area of a solar analog, the same flare competes with a substantially bigger X-ray emitting volume not involved in the flare (see Figure~8 in \citealt{reale2004}). Similar issues apply to optical flares; their bright footpoint areas produce more contrast on smaller stars; in addition, the temperature contrast between flare footpoints and the photosphere is large for M dwarfs but much smaller for G dwarfs. M dwarf flares are bright in a spectral range (e.g., the $U$ band) where the photosphere is faint. In the large statistical study of optical flares by \citet{yang2019}  the 1st percentile of all recorded flares on solar-mass stars is $10^{33}$~ergs, while for M dwarfs this value corresponds approximately to the upper bound, equivalent to the 99th percentile (see a summary in \citealt{johnstone2021a}).


The frequent observations favoring $\alpha > 2.0$ therefore open up the hypothesis that a fraction, if not 100\%, of the XUV emission is due to the sum of all flare contributions from a distribution as given in Equation~(\ref{flareenergies_diff}), depending on what the lower-energy threshold of the power-law distribution is. Often called the ``microflare'' or ``nanoflare hypothesis'', such a model has far-reaching consequences for coronal heating as it posits that stellar coronae are largely heated by flares, in particular by the large number of smaller events \citep{parker1988}. Before accepting this model for the large variety of cool stars, observational support is needed, which we summarize below: 

\begin{figure}[t]
\begin{center}
\hbox{
\includegraphics[width=0.49\textwidth]{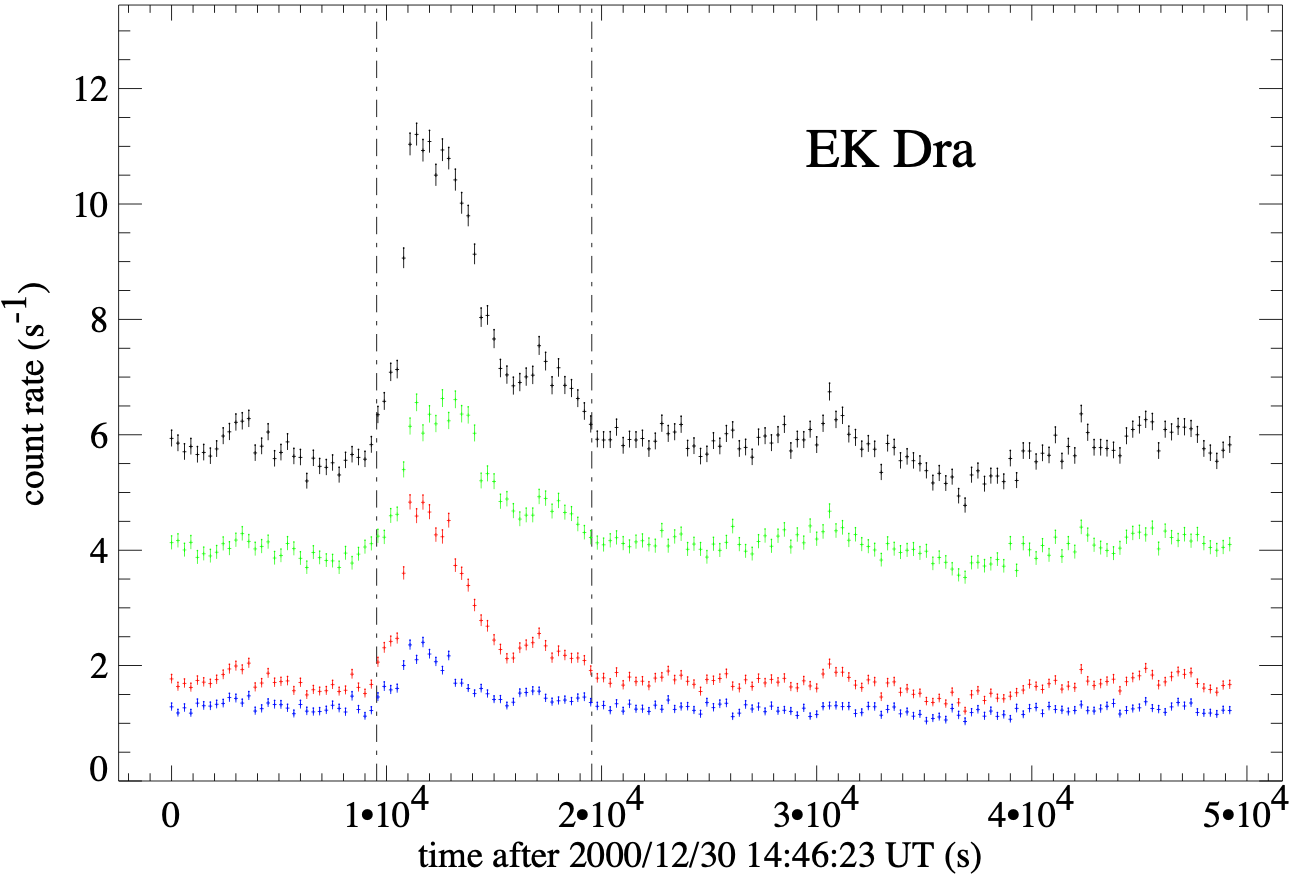} 
\includegraphics[width=0.49\textwidth]{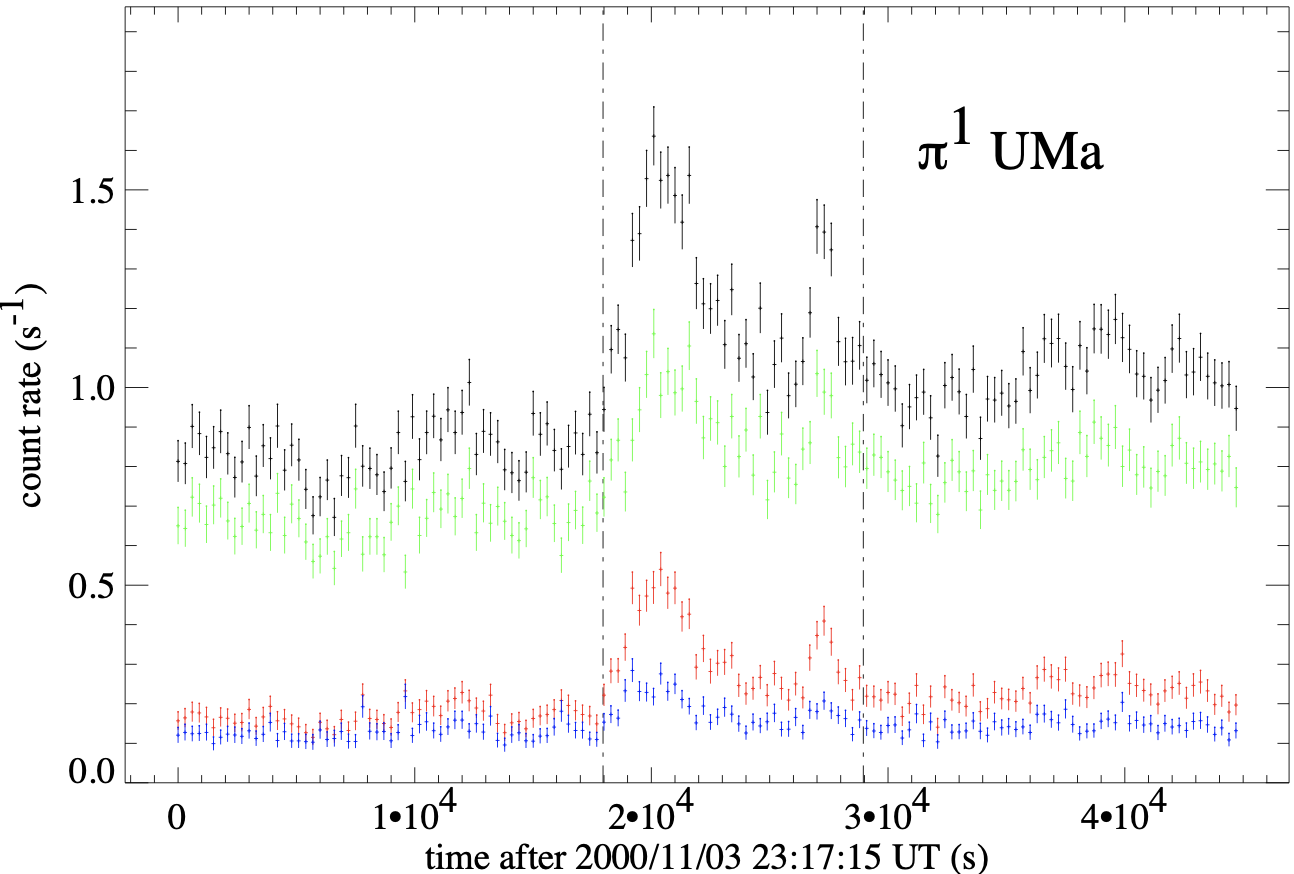} 
}\vskip 0.3cm
\hbox{
\includegraphics[width=0.49\textwidth]{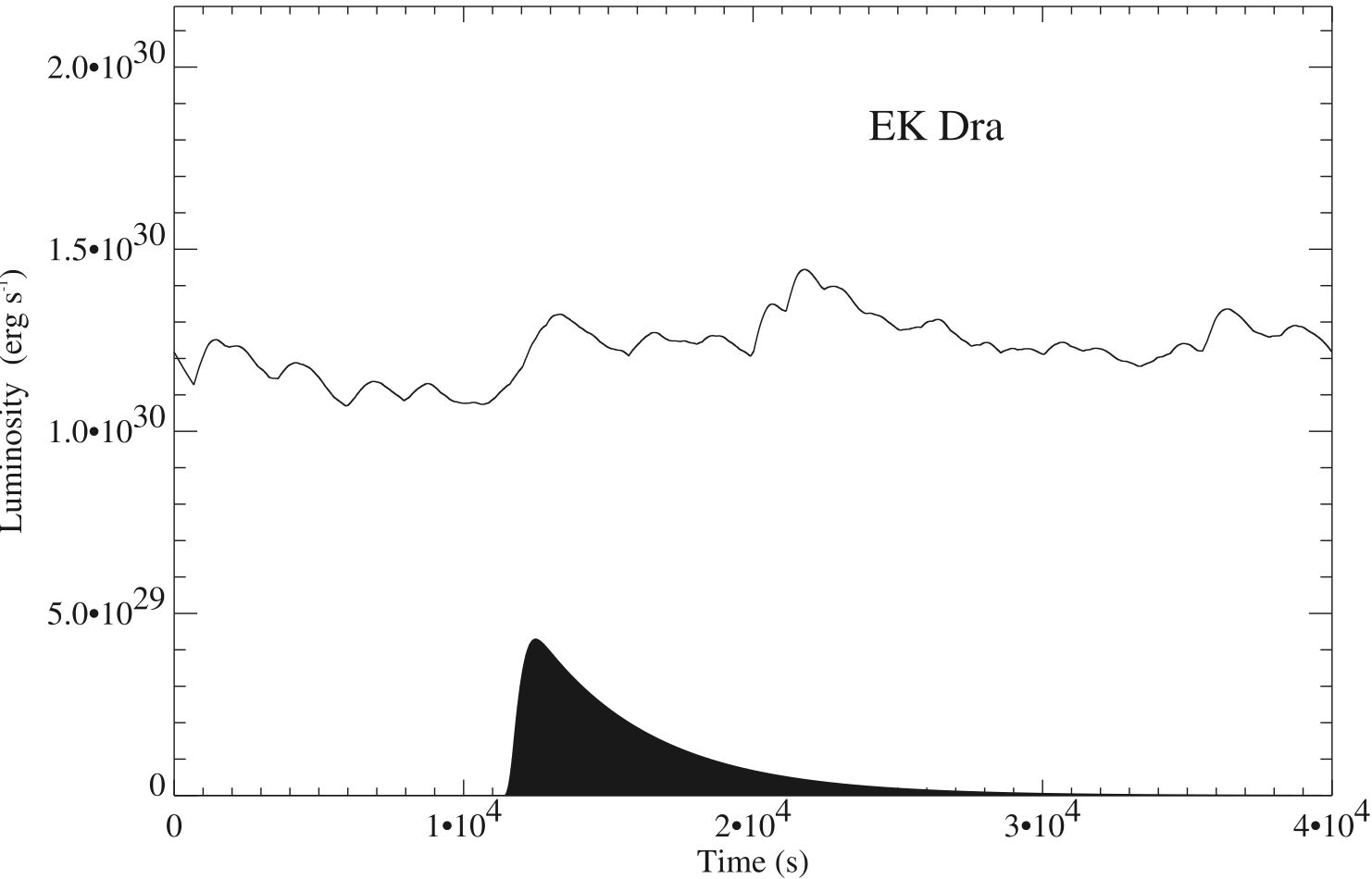} 
\includegraphics[width=0.49\textwidth]{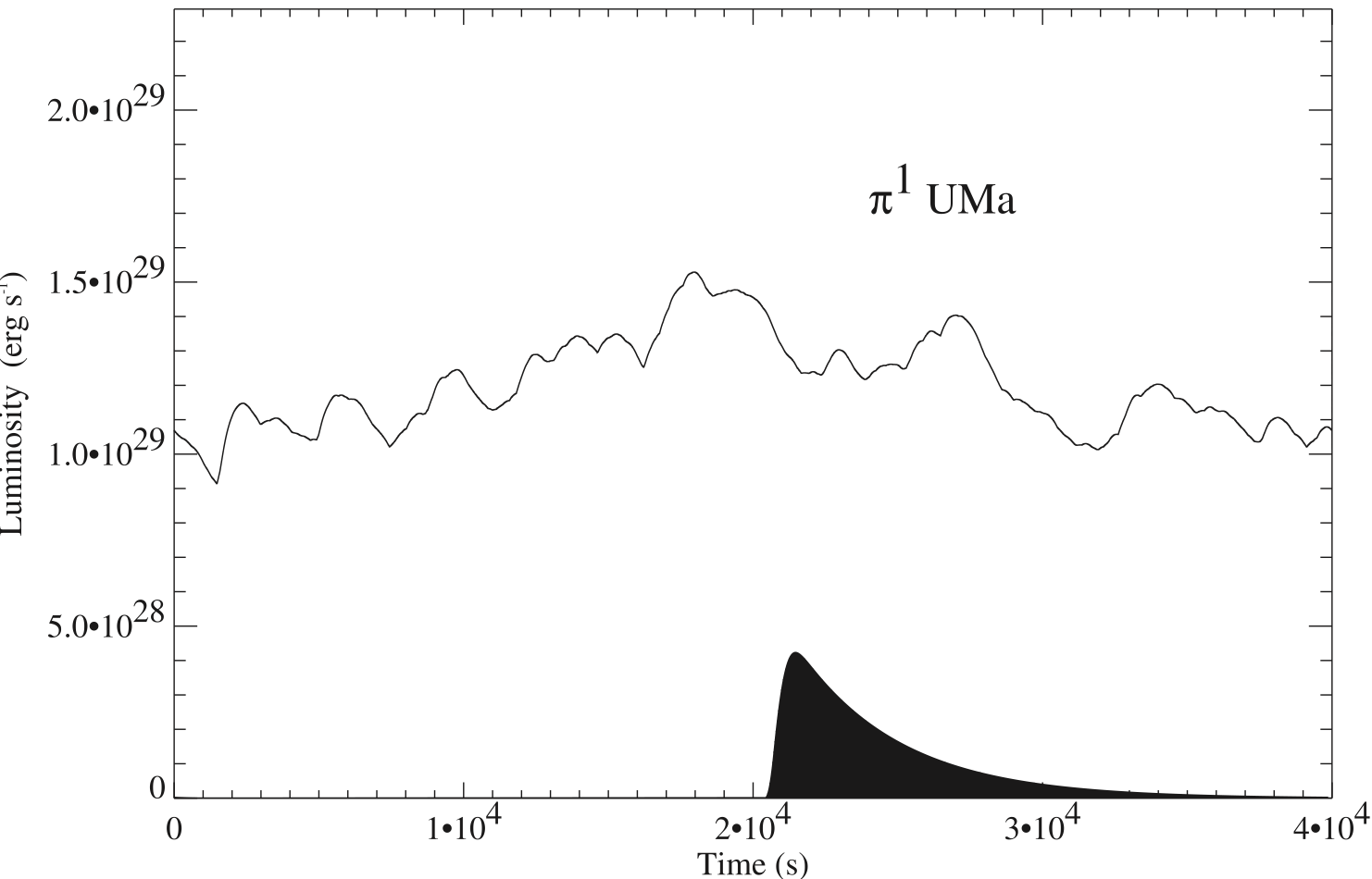} 
}
\vspace*{-0.2 cm}
\caption{Observed and synthesized X-ray count-rate light curves for two solar analogs at high (EK Dra, left) and intermediate ($\pi^1$~UMa, right) activity levels. \textit{Upper panels:} Observed X-ray light curves (from \citealt{telleschi2005}). The curves refer to the 0.2--10 keV range (black), the 0.2--1.0~keV soft band (green), the $>1$~keV hard band (red),  and the ratio of hard/soft (blue; multiplied by 3 for EK Dra and 0.5 for $\pi^1$~UMa for illustration purposes).   -- 
\textit{Lower panels}: Synthetic light curves of superposed flares following a flare energy distribution as in Equation~(\ref{flareenergies_diff}) occurring stochastically in time. A scalable shape template with constant exponential decay time as shown in black was used for all flares, where the example itself shows the amplitude of the largest flare actually used for the simulation (amplitudes reaching the singular largest flares in the observed light curves were not included in the synthetic light curves). Parameters for EK Dra (left): $\alpha = 2.28$, lowest flare energy $10^{30}$~erg; ratio highest/lowest flare energy $10^{3.1}$; and for $\pi^1$~UMa (right): $\alpha = 2.54$, lowest flare energy $10^{29}$~erg; ratio highest/lowest flare energy $10^{2}$
(from \citealt{telleschi2005}).}
\label{fig:lightcurves}
\end{center}
\end{figure}
\begin{enumerate}

\item\textbf{Light curves:} There is no doubt that many stellar XUV observations reveal no or only very few individual flares because of the limited sensitivity. There is no justification for the assumption that smaller flares are absent, in particular given the evidence of ubiquitous microflares on the Sun that could not be detected on other stars. However, the superposition of flares from an ensemble described by Equation~(\ref{flareenergies_diff}) occurring stochastically in time can also produce light curves in which the sum of all smaller flares fill up a ``quasi-steady'' emission level while only the strongest flares are detected individually (see Figure~\ref{fig:lightcurves}; \citealt{kopp1993, telleschi2005}). Assuming a fixed ``flare shape'' (e.g., a convolution of a Gaussian with an exponential function that is cut off at $t < 0$), the modulation depth of the light curve, i.e., the ratio between the light curve rms scatter and the average level, can then be used to constrain the energy range of contributing flares and the value of $\alpha$ (Figure~\ref{fig:lightcurves}, \citealt{telleschi2005}).
Observed light curves can also be directly inverted via characteristic functions (Fourier transform of the probability density) to derive the underlying $\alpha$ value \citep{arzner2004}.  Alternatively, inversion has been achieved via forward modeling using Monte Carlo methods \citep{kashyap2002, guedel2003}, generally leading to agreement between all methods. The success of such inversions lends support to the microflare hypothesis. Light curves in X-ray and other high-energy wavelengths are needed to realistically model the response of an exoplanetary atmosphere. Such light curves are possible for a few nearby stars, such as Proxima Centauri \citep{damonteetal2026}.

\begin{figure}[t]
\begin{center}
\includegraphics[width=0.55\textwidth]{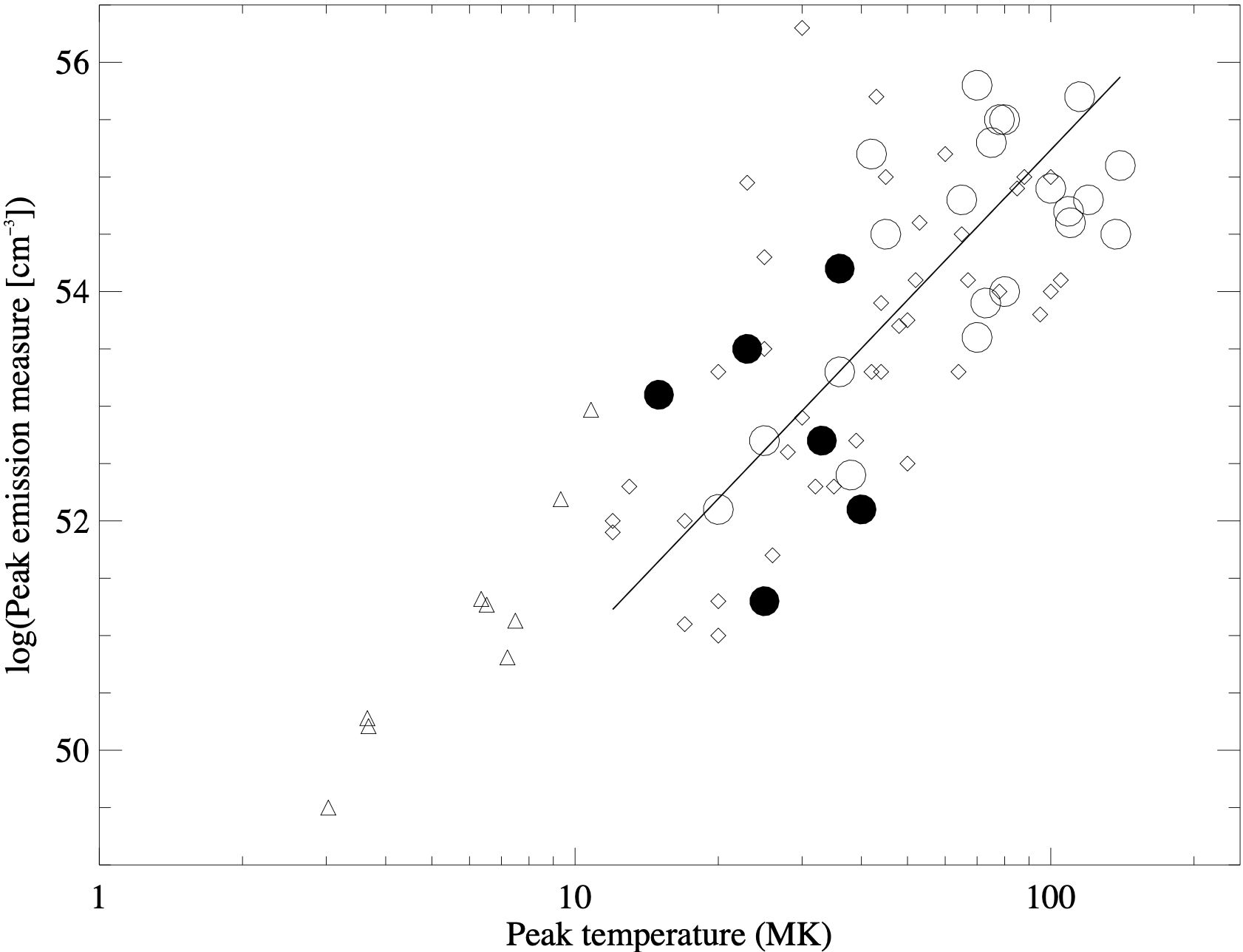} 
\caption{
Correlation between flare peak temperature and peak emission for flares from the literature; the triangles indicate temperature and total emission measure for the average emission of solar analogs at different activity levels \citep{guedel1997a}.
(From \citealt{guedel2004}.)}
\label{fig:EMT}
\end{center}
\end{figure}

\item\textbf{Flare temperatures:} Flares heat coronal plasma up to tens of MK. Larger flares are hotter: their peak temperatures scale with their peak emission measures\footnote{Stellar coronae are often described by the differential emission measure (DEM) derived from spectra. The DEM describes the amount of emission measure $n_en_i {\rm d}V$  ($n_i$ = ion density, $n_e$ = electron density, $V$ = volume) per logarithmic temperature interval ${\rm d}\ln T$} as found on the Sun \citep{feldman1995, feldman1996} and on stars \citep{guedel2004}. Interestingly, outside obvious flares the average coronal temperature of a star of given size also scales with the total average X-ray emission measure (e.g., \citealt{vaiana1983, schrijver1984, schmitt1995, guedel1997a, orlando2000, peres2000}), and the trend is the same as for flares, see Figure~\ref{fig:EMT}   \citep{guedel2004}. This suggests that at higher magnetic activity levels of stars, a larger number of higher-energy flares heat the corona to higher average temperatures \citep{guedel2004}.

\item\textbf{Coronal emission measure:} For a given flare energy distribution as in Equation~(\ref{flareenergies_diff}), smaller flares contribute lower-temperature plasma and larger flares contribute, around their peak time, hotter plasma. For stochastically occurring flares following Equation~(\ref{flareenergies_diff}), a theoretical time-averaged emission measure distribution can be analytically derived; it consists of two power laws on each side of a DEM peak. The slope of the increasing lower-$T$ part holds information on sustained heating during the decay, while the slope of the decreasing high-$T$ part depends on $\alpha$. These model DEMs agree with observationally derived DEMs (Figure~\ref{fig:DEM}, \citealt{guedel2003}). Steeply increasing lower-$T$ DEMs with $\alpha = 1-4$ (see the example of EK Dra in Figure~\ref{fig:DEM}) are also compatible with flaring coronal loops but not with steadily heated static loops \citep{guedel2004}. 

\begin{figure}[ht!]
\begin{center}
\includegraphics[width=1.0\textwidth]{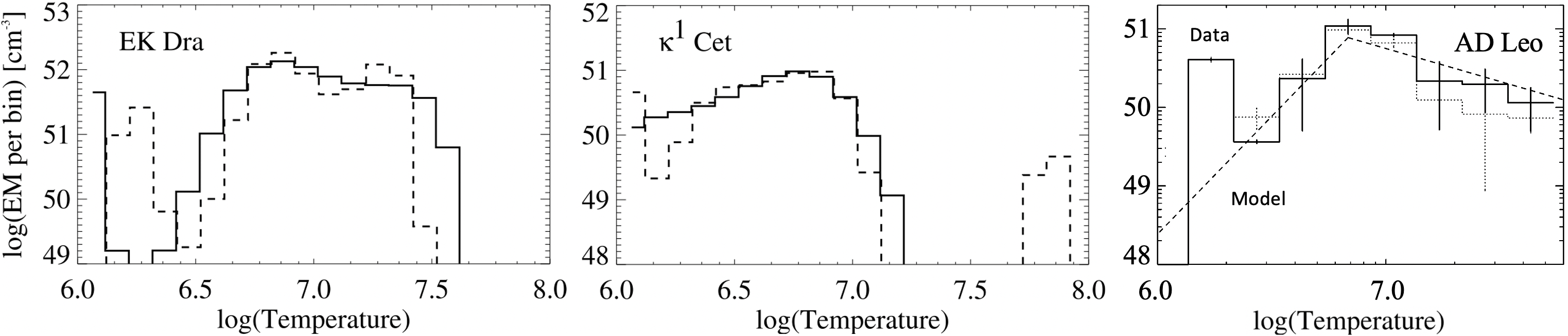} 
\caption{\textit{Left and middle}: X-ray differential emission measures derived from observations,for EK Dra (G1V) and $\kappa^1$~Cet, respectively, using two methods (solid and dashed) \citep{telleschi2005}.   \textit{Right}: Comparison between observed DEM of AD Leo from X-ray and EUV observations (solid and dotted histograms with error bars per bin for two different emission measure reconstruction methods) and a synthetic model DEM based on the flare energy distribution as in Equation~\ref{flareenergies_diff} \citep{guedel2003}.}
\label{fig:DEM}
\end{center}
\end{figure}

\item\textbf{Plasma density:} Density estimates from X-ray line ratios tend to show higher values for active stars than inactive stars 
(Section~10 in \citealt{guedel2004} and references therein). Similarly, those line ratios suggest higher densities during large stellar flares compared to quiescence (see for example \citealt{fuhrmeister2011}).

\item\textbf{X-ray luminosity:} The highest X-ray luminosities of stars exceed by up to an order of magnitude luminosities expected if the stars were entirely surrounded by a Sun-like corona \citep{vaiana1978, wood1994, vilhu1984}. A remedy may be the addition of denser, flaring regions \citep{orlando2001}. 

\item\textbf{X-ray flare rates and $L_{\rm X}$:} The rate of X-ray flares above a given radiative energy threshold is linearly correlated with the total $L_{\rm X}$, suggesting a role of flares in coronal heating, see Figure~\ref{fig:flareratevsLX} \citep{audard2000}.

\item\textbf{Correlations with other wavelengths:} The time-averaged power released from optical flares correlates with the steady X-ray emission \citep{Doyle1985, skumanich1985, whitehouse1985}; this could suggest that flares are responsible for the plasma heating although both parameters could be separate expressions of magnetic activity. Furthermore, the ratio between energy losses in X-rays and in the chromospheric Mg~\,{\sc ii} lines or between X-rays and H$\gamma$ emission is the same for flares and ``quiescent'' emission \citep{haisch1990, mathioudakis1990}, and the same holds if Mg~\,{\sc ii} is replaced by UV photometry \citep{mitrakraev2005}.

\item\textbf{Non-thermal particles:}  Magnetically active stars are strong sources of non-thermal gyrosynchrotron radiation emitted by accelerated electrons in the radio range. In view of the relatively short lifetimes of such electrons subject to collisional and radiative losses in the range of minutes to hours, frequent acceleration is required; on the Sun, the source of electron acceleration are flares, and also shocks associated with CMEs \citep{benz2010}. In magnetically active stars, the quasi-steady radio luminosity is correlated with the quasi-steady X-ray luminosity, and the same correlation holds true for flares \citep{guedel1993, benz1994}. The picture explaining this correlation is borrowed from solar flare observations; in such events, electrons are accelerated impulsively in the corona in the initial phase; the electrons stream down along the magnetic fields and collide in the chromosphere where they heat the gas to millions of K; the overpressure then drives the hot plasma into coronal magnetic loops \citep{antonucci1984, hudson1995}.
Large fluxes of non-thermal particles are often considered in RHD modeling to reproduce the observed multi-wavelength properties of stellar flares: for example,  the optical/NUV white-light continuum spectra (Figure~\ref{fig:NUV_HST_flare}, \citealt{Kowalski+2024_ApJ,Kowalski+2025_ApJ}) and line broadenings of chromospheric lines (\citealt{Namekata+2020_PASJ,Notsu+2025_ApJ}). See the review by \citet{Kowalsk+2024_LRSP} for more details.

\end{enumerate}
All evidence taken together suggests a continuous, prominent role for flares in the overall XUV emission of stars. There is an important message in this hypothesis: although we can observe flares and quasi-steady emission in the XUV, UV, optical, and radio ranges, high-energy particles always accompanying flares remain inaccessible except in the corona and the chromopshere.
However, we expect flare- and CME-related accelerated particles also throughout interplanetary space in other planetary systems. Such particles should have a profound influence on planetary atmospheric chemistry \citep{airapetian2016}.

\section{Magnetized Stellar Winds}\label{winds}
Stellar winds are a key component of any stellar system and constitute one of the main channels through which a star influences its surrounding environment. Consisting of a continuous flow of charged particles---mainly protons and electrons, with some heavier ions---they carry mass, momentum, and energy from the stellar surface out into its astrosphere. Stellar winds are a fundamental driver of stellar forcing, capable of influencing planetary magnetospheres, atmospheres, and even long-term habitability. 

In order for a stellar wind to be sustained, a non-gravitational force must overcome gravity and accelerate material outward from the stellar atmosphere. The dominant driving mechanism varies with stellar type: in hot, massive stars, radiation pressure on spectral lines---i.e., momentum transfer from photons absorbed and scattered in discrete atomic transitions---produces powerful winds, while in evolved giants and AGB stars, radiation pressure on dust grains plays the leading role. In contrast, the winds of cool, main-sequence stars are significantly weaker and cannot be explained by radiation pressure alone. Instead, they arise from a combination of thermal pressure gradients in their hot coronae and magnetic processes that provide additional heating and acceleration \citep[e.g.,][]{cranmer2019, vidotto2021}.

In this section, we review magnetized stellar wind forcing by first focusing on observational constraints. We begin with in-situ observations of our local plasma environment, i.e.\ the solar wind (Section~\ref{subsec:solarwind}), followed by the interaction of stellar astrospheres with the interstellar medium (Section~\ref{subsec:lism}), and then by observations and empirical characterization of stellar winds across a broader range of stars (Section~\ref{sec:SWinds}). We then turn to theoretical and numerical approaches, reviewing stellar wind modeling techniques (Section~\ref{sec:SW_Models}) with particular emphasis on cool main-sequence stars (Section~\ref{subsec:coolstarwinds}). Finally, we conclude with an outlook on current challenges and future prospects in stellar wind modeling (Section~\ref{subsec:sw_challenges}).

\subsection{The Solar Wind: A Window into Stellar Forcing} \label{subsec:solarwind}

As is the case for many of the mechanisms and phenomena discussed in this article, the first observational and theoretical insights into a continuous outflow of charged particles from a star originated with the Sun. Early evidence came from comet tails, which consistently point away from the Sun regardless of the comet's orbital motion, leading to the proposal of a continuous corpuscular outflow from the Sun \citep{1951ZA.....29..274B}. This idea was complemented by Alfv{\'e}n, who emphasized the role of magnetized plasmas in interplanetary space \citepads{1957Tell....9...92A}, and ultimately led to the recognition of a solar wind emanating from the hot corona \citep{1957SCoA....2....1C}. The decisive theoretical breakthrough came in the form of Parker's hydrodynamic model \citep{parker1958a}, which demonstrated that a static, million-degree corona is unstable and must expand supersonically. However, thermal pressure alone cannot account for all observed features of the solar wind, such as high-speed streams and variability linked to the magnetic cycle. Alfv{\'e}n's paradigm of magnetized outflows \citep{1947MNRAS.107..211A} introduced magnetic fields and waves as key contributors to additional heating and acceleration. This framework underpins the modern view of cool-star winds as the combined outcome of thermal expansion and magnetic driving.

These theoretical predictions were soon confirmed with the advent of the Space Age, which enabled the first in situ measurements of the solar wind between 1959 and 1962. Early observations were obtained by the Soviet Lunik and Venera missions, as well as the American Explorer 10 and Mariner 2 spacecraft. It was the latter satellite, which provided continuous solar wind measurements over several months, that revealed the presence of alternating streams of slow, dense and fast, tenuous flows---now identified as the slow and fast solar wind \citep[e.g.,][]{snyder1963}. It was also confirmed that, due to solar rotation, the interplanetary magnetic field carried away with the outflow of plasma assumes the shape of an Archimedean spiral, also known as the Parker spiral \citep[after][]{parker1958a}. Here, we first briefly review solar wind observations across the heliosphere (Section~\ref{subsubsec:solarwind_obs}) and then provide an overview of the space weather impact of the solar wind on planetary environments (Section~\ref{subsubsec:solarwind_swx}).

\subsubsection{Observing the Solar Wind in the Heliosphere} \label{subsubsec:solarwind_obs}

While the solar wind is most directly characterized via in-situ measurements, remote-sensing observations supply essential complementary information about its origin and large-scale structure, especially near the Sun. Low coronal observations at EUV and X-ray wavelengths provide insights on the morphology and properties of coronal holes \citep[known to be sources of the fast solar wind; e.g.,][]{cranmer2009} as well as the structure of the streamer belt and active region boundaries \citep[related to the slow solar wind; e.g.,][]{abbo2016}. White-light coronagraph observations further constrain the coronal context by revealing the structure and dynamics of the streamer belt at higher altitudes as well as by imaging the formation and/or release of small-scale density structures (``blobs'') believed to seed the slow solar wind variability found in the heliosphere \citep[e.g.,][]{sheeley2010, sanchez-diaz2017}. At larger elongations, heliospheric imagers provide a more ``direct'' view of denser solar wind structures such as slow--fast stream interaction regions (SIRs), which form when high-speed streams (HSSs) originating from coronal holes overtake the preceding solar wind ahead of them. Typically, SIRs do not produce observable signatures in coronagraph data but become sufficiently developed and compressed beyond the corona to be detectable in white-light imaging \citep[e.g.,][]{rouillard2008, conlon2015}.

Following the first measurements of the late 1950s--early 1960s, the solar wind has been sampled in situ by multiple interplanetary missions more or less continuously, albeit at widely separated locations throughout the heliosphere, resulting in an intrinsically sparse and nonuniform spatial coverage. One notable exception to this spatial intermittency is the Sun--Earth L1 point, where a series of spacecraft has enabled essentially continuous in-situ monitoring of the solar wind upstream of Earth since the mid-1990s, yielding a uniquely stable reference data set at a fixed heliocentric location. This has allowed the generation of catalogs and statistics on large-scale structures such as SIRs \citep[e.g.,][]{2006SoPh..239..337J} and smaller-scale properties such as solar wind turbulence \citep[e.g.,][]{2019JGRA..124.2406B} at 1~au. More sporadic measurements at other heliocentric distances have allowed investigations of the radial evolution of the solar wind, revealing how SIRs tend to steepen, broaden, and fully develop forward--reverse shock pairs as they propagate outward \citep[e.g.,][]{2018LRSP...15....1R}. Multi-point observations from spacecraft distributed between ${\sim}$0.3 and several au have shown that the SIR-associated compression regions identified near 1~au often evolve into fully developed corotating interaction regions (CIRs) beyond Earth's orbit, with increasingly pronounced forward and reverse shocks and enhanced energetic particle populations \citep[e.g.,][]{2019MNRAS.483.3730P, 2021A&A...649A..80G}. At larger radial distances, multiple SIRs/CIRs can merge, and the interaction may also include solar transient structures such as CMEs, giving rise to so-called merged interaction regions (MIRs)---believed to dominate the large-scale structure of the outer heliosphere \citep[e.g.,][]{1994JGR....9919341B}. 

A major advance in understanding the 3D structure of the solar wind was offered by the Ulysses mission, the first to venture with in-situ instrumentation out of the ecliptic plane to fly over the Sun's poles. These observations revealed that HSSs and their associated SIRs are are most prominent at high heliographic latitudes during solar minimum, when large polar coronal holes dominate the solar surface, whereas this structured flow is largely disrupted during solar maximum by the increased occurrence of transient CMEs \citep[e.g.,][]{2008GeoRL..3518103M}. Further outward, the outer heliosphere has been explored by the Pioneer 10/11 and Voyager 1/2 spacecraft as well as by New Horizons, which have provided valuable in-situ measurements of the solar wind at progressively larger heliocentric distances \citep[e.g.,][]{1994GeoRL..21.1743G, 2016ApJS..223...19E}. At the edge of the heliosphere, the twin Voyagers have provided the first in-situ measurements of the transition from supersonic solar wind to the heliosheath across the termination shock, and ultimately into interstellar space \citep[e.g.,][]{2013Sci...341.1489G, 2019NatAs...3.1019R}. In particular, Voyager 1 and 2 have revealed the highly dynamic and asymmetric nature of the heliospheric boundary regions, including pronounced temporal and spatial variability in magnetic field and plasma properties as well as energetic particle populations. These measurements have culminated in the first direct sampling of the local interstellar medium, marking the transition from solar-dominated plasma to the ambient galactic environment, and thereby completing the in-situ exploration of the heliosphere from its inner regions to its outermost boundary (see Section~\ref{subsec:lism} for further discussion on heliosphere/astrosphere interactions with the interstellar medium).

\subsubsection{Solar Wind Impact on Planetary Environments} \label{subsubsec:solarwind_swx}

The solar wind plays a fundamental role in shaping the plasma and magnetic environments of planets throughout the heliosphere. Persistent solar wind interactions govern the large-scale, long-term configuration of planetary magnetoplasma environments, from the confinement of intrinsic magnetospheres to the generation of induced magnetospheres and atmospheric escape. Superimposed on this steady forcing, shorter-timescale variations in the solar wind are able to drive space weather responses at planetary bodies. Hence, the primary factors modulating solar wind forcing on planetary environments are the presence of an atmosphere, the strength of intrinsic magnetic fields, as well as the size and structure of the resulting (full-fledged or induced) magnetospheres.

The long-term forcing of planetary and small-body environments is commonly classified according to the nature of the solar wind--body interaction \citep[e.g.,][]{2001P&SS...49.1005R, 2012EP&S...64...57B}. At the broadest level, these processes can be divided into (i) intrinsic or ``full-fledged'' magnetospheres generated by internally sustained planetary dynamos, (ii) induced magnetospheres formed through the interaction of the solar wind with conducting ionospheres or extended atmospheres, and (iii) localized or hybrid interactions associated with crustal magnetic fields, mini-magnetospheres, or weakly magnetized bodies. Among the planets, Mercury, Earth, Jupiter, Saturn, Uranus, and Neptune possess presently active internal dynamos capable of sustaining global magnetospheres. In these systems, the solar wind is primarily deflected by the planetary intrinsic magnetic fields, producing features such as bow shocks, magnetopauses, and extended magnetotails \citep[e.g.,][]{1992AREPS..20..289B, 2018SGeo...39..817B, 2022ScChD..65...25S}. In contrast, Venus and Mars lack global dipolar magnetic fields and as such their interaction with the solar wind results mainly in induced magnetospheres generated by their ionospheres and upper atmospheres \citep[e.g.,][]{2006SSRv..126...77B, 2008P&SS...56..796K}. Mars, however, also exhibits strong localized crustal magnetic fields especially in its southern hemisphere, giving rise to a hybrid structure that combines induced magnetospheric behavior with localized magnetic shielding and mini-magnetospheres \citep[e.g.,][]{2003JGRA..108.1424B, 2018GeoRL..45.4559D}.

Comets constitute another class of induced interaction whose properties depend strongly on the cometary activity phase, and therefore on heliocentric distance throughout the orbit \citep[e.g.,][]{2017RSPTA.37560256G}. During periods of strong cometary activity, interaction processes are dominated by mass loading of the solar wind, as neutral gas released from the coma becomes ionized and incorporated into the flow. Because the interaction region is large, newly generated ions are able to generate low-frequency plasma waves and turbulence, thus slowing and perturbing the solar wind and ultimately producing an induced obstacle around the comet. Small bodies and moons display a broader diversity of interaction regimes. Ganymede remains the only confirmed moon with an intrinsic internally generated magnetic field and a self-sustained magnetosphere \citep[e.g.,][]{1997GeoRL..24.2155K}. Other moons, instead, can exhibit induced and/or localized interactions analogous to those of Venus or Mars. For example, Luna possesses remanent crustal magnetic anomalies that generate localized mini-magnetospheres \citep[e.g.,][]{2010GeoRL..37.5103W}. In the case of the Galilean system, moons such as Europa and Callisto exhibit induced magnetic responses associated with conductive (salty) subsurface oceans---however, they are permanently embedded within Jupiter's magnetosphere and therefore interact primarily with this environment and the solar wind's influence can be considered more indirect. An analogous situation occurs in the Saturn system, where moons such as Titan are immersed (at least most of the time) in the Kronian magnetosphere, with plasma--atmosphere interactions governed by internal magnetospheric processes rather than direct solar wind forcing.
 
In addition to these long-term interaction regimes, planetary and small-body environments are also subject to short-term forcing associated with the large-scale structure and variability of the solar wind itself \citep[e.g.,][]{2020orep.book...19O}. In particular, HSSs and SIRs constitute major drivers of transient solar wind forcing throughout the heliosphere. As fast solar wind streams overtake slower ambient plasma, the compressed interaction regions that form can strongly perturb intrinsic and induced magnetospheres alike due to their enhanced magnetic fields, plasma densities, dynamic pressures, and turbulent fluctuations. In magnetized planets such as Earth and the giant planets, SIR/HSS forcing can trigger geomagnetic storms, magnetotail reconnection, and enhanced radiation belt dynamics \citep[e.g.,][]{2007LRSP....4....1P, 2009JGRA..114.5210C}, whereas at weakly magnetized or unmagnetized bodies such as Venus, Mars, comets, and many moons, compressed regions in the solar wind can substantially modify ion escape rates, magnetic field draping, induced currents, and plasma boundaries \citep[e.g.,][]{2017SSRv..212.1453F, 2017JGRA..12211320H}. Because SIRs and HSSs recur approximately with the solar rotation period (${\sim}27$~days) during intervals of persistent coronal holes, they provide an important source of quasi-periodic heliospheric forcing across the solar system \citep[e.g.,][]{1985JGR....90.1425S}.

\subsection{Wind Interaction with the Interstellar Medium} \label{subsec:lism}

The Sun and nearby stars reside within the Local Bubble, a region of very low density, which contains scattered molecular clouds, including the Local Leo Cold Cloud and warm, partially ionized clouds, extending to approximately 100~parsecs \citep{Zucker22}. These clouds are collectively known as the local interstellar medium (LISM) and mark the outer boundary of the influence exerted by solar and stellar magnetic fields. Observations made by the Hubble Space Telescope (HST) have identified distinct LISM absorbers \citep{redfield2015}, allowing measurements of their temperature \citep{redfield2004} and turbulent velocity \citep{frisch2011}, as well as detection of dust inhomogeneities and characteristics \citep{linsky2022,linsky2025}.

The wind--LISM interaction creates a protective plasma bubble---known as the heliosphere (for the Sun) and astrosphere (for other stars)---where the stellar wind pressure balances that of the surrounding medium \citep{wood2004}. From a hydrodynamic perspective, in the rest frame of a star moving uniformly through the surrounding interstellar medium (ISM), the ISM appears as a uniform flow whose velocity corresponds to the star's relative motion and is often supersonic. Since the solar wind is also supersonic and propagates radially outward, both the ISM and the wind experience a transition to subsonic velocities, resulting in a potential bow shock for the ISM and a termination shock for the solar wind. Because a merging of these two flows is impossible, the heliopause, a tangential discontinuity where pressure equilibrium is maintained, is formed.  

Both Voyager spacecraft, launched in 1977, by now have passed the outer boundary of our solar system providing unique insights into the shape and structure of the heliosphere. According to observations, Voyager~1 passed the termination shock at 94~au and the heliopause at 122~au, while Voyager~2 crossed both discontinuities at 84~au and 119~au, respectively \citep[e.g.,][]{stone2019}. These observations revealed a compression of the heliosphere in the Voyager~2 direction, which indicated that the ISM is not aligned with the ISM inflow vector, resulting in the heliosphere being asymmetric. In addition, no clear signatures of a bow shock were found, thus it is commonly accepted that the the heliosphere has a bow wave rather than a full-fledged shock. Thus, the propagation speed is no longer defined by the sound speed but rather by the fast-magnetosonic speed---numerically speaking, this necessitates the use of full 3D MHD modeling (in contrast to hydrodynamics). 

\subsubsection{Modeling the Heliosphere}

Numerical models of the heliosphere often make use of (multi-)fluid (Euler) equations---a collection of interrelated differential equations that streamline the Navier--Stokes equations. Assuming that the permeability $\mu$ is equal to 1, the conservative formulation of the integrated continuity, momentum, and energy equations, in conjunction with the Euler--Maxwell equations, is given as:

\begin{gather}
 \frac{\partial}{\partial t}\begin{bmatrix} \rho_j \\ \rho_j \vec{v}_{j} + P_1 \vec{F}_{{\mathrm{rad}}} \\ E_j + P_2 E_{\mathrm{rad}} \\ \vec{B} \end{bmatrix} + \vec{\nabla} \cdot 
 \begin{bmatrix} 
 \rho_j \vec{v}_{j} \\ 
 \rho_j \vec{v}_{j} \otimes \vec{v}_{j} + \left(p_j + \frac{B^2}{2}\right)\widetilde{I} + P_3\vec{F}_{{\mathrm{rad}}} - \vec{B}\otimes\vec{B}\\ 
 \left(E_j+p_j+\frac{B^2}{2}\right)\vec{v}_{j} + P_4 \vec{F}_{{\mathrm{rad}}} - \left(\vec{B}\cdot \vec{v}_{j}\right)\vec{B}\\ 
 \vec{v}_{j}\otimes\vec{B} - \vec{B}\otimes\vec{v}_{j}
  \end{bmatrix}
 = \nonumber\\
 \begin{bmatrix} 0 \\ \rho_j \vec{F} + \vec{\nabla}\cdot \hat{\sigma} - \vec{\nabla} p_{\mathrm{CR}} \\ \rho_j \vec{v}_{j}\cdot \vec{F} + \vec{\nabla}\cdot \left(\vec{v}_{j}\cdot \hat{\sigma}\right) - \vec{\nabla}\cdot \vec{Q} - R_{\mathrm{L}} - \vec{v}_{j}\cdot \vec{\nabla}p_{\mathrm{CR}}\\ 0 \end{bmatrix} + \begin{bmatrix}S_j^{\mathrm{c}} \\ \vec{S}_j^{\mathrm{m}} \\ S_j^{\mathrm{e}} \\ \vec{A}_{j}\end{bmatrix},
 \label{eq:MHD}
\end{gather}

\noindent with $j$ as the particle species, $\vec{v_j}$ its velocity, $\rho_j$ its mass density, and $p_j$ its pressure. The total energy is given by $E_j = e_j + \frac{1}{2} \rho_j v_j^2 + \frac{B^2}{2}$, where $e_j$ is the inner energy, $\widetilde{I}$ represents the unit tensor, $\hat{\sigma}$ the stress tensor, $\vec{F}$ the external force per unit mass and volume, $\vec{Q}$ gives the heat flow, $S_j^{x}$ the sources and sinks caused by charge exchange in the continuity equation ($x = \mathrm{c}$), energy equation ($x = \mathrm{e}$) and momentum equation ($x = \mathrm{m}$), while $\otimes$ represents the dyadic/tensorial product. $R_{\mathrm{L}}$ is a function that accounts for cooling effects, the subscript $\mathrm{rad}$ accounts for the radiation transport of momentum and energy coupling, and $\mathrm{CR}$ accounts for cosmic rays.  $P_1 - P_4$ are constants and $\vec{A}_j$ describes the ambipolar diffusion between neutrals and ions. Note that, in general, the ideal gas law ($p_j = (\gamma - 1) e_j$) is used instead of the heat flux, which is often assumed to be zero. Single- and multi-fluid MHD studies often are based on solving the ideal (M)HD equations where the following conditions must be met: $\rho \neq 0$, $\vec{v} \neq 0$, $E \neq 0$, $P_1 - P_4 = 0$, $R_L = S_j^x = \vec{F_{\mathrm{rad}}} = \vec{E}_{{\mathrm{rad}}} = 0$, and $\vec{B} = 0$ ($\vec{B} \neq 0$).
%

\begin{figure}[!t]
    \centering
    \includegraphics[width=\textwidth]{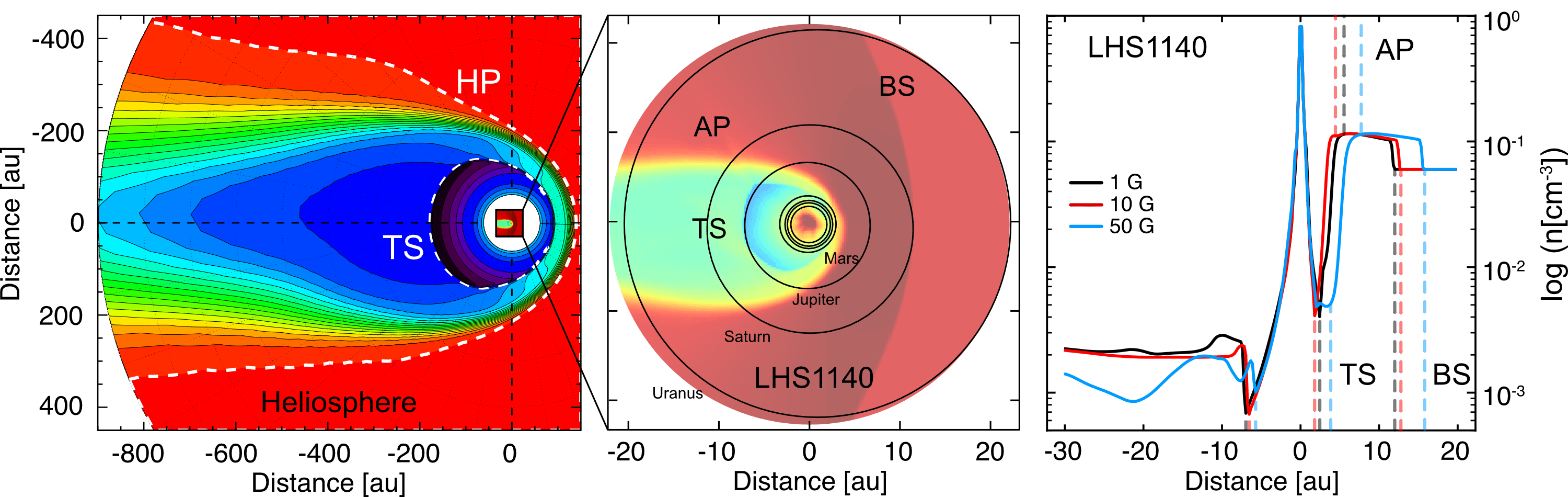}
    \caption{Left panel: Single-fluid 3D MHD model of the heliosphere including a direct size-comparison to the small astrosphere of the M dwarf LHS1140 . Middle panel: Zoom in to LHS1140 including the planetary orbits of our solar system planets. Right panel: TS, AP, and BS distances for varying stellar magnetic field strengths of LHS1140. Figures from \citet{HerbstEA22} (left panel) and \citet{Scherer-etal-2025} (middle and right panel).}
    \label{fig:HelioAstrosphere}
\end{figure}

Several heliospheric models exist in the literature. Often used are the Multi-Scale Fluid-Kinetic Simulation Suite \citep[MS-FLUKSS, e.g.][]{2009ASPC..406..149P, Pogorelov_2025}, the multi-fluid code CRONOS \citep[e.g.][]{Kissmann-etal-2018}, and a BATS-R-US-based multi-fluid model \citep[e.g.][]{Opher_2021}. While the different models are in agreement with the nose-ward Voyager observations toward the LISM, severe differences in shape and size of the tail region, varying from bullet-shaped \citep[e.g.,][]{PogorelovEA17, HerbstEA22} to croissant-like \citep{Opher_2021} structures. As an example, the left panel of Figure~\ref{fig:HelioAstrosphere} shows single-fluid 3D MHD results using CRONOS.
\subsubsection{Modeling Cool Star Astrospheres} \label{subsubsec:stellarwind_models}

Using the CRONOS code, the astrospheres of planet-hosting cool stars, particularly M dwarfs, have been recently modeled. Thereby, the set of relations given in Equation~\ref{eq:MHD} is integrated over time on a fixed 3D grid until a steady state is reached. According to the model attempt discussed in \citet{Scherer-etal-2020}, simulation runs were performed in a star-centred spherical grid with a resolution of $N_r \times N_{\theta} \times N_{\phi}$ cells. Besides the information on the stellar surroundings (i.e., ISM), the most crucial input parameters for modeling astrospheric environments are the stellar magnetic field strength, the stellar wind speed and density, the stellar mass loss rate, and the stellar magnetic field. However, for most cool stars, only little information on these parameters can be found in the literature. 

The first results based on the single-fluid 3D MHD version of CRONOS were discussed in \citet{HerbstEA20} and \citet{HerbstEA22} (i.e., V374 Peg, Prox Cen, and LHS1140), while the first multi-fluid 3D MHD model efforts for LHS1140 are presented in \citet{Scherer-etal-2025}. The middle and right panels of Figure~\ref{fig:HelioAstrosphere} give insight into the numerical modeling of the small-scale astrosphere of LHS1140 compared to the solar system (middle) and the TS, AP, and BS distances depending on different stellar magnetic field strength (right).

Charge exchange between ionized stellar winds and cold ISM hydrogen further generates hot atoms that accumulate at the astrospheric outer boundary, forming a detectable ``hydrogen wall'' seen as excess H\,I Lyman-$\alpha$ (Ly-$\alpha$) absorption in UV spectra of cool stars (see Section~\ref{sec:Ly-a_Astrospheres} for further details). Charge exchange between heavy ions of the stellar wind and the neutral ISM can also be detected via X-ray observations of stellar astrospheres (see Section~\ref{subsubsec:astroCX}).

\subsection{Quantifying Stellar Winds: Observations and Characterization}\label{sec:SWinds}

Today, the modern picture of winds in cool main-sequence stars combines Parker's thermal expansion with Alfv\'en’s magnetically mediated acceleration. This paradigm explains the dual nature of the solar wind (slow and fast components) and is expected to apply universally to other cool stars with outer convection zones and magnetized coronae. As mentioned in Section~\ref{subsec:solarwind}, in the solar case, in-situ spacecraft observations since the 1960s have directly confirmed the existence of a supersonic wind and revealed its complex structure \citep{1962Sci...138.1095N, 1998GeoRL..25....1M, 2021PhRvL.127y5101K}. By analogy, other cool stars are also expected to sustain winds governed by the same coupled processes of coronal heating, pressure-driven expansion, and magnetic acceleration.  




While we know physical characteristics, composition, dynamic, magnetic, and thermodynamic properties of the solar wind from in-situ measurements from the outermost coronal layers to the edge of the heliosphere (see Section~\ref{subsubsec:solarwind_obs}), magnetized winds of other cool main-sequence stars are much more difficult to characterize. This is not so much a consequence of the unavailability of observing techniques, but rather of the intrinsic faintness of wind signatures and the low densities of wind plasma. Nevertheless, over the past decades a multitude of detection methods have been developed that address very different physical properties of a wind and therefore open a window into the characterization of different types of wind and features in a wind.

Stellar wind detection methods can be generally classified into two main groups, apart from in-situ measurements in the solar system: direct and indirect \citep{guedel2020}. A direct method records signatures of emission, absorption, or scattering of electromagetic radiation due to the particles (electrons, ions, neutrals, and potentially dust) of the wind itself; the relevant physical process takes place at the location of these particles in the wind. In the case of scattering and absorption, external or background radiation may be involved. An indirect method instead measures properties or actions of features that are generated, influenced, or modified by the stellar wind but are not part of it. This opens the window to observations that are less challenging than direct observations; however, extensive theoretical modeling is required to infer physical wind properties from such measurements. Historically, indirect methods have prevailed and they have led to the most extensive evidence and systematics of solar-type winds. We therefore start our presentation with this group of methodologies. Similar presentations can be found in \citet{guedel2020} and \citet{vidotto2021}.

\subsubsection{Stellar Rotation, Angular Momentum Transport, and Spin-Down}\label{rotation}

Evidence for the presence of solar-like, magnetized winds around other stars emerged from the long-term evolution of stellar rotation inferred from large stellar samples. 
The basic principle is relatively straightforward. Because stellar winds are strongly ionized, their particles are effectively ``attached'' to the star's open magnetic field lines as they flow outward, a phenomenon known as the frozen-in condition. As long as the field lines rotate approximately rigidly with the star, the particle gains angular momentum with increasing distance, removing it from the star. As the particle reaches the Alfv\'en radius, the angular momentum of the particle at that distance is essentially lost from the stellar surface, contributing to the spin-down. The transition from the sub-Alf\'enic (inside the Alfv\'en radius) to the super-Alfv\'enic regime (outside) is gradual; the first comprehensive theory of angular momentum transport in a magnetized solar-type wind was presented by \citet{weber1967}. \citet{Kraft1967} suggested that spin-down is due to angular momentum loss in a magnetized wind, and subsequently the connection to stellar rotation and also stellar magnetic activity was developed in the seminal paper by \citet{skumanich1972}. 

Large samples of stellar rotation periods $P$ and ages were later presented e.g.\ by \citet{soderblom1993}, \citet{irwin2011}, and many others, in particular relying on open-cluster samples. The main observational features that a wind-angular momentum-rotation theory must be able to explain are: i) the general trend of decreasing rotation rates (longer rotation periods) toward higher stellar main-sequence ages although there is no simple correlation; ii) the large scatter of rotation periods even for a given stellar mass at a given stellar age, at least for younger stars; and iii) the observation that at ages of ${\sim}0.5$--2~Gyr (depending on stellar mass), rotation periods converge to typical values that are a function of age  for a given mass
(we exclude outliers here that may be related to angular momentum exchange with a companion such as a massive planet or a close stellar/substellar secondary).

The convergence of rotation periods is a consequence of a fundamental negative feedback loop between winds, stellar activity, and rotation. The stellar magnetic dynamo is rotationally driven, and the stellar wind is a consequence of stellar surface magnetic activity; angular momentum removal thus weakens the surface field (due to a weakened dynamo) and consequently the rate of angular momentum loss itself. 

\begin{figure}[t]
\begin{center}
\hbox{\hspace{-0.5cm}
\includegraphics[width=0.55\textwidth]{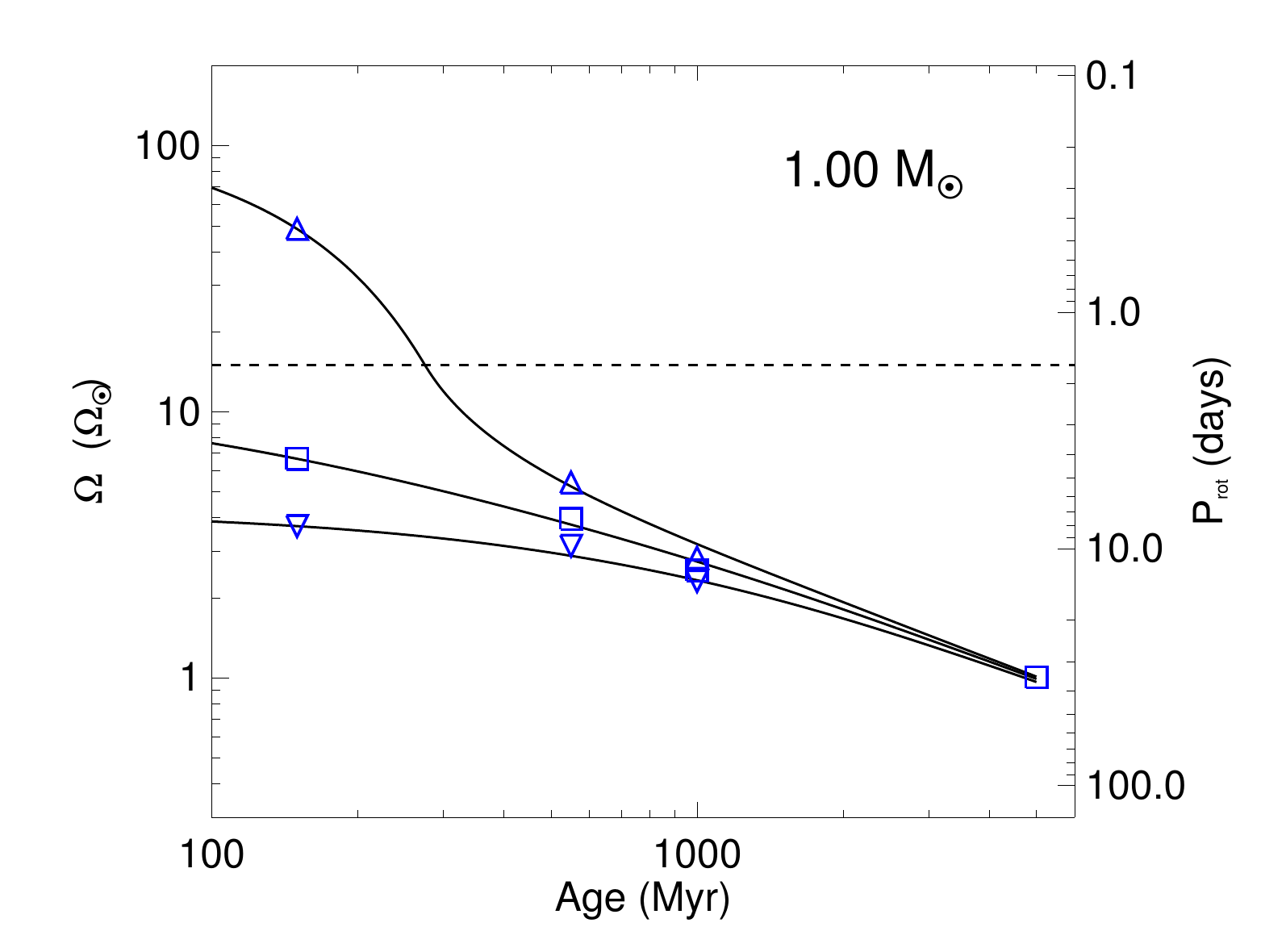} 
\hspace{-0.3cm}
\includegraphics[width=0.5\textwidth]{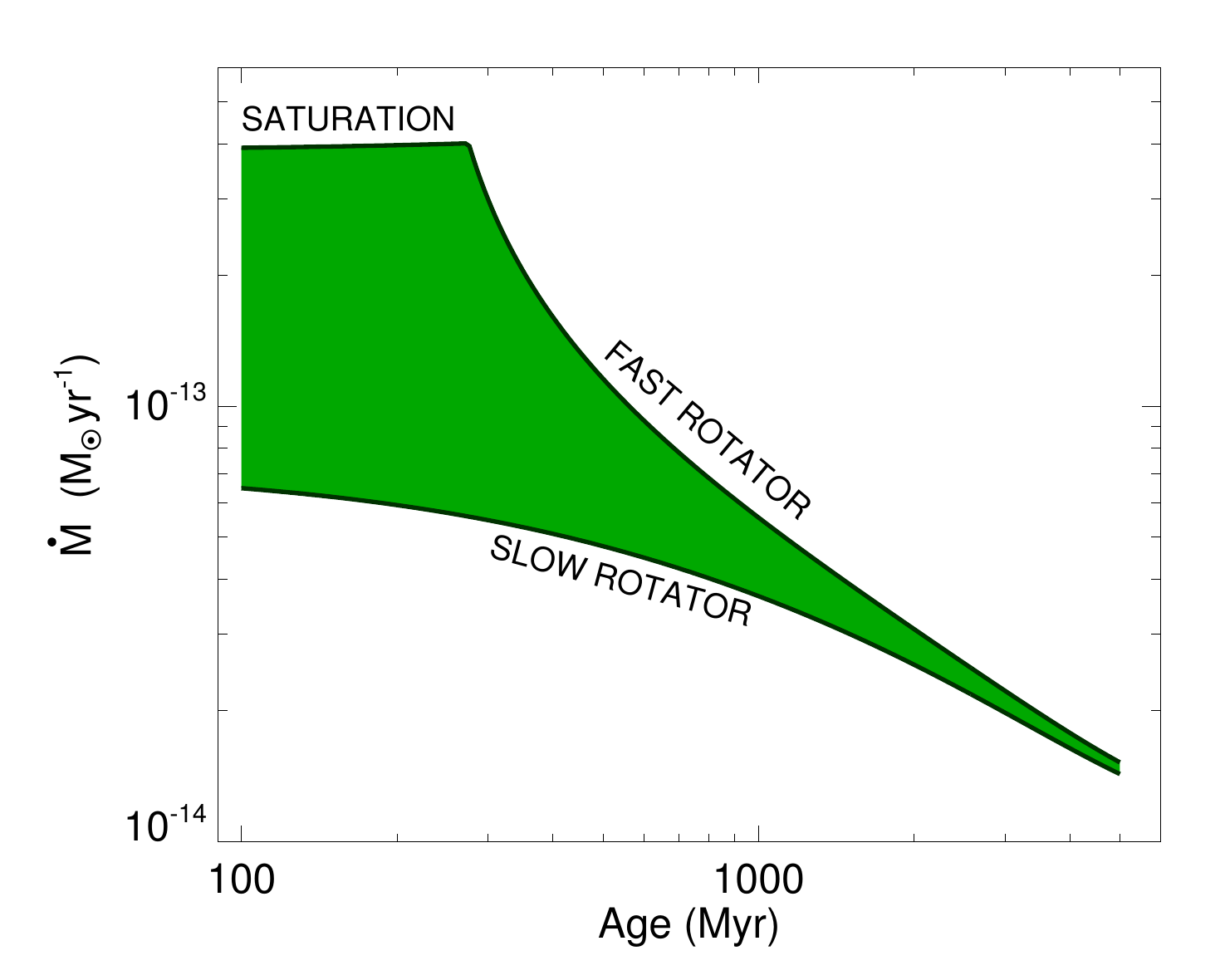} 
 }
\vspace*{-0.2 cm}
\caption{\textit{Left:} { Modeled rotational evolution of a 1$M_{\odot}$ main-sequence star at the 10th, 50th, and 90th percentiles of the observed, ``initial'' rotational distribution at 100~Myr, shown for ages of 100~Myr to 5~Gyr. The blue symbols mark the same percentiles from observational distributions at different ages. Above the dashed line, the wind mass-loss rate and the surface magnetic field saturate. -- 
\textit{Right}: The corresponding evolution of the wind mass-loss rate with age on the main-sequence. The green area 
includes all possible $\dot{M}$ evolutionary tracks for different initial rotation periods at the 10th
percentile (bottom edge) and the 90th percentile (top edge). The fastest rotators remain saturated
at a maximum $\dot{M}$ level during the first 300~Myr.
(From \citealt{johnstone2015a}.)}}
   \label{windloss}
\end{center}
\vskip -0.5cm
\end{figure}

A sufficient, approximate theory of the observed rotation period distributions (see, e.g., \citealt{gallet2013, johnstone2015a}) will therefore need to explain the rate of change of the rotation period $P$ as a function of stellar mass $M_*$, surface magnetic field $B$, and surface wind speed and acceleration. It will require a starting value $P(t=0)$ as an initial condition at a given initial age $t$. In the following, we prefer to use the stellar angular rotation rate $\Omega_* = 2\pi/P$. Further parameters involve the interior structure of the star and core-envelope decoupling. 
The fundamental equations for the problem include an equation for the wind torque $\tau$ acting on the star (we present the equations in \citealt{johnstone2015a} in what follows):
\begin{equation}\label{tau}
\tau  = f(B, \dot{M}, R_*, \Omega_*) \propto B^{0.87}\dot{M}^{0.56}R_*^{2.87}\Omega_*\, ,
\end{equation}
where $\dot{M}$ is the total mass loss rate through the wind and $R_*$ is the stellar radius. The approximation on the right-hand side is an example torque law from magnetohydrodynamic simulations \citep{Matt2012}. Given this expression, we can then formulate the spin-down rate as
\begin{equation}\label{torque}
\frac{{\rm d}\Omega_*}{{\rm d}t} = \frac{1}{I_*}\left(\tau - \frac{{\rm d}I_*}{{\rm d}t}\right)\Omega_*\, , 
\end{equation}
where $I_*$ is the moment of inertia of the star from stellar structure models.
Further equations needed to close the system are 
\begin{itemize}
   \item $B = B(\Omega_*, \tau) \propto (\Omega_*\tau)^{1.32}$, an empirical relation expressing the surface magnetic field strength $B$ as a function of  $\Omega_*$ and the theoretically known convective turnover time $\tau_*$ (see, e.g., \citealt{vidotto2014});
   a complication is added at mass-dependent high rotation rates of typically a few days: there, the magnetic field (and other expressions of magnetic activity) ``saturates'' at a characteristic value, i.e., stays constant if the rotation rate increases further;
   \item $\dot{M} = f(R_*, \Omega_*, M_*) = cR_*^2\Omega_*^aM_*^b$, an adopted mass-loss rate function; $a, b,$ and $c$ are constants to be determined from fits to observational samples.
   \end{itemize}
Solving these equations and fitting the free parameters (e.g., $a, b, c$) to observational samples, we find evolutionary wind mass-loss models given an initial stellar rotation rate. Such models can be extended to include the pre-main sequence stellar contraction phase where the star spins up due to angular momentum conservation. Example results are shown in Figure~\ref{windloss}.

\subsubsection{Ly\texorpdfstring{$\alpha$}{α} Absorption in Hydrogen Walls} \label{sec:Ly-a_Astrospheres}

If the ISM contains neutrals, then they may enter the heliosphere directly and subsequently undergo charge exchange with the solar wind ions. In a complicated process, a neutral but hot hydrogen wall builds up outside the heliopause \citep{zank1996, wood2000}. This so-called neutral hydrogen wall will sensitively absorb stellar Ly$\alpha$ radiation, which can be seen as excess absorption in the stellar Ly$\alpha$ profile whose center is usually already heavily absorbed by the neutral hydrogen in the ISM. 

Numerical model simulations show that the depth of the excess Ly$\alpha$ absorption is related to the stellar wind momentum rate. Through calibration with such simulations, it should therefore be possible to derive the stellar wind mass loss rate given reasonable assumptions for the wind velocity \citep{wood2004}. The method uses the profiles of the H\,{\sc i} and D\,{\sc i} lines at $\sim$1216~\AA, formed in the stellar chromosphere. The excess absorption by the hydrogen wall is found in the wings of the ISM-absorbed Ly$\alpha$ profile.
A closer analysis using relative motions between the star and our Sun shows that neutral hydrogen around the 
(solar) heliosphere produces excess absorption on the red side of the Ly$\alpha$ absorption profile \citep{gayley1997} while the (stellar) astrospheric hydrogen wall produces equivalent absorption on the blue side.

This method has been successful in collecting  a significant number of stellar wind mass-loss rate estimates \citep{wood2018}, but it strongly depends on the model calibration and requires interstellar neutrals to be present. Th absence of excess absorption may therefore not be an indication of the absence of winds but could indicate the absence of ISM neutrals. Generally, it is thought that owing to the presence of partially neutral ISM out to about 10~pc from the Sun, the method applies out to such distances but may fail beyond. Also, the wind speed needs to be assumed, and normally a speed equal to the solar wind speed is adopted. We note, however, that rapidly rotating star may eject winds at significantly higher speeds due to centrifugal acceleration \citep{holzwarth2007, johnstone2015a,Johnstone2017}. 

The success of the method is owed to the systematic trends seen in samples of stars at different activity levels. The latter are conveniently expressed by the stellar surface X-ray flux $F_{\rm X}$, while the wind mass-loss rate is similarly normalized to a unit surface area. Empirically, the hydrogen wall Ly$\alpha$ method indicates a correlation between $F_{\rm X}$ and the mass loss per unit surface area, $\dot{M}_{\rm s}$, of the form 
\begin{equation}\label{mdot}
\dot{M}_{\rm s} \propto F_{\rm X}^{1.34\pm 0.18}
\end{equation}
\citep{wood2005}. This relation seems to apply only to F, G, and K-type stars with activity levels up to $\approx$20$\times$ the solar $F_{\rm X}$. For stars at higher activity levels, the trend appears to break down, indicating much lower mass-loss rates. This is problematic, however, because such stars are subject to strong spin-down as statistics on rotation periods shows. A possibility is that the mass loss is more directional (e.g., along the stellar poles) and that the hydrogen wall may not often not affect the line of sight
toward the Sun.

Using relations for between rotational velocity and evolutionary age and between the former and $F_{\rm X}$,
we can derive
\begin{equation}\label{mdotage}
\dot{M} \propto t^{-2.33\pm 0.55}.
\end{equation}
but as already discussed in Section~\ref{rotation}, stellar spin-down is non-unique between $\sim 700$~Mys and $\approx 2$~Gyr, depending on stellar mass, therefore relation~(\ref{mdotage}) applies only past such ages. \citet{wood2021} also added M dwarfs to the picture, which however statistically make relation~\ref{mdot} considerably flatter and add much scatter to it; potentially other parameters are also relevant for such a relation.

\subsubsection{Wind--Planet Interactions}

Planetary atmospheres and magnetospheres react to stellar winds in various ways, for example in the form of mass loss that can be measured in various ways, or the generation of detectable electromagnetic radiation. In particular, a plethora of atmospheric mass-loss processes relate to stellar wind interactions, the so-called non-thermal loss mechanisms. We also note that stellar winds are the medium in which CMEs and high-energy particles propagate; these will be addressed below in Section~\ref{CMEs} and Section~\ref{particles}, respectively.
We will not describe these mechanisms in detail here because the review by \citet{Chapter3}, part of this collection, is devoted to star--planet interactions. Nevertheless, we briefly summarize a few approaches in which observations of star--planet interactions have been used
to derive stellar wind properties.

During a planetary transit, a bow shock forming between the mass-loss induced planetary ``wind'' and the stellar wind may become detectable. The shock itself depends on the unknown stellar wind temperature that also control the stellar wind velocity, while the pre-shock wind density follows from the observation of the absorption depth indicting the shock density. \citet{vidotto2010} and \citet{lai2010} observed a planetary transit in WASP-12 to infer that the wind temperature must be $<4.2\times 10^6$~K and the pre-shock wind density 
$n_{\rm w} \approx 1.5\times 10^{6}$~cm$^{-3}$. $\dot{M}$ could be determined if $T$ could be confined.

\begin{figure}
\begin{center}  
\includegraphics[width=0.9\textwidth]{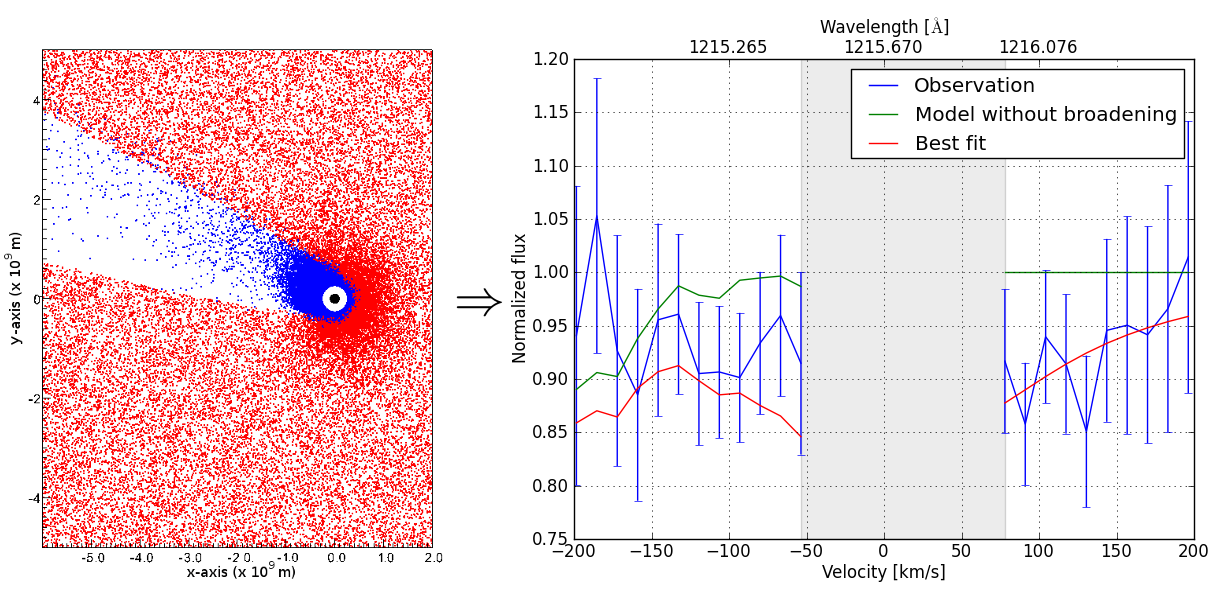} 
\caption{\textit{Left:} Slice of modeled 3D atomic H corona around HD 209458b. Blue and red dots  correspond to neutral H atoms and H ions, including stellar wind protons, respectively. The black dot represents the planet. The white area around the planet corresponds to the thermosphere below the inner simulation boundary. The star is on the right. The area dominated by the neutral H particles represented with the blue dots depends on the dynamic pressure of the stellar wind. \textit{Right:} Comparison of modeled \citep{kislyakova2014} and observed \citep{Ben-Jaffel10} Ly$\alpha$ spectra of HD~209458b mid-transit. Blue, the observed in-transit profile, normalized by the out-of-transit profile. Red, the modeled profile, computed including charge exchange, radiation pressure with self-shielding, photo- and electron impact ionization, and natural broadening.
From \citet{kislyakova2014}.}
\label{Lya_hd209458b}
\end{center}
\vskip -0.5cm
\end{figure}

Inflated planetary atmospheres may induce excess Ly$\alpha$ absorption during transits \citep{benjaffel2007}. A full model requires consideration of
i) Ly$\alpha$ radiation pressure, ii) natural line broadening, iii) line broadening due to the velocity distribution of particles along the line of sight, iv) photoionization, v) electron impact ionization, and vi) exospheric energetic charge exchange to form energetic neutral atoms \citep{kislyakova2014}. 
In the case of HD~209458 (F9V), \citet{kislyakova2014} derived, from the Ly$\alpha$ profile,  a wind velocity of 400~km~s$^{-1}$, a wind particle density of $5\times 10^{3}$~cm$^{-3}$; at the the orbital distance of 0.047~au, the mass-loss rate is
$\dot{M}= 4\times 10^{-14}~M_{\odot}$~yr$^{-1}$ for this F9~V star, or twice the solar $\dot{M}$. Figure~\ref{Lya_hd209458b} illustrates the method. The left panel shows the slice of the modeled 3D atomic hydrogen corona surrounding HD~209458b. The right panel shows the best fit of the Ly$\alpha$ transit observations by the Hubble Space Telescope (the red and green lines show the absorption calculated with and without spectral line broadening, respectively). The depth of in-transit absorption in the Ly$\alpha$ line depends on the size and density of the neutral cloud of atomic hydrogen surrounding the planet. Close-in giant planets are inflated due to their proximity to their host stars. Their upper atmospheres contain a large amount of atomic H due to photodissociation by the host star's short wavelength radiation, which leads to their Ly$\alpha$ transits being much deeper than the optical transits (e.g., \citealp{Ehrenreich15,Lavie17,ben-jaffeletal2022}). The extension of the atomic hydrogen corona around a planet is sensitive to stellar activity level and the temperature of the upper atmosphere, and is also regulated by the size of the planet's magnetosphere \citep{garciamunozetal2020}. The size of the magnetosphere is determined by the balance between the pressure of the planetary magnetic field and the dynamic pressure of the stellar wind \citep{Baumjohann12}. If the stellar wind is dense and fast, the planetary magnetosphere gets more compressed and the atomic hydrogen cloud around the planet gets more easily ionized by incoming stellar wind particles. Therefore, planets orbiting stars with lower mass loss rates tend to have deeper Ly$\alpha$ transits compared to planets orbiting stars with lower mass loss rates. In addition to stellar mass loss rates, one can draw conclusions about the magnetic field strength of the planet and the mass loss rate of the host star based on the Ly$\alpha$ transit observations of close-in gas giants \citep{kislyakova2014}. Although this method is indirect and some parameters are degenerate, it can still provide important insight into exoplanetary magnetism and stellar winds.

The Ly$\alpha$ profile was analogously analyzed for  HD~189733 (K2~V) using radiation pressure and charge exchange to find a wind temperature $T \approx 10^5$~K, a wind density $n=3\times 10^3$~cm$^{-3}$,
and a wind velocity $v = 190$~km~s$^{-1}$. At the know orbital radius (0.031~au) this implies $\dot{M} \approx 5\times 10^{-15}~M_{\odot}$~yr$^{-1}$ or 25\% of the present-day Sun's wind rate \citep{lecavelier2012}.

And the same method was applied to GJ~436 (M2.5V, with a warm Neptune) \citep{vidotto2017b}. Here, the authors found a wind temperature of $4.1\times 10^5$~K, a local wind velocity at 0.029~au of 85~km~s$^{-1}$ although the wind is still accelerating to reach a terminal velocity of 370~km~s$^{-1}$. Furthermore, the local proton density is $2\times 10^3$~cm$^{-3}$, implying $\dot{M} \approx 1.2_{-0.75}^{+1.3}\times 10^{-15}~M_{\odot}$~yr$^{-1}$, or 5.9\% of $\dot{M}_{\odot}$.

In a different type of application, \citet{2021MNRAS.504.1511K} studied the generation of Jupiter-like radio emission around planets orbiting in the stellar winds of Proxima Centauri (M5.5V) and AU Mic (M1V). For AU Mic, they estimate, depending on the wind mass-loss rate, that planet b could emit up to 10~mJy at radio wavelengths of 10~MH to 3~GHz. Such fluxes were observed from AU Mic indeed \citep{tristan2026}, although it remains to be seen whether they are purely coronal or have a planetary component. Such estimates would, vice versa, provide a pathway to estimate stellar wind mass loss.


\subsubsection{Slingshot Prominences}

Active, rapidly rotating stars occasionally reveal periodic absorption in H$\alpha$ ascribed to transiting prominences \citep{collier1989}. This relatively cool material is trapped in stellar magnetic fields far above the surface and corotates with the star. It is thought that prominences form in the apex region of high magnetic loops. 

\citet{jardine2019} developed a theoretical framework 
for these features that suggests that coronal plasma cools most efficiently in region of magnetic loops that reach the co-rotation radius. If this region is above the sonic point where the flow speed in the coronal loop equals the sound speed, then material accumulates there because hydrostatic equilibrium cannot be established anymore across the entire loop. Because the pressure increases, the loop will eventually break open, releasing the material that will (if located above the corotation radius) flow away from the star. If this material alone is taken as a stellar wind, its mass loss can be estimated from the H$\alpha$ absorption of transiting prominences and a geometric extrapolation to the entire stellar surface. 
One then finds 
\begin{equation}
{\dot M} = \frac{4\pi R_*^2}{2A_0} \dot{m_{\rm p}} \approx 100 \frac{m_{\rm p}}{\tau}\, ,
\end{equation}
where $A_0$ is the footpoint area of one loop with an observed prominence, and $\dot{m}_{\rm p}$ is the rate of chance of the mass $m_{\rm p}$ in the prominence, depending on its lifetime $\tau$. The assumption here is that the ``wind'' has its origin in flows inside magnetic loops that break open, and that the entire stellar surface is covered with loops containing prominences of this type. This methodology works only for very active, rapidly rotating stars. \citet{jardine2019} find up to 3000 times the solar wind mass-loss rate per unit surface area in some objects with X-ray surface fluxes of order 100 times the solar value and more (Figure~\ref{windplot}--left).
 
\subsubsection{Wind Contamination of White Dwarfs}

Stellar winds from companions to white dwarfs (WD) may contaminate the latters' surfaces with heavy elements. Without contamination, WD surfaces are supposed to be metal-poor because metals should settle to layers below the photosphere in a matter of less than a million years.
But metals have been found on WDs with companions \citep{debes2006}. Companion winds can accrete onto the WD via Bondi-Hoyle accretion with an accretion rate of $\dot{M}_{\rm acc}$ which is a function of the density and the relative velocity $v$  (including orbital motion) between wind and the WD at a distance of $R$ from the companion, and the WD mass $M_{\rm WD}$ (\citealt{bondi1944}; see \citealt{debes2006}). The metallicity will settle where accretion and diffusion into the WD balance. Using the observed [Ca/H] abundance in the WD, one finds for the wind mass-loss rate of the companion
\begin{equation}
\dot{M} = \frac{q}{\tau} 
\frac{{\rm {[Ca/H]\phantom{\odot}}}}
{\rm [Ca/H]_{\odot}}\frac{R^2v^4}{G^2M_{\rm WD}}\ .
\end{equation}
Here, $G$ is the gravitational constant, and $v$ is usually dominated by the wind velocity which itself is of order the escape speed from the companion star. The constants $q$ and $\tau$ are, respectively,
the fractional mass of Ca at the base of the convection zone and the  $e$-folding time scale for Ca to settle out of the atmosphere, respectively. \citet{debes2006} derived order-of-magnitude mass loss  rates of  $\dot{M} \approx 10^{-16} - 6\times 10^{-15}~M_{\odot}$~yr$^{-1}$ for mid
M dwarfs with WDs in close orbit ($<$0.015~au). \citet{parsons2012} reported even smaller $\dot{M}$.

\subsubsection{Wind Evidence from Debris Disk Dynamics}
Debris disks are made of leftover material from the planet formation process after protoplanetary disks have dispersed.The consist of dust particles in a distribution of sizes, and in cases there is also a thin gas component that may be leftover primordial gas or be released after collisions of particles. In analogy to radiation pressure due to stellar luminosity, dust grains are also subject to momentum transfer from stellar-wind particles. Dust particles produced during  collisions of larger bodies can therefore be pushed to larger orbital distances or even be ``blown out'' of the system. Collisions between wind and dust particles also generate a ``wind drag'' that is analogous to the Poynting-Robertson (PR) drag \citep{plavchan2005}. It originates from an ``aberration'' effect that, due to the orbital velocity of a dust particle, causes wind ions and stellar radiation to hit the dust particle not precisely radially but at a small angle toward the direction of motion. These effects lead to a deceleration of the dust particle and therefore inward motion.

With a sufficiently detailed theory in hand considering PR and wind drag as well as radiation pressure and wind momentum transfer, the particle size distribution and the radial brightness profile of a debris disk can therefore be used to estimate the stellar wind momentum less rate \citep{plavchan2005, augereau2006}. \citet{strubbe2006} constrained the wind mass-loss rate of the early M dwarf AU Mic to $\dot{M} \lapprox 10\dot{M}_{\odot}$,
while \citet{schueppler2015} find a best-fitting 
$\dot{M} \approx 50\dot{M}_{\odot}$.

\centerline{---------------------}

\noindent We now turn to direct methods, in which the wind itself is observed through its emission and absorption mechanisms.

\subsubsection{Radio Bremsstrahlung from Ionized Winds}\label{radiobrems}

The solar wind is an essentially fully ionized plasma and is therefore subject to bremsstrahlung in  wavelength ranges from radio to X-rays. X-ray bremsstrahlung is a fraction of the wind's X-ray emission but not all of it; see Section~\ref{xrays} for more details. 

The theory for radio bremsstrahlung of a wind boils down to (approximately) deducing the total radio flux from a spherical bubble of plasma based on radiative transport toward an observer at infinity (see, e.g., \citealt{2021MNRAS.504.1511K}). The theory was developed by \citet{panagia1975}, \citet{wright1975}, and \citet{olnon1975}. The emission itself is proportional to the electron density $n_{\rm e}$ squared (as collisions are required) and the square root of the electron temperature, $T_{\rm e}$.

The spectral shape for a wind that become optically thick at some geometrical depth above the stellar surface is given by (in flux units of milliJansky)
\begin{equation}
 S_{\nu} = 9 \times 10^{10} \ \left(\frac{\dot{M}}{v}\right)^{4/3} T_{\rm e}^{0.1} \ \nu^{0.6} \ d^{-2} \ \rm mJy \, ,	
	\label{eq:fluxiso}
\end{equation}
where $\nu$ is the frequency in Hz, $v$ is the wind velocity in km~s$^{-1}$, $\dot{M}$ is the wind mass loss rate in units of $M_{\odot}$~yr$^{-1}$, and $d$ is the distance to the star in units of pc. If the wind remains \textit{optically thin} througout, the spectral flux becomes 
\begin{equation}
	S_{\nu} =  5\times 10^{39}\left(\frac{\dot{M}}{v}\right)^{2}T_{\rm r}^{-0.35}R_*^{-1}
	\nu^{-0.1}d^{-2}\ \rm mJy\, ,  \label{eq:fluxthin}
\end{equation}
where $R_*$ is the stellar radius. We thus see that in the optically thick regime (typically at longer radio wavelengths), the spectrum is rising with $\nu$ with a slope of +0.6, while in the optically thin regime at shorter wavelengths, it is nearly flat, with a power-law index of $\sim -0.1$. These two equations are valued for an isothermal wind with a constant expansion velocity. If this is not the case, then numerical integration is needed. Non-spherical winds were treated in \citet{reynolds1986}.

Many searches have been conducted at centimeter and millimeter radio wavelengths but only upper limits have been deduced \citep{doyle1991, mullan1992, drake1993}. \citet{lim1996a}, \citet{lim1996b}, and \citet{vandenoord1997} derived upper limits for M-dwarf winds as low as 10--100 times the solar wind mass-loss rate, or specifically
$\dot{M} < 7\times 10^{-12} M_{\odot}$~yr$^{-1}$ for Proxima Centauri
for a wind velocity of 300~km~s$^{-1}$. Both \citet{gaidos2000} and \citet{fichtinger2017}  studied different evolutionary stages of solar analogs to derive potential evolutionary trends.  Although these studies still resulted in upper limits, they were sensitive enough to exclude a significantly higher total mass of the young Sun as an explanation of the \textit{Faint Young Sun Paradox} \citep{fichtinger2017}, i.e.\ the observation that the early climate on Earth was mild even though the Sun's luminosity in the first Gyr of its main-sequence life was as low as 70--75\% of the present luminosity. The limit of this method comes when the flux level is lower than that of the ubiquitous chromospheric radio emission from the full stellar disk, a level that was reached in observations by \citet{drake1993} of Procyon and later
in further Sun-like stars \citep{villadsen2014}.

\subsubsection{Free-Free Optical Depth of Winds at Radio Wavelengths}\label{sec:Wind_Radio2}

In continuation of the bremsstrahlung radiation theory in SubSection~\ref{radiobrems}, we can also study the effect of free-free absorption of background radiation. Radio emission emitted for example by coronal flares or close-in exoplanets can be continuously absorbed by the wind free-free opacity. If the wind is optically thick along the line of sight from a radio source to the observer, then we can estimate its density and, given a wind velocity, the mass-loss rate. In a spherically symmetric, stationary wind
with constant velocity, optical depth unity along a radial line of sight between the observer and the stellar center
is reached at a radius of
\begin{equation}\label{windsurface}
R_{\mathrm{thick}} = 8\times 10^{28}\left(\frac{\dot{M}}{v}\right)^{2/3} T^{-0.45}\nu^{-0.7}\ \ \rm cm
\end{equation}
(e.g., \citealt{lim1996a}; $T$ is in K, $\nu$ in Hz, $v$ in km~s$^{-1}$, and $\dot{M}$ in $M_{\odot}$~yr$^{-1}$). If one knows that  radio emission should be present but remains undetected, then solving this equation for $\dot{M}$ indicates lower limits for $\dot{M}$. If radio emission \textit{is} seen, then one obtains an upper limit for $\dot{M}$. The method works best for low radio frequencies. This method was used by \citet{lim1996a} to derive $\dot{M} \lapprox 5\times 10^{-14} - 10^{-12} M_{\odot}$~yr$^{-1}$ assuming
$v = 300-600$~km~s$^{-1}$ and wind temperatures of $10^4-10^6$~K based on flare observations of the M dwarf YZ CMi.

Because the free-free absorption coefficient increases at lower frequencies as $\nu^{-2}$, meter-wavelength observations could, in principle, provide better constraints. Current radio telescopes do not have the sensitivity to detect even optically thick free-free emission from coronae. For instance, the Rayleigh-Jeans flux density of an optically thick corona at temperature of $T=10^6\,{\rm K}$ and size of $R_\odot$ at a distance of $d = 10\,{\rm pc}$ is $\approx 5\,(\nu/100\,{\rm MHz})^2\,$nanoJansky --- a depth that is unattainable with current telescope. However, around $0.5\%$ of stars within $d\approx 50\,{\rm pc}$ are detected at milliJansky-level flux densities around 150\,MHz \citep{callingham2021}. These emissions are nearly all coherent in nature. The lack of catastrophic free-free absorption of the coherent emission could be used to place upper limits on the wind mass-loss rate assuming some model for the radial evolution of wind temperature and density. The mass loss upper limits thus derived for stars detected in the LOFAR Two Metre Sky Survey at 150\,MHz are comparable (within an order of magnitude) to that obtained using astrospheric Ly$\alpha$ observations \citep{bloot2025}. The advantages of this method is that, unlike the astrospheric method, it does not require intimate knowledge of the interstellar conditions. On the other hand, it does ideally requite the magnetic field strength of the stars to be known so that the emitter radial distance can be determined and its mass loss constraints are quite sensitive to the assumed density structure in the wind acceleration region (where most of the free-free opacity originates).

\subsubsection{Propagation of Radio Waves}\label{sec:Wind_Radio3}
For radio waves to propagate through a plasma, they must exceed the local plasma frequency along their path. Observing a radio source through a stellar wind requires the plasma frequency to be lower than the wave frequency along the entire line of sight.
The radio source could be near the stellar surface (e.g., the chromosphere or coronal flares, or radio-emitting mass ejections), or planets emitting radio waves. Using this requirement for a stationary, isotropic wind at constant velocity $v$ (expressed in kilometers per second: $v_{\rm km/s}$), one can derive an upper limit to the wind mass-loss rate $\dot{M}$ if the observed radio emission is emitted at a radius $r$ (in cm) in front of the star relative to the observer \citep{guedel2020}: 
\begin{equation}
\dot{M} < \frac{r^2 m_{\rm e} m_{\rm i} v \nu^2}{1.09 e^2} 
        \approx 1.2\times 10^{-53}r_{\rm cm}^2 v_{\rm km/s} \nu_{\rm Hz}^2~[M_{\odot}~{\rm yr}^{-1}]\, , \label{propagation}
\end{equation}
where $e$ is the unit electric charge, $m_{\rm e}$ is the mass of an electron, and $m_{\rm i}$ is the mass of an (average) ion $\approx 1.25m_p$, $m_p$ being the mass of a proton. An analogous methodology was applied in \citet{vidotto2017} for potential radio emission of a young Jupiter-mass planet orbiting its star at 0.057~au. A reasonable estimate of the radio frequency for planetary radio emission is the gyrofrequency, which requires knowledge of the magnetic field strength in the planetary magnetosphere. For $B = 10$~G, the authors set a limit of $\dot{M} \lapprox 10^{-10}$~M$_{\odot}$~yr$^{-1}$.
In a similar way, the fact that low-frequency stellar radio flares are observed
can be used to set limits on $\dot{M}$. 

The method outlined above sets an upper limit to the upper limit of $\dot{M}$ given in Equation~(\ref{propagation}). 
Because stationary winds are only accelerating close to the star, their densities in the regions close to the star are higher than if $v$ were constant, requiring lower $\dot{M}$ for propagation through the wind; this also applies if a radio planet orbits in the wind acceleration region.
Also, if the radio source is not in front of the star, the maximum density needs to be found along the line of sight, which may require lower $\dot{M}$ for radio wave propagation to the observer.

\subsubsection{X-Ray Emission from Winds}\label{xrays}

A coronal plasma at temperature around 1~million K (MK) and higher emits a combination of line radiation and bremsstrahlung (apart from some free-bound emission). A stellar wind is roughly isotropic; if the wind velocity is constant, then the density decreases with $d^{-2}$ where $d$ is the distance from the stellar center. If we assume a \textit{coronal approximation} for the innermost stellar wind that dominates the total luminosity (e.g., gas is in collisional equilibrium, which is not correct in the extended wind), then we can easily integrate the emission across the wind because the wind can be assumed to be optically thin to X-rays, in contrast to the situation at radio wavelengths (SubSection~\ref{radiobrems}). In this case, one obtains for the mass-loss rate \citep{guedel2020}
\begin{align}
\dot{M} &\approx 7.1\times 10^{-13}\left(L_{\rm X}R_* \right)^{1/2}v~\textrm{[g~s}^{-1}]\nonumber\\
&\approx 3.75\times 10^{-12}\left(\frac{L_{\rm X}}{10^{27}~{\rm erg/s}}\right)^{1/2}
                          \left(\frac{v}{400~{\rm km/s}}\right)
                          \left(\frac{R_*}{7\times 10^{10}~{\rm cm}}\right)^{1/2}
                          M_{\odot}~{\rm yr^{-1}}\, .
\label{mdotLx}
\end{align}
However, in the inner wind zone, the wind continuously accelerates from sub-Alfv\'enic to super-Alfv\'enic velocities, which implies a modified density profile. This was discussed further in \citet{vidotto2021} using some approximations, namely X-ray emission limited to bremsstrahlung (although in fact lines dominate at typically expected wind temperatures), and adopting a hydrostatic density structure below the sonic point, and a wind plasma that is isothermal. The approximate solution then is a rough estimate and in fact a lower limit and can be found as follows:
\begin{align}
    x &= \frac{r}{R} \, , \\
    \mathrm{EM} &\approx \pi R^3  n_0^2 \int_1^2 e^{2R(1/x - 1)/H}x^2 \textrm{d}x \, , \\
    L_{\rm X} &\approx 1.4\times 10^{-27}T^{1/2}\mathrm{EM}\left(e^{-h\nu_1/kT} - e^{-h\nu_2/kT}\right) \, ,
\end{align}
where $r$ is the distance from the stellar center, $R$ is the stellar radius, EM is the total emission measure, $k$ is the Boltzmann constant, and $[\nu_{1},\nu_{2}]$ is the X-ray radiation frequency range considered for observations. This relates $L_{\rm X}$ to the emission measure of the wind which depends on density and volume. To obtain the mass-loss rate, one needs a wind velocity and density profile, for example given by a standard Parker wind profile for a non-magnetic wind \citepads{parker1958a}.  Using a solar-wind temperature of $1.5\times 10^6$~K and a base density of $n_0 = 10^8$~cm$^{-3}$, one finds \citep{vidotto2021} $\mathrm{EM} = 10^{48}$~cm$^{-3}$ and a 0.2--10~keV luminosity of order $10^{-10}L_{\odot}$ while the observed solar X-ray luminosity is of order $\sim (10^{-7}-10^{-6})L_{\odot}$. This demonstrates that we expect stellar wind X-ray luminosities to be orders of magnitude lower than the coronal X-ray luminosity, in particular also because coronal X-ray luminosity and the mass-loss rate in winds are expected to be correlated (see Section~\ref{sec:Ly-a_Astrospheres}).

The severe limitation of this method is thus given by the usually much larger $L_{\rm X}$ from the stellar corona. 
Typically, therefore, $L_{\rm X}$ measurements will provide upper limits to the wind mass loss rate. \citet{lim1996a} mention an estimate for the mid M dwarf YZ CMi of $\dot{M} \lapprox 6\times 10^{-11} M_{\odot}$~yr$^{-1}$.

\subsubsection{Astrospheric Charge Exchange}
\label{subsubsec:astroCX}

\begin{figure}
\begin{center}
\hbox{
\includegraphics[width=0.40\textwidth]{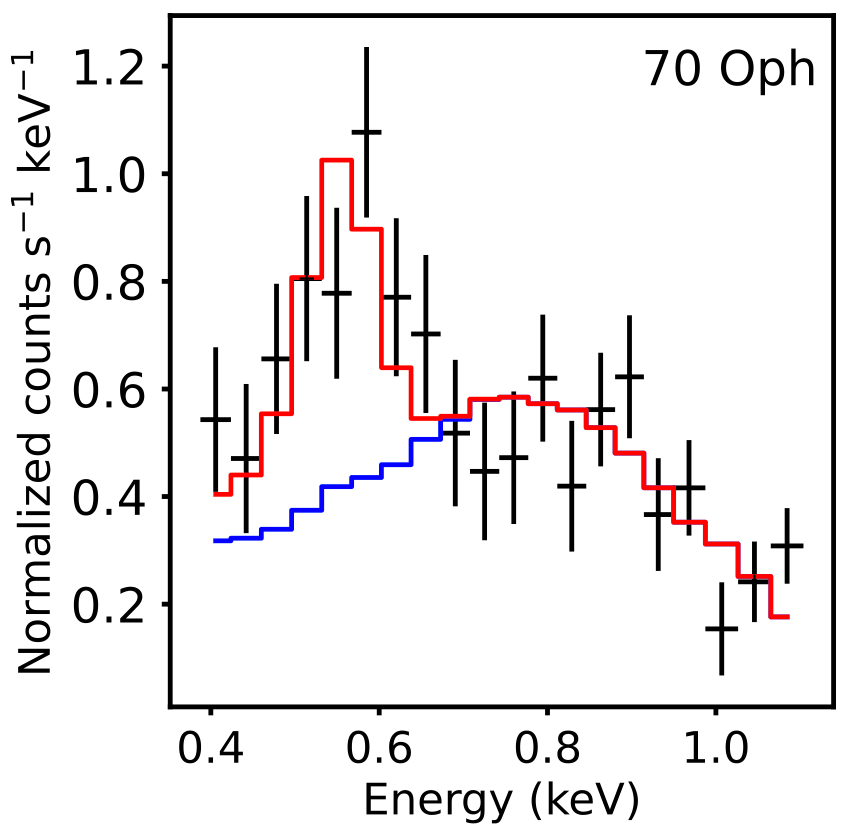} 
\hskip 0.7cm\includegraphics[width=0.54\textwidth]{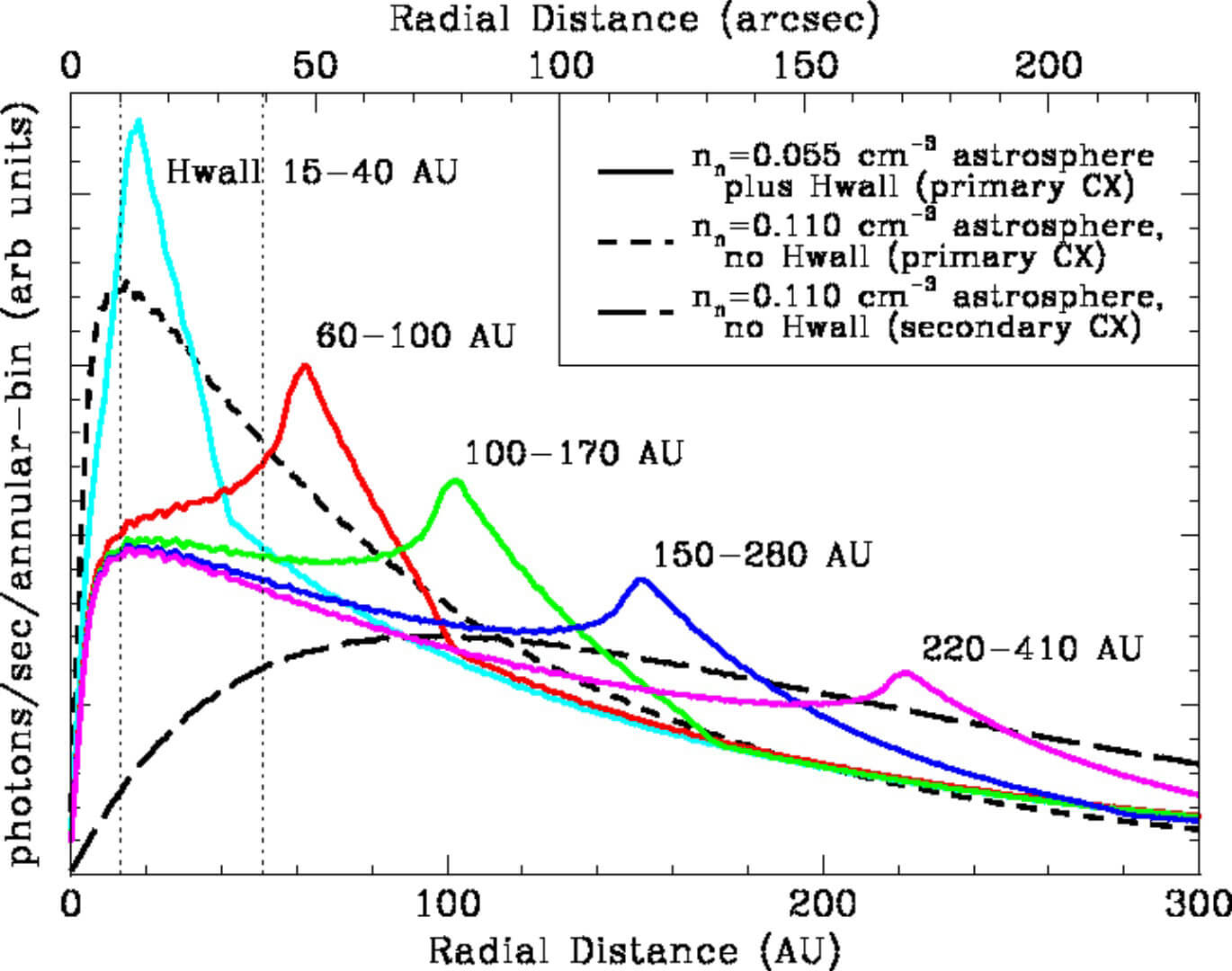} 
}
\caption{\textit{Left:} Stellar-wind charge exchange signal in the 0.56~keV oxygen K$\alpha$ triplet. The black crosses show the observed spectrum (from the \textit{XMM-Newton} PN camera), the blue solid lines show the scaled-down stellar models extracted from the stellar point source on the detector, and the red solid lines shown the same model with an added Gaussian line at 0.56 keV.
From \citet{kislyakova2024}.
-- \textit{Right:} Annular X-ray CX photon rates from the astrosphere of Proxima Centauri simulated for Proxima Centauri observed with the Chandra ACIS CCD detector. Solid lines: models with a hemispherical hydrogen wall with twice the ISM neutral density, and half the ISM neutral density inside the astrosphere. Short-dashed line: model with no hydrogen wall, adopting a neutral H density equal to the ISM. Long-dashed line: also without H walls but for secondary CX emission. From \citet{wargelin2002}. }
\label{CXmodels}
\end{center}
\vskip -0.5cm
\end{figure}
Charge exchange (CX) can occur between solar/stellar wind ions and neutral penetrating into the wind region from the interstellar medium \citep{cox1998}. In He-like ions CX populates at high $n$, followed by cascades of decays; in X-rays, most of the CX emission comes from $2\rightarrow 1$ transitions \citep{wargelin2001}. An obvious condition for such line emission to be observable is that the astrosphere is embedded in a partly neutral insterstellar medium.

\citet{wargelin2001} modeled the spatial distribution of neutrals inside an astrosphere (Figure~\ref{CXmodels}-right) and calculated charge exchange reactions of O$^{7+}$ and O$^{8+}$ to O$^{6+}$ and subsequent emission O~\,{\sc vii} CX line emission in the X-ray range (Figure~\ref{CXmodels}-left). The model parameters are the wind velocity $v$, the neutral density outside the astrosphere, adopting an exponentially decreasing neutral density toward the star on a length scale depending on $\dot{M}$;  also, a reasonable assumption for the fraction of O$^{7+}$ and O$^{8+}$ ions injected into the wind per second has to be made. The emissivity of CX can then easily be calculated from the ion density, the neutral density, the wind velocity, and the CX cross section. Observing a CX line flux can then be used to infer the density of O$^{7+}$ and O$^{8+}$ ions and by implication the wind density and therefore $\dot{M}$. A reasonable estimate must be made for the emitting astrospheric size and therefore its volume.
The total X-ray line luminosity of a spherical shell of radius $r$ around the star then is
\begin{equation}
\Delta L_{\rm X} = 4\pi r^2\Delta r\,n_{\rm H}n_{\rm i}v_{\rm w}\sigma_{\rm CX}\epsilon_{\rm CX}\, , \label{CXshell}
\end{equation}
where $n_{\rm H}$ is the neutral H density in that shell, $n_{\rm i}$ is the wind ion density at the same location, $\Delta r$ is radial width of the shell, $v_{\rm w}$ is the wind velocity (assuming that the neutrals are slow relative to the star), $\sigma_{\rm CX}$ is the CX cross section ($4\times 10^{-15}$~cm$^2$ for O~\,{\sc vii}, \citealt{koutroumpa2007}), and $\epsilon_{\rm CX}$ is the photon energy for the CX line, i.e., 0.56~keV $\approx 9\times 10^{-10}$~erg in the case of O~\,{\sc vii} emission.
Because the wind mass-loss rate for a spherical, stationary wind is $\dot{M} = 4\pi r^2 v_{\rm w}m_{\rm i}n_{\rm i}/f$ with $m_{\rm i}$ being the mean mass of a wind ion and $f$ the fractional abundance of O$^{7+}$ and O$^{8+}$ ions in the wind ($1.272\times 10^{-4} + 4.77\times 10^{-5}$ for the solar wind, \citealt{wargelin2001}),
we can invert Equation~(\ref{CXshell}) to find
\begin{equation}
\dot{M} = \frac{\Delta L_{\rm X}}{\Delta r}\frac{m_{\rm i}}{f n_{\rm H}\sigma_{\rm CX}\epsilon_{\rm CX}}\, .
\end{equation}
\citet{wargelin2002} applied this method to a deep X-ray observation of Proxima Centauri and could report a meaningful upper limit of $\dot{M} \lapprox 3\times 10^{-13}~M_{\odot}$~yr$^{-1}$. Using the \textit{XMM-Newton} CCD cameras, \citet{kislyakova2024} succeeded with detections of CX lines from around three nearby cool stars plus sensitive upper limits for several more (Figure~\ref{windplot}-left). Like for the hydrogen wall method, CX measurements reveal a correlation between $\dot{M}$ and stellar activity. \citet{kislyakova2024} have searched for the CX signal in a wide annulus surrounding several MS stars, and have detected X-ray emission likely originating from charge exchange from three of their targets. Similar to the hydrogen wall method, estimating stellar mass-loss rates based on the observed CX emission can be challenging. The interpretation of the results and the mass-loss rate estimates depend on the preliminary knowledge of the properties of the ISM surrounding the stars. A star with a high mass loss rate might produce no detectable astropheric signal, if the neutral ISM density around it is low. However, when an astrospheric signal is present, this method provides important information not only about stellar mass loss rates, but also potentially about the presence of specific highly ionized heavy elements. Recently, \citet{Lisse2026} has also successfully used this method to detect the CX halo around the young star HD~61005 with \textit{Chandra}. 


\subsubsection{Summary of Results from Wind Measurements}\label{sec:Wind_Overview}

In Table~\ref{windmethods}, we summarize the methodologies described above, their classification as direct or indirect, and caveats; the latter include limitations, contamination in data from other sources, the dependence on models, or requirements for successful applications.
\begin{table}
  \begin{center}
  \caption{Summary of methods for wind mass-loss rate measurements}\label{windmethods}
  \begin{tabular}{lll}\hline
   Method                          &           Type            & Caveats, requirements, limitations                                      \\
                                   &                           &                                              \\
   \hline               
   Spin-down                       &          indirect         & torque formula                               \\
   H walls, Ly$\alpha$ absorption  &          indirect         & hydrodynamic model, presence of neutral ISM  \\
   Astrospheric CX (X-rays)        &          direct           & presence of neutral ISM                                  \\
   X-ray emission                  &          direct           & assumed wind $T$; contaminating stellar X-rays           \\
   Radio bremsstrahlung            &          direct           & other radio emission; wind $T$ and velocity     \\
   Radio free-free absorption      &          direct           & presence coronal emission; wind $T$ and velocity      \\
   Radio wave propagation          &          direct           & planetary magnetic field strength and radio emission       \\
   Planet. Ly$\alpha$/radio suppression &          indirect         & complex models, planetary atmosphere                  \\
   Slingshot prominences           &          indirect         & model assumptions                           \\
   WD metal contamination          &          indirect         & orbital separation, wind velocity, diffusion parameters \\
   Debris disk dynamics            &          indirect         & presence of debris disk \\
   \hline
  \end{tabular}
 \end{center}
\end{table}

\begin{figure}
\begin{center}
\hbox{\hspace{-0.2cm}
\includegraphics[width=0.5505\textwidth]{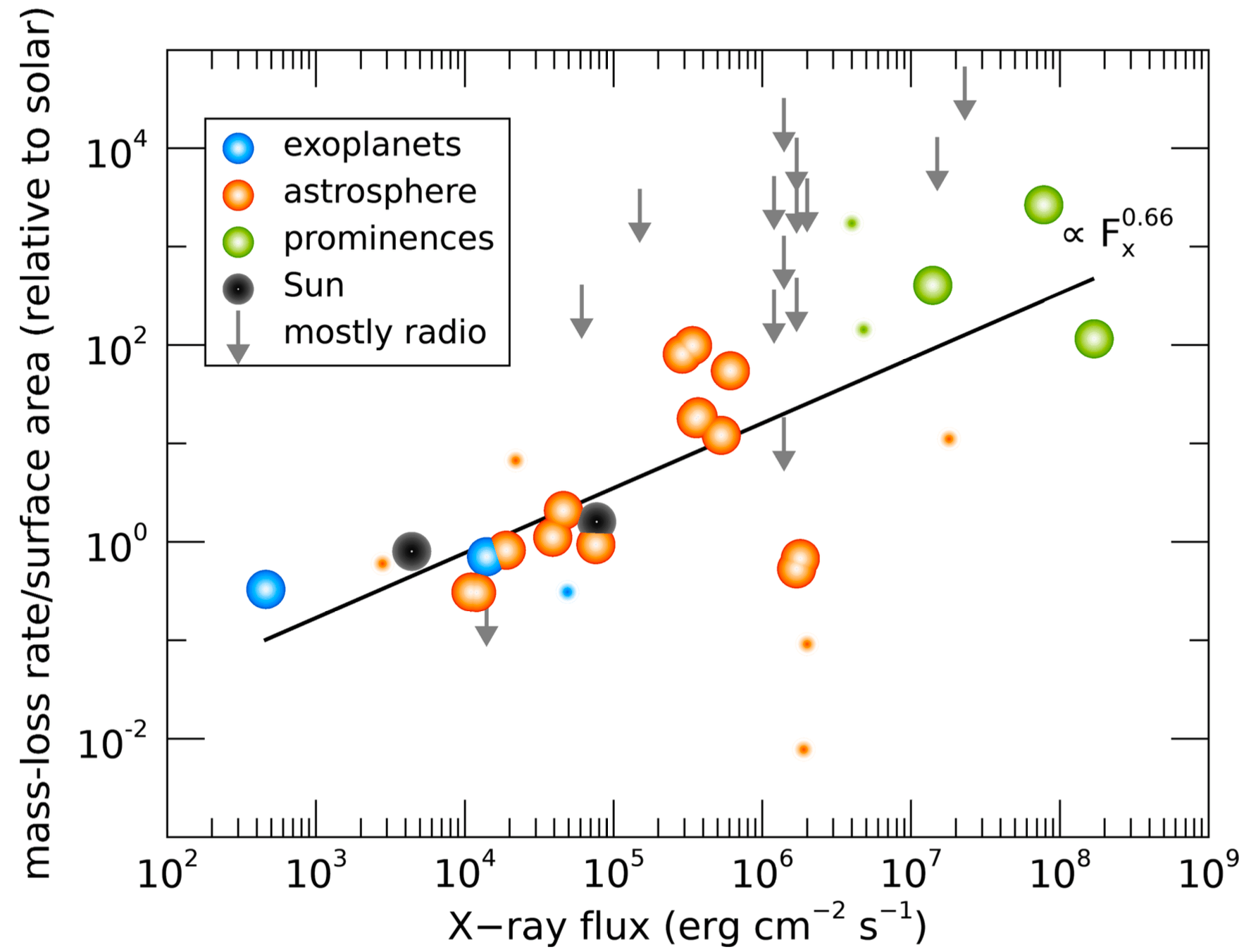} 
\includegraphics[width=0.4445\textwidth]{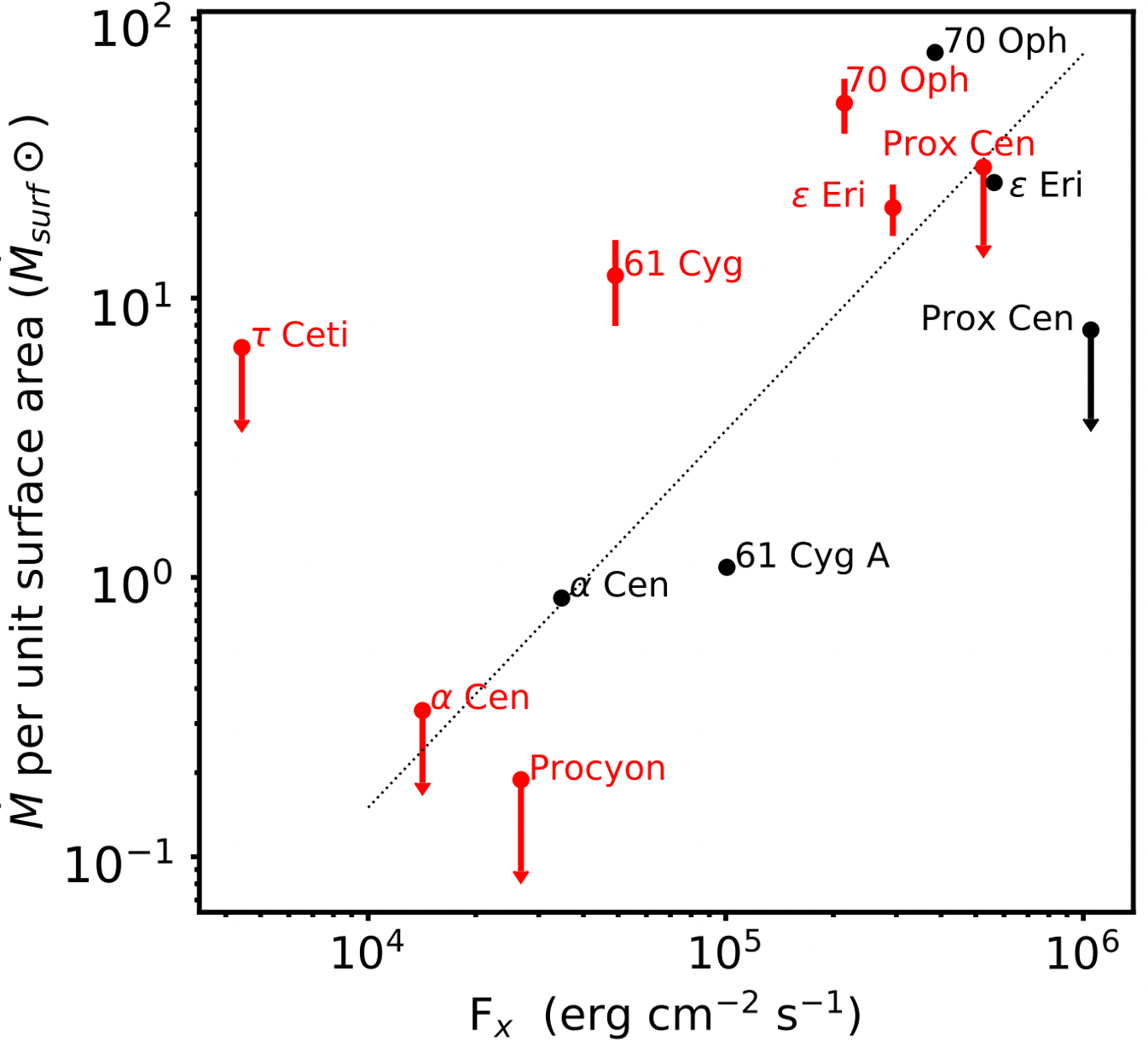} 
}
\caption{\textit{Left:} Selection of wind mass-loss estimates using different methods, plotted as mass-loss rate per unit stellar surface area versus the X-ray surface flux. Results focus on radio bremsstrahlung, astrospheric hydrogen wall  Ly$\alpha$ absorption, slingshot prominences, and inferences from exoplanets. The solid line is a power-law fit through the larger circles, not including M dwarfs or evolved stars (smaller circles) or the upper limits.
From \citet{vidotto2021}.  -- \textit{Right}: Wind mass-loss estimates from astrospheric X-ray charge exchange X-ray lines, the only so far directly detected cool-star winds apart from the solar wind, compared with estimates from the Ly$\alpha$ hydrogen wall method. 
From \citet{kislyakova2024}.}
\label{windplot}
\end{center}
\vskip -0.5cm
\end{figure}

We summarize order-of-magnitude estimates and upper limits from different methods separately for F/G/K dwarfs and for M dwarfs in Table~\ref{resultstable}. It is instructive to compare results from the different methods. Each detection method teaches us something about the physical conditions in the wind, but this also makes the methods vulnerable to simplifications and biases. Also, the sensitivity range vary widely between the methods. Figure~\ref{windplot} shows two versions of plots of $\dot{M}$ per stellar surface area versus the X-ray surface flux as an activity indicator.

\begin{table}
\begin{center}
  \caption{Selected published estimates and upper limits of $\dot{M}$ using different methodologies}\label{resultstable}
  \begin{tabular}{lrrr}\hline
   Method              &            $\dot{M}$ (F/G/K Dwarfs)               & $\dot{M}$ (M Dwarfs)                              &  R\textsuperscript{\textdagger}\\
                       &            $[M_{\odot}$~yr$^{-1}]$                & $[M_{\odot}$~yr$^{-1}]$                           &  \\
   \hline                 
   Spin-down           &            $10^{-14}-\, $4$\times 10^{-13}$       & $10^{-14}-10^{-13}$                               &  1     \\
   H walls, Ly$\alpha$ absorption&  $<$10$^{-15}-\,$2$\times 10^{-12}$     & $<$2$\times 10^{-15}$                             &  2      \\
   Astrospheric CX (X-rays) &       (1.9--13)$\times 10^{-13}$, 
                                               $<$1.4$\times 10^{-14}$     & 1.4$\times 10^{-14}$, $<$3$\times 10^{-13}$       &  3      \\
   X-ray emission      &            $<$3.5$\times 10^{-12} - <$10$^{-10} $ & $<$7$\times 10^{-13} - <$2$\times 10^{-11}$       &  4      \\
   Radio bremsstrahlung&            $<$(2--3)$\times 
                                    10^{-11}$\textsuperscript{\textdagger\textdagger} & $<$(6--9)$\times 10^{-10}$, $<$7$\times 10^{-12}$ &  5      \\
   Radio free-free absorption &     $<$7$\times 10^{-10} $                 & $<$5$\times 10^{-14}- <$10$^{-11} $               &  6      \\
   Radio wave propagation     &     $<$10$^{-12} - <$3$\times 10^{-9} $    & ...                                               &  7      \\
   Planet. Ly$\alpha$/radio suppression& 5$\times 10^{-15}-\,$4$\times 10^{-14}$& 1.2$\times 10^{-15}$                         &  8      \\ 
   Slingshot prominences &        2.6$\times 10^{-12}-\,$9$\times 10^{-11}$& $(1-4)\times 10^{-12}$                            &  9      \\ 
   WD  metal contamination  &       ...                                    & $10^{-16} -\, $6$\times 10^{-15}$                 &  10     \\
   Debris disk dynamics  &          ...                                    & $<$2$\times 10^{-13}$, $10^{-12}$                 &  11     \\    
   \hline
  \end{tabular}
 \end{center}
 {\it Notes:} The presented values or value ranges represent published estimates subject to selection bias  (including stellar activity levels, rotation periods, or age) and methodological limitations, not the true distributions of $\dot{M}$. \\
  \textsuperscript{\textdagger} References: 1: \citet{johnstone2015a};
          --- 2: \citet{wood2004};
          --- 3: \citet{wargelin2002}, \citet{kislyakova2024};
          --- 4: \citet{guedel2020};
		  --- 5: \citet{fichtinger2017} (G dwarfs), \citet{lim1996a} (M dwarfs AD Leo, YZ CMi), \citet{lim1996b} Prox. Cen. $(<7\times 10^{-12}$);
		  --- 6: \citet{fichtinger2017} (G dwarfs), \citet{lim1996a} (M dwarf YZ CMi);
          --- 7: \citet{vidotto2017}  (V830 Tau, T Tauri star);
          --- 8: \citet{lecavelier2012} for HD~189733 (K2~V), 
		     \citet{kislyakova2014} for HD 209458 (F9~V), 
		     \citet{vidotto2017b} for GJ~436 (M2.5);
          --- 9: \citet{jardine2019};  
		  --- 10: \citet{debes2006}, also lower values in \citet{parsons2012};  
		  --- 11: \citet{strubbe2006}, \citet{schueppler2015}       \\
     \textsuperscript{\textdagger\textdagger} Only for spherical winds; lower upper limits possible for conical winds.
\end{table}



The best-studied cases across all methods are Proxima Cen (M5.5V), AU Mic (M1V), and $\pi^1$~UMa (G0.5V), summarized in Table~\ref{windcomparison}. Estimates of $\dot{M}$ using stellar spin-down, Ly$\alpha$ absorption, astrospheric CX, X-ray thermal emission, and radio bremsstrahlung are available
for Proxima Centauri; stellar spin-down, Ly$\alpha$ absorption, X-ray upper limits, and radio bremsstrahlung for the solar analog 
$\pi^1$~UMa; and spin-down, X-ray upper limits, and debris disk dynamics for AU Mic. While there is some agreement between different measurements or upper limits, there are also contradictions.
The last row of the Table predicts $\dot{M}$ per unit stellar surface area for the observed X-ray surface flux based on the correlation fit shown in Figure~\ref{windplot}-left (from \citealt{vidotto2021}).
It is noteworthy that the $\dot{M}$ estimates from spin-down for $\pi^1$~UMa and AU Mic give values close to the prediction from the correlation although the latter did not use estimates from the spin-down method. On the other hand, spin-down modeling predicts 30 times stronger winds for Proxima Cen than predicted 
from the correlation.

It is obvious that obtaining multiple $\dot{M}$ determinations are important.
We also note that some of the direct methods, in particular using X-ray luminosity and radio bremsstrahlung, provide only very high upper limits. In the former case, this is because the coronal X-ray emission usually outshines a wind contribution by probably up to orders of magnitude while the two components (wind X-rays are emitted mostly close the the corona) cannot be spatially separated. The upper limits are therefore based on the coronal X-ray luminosity and not instrumental sensitivity limits. In the case of radio bremsstrahlung, the high upper limits can be related to  instrumental sensitivity but also to coronal radio emission that outshines wind emission. If coronal radio emission is weak, a practical limit will be set by the ubiquitous  chromospheric thermal emission \citep{drake1993, villadsen2014}.

\begin{table}
  \begin{center}
  \caption{Comparison of wind measurements and models for three stars, in units of the solar-wind mass-loss rate $\dot{M} = 2\times 10^{-14}~M_{\odot}$~yr$^{-1}$}\label{windcomparison}
  \begin{tabular}{lllll}
  \hline
   Method                           &           $\pi^1$ UMa     & AU Mic                     & Proxima Cen                & Ref.\textsuperscript{\textdagger}         \\
   Spectral type                    &           G0.5V           & M1V                        & M5.5V                      &        \\
   \hline               
   Spin-down                        &    6.6                    & 34.6                       & 12.2                        &  1     \\
   H walls, Ly$\alpha$ absorption   &    0.5                    & --                         & $<$0.2                     &  2      \\
   Astrospheric CX (X-rays)         &    --                     & --                         & $<$0.7                     &  3      \\
   X-ray emission                   &    $<$2160                & $<$2970                    & $<$90                      &  4      \\
   Radio bremsstrahlung             &    $<$1450                & $<$30,000                  & $<$350                     &  5      \\
   Planetary radio emission        &          --                 &                         30--600   &          --            & 6      \\ 
   Debris disk dynamics             &  --                       & $\le$10; 50                   &  --                     &  7      \\                          
   Correlation Figure~\ref{windplot}--left &  27                  & 38                         &  0.4                    &  8      \\                          
   \hline
  \end{tabular}
 \end{center}
  \textsuperscript{\textdagger} References: 
         1: Determined from Equation~(4) of \citet{johnstone2015a} using $a=1.33, b=-3.36$, and using  
             radii from \citet{wright2012} ($\pi^1$~UMa, 1.02$R_{\odot}$), \citet{white2019} (AU Mic, 0.75$R_{\odot}$), and \citet{wright2012} (Prox. Cen, 0.14$R_{\odot}$), 
             masses from \citet{wright2012} (1.1$M_{\odot}$ for $\pi^1$~UMa, 0.6$M_{\odot}$ for AU Mic, 0.12$M_{\odot}$ for Prox. Cen),
             and rotation rates from rotation periods in \citet{wright2012} (4.9~d for $\pi^1$~UMa, 4.85~d for AU Mic, 42~d for Prox. Cen). 
      --- 2: \citet{wood2018},
      --- 3: \citet{kislyakova2024},
      --- 4: calculated from $L_{\rm X}$ and radius using Equation~
             (\ref{mdotLx}); $L_{\rm X}$ from \citet{fichtinger2017} ($\pi^1$~UMa, $1.3\times 10^{29}$~erg~s$^{-1}$),
            \citet{mitrakraev2005} (AU Mic, $3\times 10^{29}$~erg~s$^{-1}$), and \citet{wood2018} (Prox. Cen, $1.7\times 10^{27}$~erg~s$^{-1}$), and radii as in 1.
      --- 5: \citet{fichtinger2017}, \citet{tristan2026},     
             \citet{lim1996b}; note that AU Mic was detected  as a coronal X-ray source ($\sim$1~mJy at 21~GHz, \citealt{tristan2026}).
     ---  6: \citet{2021MNRAS.504.1511K} for AU Mic and
             its planets b and c for radio fluxes of 1--10~mJy
      --- 7: \citet{strubbe2006}, \citet{schueppler2015} 
      --- 8: using Equation~(17) of \citet{vidotto2021} for the regression curve in Figure~\ref{windplot}-left, with $L_{\rm X} =3\times 10^{29}$~erg~s$^{-1}$  \citep{mitrakraev2005}.  
\end{table}
What do we learn from the various methodologies? The discussed approaches to detecting stellar winds are not merely reflecting attempts to pin them down. Each method has its own strength that teaches us specific properties of the winds themselves. Stellar spin-down shows that cool-star magnetized winds are an expression of dynamo action in the stellar interior and are coupled with ``magnetic activity''. Methodologies using Ly$\alpha$ absorption teach us about the momentum loss rate in a wind and the interaction with the local interstellar medium, and this holds similarly for studies of white-dwarf surface contamination. The slingshot prominence interpretation links winds to the dynamics in the corona itself. The radio bremsstrahlung methods prove, in the first place, that the winds are ionized (although detections are yet to be reported), and give information about the ratio of $\dot{M}/v$; thus, with additional information, e.g. from the Ly$\alpha$ method that provides information on $\dot{M}v$, it could potentially give us information on the wind velocity. X-ray emission from the winds, including charge exchange emission in the winds, provides the most direct information on the kinetic energy and thermal and density structure of the wind particles and also links to the presence of a neutral interstellar medium. Taken together, the various methodologies hold promise to converge on a unique description of Sun-like ionized and magnetized stellar winds throughout the main-sequence evolution of a star.

\FloatBarrier 

\subsection{Modeling of Stellar Winds}\label{sec:SW_Models}

 \subsubsection{Solar Wind Models: A Brief Historical Summary} \label{sec:SolW-Models}

The development of solar wind models began with a progressive shift from the concept of a static corona to that of a continuously expanding and dynamic corona--heliosphere interconnected system. A summary of the most important historical milestones is provided below, but interested readers are referred to \citetads{2018LRSP...15....4G} for a comprehensive review.  


As recounted in Section~\ref{subsec:solarwind}, the \citepads{parker1958a} prediction of a supersonically expanding solar corona established the theoretical foundation of the solar wind paradigm, later extended by magnetized wind models incorporating magnetic fields and wave-driven acceleration. With the existence of the solar wind established, modeling efforts turned toward quantitative descriptions of the expanding corona. Early numerical work incorporated the full set of hydrodynamic equations, including energy transport processes such as thermal conduction and viscosity, allowing transonic solutions that reproduced observed coronal densities with reasonable accuracy (\citeads{1963ApJ...138.1169N}, \citeads{1965ApJ...141.1479S}). At the same time, the recognition that electrons and ions may not remain thermally coupled led to the development of two-fluid models, in which separate energy equations for each species captured deviations between electron and proton temperatures in the outer corona \citepads{1966PhRvL..16..628S}. These advances marked the transition from purely analytic models to physically more complete numerical descriptions of the solar wind.  
 
A major step forward was the inclusion of magnetic field structure in global models. Potential field formulations, based on the assumption of a current-free corona, enabled the reconstruction of large-scale magnetic topologies from photospheric observations and introduced the concept of a source surface beyond which field lines become radial (\citeads{1968Natur.220.1211S}, \citeads{1969SoPh....6..442S}, \citeads{1969SoPh....9..131A}, \citeads{1970SoPh...13..131N}). At the same time, models of an expanding magnetized corona demonstrated that plasma outflows modify the magnetic configuration itself, producing helmet streamers, neutral points, and current sheets as natural consequences of the coupled plasma-field system \citepads{1971SoPh...18..258P}. This established that realistic solar wind modeling requires a self-consistent treatment of both plasma dynamics and magnetic structure.  
From the 1970s onward, the field evolved toward global magnetohydrodynamic descriptions of the corona and heliosphere. Early MHD simulations reproduced the large-scale bimodal structure of the solar wind, with fast streams originating in coronal holes and slower wind associated with streamer regions, while also capturing the formation of corotating interaction regions and large-scale current sheets (e.g., \citeads{1975JGR....80.1223S}, \citeads{1978JGR....83.5563P}, \citeads{1988JGR....9314269S}, \citeads{1994GeoRL..21.2063P}). Subsequent developments introduced semi-empirical relations linking solar wind speed to magnetic flux tube expansion, enabling practical coupling between coronal and heliospheric models (e.g., Wang \& Sheeley~\citeyearads{1990ApJ...355..726W}, \citeyearads{1995ApJ...447L.143W}, \citeads{1999JGR...10428225O}, \citeads{2000JGR...10510465A}). These approaches formed the basis of modern global simulations that connect the inner corona to interplanetary space.  
 
Contemporary modeling efforts focus on incorporating increasingly realistic energetics and microphysics. Thermodynamic models improved upon earlier polytropic treatments by including parameterized heating and energy transport, while more recent frameworks explicitly account for Alfvén wave turbulence as a key mechanism for coronal heating and wind acceleration (e.g.~\citeads{1993SoPh..146..377U}, \citeads{1999SSRv...87..193G}, \citeads{2001ApJ...546..542L}, \citeads{2007ApJ...654L.163C}). These developments have culminated in advanced three-dimensional MHD models that self-consistently couple the chromosphere, corona, and solar wind, representing the current state of the art in global solar and stellar wind modeling (\citeads{2013ApJ...764...23S}, van der Holst et al.~\citeyearads{2014ApJ...782...81V}, \citeyearads{2022ApJ...925..146V}). More recently, efforts have increasingly shifted toward fully time-dependent coronal and heliospheric modeling, in which the boundary conditions---typically photospheric magnetic maps---are continuously updated to capture the evolving magnetic structure and transient dynamics of the solar wind \citep[e.g.,][]{2025Sci...388.1306D, 2025ApJ...994..236S, 2026ApJ..1001L..23L}.

\subsubsection{From the Sun to the Stars: Stellar Wind Models}\label{sec:Sun-to-Stars_Models}

Building on the solar wind modeling framework outlined in the previous section, stellar wind studies have progressively moved from idealized, largely 1D descriptions toward fully 3D MHD simulations constrained by observed surface magnetic fields. Early analytic work, such as the Weber--Davis model \citepads{weber1967}, extended the Parker wind \citepads{parker1958a} by incorporating the Lorentz forces and thereby established the basic theory of magnetized winds acceleration and angular momentum losses (e.g., \citeads{1984LNP...193...49M}, \citeads{1988ApJ...333..236K}). However, the radial magnetic geometry assumed in those early treatments was too restrictive to represent realistic stellar topologies, particularly for young and active stars whose surface fields are strongly non-axisymmetric (see \citeads{2009ARA&A..47..333D}). In parallel, the importance of Alfv\'en waves for coronal heating and wind acceleration motivated the development of wave-driven models once numerical methods became sufficiently mature to treat wave transport, reflection, and turbulent dissipation. 1D and 2D formulations (e.g., \citeads{2006JGRA..111.6101S}, \citeads{2005ApJS..156..265C}, \citeads{2007ApJS..171..520C}, \citeads{2012ApJ...749....8M}) provided the first physically motivated descriptions of fast and slow wind components. Advancing such approaches with explicit time-dependent 1D MHD simulations that couple the photosphere to the wind, \citetads{2013PASJ...65...98S} investigated the effect of increasing the photospheric Poynting flux over a wide range, finding that the mass-loss rate first rises rapidly as Alfvén wave reflection is suppressed, before saturating as the enhanced wave energy is increasingly radiated away from a denser, extended chromosphere rather than escaping to drive the wind. \citetads{2020ApJ...896..123S} further refined this 1D framework by incorporating plasma compressibility and a phenomenological treatment of Alfvén-wave turbulence, and applied it to low-mass stars across a range of rotation rates to simultaneously constrain mass-loss rates and magnetic braking; yet this model focused on the unsaturated regime of magnetic activity. To address this shortcoming, \citetads{2023ApJ...957...71S} proposed a hybrid model combining Alfvén wave dynamics with flux emergence-driven interchange reconnection, finding that this additional energy injection channel may be key to explaining the enhanced winds of magnetically active low-mass stars. Directly targeting M-dwarf atmospheres, \citetads{2021ApJ...919...29S} performed 1D compressive MHD simulations for TRAPPIST-1, Proxima Centauri, YZ CMi, and other M dwarfs, finding that M-dwarf coronae tend to be cooler and denser than the solar corona, and that M-dwarf winds are characterized by relatively faster velocities but much smaller mass-loss rates — tendencies attributed to the stronger atmospheric stratification and smaller Alfvén wave energy input from M-dwarf photospheres. On a broader parametric footing, \citeads{2020MNRAS.494.1297M} conducted a systematic parameter survey of M-dwarf winds using a 1D Alfvén-wave-driven model, characterizing the dependences of mass-loss rates, terminal velocities, and related properties on stellar parameters across the M-dwarf sequence. More recently, \citetads{2024ApJ...976...65R} applied the WindPredict-AW model to M dwarfs, demonstrating that the characteristics of the turbulence — particularly the turbulent correlation length — play a key role in setting the terminal wind speed, and highlighting the relevance of the resulting wind conditions for star-planet magnetic interactions in systems like TRAPPIST-1. An overarching challenge for all these wave-driven frameworks remains the parametrization of chromospheric turbulence spanning spectral type from from F-type to M-type stars. 

Despite these uncertainties, the advent of global MHD codes such as BATS-R-US \citepads{1999JCoPh.154..284P} marked a major transition toward realistic stellar wind modeling. These models enabled fully 3D simulations of thermally-driven Parker-type stellar outflows rooted in a prescribed hot corona, using surface radial magnetic fields inferred from Zeeman--Doppler Imaging (ZDI) as boundary conditions (e.g., \citeads{2010ApJ...721...80C}, \citeads{2011MNRAS.412..351V}, \citeads{2013ApJ...764...32G}). Initially based on a polytropic approach, these models use a reduced $\gamma$ (typically between 1.05 and 1.2) to ensure energy deposition in the accelerating wind, and are piloted by setting the density ($n_0$) and temperature ($T_0$) at the base of the corona. In the context of theses models, \citetads{2020A&A...635A.170A} developed a systematic approach to put constraints on the acceptable $(n_0,T_0)$ for a given cool star. Theses models, nevertheless, generally have difficulties reproducing the bi-modality of slow and fast winds such as those observed for the Sun. As mentioned in Section~\ref{sec:SolW-Models}, more advanced implementations incorporated Alfv\'en-wave physics, culminating in the three-dimensional Alfv\'en Wave Solar Model (AWSoM), which self-consistently treats wind heating and acceleration through wave propagation and turbulent dissipation (\citeads{2013ApJ...764...23S}, van der Holst et al.~\citeyearads{2014ApJ...782...81V}, \citeyearads{2022ApJ...925..146V}). 

In recent years, two modified versions of AWSoM, known as AWSoM-R and stream-aligned AWSoM-R, were introduced. AWSoM-R is a threaded-field-line extension of AWSoM designed to improve the run speed of the code by simplifying the treatment of the upper chromosphere while retaining the global 3D MHD context (see \citeads{2021ApJ...908..172S}). Instead of generating a fully volumetric description near the top of the chromosphere and transition region, it resolves each magnetic flux tube with a dedicated one-dimensional structure embedded in the global magnetic environment, which allows much sharper gradients in temperature and density to be captured along individual field lines as well as a reduction on the computational cost. The model keeps the Alfv\'en-wave-driven heating and acceleration physics of AWSoM, including upward wave propagation, reflection, mode coupling, and turbulent dissipation, but does so in a way that is better suited to the strongly stratified chromosphere-corona interface. Compared with AWSoM, this gives a more efficient and rapidly converging representation of the corona and wind structure, although the threaded approximation is less general than a full three-dimensional treatment. This is reflected in a slight reduction of the covered computational domain ($R \geq 1.05~R_{\odot}$), with the corresponding reduction of the thermodynamic structure coverage at upper chromosphere temperatures. 
  
The stream-aligned (SA) variant of AWSoM-R imposes $ \mathbf{u} \parallel \mathbf{B} $ in the co-rotating frame, so that $ \mathbf{E} = -\mathbf{u}\times\mathbf{B} = 0 $ and the associated Poynting flux drops out of the induction and energy equations, while the Lorentz force remains active in the momentum balance. Apart from the trade-offs associated with its AWSoM-R foundation, the SA variant is intentionally more restrictive than AWSoM, since the enforced alignment suppresses some fully dynamical MHD behavior in exchange for cleaner steady-state structure. Nevertheless, the simplifications remove numerical misalignment artifacts, including the spurious ``u-shape'' and ``v-shape'' distortions of magnetic field lines that can arise from diffusion in full MHD, and still admits physically consistent steady-state solutions such as Parker-spiral-like configurations. Its main advantage over AWSoM is therefore robustness for field-line tracing, magnetic connectivity studies, and space-weather applications, especially when extrapolating back to the Sun from spacecraft measurements or constructing a steady background wind for CME and SEP calculations. More details concerning its implementation and validation can be found in \citetads{2022ApJ...926..102S}. 

The open-source release of the Space Weather modeling Framework (SWMF; \citeads{2021JSWSC..11...42G}), incorporating AWSoM and its variants as some of the possible coupled coronal and heliospheric models, opened the path for its application and exploitation in the stellar regime. In practice, SWMF/AWSoM-based stellar wind models typically employ ZDI observations as magnetic boundary conditions, adopt a potential field extrapolation as the initial 3D magnetic field configuration, apply Adaptive Mesh Refinement (AMR) to resolve the coronal structure and the astrospheric current sheet, and link together the driving magnetic topology to the resulting wind density, speed, and mass-loss rate. This framework has been applied extensively to cool stars across the F--M spectral types to investigate a broad range of problems in stellar astrophysics. These include the characterisation of coronal environments and wind structures in connection with stellar magnetism (e.g., \citeads{2016A&A...588A..28A}, \citeads{2017ApJ...834...14C}, \citeads{2020A&A...635A.178B}, \citeads{2023MNRAS.522..792M}), as well as the quantification of stellar mass and angular momentum loss rates (e.g., Garraffo et al.~\citeyearads{2015ApJ...807L...6G}, \citeyearads{2016A&A...595A.110G}, \citeads{2023MNRAS.524.2042E}). It has also been instrumental in exploring the relation between stellar wind mass loss and magnetic activity levels (\citeads{2016A&A...594A..95A}, \citeads{2023MNRAS.524.5060C}), and in modeling the rotational evolution of cool stars (\citeads{2018ApJ...862...90G}, \citeads{2024MNRAS.529L.140E}). In addition, these data-driven models have been employed to study a variety of star--planet interaction scenarios governed by the ambient stellar wind (e.g., \citeads{2014ApJ...790...57C}, \citeads{2021ApJ...913..130H}, \citealt{2021MNRAS.504.1511K}, \citeads{2023A&A...678A.152V}). 
 
A complementary effort is the \textsc{WindPredict} framework (\citeads{2012ApJS..198....7M}), which builds on the open-source general purpose MHD code \textsc{PLUTO}. The WindPredict suite can be run in 1D, 2D and 3D. It is comprised of a polytropic-based version (\citeads{2012ApJS..198....7M}, \citeads{2015ApJ...815..111S}, \citeads{2016ApJ...832..145R}) and an alfv\'en-wave driven version (\textsc{WindPredict-AW}) inspired from the original derivations of the AWSoM code (\citeads{2020ApJS..246...24R}). They are all available in cartesian and spherical geometry. The \textsc{WindPredict-AW} version differs from the AWSoM approach in some of its details. On the physics part, one can mention the choice of the radiative cooling function and the choice of correlation scale of Alfv\'en wave turbulence (\citeads{2020ApJS..246...24R}, \citeads{2021ApJ...910...90H}). On the numerical part, both codes differ in the solver they use, the strategy they use to control $\nabla\cdot{\bf B}=0$, or in the grid construction. The polytropic version has been extensively used for generic studies on wind mass loss and torque (e.g. \citeads{2017ApJ...850...45R}, \citeads{2018ApJ...854...78F}), both for the Sun and for cool stars. \textsc{WindPredict-AW} has been mostly validated and used so far for solar applications (\citeads{2022ApJ...929...75P}), with the exception of an application to Proxima Centauri and Trappist-1 in 2D (\citeads{2024ApJ...976...65R}). Both versions of the code have the capabilities to use analytical magnetic field configurations or to be be informed by solar synoptic and stellar ZDI maps at the inner boundary (\citeads{2022MNRAS.512.4556S}). Finally, the same code base has been used extensively to study global star-planet magnetic interactions, adding a planet as an internal boundary in the computational domain (e.g. \citeads{2015ApJ...815..111S}, \citeads{2016ApJ...833..140S}, \citeads{2019ApJ...881..136S}, \citeads{2026A&A...705A..12P}). Together with the AWSoM suite, these approaches now form the backbone of state-of-the-art stellar wind modeling.
 
Modern 3D MHD simulations constrained by ZDI magnetograms of cool main-sequence stars provide the most physically robust description of stellar winds including estimates of their associated mass-loss rates. In general, the resulting wind properties depend sensitively on magnetic topology, the coronal base conditions as well as the accelation mechanism. Consequently, different numerical implementations such as the polytropic version of \textsc{BATS-R-US}, the Alfv\'en-wave-driven \textsc{AWSoM} model and its variants, or the \textsc{WindPredict} suite (polytropic and Alfv\`en-wave driven versions), can yield mass-loss rates that differ by factors of a few even when applied to the same ZDI magnetogram. Nevertheless, several benchmark stars have now been modeled by multiple groups, enabling increasingly robust cross-comparisons and providing an empirical basis for assessing model dependencies. Note that a proper, coordinated benchmark comparing \textsc{WindPredict} and AWSoM is yet to be done. In the following Section, we provide a summary of recent main-sequence 3D stellar wind modeling studies across F to M spectral types. We will focus mostly on the predicted mass loss rate values given their prevalence and importance in connection with the currently available observational constraints (see Section~\ref{sec:Wind_Overview}). 

\subsection{Winds of Cool Main-Sequence Stars: 3D Numerical Results} \label{subsec:coolstarwinds}


\subsubsection{F-type Stars.}\label{sec:SWModels_F}

The F7V planet host $\tau$~Boo is one of the best-studied cases among F-type stars. Using a \textsc{BATS-R-US} polytropic wind model driven by observed ZDI maps, \citetads{2012MNRAS.423.3285V} derived a mass-loss rate of $\dot M \approx 2.7\times10^{-12}~M_\odot~{\rm yr^{-1}}$, while \citetads{2016MNRAS.459.1907N} reported a similar value of $\sim 2.3\times10^{-12}~M_\odot~{\rm yr^{-1}}$, with variations of less than $4\%$ over the magnetic cycle.

In contrast, simulations using the AWSoM framework generally yield lower mass-loss rates, typically in the range of $\sim10^{-14}$–a few $\times10^{-13}~M_\odot~{\rm yr^{-1}}$. For instance, \citeads{2023MNRAS.524.5060C} reported $\dot M = 9.8\times10^{-14}~M_\odot~{\rm yr^{-1}}$, while \citetads{2026MNRAS.546f2070E} obtained values of $\dot M = 3.3\times10^{-14}~M_\odot~{\rm yr^{-1}}$ and $\dot M = 4.5\times10^{-13}~M_\odot~{\rm yr^{-1}}$ for a case where the $S_{\rm A}/B$ ratio was set to 10 times the solar value ($1.1 \times 10^{6}$~W~m$^{-2}$~T$^{-1}$). These variations highlight the strong sensitivity of the inferred mass-loss rates to the adopted input parameters and numerical approach, as well as the critical role of the chosen ZDI maps, particularly for stars with evolving magnetic topologies such as $\tau$~Boo.

Other ZDI-driven, AWSoM-based stellar wind models of F-type stars include $\beta$~Vir (F9V; \citeads{2022MNRAS.509.5117S}), HD~179949 (F7V; \citeads{2023MNRAS.524.5060C}), and $\chi$~Dra A (F7V; \citeads{2023MNRAS.522..792M}), for which simulations yield mass-loss rates of $\sim2$–$5$ times the solar value (i.e.~$\dot M \approx 5.0$–$9.0 \times 10^{-14}~M_\odot~{\rm yr^{-1}}$). Additional targets include HD~35296 (F8V; \citeads{2023MNRAS.524.5060C}) and $\tau$~Draconis (F8IV; \citeads{2023MNRAS.522..792M}), both exhibiting mass-loss rates higher than solar value, with $\dot M \approx 1.8\times10^{-13}$ and $3.6\times10^{-13}~M_\odot~{\rm yr^{-1}}$ respectively.

Overall, these studies show that F-type stars span a wide range of mass-loss rates, from values comparable to the solar wind ($\dot{M}_{\odot} \approx 2\times10^{-14}~M_{\odot}~{\rm yr^{-1}}$) up to more than two orders of magnitude higher. This broad spread reflects both intrinsic variability and the strong dependence of inferred rates on the adopted magnetic field maps and modeling approaches. However, wind studies of F-type stars remain relatively limited, as surface magnetic fields are detected in relatively few stars of this spectral type. This is consistent with their comparatively weak coronal activity and the limited number of successful ZDI reconstructions. Consequently, the observational constraints available to drive and validate ZDI-based wind models are still sparse (see \citet{Chapter1} in this collection; see also \citeads{2020MNRAS.494.5682S}).

\begin{figure}[p!]
    \centering
    \includegraphics[width=0.7\linewidth]{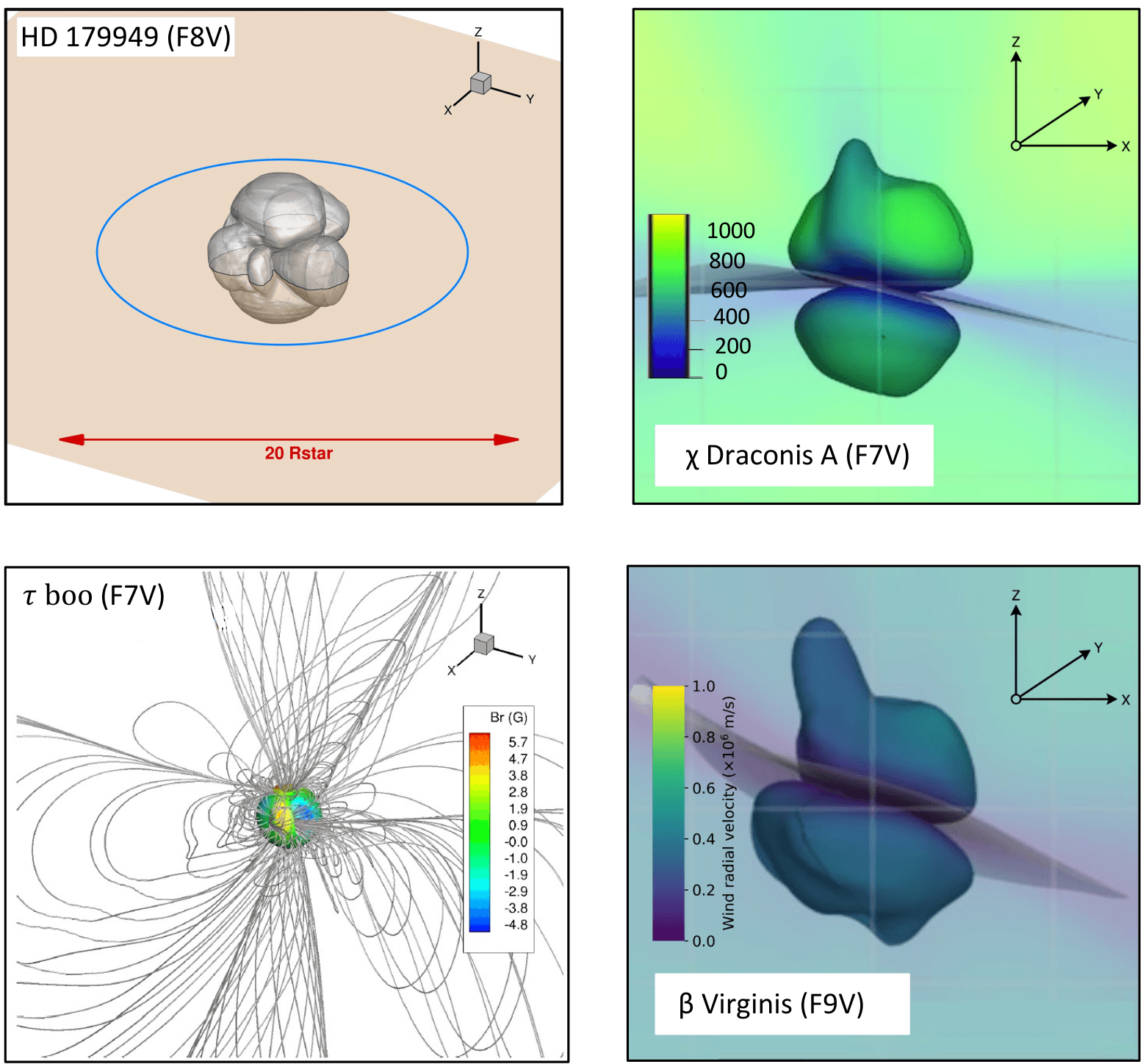}
    \caption{Stellar wind steady-state solutions for F-type stars obtained using polytropic and Alfv\'en wave-driven models.\textbf{First column (polytropic; BATS-R-US).}
\textit{Top:} HD~179949 (F8V; \citeads{2015MNRAS.449.4117V}). The Alfv\'en surface is shown in grey and exhibits an irregular, asymmetric morphology due to the complex observed magnetic field distribution. The stellar equatorial plane ($xy$), assumed to host the planetary orbit, is displayed along with its intersection with the Alfv\'en surface (thin black contour) and the planetary orbital radius (thick blue contour).
\textit{Bottom:} $\tau$~Boo (F7V; \citeads{2016MNRAS.459.1907N}). The stellar surface is color-coded by the radial magnetic field reconstructed from ZDI maps (red/blue for positive/negative polarity). Magnetic field lines are shown in grey, illustrating the large-scale magnetic topology. 
\textbf{Second column (Alfv\'en wave-driven; AWSoM).}
\textit{Top:} $\chi$~Dra~A (F7V; \citeads{2023MNRAS.522..792M}). The Alfv\'en surface is shown as a solid surface, while the current sheet ($B_r = 0$) appears as a translucent grey structure. Both the Alfv\'en surface and the plane of the sky ($xz$ plane) are color-coded by the radial wind velocity.
\textit{Bottom:} $\beta$~Vir (F9V; \citeads{2022MNRAS.509.5117S}). The Alfv\'en surface and inner current sheet are displayed, with the current sheet shown edge-on to highlight the characteristic two-lobed structure. The $z$-axis corresponds to the stellar rotation axis. Both the Alfv\'en surface and the $xz$ plane are color-coded by the radial wind velocity $u_r$. The current sheet is truncated at $30\,R_\star$. 
In all panels, the star is located at the center of the computational domain, and the rotation axis is aligned with the $z$-axis. The 3D orientation is indicated in each panel. \textbf{Note:} Visualizations are adapted from the referenced works and may differ slightly from the originals for consistency and clarity.}
    \label{F_stars_mosaic}
\end{figure}

\begin{table*}[t]
\centering
\begin{threeparttable}
\caption{Summary of 3D Stellar Wind Models for F-type Stars}\label{tab:FstarWinds}
\begin{scriptsize}
 \begin{tabular}{l c c c c c c c}
    \hline
  & & & & & & & \\[-0.2cm]
  \textbf{Star (S. Type)} & \textbf{Model} & \multicolumn{3}{c}{\textbf{Input Parameters}\tnote{$\dagger$}} & \textbf{Magnetic} & $\mathbf{\dot M}$ & \textbf{Ref.\tnote{$\S$}}\\
  & (Polytropic) & $\gamma$ & $n_0$ [cm$^{-3}$] & $T_0$ [K]& \textbf{Boundary} & \textbf{[}$\mathbf{M_\odot\,{\rm \textbf{yr}\mathbf{^{-1}}}}$\textbf{]} &\\[0.1cm]
  \hline\\[-0.25cm]
$\tau$~Boo (F7V) & BATS-R-US & 1.1 & $1.0\times10^{9}$ & $2.0\times10^{6}$ & ZDI\tnote{*} & $2.7\times 10^{-12}$ & 1\\
--- & BATS-R-US & 1.1 & $5.0\times10^{8}$ & $2.0\times10^{6}$ & ZDI\tnote{*} & $2.3\times 10^{-12}$ & 2\\
HD 179949 (F8V) & BATS-R-US & 1.1  & $1.0\times10^{9}$ & $2.0\times10^{6}$ & ZDI & $8.0 \times 10^{-13}$ & 3\\
  \hline\\[-0.2cm]
  \textbf{Star (S. Type)} & \textbf{Model} & \multicolumn{3}{c}{\textbf{Input Parameters}\tnote{$\ddagger$}} & \textbf{Magnetic} & $\mathbf{\dot M}$ & \textbf{Ref.\tnote{$\S$}}\\
  & (AW-driven) & & $(S_{\rm A}/B)_{\bigstar}$ & $(L_{\perp}\sqrt{B})_{\bigstar}$ & \textbf{Boundary} & \textbf{[}$\mathbf{M_\odot\,{\rm \textbf{yr}\mathbf{^{-1}}}}$\textbf{]} &\\
  & & & [W m$^{-2}$ T$^{-1}$] & [m $\sqrt{\rm T}$] & & &\\[0.1cm]
  \hline\\[-0.2cm]
$\tau$~Boo (F7V)        & AWSoM & & $1.1\times10^{6}$ & $1.5\times10^{5}$ & ZDI & $9.8\times10^{-14}$ & 4\\
---                     & AWSoM & & $1.1\times10^{6}$ & $1.5\times10^{5}$ & ZDI & $3.3\times10^{-14}$ & 5\\
---                     & AWSoM & & $1.1\times10^{7}$ & $1.5\times10^{5}$ & ZDI & $4.5\times 10^{-13}$ & 5\\
$\chi$~Draconis A (F7V) & AWSoM & & $1.1\times10^{6}$ & $1.5\times10^{5}$ & ZDI\tnote{*} & $8.5 \times 10^{-14}$& 6\\
HD 179949 (F8V)         & AWSoM & & $1.1\times10^{6}$ & $1.5\times10^{5}$ & ZDI & $5.4\times10^{-14}$ & 4\\
HD 35296 (F8V)          & AWSoM & & $1.1\times10^{6}$ & $1.5\times10^{5}$ & ZDI & $1.8\times10^{-13}$ & 4\\
$\beta$ Virginis (F9V)  & AWSoM & & $1.1\times10^{6}$ & $1.5\times10^{5}$ & ZDI\tnote{*} & $5.5\times10^{-14}$ & 7\\
$\theta$ Draconis (F8IV) & AWSoM & & $1.1\times10^{6}$ & $1.5\times10^{5}$ & ZDI\tnote{*} & $3.6\times 10^{-13}$& 6\\
\hline 
\end{tabular}
\end{scriptsize}
\begin{tablenotes}
\footnotesize
\item[$\dagger$] Typical parameters for 3D polytropic solar wind models are $\gamma = 1.05$, $n_{0} = 1.0\times10^8$~cm$^{-3}$ and $T_{0} = 2\times10^{6}$~K \citepads{2017ApJ...835..220C}.
\item[$\ddagger$] The solar-validated AWSoM parameters correspond to $(S_{\rm A}/B)_{\odot} = 1.1\times10^{6}$~W m$^{-2}$ T$^{-1}$ and $(L_{\perp}\sqrt{B})_{\odot} = 1.5\times10^{5}$~m~$\sqrt{\rm T}$ (van der Holst et al.~\citeyearads{2014ApJ...782...81V}, \citeyearads{2022ApJ...925..146V}).  
\item[*] Different ZDI maps were employed. The listed $\dot M$ value correspond to the average of all models.
\item[$\S$] \textbf{References:} ---1: \citetads{2012MNRAS.423.3285V} ---2: \citetads{2016MNRAS.459.1907N} ---3: \citetads{2015MNRAS.449.4117V} ---4: \citetads{2023MNRAS.524.5060C} ---5: \citetads{2026MNRAS.546f2070E} ---6: \citetads{2023MNRAS.522..792M} ---7: \citetads{2022MNRAS.509.5117S} 
\end{tablenotes} 
\end{threeparttable}
\end{table*}

\subsubsection{G-type Stars.}\label{sec:SWModels_G} 

Several studies have generated stellar wind models for G-type (Sun-like) stars. The more direct applicability of the different modeling frameworks (originally developed for the Sun; see Section~\ref{sec:Sun-to-Stars_Models}) as well as the relatively high magnetic field detection rates (\citeads{2014MNRAS.444.3517M}, \citeads{2017AN....338..428K}) allowing the reconstruction of the required ZDI input data, favors the investigation within this spectral type. 

The ensemble of three-dimensional polytropic wind models for G-type stars reveals a relatively narrow parameter space for polytropic models, typically clustered around $\gamma \sim 1.05$--$1.10$, base densities $n_0 \sim 10^{7}$--$10^{9}$~cm$^{-3}$, and coronal temperatures $T_0 \sim 1$--$2\times10^{6}$~K, broadly consistent with canonical solar-like assumptions \citepads{2017ApJ...835..220C}. Within this framework, \textsc{BATS-R-US} and \textsc{WindPredict} simulations systematically yield mass-loss rates spanning $\dot{M} \sim 10^{-14}$--$10^{-13}~M_\odot~{\rm yr}^{-1}$. Notable examples include $\kappa^1$~Cet and HD~73256, which exhibit comparatively elevated mass-loss rates of $\sim 10^{-13}~M_\odot,{\rm yr}^{-1}$ (\citeads{2016ApJ...820L..15D}, \citeads{2015MNRAS.449.4117V}), while more quiescent systems such as 55~Cnc and Kepler-78 fall below $10^{-13}~M_\odot~{\rm yr}^{-1}$ (\citeads{2020A&A...633A..48F}, \citeads{2019ApJ...881..136S}). 

Naturally AWSoM modeling efforts adopt coronal heating and stellar wind acceleration parameters calibrated to solar values (van der Holst et al.~\citeyearads{2014ApJ...782...81V}, \citeyearads{2022ApJ...925..146V}). The various studies consistently employ ZDI-derived magnetic field maps, with occasional use of dynamo-generated magnetic field distributions (e.g.~\citeads{2024ApJ...971..153X}), and predict a broader range of mass-loss rates, extending from $\sim 10^{-14}$ up to $\sim 10^{-12}~M\odot~{\rm yr}^{-1}$. Enhanced mass loss is observed in young or magnetically active stars such as HII,739 and DS~Tuc~A, reaching $\dot{M} \gtrsim 4\times10^{-13}~M_\odot~{\rm yr}^{-1}$ (\citeads{2023MNRAS.524.2042E}, \citeads{2026A&A...705A..40B}), while the highest value in the sample corresponds to $\kappa^1$~Cet with $\dot{M} \approx 2.6\times10^{-12}~M_\odot~{\rm yr}^{-1}$ \citepads{2021ApJ...916...96A}. Conversely, more evolved or less active stars such as those in open clusters (e.g. Melotte systems) exhibit weaker winds with $\dot{M} \lesssim 10^{-13}~M_\odot~{\rm yr}^{-1}$ \citepads{2021MNRAS.506.2309E}. 

In general, the large majority of stellar wind models for G-type stars produce mass-loss rates in the range of $10^{-13}$--$10^{-14}\,M_\odot~{\rm yr^{-1}}$ depending on activity level. For reference, the accepted value for the present-day solar wind mass-loss rate is $\dot M_\odot \approx 2\times10^{-14}~M_\odot~{\rm yr^{-1}}$.

\begin{table*}[t]
\centering
\begin{threeparttable}
\caption{Summary of 3D Stellar Wind Models for G-type Stars}\label{tab:placeholder2}
\begin{scriptsize}
 \begin{tabular}{l c c c c c c c}
  \hline
  & & & & & & & \\[-0.2cm]
  \textbf{Star (S. Type)} & \textbf{Model} & \multicolumn{3}{c}{\textbf{Input Parameters}\tnote{$\dagger$}} & \textbf{Magnetic} & $\mathbf{\dot M}$ & \textbf{Ref.\tnote{$\S$}}\\
  & (Polytropic) & $\gamma$ & $n_0$ [cm$^{-3}$] & $T_0$ [K]& \textbf{Boundary} & \textbf{[}$\mathbf{M_\odot\,{\rm \textbf{yr}\mathbf{^{-1}}}}$\textbf{]} &\\[0.1cm]
  \hline\\[-0.25cm]
$\kappa^1$ Cet  (G5V)       & BATS-R-US   & 1.10 & $1.0\times10^{9}$ & $2.0\times10^{6}$ & ZDI & $9.7\times10^{-13}$ & 1\\
HII\,296        (G8)        & WindPredict & 1.05 & $4.2\times10^{8}$ & $1.9\times10^{6}$ & ZDI & $1.1\times10^{-13}$ & 2\\
55 Cnc          (G8V)       & BATS-R-US   & 1.05 & $5.9\times10^{7}$ & $1.0\times10^{6}$ & ZDI & $2.2\times10^{-14}$ & 3\\
Kepler-78       (G8V)       & WindPredict & 1.05 & $1.6\times10^{8}$ & $1.6\times10^{6}$ & ZDI & $1.7\times10^{-14}$ & 4\\
TYC 1990-108-1  (G8)        & WindPredict & 1.05 & $2.1\times10^{8}$ & $1.7\times10^{6}$ & ZDI & $6.0\times10^{-14}$ & 2\\
HD\,73256       (G9V)       & BATS-R-US   & 1.10 & $1.0\times10^{9}$ & $2.0\times10^{6}$ & ZDI & $2.1\times10^{-13}$ & 5\\
DX Leo          (G9V)       & WindPredict & 1.05 & $2.7\times10^{8}$ & $1.8\times10^{6}$ & ZDI & $6.2\times10^{-14}$ & 2\\
BD- 16351       (PMS)       & WindPredict & 1.05 & $3.6\times10^{8}$ & $1.9\times10^{6}$ & ZDI & $1.1\times10^{-13}$ & 2\\
  \hline\\[-0.2cm]
  \textbf{Star (S. Type)} & \textbf{Model} & \multicolumn{3}{c}{\textbf{Input Parameters}\tnote{$\ddagger$}} & \textbf{Magnetic} & $\mathbf{\dot M}$ & \textbf{Ref.\tnote{$\S$}}\\
  & (AW-driven) & & $(S_{\rm A}/B)_{\bigstar}$ & $(L_{\perp}\sqrt{B})_{\bigstar}$ & \textbf{Boundary} & \textbf{[}$\mathbf{M_\odot\,{\rm \textbf{yr}\mathbf{^{-1}}}}$\textbf{]} &\\
  & & & [W m$^{-2}$ T$^{-1}$] & [m $\sqrt{\rm T}$] & & &\\[0.1cm]
  \hline\\[-0.2cm]
HN\,Peg (G0V)               & AWSoM & & $1.1\times10^{6}$ & $1.5\times10^{5}$ & ZDI & $1.6\times10^{-13}$ & 6\\
$\iota$ Hor (G0V)           & AWSoM & & $1.1\times10^{6}$ & $1.5\times10^{5}$ & Dynamo\tnote{$\sharp$} & $1.7\times10^{-13}$ & 7\\
GJ\,504 (G0V)               & AWSoM & & $1.1\times10^{6}$ & $1.5\times10^{5}$ & ZDI & $1.1\times10^{-13}$ & 8\\
HD\,190771 (G2V)            & AWSoM & & $1.1\times10^{6}$ & $1.5\times10^{5}$ & ZDI & $1.1\times10^{-13}$ & 6\\
HII\,739 (G3)               & AWSoM & & $1.1\times10^{6}$ & $1.5\times10^{5}$ & ZDI & $4.9\times10^{-13}$ & 9\\
HD\,63433 (G5V)             & AWSoM & & $1.1\times10^{6}$ & $1.5\times10^{5}$\tnote{$\parallel$} & ZDI\tnote{*} & $1.3\times10^{-13}$ & 10\\
TOI-1860 (G5V)              & AWSoM & & $1.1\times10^{6}$ & $1.5\times10^{5}$\tnote{$\parallel$} & ZDI & $1.5\times10^{-13}$ & 10\\
DS\,Tuc\,A (G5V)            & AWSoM & & $1.1\times10^{6}$ & $1.5\times10^{5}$\tnote{$\parallel$} & ZDI & $4.4\times10^{-13}$ & 10\\    
HD\,147513 (G5V)            & AWSoM & & $1.1\times10^{6}$ & $1.5\times10^{5}$ & ZDI & $1.1\times10^{-13}$ & 11\\
$\kappa^1$ Cet (G5V)        & AWSoM & & $3.0\times10^{7}$ & ---\tnote{**}   & ZDI\tnote{*} & $2.6\times10^{-12}$ & 12\\
Cl Melotte 25-21 (G5)       & AWSoM & & $1.1\times10^{6}$ & $1.5\times10^{5}$ & ZDI & $5.7\times10^{-14}$ & 13\\
Cl Melotte 111 AV 2177 (G6) & AWSoM & & $1.1\times10^{6}$ & $1.5\times10^{5}$ & ZDI & $2.1\times10^{-14}$ & 13\\
TYC\,1987-509-1 (G7)        & AWSoM & & $1.1\times10^{6}$ & $1.5\times10^{5}$ & ZDI & $2.8\times10^{-13}$ & 6\\
---                         & AWSoM & & $1.1\times10^{6}$ & $1.5\times10^{5}$ & ZDI & $5.4\times10^{-14}$ & 13\\
HD\,1237 (G8V)              & AWSoM & & $1.1\times10^{6}$ & $1.5\times10^{5}$ & ZDI\tnote{*} & $9.3\times10^{-14}$ & 11\\
HII\,296 (G8)               & AWSoM & & $1.1\times10^{6}$ & $1.5\times10^{5}$ & ZDI & $3.5\times10^{-13}$ & 9\\
Cl Melotte 111 AV 1693 (G8) & AWSoM & & $1.1\times10^{6}$ & $1.5\times10^{5}$ & ZDI & $6.9\times10^{-14}$ & 13\\
HH Leo (G8)                 & AWSoM & & $1.1\times10^{6}$ & $1.5\times10^{5}$ & ZDI & $5.1\times10^{-14}$ & 13\\
HD\,73256 (G9V)             & AWSoM & & $1.1\times10^{6}$ & $1.5\times10^{5}$ & ZDI & $7.7\times10^{-14}$ & 6\\
Cl Melotte 111 AV 1826 (G9) & AWSoM & & $1.1\times10^{6}$ & $1.5\times10^{5}$ & ZDI & $4.6\times10^{-14}$ & 13\\
DX Leo (G9)                 & AWSoM & & $1.1\times10^{6}$ & $1.5\times10^{5}$ & ZDI & $6.6\times10^{-14}$ & 13\\
\hline 
\end{tabular}
\end{scriptsize}
\begin{tablenotes}
\footnotesize
\item[$\dagger$] Typical parameters for 3D polytropic solar wind models are $\gamma = 1.05$, $n_{0} = 1.0\times10^8$~cm$^{-3}$ and $T_{0} = 2\times10^{6}$~K \citepads{2017ApJ...835..220C}.
\item[$\ddagger$] The solar-validated AWSoM parameters correspond to $(S_{\rm A}/B)_{\odot} = 1.1\times10^{6}$~W m$^{-2}$ T$^{-1}$ and $(L_{\perp}\sqrt{B})_{\odot} = 1.5\times10^{5}$~m~$\sqrt{\rm T}$ (van der Holst et al.~\citeyearads{2014ApJ...782...81V}, \citeyearads{2022ApJ...925..146V}).  
\item[$\sharp$] This model employed a Dynamo-generated magnetic field distribution as boundary condition from the EULAG-MHD code \citepads{2019ApJ...880....6G}.  
\item[*] Different ZDI maps were employed. The listed $\dot M$ value corresponds to the average of all models.
\item[$**$] The $(L_{\perp}\sqrt{B})_{\bigstar}$ value was not specified in the publication.
\item[$\parallel$] The $(L_{\perp}\sqrt{B})_{\bigstar}$ value listed in \citetads{2026A&A...705A..40B} has an unit conversion error (Priv. Comm). The correct value has been added to the table.
\item[$\S$] \textbf{References:} ---1: \citetads{2016ApJ...820L..15D} ---2: \citetads{2016ApJ...832..145R} ---3: \citetads{2020A&A...633A..48F} ---4: \citetads{2019ApJ...881..136S} ---5: \citetads{2015MNRAS.449.4117V} ---6: \citetads{2023MNRAS.524.5060C} ---7: \citetads{2024ApJ...971..153X} ---8: \citetads{2026A&A...707L..12B} ---9: \citetads{2023MNRAS.524.2042E} ---10: \citetads{2026A&A...705A..40B} ---11: \citetads{2016A&A...594A..95A} ---12: \citetads{2021ApJ...916...96A} ---13: \citetads{2021MNRAS.506.2309E} 
\end{tablenotes} 
\end{threeparttable}
\end{table*}
%

\begin{figure}[p!]
    \centering
    \includegraphics[width=0.9\linewidth]{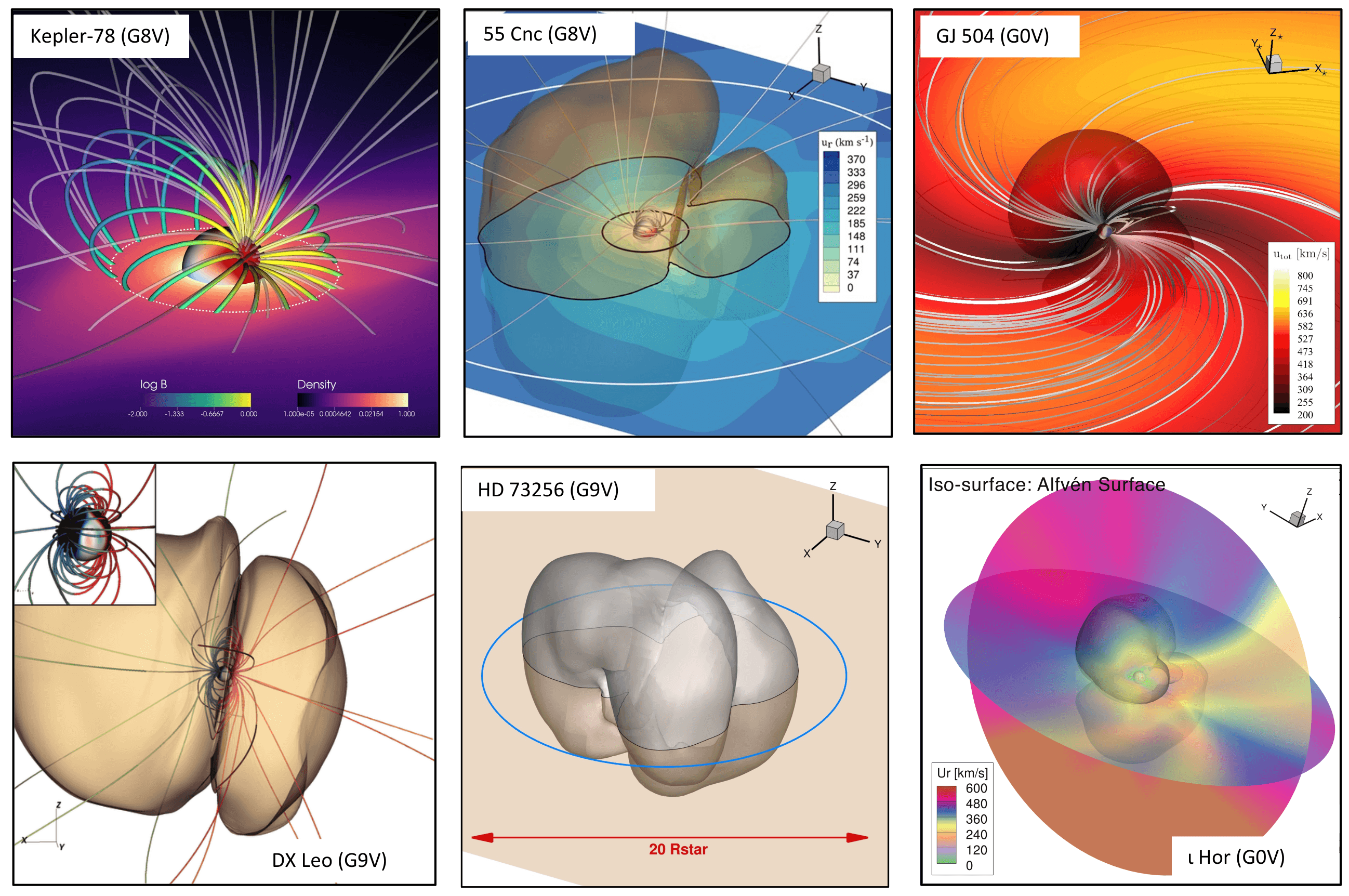}
   \caption{
Stellar wind steady-state solutions for G-type stars obtained using polytropic and Alfv\'en wave-driven models.
\textbf{First column (polytropic; WindPredict).}
\textit{Top:} Kepler-78 (G8V; \citeads{2019ApJ...881..136S}). The equatorial plane is color-coded by wind density, with magnetic field lines shown in grey. The stellar surface is colored by the radial magnetic field (blue/red for negative/positive polarity). The planetary orbit is indicated by a dashed circle, with selected star–planet connecting field lines highlighted.
\textit{Bottom:} DX~Leo (G9V; \citeads{2016ApJ...832..145R}). Steady-state solution showing selected magnetic field lines and the Alfv\'en surface (translucent beige). The surface radial magnetic field is mapped onto the star (red/blue for polarity), and the Alfv\'en surface reflects the large-scale magnetic topology, including dipolar and higher-order components.
\textbf{Second column (polytropic; BATS-R-US).}
\textit{Top:} 55~Cnc (G8V; \citeads{2020A&A...633A..48F}). The equatorial plane is color-coded by radial wind velocity, with the Alfv\'en surface shown as a translucent structure. Open magnetic field lines are shown in grey and closed field lines in red. Planetary orbits are indicated in the equatorial plane.
\textit{Bottom:} HD~73226 (G9V; \citeads{2015MNRAS.449.4117V}). The Alfv\'en surface (grey) exhibits irregular, asymmetric structure due to the observed magnetic field distribution. The equatorial plane (xy) and its intersection with the AS (thin black contour) and planetary
orbit (thick blue contour) are shown.
\textbf{Third column (Alfv\'en wave-driven; AWSoM).}
\textit{Top:} GJ~504 (G0V; \citeads{2026A&A...707L..12B}). The equatorial plane is color-coded by radial wind velocity, with the Alfvén surface shown as a translucent, approximately dipolar structure.
\textit{Bottom:} $\iota$~Hor (G0V; \citeads{2024ApJ...971..153X}). The Alfv\'en surface is represented as a translucent isosurface, with slices of radial wind velocity shown in the $X$–$Y$ and $Y$–$Z$ planes.
In all panels, the star is located at the center of the computational domain, and the rotation axis is aligned with the $z$-axis. \textbf{Note:} Visualizations are adapted from the referenced works and may differ slightly from the originals for consistency and clarity.
}
    \label{G_star_mosaic}
\end{figure}

\begin{table*}[p!]
\centering
\begin{threeparttable}
\caption{Summary of 3D Stellar Wind Models for K-type Stars}\label{tab:placeholder3}
\begin{scriptsize}
 \begin{tabular}{l c c c c c c c}
  \hline
  & & & & & & & \\[-0.2cm]
  \textbf{Star (S. Type)} & \textbf{Model} & \multicolumn{3}{c}{\textbf{Input Parameters}\tnote{$\dagger$}} & \textbf{Magnetic} & $\mathbf{\dot M}$ & \textbf{Ref.\tnote{$\S$}}\\
  & (Polytropic) & $\gamma$ & $n_0$ [cm$^{-3}$] & $T_0$ [K]& \textbf{Boundary} & \textbf{[}$\mathbf{M_\odot\,{\rm \textbf{yr}\mathbf{^{-1}}}}$\textbf{]} &\\[0.1cm]
  \hline\\[-0.25cm]
AB Dor (K0V)          & BATS-R-US & 1.10  & $2.0\times10^{8}$ & $5.0\times10^{6}$ & ZDI & $4.5 \times 10^{-13}$ & 1\\
---                   & BATS-R-US & 1.10  & $1.0\times10^{9}$ & $5.0\times10^{6}$ & ZDI & $2.1 \times 10^{-12}$ & 1\\
---                   & BATS-R-US & 1.10  & $1.0\times10^{10}$ & $5.0\times10^{6}$ & ZDI & $1.1 \times 10^{-11}$ & 1\\
HD 130322 (K0V)       & BATS-R-US & 1.10  & $1.0\times10^{9}$ & $2.0\times10^{6}$ & ZDI & $5.8 \times 10^{-13}$ & 2\\
HD 46375 (K1IV)       & BATS-R-US & 1.10  & $1.0\times10^{9}$ & $2.0\times10^{6}$ & ZDI & $1.9 \times 10^{-13}$ & 2\\
TYC 51640567-1 (K1V)  & WindPredict & 1.05 & $2.9\times10^{8}$ & $1.8\times10^{6}$ & ZDI & $5.5\times10^{-14}$ & 3\\
HD 189733~(K2V)       & BATS-R-US & 1.10 & $1.0\times10^{9}$ & $5.0\times10^{6}$ & ZDI & $8.3 \times 10^{-14}$ & 4\\
---                   & BATS-R-US & 1.10 & $1.9\times10^{9}$ & $2.0\times10^{6}$ & ZDI & $4.5 \times 10^{-13}$ & 5\\
---                   & BATS-R-US & 1.01 & $1.0\times10^{8}$ & $3.0\times10^{6}$ & ZDI\tnote{*} &  $4.1 \times 10^{-14}$ & 6\\
---                   & BATS-R-US & 1.10 & $1.0\times10^{10}$ & $2.0\times10^{6}$ & ZDI\tnote{*} &  $3.0 \times 10^{-12}$ & 7\\
---                   & WindPredict & 1.05 & $1.8 \times 10^{10}$ & $2.0\times10^{6}$& ZDI & $8.7 \times 10^{-12}$ & 8\\
---                   & WindPredict & 1.05 & $2.6 \times 10^{9}$ & $2.0\times10^{6}$& ZDI & $1.2\times 10^{-12}$ & 8\\
---                   & WindPredict & 1.05 & $1.0 \times 10^{9}$ & $2.0\times10^{6}$& ZDI & $4.4\times 10^{-13}$ & 8\\
---                   & WindPredict & 1.05 & $2.4\times10^{7}$ & $1.6\times10^{6}$ & ZDI & $1.5 \times 10^{-15}$ & 9\\
---                   & WindPredict & 1.05 & $9.9\times10^{9}$ & $2.0\times10^{6}$ & ZDI & $1.5 \times 10^{-12}$ & 9\\
BD-16351 (K5V)        & BATS-R-US & 1.05 & $1.0\times10^{8}$ & $1.5\times10^{6}$& ZDI& $8.9\times 10 ^{-14}$ & 7\\

\hline\\[-0.2cm]
  \textbf{Star (S. Type)} & \textbf{Model} & \multicolumn{3}{c}{\textbf{Input Parameters}\tnote{$\ddagger$}} & \textbf{Magnetic} & $\mathbf{\dot M}$ & \textbf{Ref.\tnote{$\S$}}\\
  & (AW-driven) & & $(S_{\rm A}/B)_{\bigstar}$ & $(L_{\perp}\sqrt{B})_{\bigstar}$ & \textbf{Boundary} & \textbf{[}$\mathbf{M_\odot\,{\rm \textbf{yr}\mathbf{^{-1}}}}$\textbf{]} &\\
  & & & [W m$^{-2}$ T$^{-1}$] & [m $\sqrt{\rm T}$] & & &\\[0.1cm]
  \hline\\[-0.2cm]
HD 130322 (K0V)              &  AWSoM & & $1.1\times10^{6}$ & $1.5\times10^{5}$ & ZDI & $1.6 \times 10^{-14}$ & 10\\
BD-07 2388 (K0V)             &  AWSoM & & $1.1\times10^{6}$ & $1.5\times10^{5}$ & ZDI & $5.5\times 10 ^{-13}$ & 11\\
Cl Melotte 25-5~(K0V)        & AWSoM & & $1.1\times10^{6}$ & $1.5\times10^{5}$ & ZDI & $4.9 \times 10^{-14}$ & 12\\
Cl Melotte 25-43~(K0V)       & AWSoM & & $1.1\times10^{6}$ & $1.5\times10^{5}$ & ZDI & $3.2 \times 10^{-14}$ & 12\\
Cl Melotte 25-179~(K0V)      & AWSoM & & $1.1\times10^{6}$ & $1.5\times10^{5}$ & ZDI & $6.5 \times 10^{-14}$ & 12\\
V439 And~(K0V)               & AWSoM & & $1.1\times10^{6}$ & $1.5\times10^{5}$ & ZDI & $5.2 \times 10^{-14}$ & 13\\
HD~6569~(K1V)                & AWSoM & & $1.1\times10^{6}$ & $1.5\times10^{5}$ & ZDI & $7.7 \times 10^{-14}$ & 10\\
---                          & AWSoM & & $1.1\times10^{6}$ & $1.5\times10^{5}$ & ZDI & $1.2\times10^{-13}$ & 11\\
BD-16 351 (K1V)              & AWSoM & & $1.1\times10^{6}$ & $1.5\times10^{5}$ & ZDI & $3.8\times10^{-13}$ & 11\\
TYC 5164-567-1 (K1V)         & AWSoM & & $1.1\times10^{6}$ & $1.5\times10^{5}$ & ZDI & $4.8\times10^{-13}$ & 11\\
HIP 10272 (K1V)              & AWSoM & & $1.1\times10^{6}$ & $1.5\times10^{5}$ & ZDI &$8.1\times10^{-14}$ & 11\\
EP Eri (K1V)                 & AWSoM & & $1.1\times10^{6}$ & $1.5\times10^{5}$ & ZDI &$3.3\times10^{-14}$ & 13\\
V447 Lac (K1V)               & AWSoM & & $1.1\times10^{6}$ & $1.5\times10^{5}$ & ZDI &$3.6\times10^{-14}$ & 13\\
PW And (K2V)                 & AWSoM & & $1.1\times10^{6}$ & $1.5\times10^{5}$ & ZDI & $4.0\times10^{-13}$ & 11\\

$\varepsilon$~Eri~(K2V)   & AWSoM & & $1.1\times10^{6}$ & $1.5\times10^{5}$ & ZDI & $4.7 \times 10^{-14}$ & 10\\
HD 189733~(K2V)              & AWSoM & & $1.1\times10^{6}$ & $1.5\times10^{5}$ & ZDI & $1.6 \times 10^{-14}$& 6\\
---                          & AWSoM & & $1.1\times10^{6}$ & $1.5\times10^{5}$ & ZDI & $1.4 \times 10^{-13}$& 10\\
Cl Melotte 25-151~(K2V)      & AWSoM & & $1.1\times10^{6}$ & $1.5\times10^{5}$ & ZDI & $3.9 \times 10^{-14}$ & 12\\
Cl Melotte 111 AV 523~(K2V)  & AWSoM & & $1.1\times10^{6}$ & $1.5\times10^{5}$ & ZDI & $3.0 \times 10^{-14}$ & 13\\
Cl Melotte 22 PELS 31 (K2V)  & AWSoM & & $1.1\times10^{6}$ & $1.5\times10^{5}$ & ZDI & $2.7 \times 10^{-13}$ & 11\\
HD 22049 (K2V) & AWSoM & & $1.1\times10^{6}$ & $1.5\times10^{5}$ & ZDI & $2.8\times 10^{-14}$ & 14\\
HD 219134~(K3V)              & AWSoM & & $1.1\times10^{6}$ & $1.5\times10^{5}$ & ZDI & $1.8 \times 10^{-14}$& 10\\
LO Peg (K3V)                 & AWSoM & & $1.1\times10^{6}$ & $1.5\times10^{5}$ & ZDI & $3.5\times 10 ^{-13}$ & 11\\
TYC 6878-0195-1~(K4V)        & AWSoM & & $1.1\times10^{6}$ & $1.5\times10^{5}$ & ZDI & $1.0 \times 10^{-13}$  & 10\\
---                          & AWSoM & & $1.1\times 10^{6}$ & $1.5\times10^{5}$ & ZDI & $3.5 \times 10^{-13}$ & 11\\
TYC 0486-4943-1 (K4V)        & AWSoM & & $1.1\times 10^{6}$ & $1.5\times10^{5}$ & ZDI & $1.4\times 10^{-13}$ & 11\\
61 Cyg A~(K5V)               & AWSoM & & $1.1\times 10^{6}$ & $1.5\times10^{5}$ & ZDI & $3.1 \times 10^{-14}$ & 10\\
HIP 76768 (K5V)              & AWSoM & & $1.1\times 10^{6}$ & $1.5\times10^{5}$ & ZDI & $4.8\times10^{-13}$ & 11\\
HIP 12545~(K6V)              & AWSoM & & $1.1\times 10^{6}$ & $1.5\times10^{5}$ & ZDI & $1.3 \times 10^{-13}$& 10\\
---                          & AWSoM & & $1.1\times 10^{6}$ & $1.5\times10^{5}$ & ZDI & $3.5 \times 10^{-13}$ & 11\\
TYC 6349-0200-1 (K7V)        & AWSoM & & $1.1\times 10^{6}$ & $1.5\times10^{5}$ & ZDI & $9.7 \times 10^{-14}$ & 10\\
---                          & AWSoM & & $1.1\times 10^{6}$ & $1.5\times10^{5}$ & ZDI & $6.3\times 10^{-13}$ & 11\\
\hline
\end{tabular}
\end{scriptsize}
\begin{tablenotes}
\footnotesize
\item[$\dagger$] Typical parameters for 3D polytropic solar wind models are $\gamma = 1.05$, $n_{0} = 1.0\times10^8$~cm$^{-3}$ and $T_{0} = 2\times10^{6}$~K \citepads{2017ApJ...835..220C}.
\item[$\ddagger$] The solar-validated AWSoM parameters correspond to $(S_{\rm A}/B)_{\odot} = 1.1\times10^{6}$~W m$^{-2}$ T$^{-1}$ and $(L_{\perp}\sqrt{B})_{\odot} = 1.5\times10^{5}$~m~$\sqrt{\rm T}$ (van der Holst et al.~\citeyearads{2014ApJ...782...81V}, \citeyearads{2022ApJ...925..146V}).  
\item[*] Different ZDI maps were employed. The listed $\dot M$ value correspond to the average of all models.
\item[$\S$] \textbf{References:} ---1: \citetads{2010ApJ...721...80C} ---2: \citetads{2015MNRAS.449.4117V} ---3: \citetads{2016ApJ...832..145R} ---4: \citetads{2011ApJ...733...67C} ---5: \citetads{2013MNRAS.436.2179L} ---6: \citetads{2017ApJ...835..220C} ---7: \citetads{2019MNRAS.485.4529K} ---8: \citetads{2014sf2a.conf..279S} ---9: \citetads{2022MNRAS.512.4556S} ---10: \citetads{2023MNRAS.524.5060C} ---11: \citetads{2023MNRAS.524.2042E} ---12: \citetads{2021MNRAS.506.2309E} ---13: \citetads{2022MNRAS.510.5226E} ---14: \citetads{2016A&A...594A..95A}
\end{tablenotes} 
\end{threeparttable}
\end{table*}



\begin{figure}[p!]
    \centering
    \includegraphics[width=\linewidth]{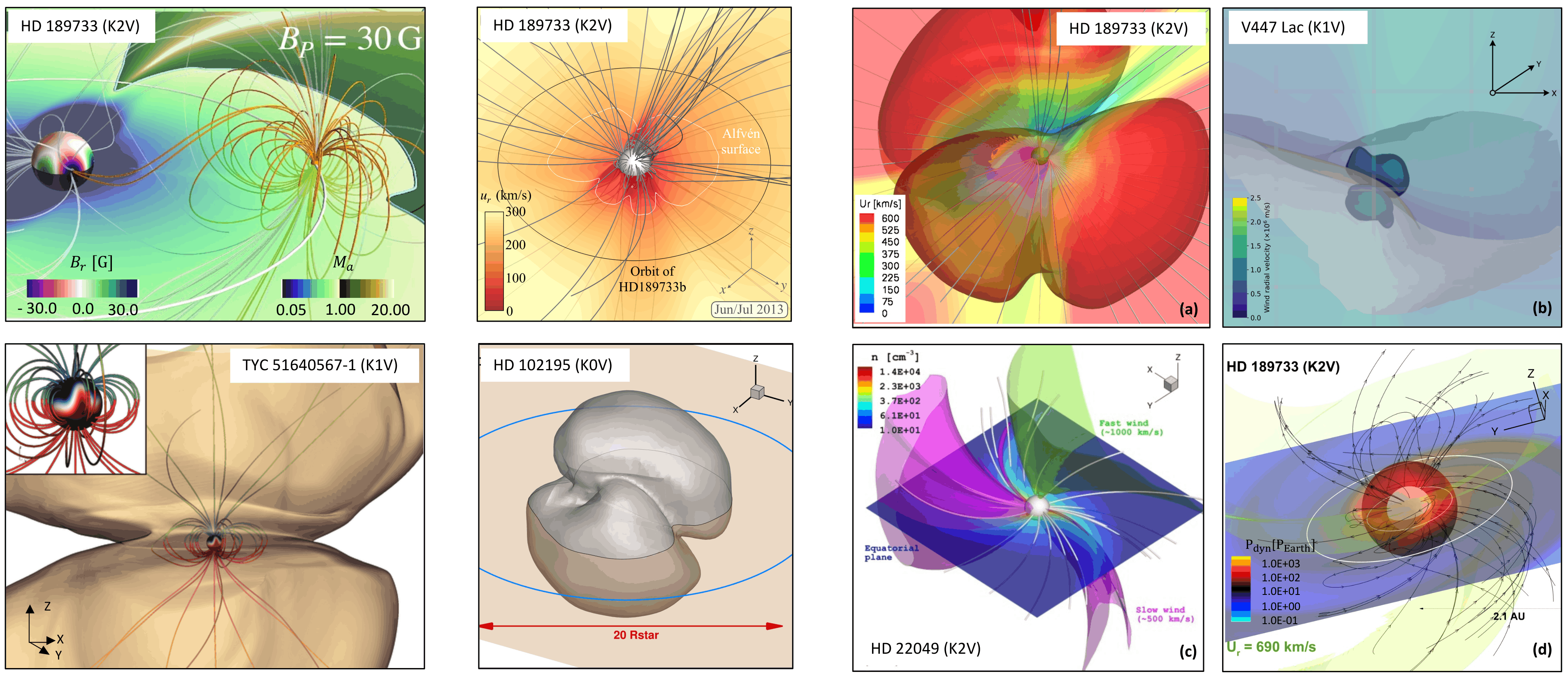}
    \caption{
Stellar wind steady-state solutions for K-type stars obtained using polytropic and Alfv\'en wave-driven models.
\textbf{First column (polytropic; WindPredict).}
\textit{Top:} HD 189733 (K2V) from \citetads{2022MNRAS.512.4556S}. Three-dimensional rendering of the corona and stellar wind (model HJ07, $\alpha\beta$-NP). Grey tubes represent magnetic field lines. The Alfv\'enic Mach number ($M_{\mathrm{A}}$) is shown in the orbital plane (log scale), with blue tones for $M_{\mathrm{A}}<1$ and yellow tones for $M_{\mathrm{A}}>1$. The surface radial magnetic field is shown in green (negative $B_r$) and red (positive $B_r$). The planetary orbit is indicated by the white circle. A hypothetical planetary magnetic field (30 G at the surface), added in post-processing, is shown in orange and does not affect the MHD solution.
\textit{Bottom:} TYC 5164-0567-1 (K1V) from \citetads{2016ApJ...832..145R}. Steady-state 3D solution showing the Alfvén surface (beige). Surface radial magnetic field and polarity are shown (red: positive, blue: negative). The Alfv\'en surface~(AS) reflects the dominant dipolar structure. A zoom highlights small-scale surface features.
\textbf{Second column (BATS-R-US).}
\textit{Top:} HD 189733 (K2V) from \citetads{2019MNRAS.485.4529K}. Stellar wind simulation showing large-scale magnetic field structure (grey lines) and radial wind velocity ($u_r$) in the orbital plane. The planetary orbit (8.8 $R_\star$) is shown in black, and the AS in white.
\textit{Bottom:} HD 102195 (K0V) from \citetads{2015MNRAS.449.4117V}. The Alfv\'en surface (grey) exhibits irregular, asymmetric structure due to the observed magnetic field distribution. The equatorial plane (xy) and its intersection with the AS (thin black contour) and planetary orbit (thick blue contour) are shown.
\textbf{Third and fourth columns (AWSoM).}
\textit{(a)} HD 189733 (K2V) from \citetads{2017ApJ...835..220C}. Wind speed in the equatorial plane; AS is shown as a translucent red isosurface.
\textit{(b)} V447 Lac (K1V) from \citetads{2022MNRAS.510.5226E}. The rotation axis ($\hat{z}$) is indicated. The current sheet ($B_r=0$) appears as a translucent grey surface. The plane of the sky (xz) and AS are colored by radial wind velocity.
\textit{(c)} HD 22049 (K2V) from \citetads{2016A&A...594A..95A}. Inner heliosphere simulation driven by ZDI maps. The white sphere marks the boundary with the SC domain (25 $R_\star$). Density is shown in the equatorial plane, along with radial velocity structure (fast: green; slow: magenta) and selected magnetic field lines (white).
\textit{(d)} HD 189733 (K2V) from \citetads{2023MNRAS.524.5060C}. Multidomain wind solution extending to 1200 $R_\star$. The green isosurface represents averaged wind velocity at 1 au. Dynamic pressure ($P_{\mathrm{dyn}}=\rho u^2$), normalized to the Sun–Earth value ($\sim$1.5 nPa), is shown in the equatorial plane and on a translucent sphere (0.5 au). Magnetic field lines are shown in black. White circles indicate habitable zone limits, and the translucent sphere marks the coupling region (67 $R_\star$). \textbf{Note:} The visualizations are adapted from the referenced works, but may differ slightly from the original figures for consistency and clarity.
}
    \label{K_star_mosaic}
\end{figure}

\subsubsection{K-type stars.}\label{sec:SWModels_K}  



Three-dimensional polytropic wind models have been applied across a broad range of K-type stellar subtypes, from K0V through K5V, using either \textsc{BATS-R-US} or \textsc{WindPredict}. The adopted base thermodynamic parameters are generally close to solar-like values, with $\gamma \approx 1.05$--$1.10$, $n_0 \sim 10^{8}$--$10^{10}$~cm$^{-3}$, and $T_0 \sim 1.5$--$5.0\times10^{6}$~K. The resulting mass-loss rates span nearly four orders of magnitude, from $\sim 10^{-15}$ to $\sim 10^{-11}~M_\odot {\rm yr}^{-1}$, with the largest values found for the very active K0V star AB~Dor under increasing base density assumptions \citepads{2010ApJ...721...80C}. By contrast, more moderate cases such as HD~189733, TYC~5164-567-1, and BD-16351 yield $\dot{M}$ values in the $10^{-14}$--$10^{-13}~M_\odot~{\rm yr}^{-1}$ range (\citeads{2011ApJ...733...67C}, \citeads{2016ApJ...832..145R}, \citeads{2017ApJ...835..220C}, \citeads{2019MNRAS.485.4529K}).

Similarly, published AWSoM models adopt the same solar-calibrated wave-energy parameters across sub-spectral types (K0V to K7V). Compared with the polytropic cases, the AWSoM mass-loss rates are more uniformly distributed and typically fall between $\sim 10^{-14}$ and a few $\times 10^{-13}~M\odot~{\rm yr}^{-1}$, although stronger winds are reported for the most active stars, such as LO~Peg, HIP~76768, $\varepsilon$~Eri and TYC~6349-0200-1, which reach several $10^{-13}~M_\odot~{\rm yr}^{-1}$ (\citeads{2016A&A...594A..95A}, Evensberget et al.~\citeyearads{2021MNRAS.506.2309E}, \citeyearads{2022MNRAS.510.5226E}, \citeyearads{2023MNRAS.524.2042E}). The lowest values are found for relatively quiet or less magnetically extreme systems, including HD~130322, HD~219134, and 61~Cyg~A, where $\dot{M}$ remains near or below $10^{-14}$--$10^{-13}~M_\odot~{\rm yr}^{-1}$ (\citeads{2023MNRAS.524.5060C}). 

Overall, the majority of ZDI-driven 3D stellar wind models of main-sequence K-type stars yield $\dot{M}$ values between $10^{-12}$ to $10^{-14}~M_\odot~{\rm yr^{-1}}$.

\subsubsection{M-type stars.}\label{sec:SWModels_M} 


\begin{figure}[p!]
    \centering
    \includegraphics[width=1\linewidth]{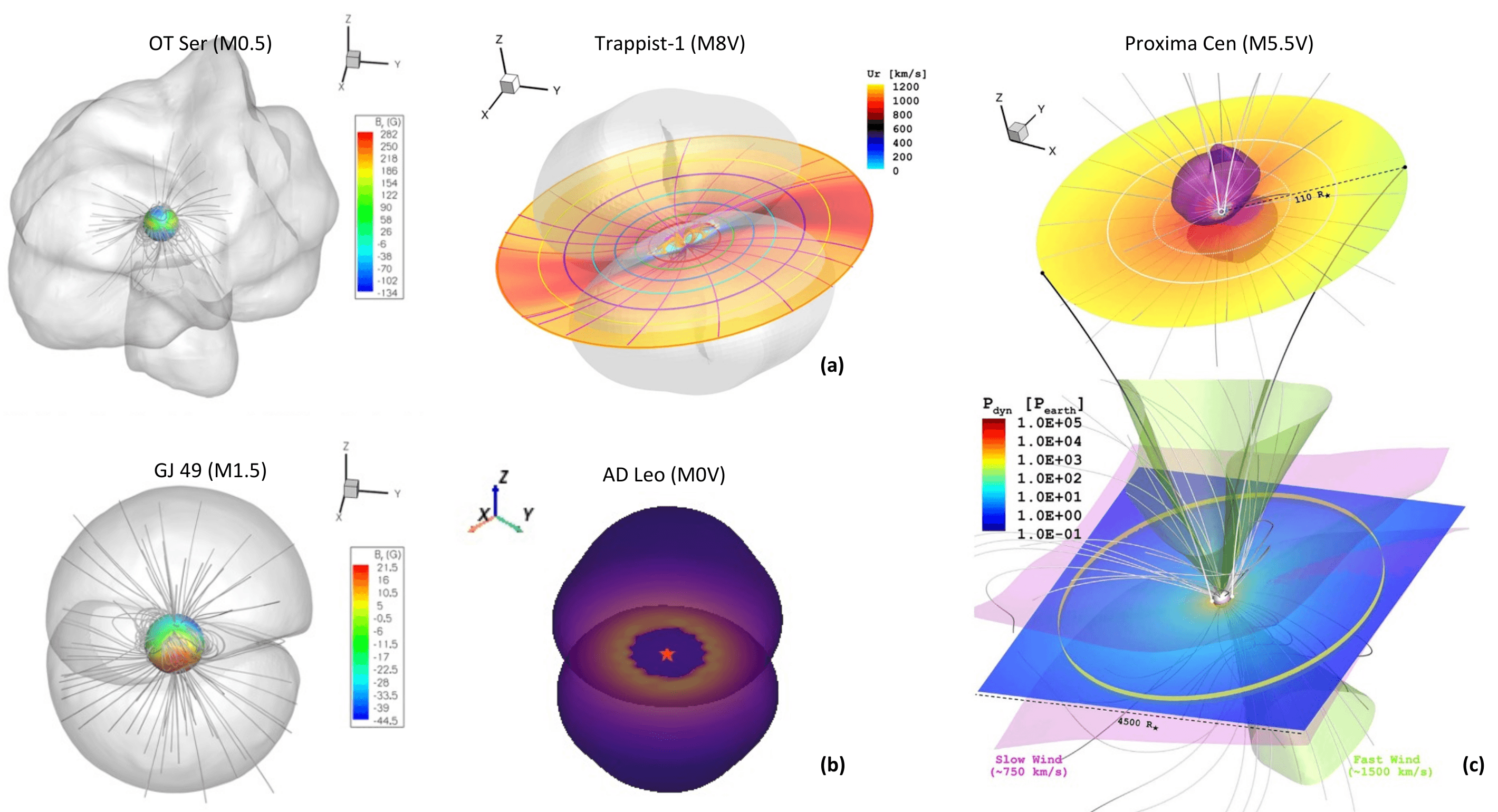}
\caption{
Stellar wind steady-state solutions for M-type stars obtained using different wind models.
\textbf{First column (polytropic; BATS-R-US).}
Adapted from \citetads{2014MNRAS.438.1162V}. 
\textit{Top:} OT~Ser (M0.5V). 
\textit{Bottom:} GJ~49 (M1.5V). 
The panels show the relaxed magnetic field configuration after convergence of the wind solution. The stellar surface is overplotted with observationally reconstructed magnetic field maps (ZDI), used as boundary conditions for the radial magnetic field. The Alfv\'en surface~(AS) is shown in grey and exhibits irregular, asymmetric shapes reflecting the complex surface magnetic field distribution.
\textbf{Second and third columns (Alfv\'en wave-driven; AWSoM).}
\textit{(a)} TRAPPIST-1 (M8V; \citeads{2017ApJ...843L..33G}). Three-dimensional stellar wind and magnetosphere computed using a proxy magnetogram (GJ~3622, $\sim$600~G average field strength). The orbital plane (top) and two meridional cuts ($x=0$, $y=0$) are shown. The wind speed is color-coded, and AS appears as a translucent grey structure. Selected closed (orange) and open (purple) magnetic field lines are shown, with planetary orbits indicated.
\textit{(b)} AD~Leo (M0V; \citeads{2026A&A...707A.397S}). The Alfvén surface is shown as a purple iso-surface.
\textit{(c)} Proxima~Centauri (M5.5V; \citeads{2020ApJ...902L...9A}). The stellar corona domain includes the orbits of Proxima~b (solid white) and Proxima~d (dashed white), while the extended heliospheric domain includes Proxima~c (yellow). The Alfv\'en surface ($M_{\mathrm{A}}=1$) is shown as a purple iso-surface. Wind dynamic pressure ($P_{\mathrm{dyn}}$) is color-coded in the equatorial plane, normalized to the Sun–Earth value. Iso-surfaces indicate slow ($u_r < 750$~km~s$^{-1}$) and fast ($u_r > 1500$~km~s$^{-1}$) wind regions. Selected magnetic field lines are shown in white.
In all panels, the star is located at the center of the computational domain, and the rotation axis is aligned with the $z$-axis. \textbf{Note:} Visualizations are adapted from the referenced works and may differ slightly from the originals for consistency and clarity.
}
    \label{fig:M_star_model}
\end{figure}

\noindent The set of polytropic stellar wind simulations for M dwarfs is characterised by significantly more extreme base conditions than those typically adopted for solar-like stars, with densities reaching $n_0 \sim 10^{10}$--$10^{12}$~cm$^{-3}$ and temperatures around $T_0 \sim 2\times10^{6}$~K, alongside slightly higher polytropic indices $\gamma \approx 1.1$--$1.2$. All models employ ZDI magnetic boundary conditions within the \textsc{BATS-R-US} framework (no M-dwarf 3D models have been published with the \textsc{WindPredict} code so far), and consistently predict very strong stellar winds, with mass-loss rates spanning $\sim 10^{-11}$ up to $\sim 10^{-10}~M_\odot~{\rm yr}^{-1}$ (Vidotto et al.~\citeyearads{2011MNRAS.412..351V}, \citeyearads{2014MNRAS.438.1162V}). Studies on early-M dwarfs such as DT~Vir, DS~Leo, and GJ~182 exhibit $\dot{M} \gtrsim 10^{-11}~M_\odot~{\rm yr}^{-1}$, while the rapidly rotating and highly magnetized V374~Peg reaches extreme values approaching $\sim 5\times10^{-10}~M_\odot~{\rm yr}^{-1}$ under high-density or high-temperature base conditions \citepads{2011MNRAS.412..351V}. These results highlight the strong dependence of mass loss on assumed coronal base density and temperature, as well as on the large-scale magnetic topology in fully convective or near-convective stellar regimes.

In contrast, published AWSoM models for M dwarfs adopt solar-calibrated Alfvén wave input parameters, $(S_{\rm A}/B)_{\odot}$ and $(L_{\perp}\sqrt{B})_{\odot}$, while exploring variations around these values by factors of a few. The resulting mass-loss rates are substantially lower and span a wide range from $\sim 10^{-15}$ to $\sim 10^{-13}~M\odot~{\rm yr}^{-1}$ for most stars, including systems such as GJ~205, TOI~700, EV~Lac, and Barnard’s Star (\citeads{2019ApJ...875L..12A}, \citeads{2020ApJ...897..101C}, \citeads{2023MNRAS.524.5060C}). More active stars, such as AU~Mic, AD~Leo, and YZ~CMi, can reach enhanced values up to $\sim 10^{-13}$--$10^{-11}~M_\odot~{\rm yr}^{-1}$ depending on the adopted wave energy flux and transverse correlation length (\citeads{2021MNRAS.504.1511K}, \citeads{2022ApJ...928..147A}, \citeads{2026A&A...707A.397S}). Additionally, the use of dynamo-generated or proxy magnetic maps (e.g. for Proxima~Cen or TRAPPIST-1) introduces further variability in $\dot{M}$ predictions (Garraffo et al.~\citeyearads{2016ApJ...833L...4G}, \citeyearads{2017ApJ...843L..33G}, \citeads{2020ApJ...902L...9A}). Overall, compared to the polytropic approach, AWSoM-based models yield systematically lower mass-loss rates for M dwarfs which in turn appear more consistent with the current observational constraints for this spectral type (see Section~\ref{sec:Wind_Overview}).

\begin{table*}[p!]
\centering
\begin{threeparttable}
\caption{Summary of 3D Stellar Wind Models for M dwarf Stars}\label{tab:MstarWinds}
\begin{scriptsize}
 \begin{tabular}{l c c c c c c c}
  \hline
  & & & & & & & \\[-0.2cm]
  \textbf{Star (S. Type)} & \textbf{Model} & \multicolumn{3}{c}{\textbf{Input Parameters}\tnote{$\dagger$}} & \textbf{Magnetic} & $\mathbf{\dot M}$ & \textbf{Ref.\tnote{$\S$}}\\
  & (Polytropic) & $\gamma$ & $n_0$ [cm$^{-3}$] & $T_0$ [K]& \textbf{Boundary} & \textbf{[}$\mathbf{M_\odot\,{\rm \textbf{yr}\mathbf{^{-1}}}}$\textbf{]} &\\[0.1cm]
  \hline\\[-0.25cm]  
DT Vir (M0V)  & BATS-R-US & 1.1 & $1.0\times10^{11}$ & $2.0\times10^{6}$ & ZDI & $1.2\times10^{-11}$ & 1 \\
DS Leo (M0)   & BATS-R-US & 1.1 & $1.0\times10^{11}$ & $2.0\times10^{6}$ & ZDI & $7.8\times10^{-12}$ & 1 \\
GJ 182 (M0.5) & BATS-R-US & 1.1 & $1.0\times10^{11}$ & $2.0\times10^{6}$ & ZDI & $6.7\times10^{-11}$ & 1 \\
OT Ser (M0.5) & BATS-R-US & 1.1 & $1.0\times10^{11}$ & $2.0\times10^{6}$ & ZDI & $1.9\times10^{-11}$ & 1 \\
GJ 49 (M1.5)  & BATS-R-US & 1.1 & $1.0\times10^{11}$ & $2.0\times10^{6}$ & ZDI & $7.2\times10^{-12}$ & 1\\
CE Boo (M1.5) & BATS-R-US & 1.1 & $1.0\times10^{11}$ & $2.0\times10^{6}$ & ZDI & $1.0\times10^{-11}$ & 1 \\
V374 Peg (M4) & BATS-R-US & 1.2 & $1.0\times10^{10}$ & $2.0\times10^{6}$ & ZDI & $4.2\times10^{-11}$ & 2\\
---           & BATS-R-US & 1.2 & $1.0\times10^{11}$ & $2.0\times10^{6}$ & ZDI & $1.4\times10^{-10}$ & 2\\
---           & BATS-R-US & 1.2 & $1.0\times10^{12}$ & $2.0\times10^{6}$ & ZDI & $5.0\times10^{-10}$ & 2\\
---           & BATS-R-US & 1.2 & $1.0\times10^{11}$ & $1.0\times10^{7}$ & ZDI & $2.6\times10^{-10}$ & 2\\
\hline
  \textbf{Star (S. Type)} & \textbf{Model} & \multicolumn{3}{c}{\textbf{Input Parameters}\tnote{$\dagger$}} & \textbf{Magnetic} & $\mathbf{\dot M}$ & \textbf{Ref.\tnote{$\S$}}\\
  & (AW-driven) &  & $(S_{\rm A}/B)_{\bigstar}$ & $(L_{\perp}\sqrt{B})_{\bigstar}$ & \textbf{Boundary} & \textbf{[}$\mathbf{M_\odot\,{\rm \textbf{yr}\mathbf{^{-1}}}}$\textbf{]} &\\
  & & & [W m$^{-2}$ T$^{-1}$] & [m $\sqrt{\rm T}$] & & &\\[0.1cm]
  \hline\\[-0.2cm] 
DT Vir (M0V)     & AWSoM & & $1.1\times10^{6}$ & $1.5\times10^{5}$ & ZDI & $2.1\times10^{-14}$ & 3 \\
AU Mic (M1V)     & AWSoM & & $1.1\times10^{6}$ & $1.5\times10^{5}$ & ZDI & $5.4\times10^{-13}$ & 4 \\
---              & AWSoM & & $6.0\times10^{5}$ &  $1.5\times10^{5}$ & ZDI & $1.2\times10^{-11}$ & 4 \\
---     & AWSoM & & $1.1\times10^{6}$ & $6.0\times10^{5}$ & ZDI\tnote{$*$} & $1.5\times10^{-13}$ & 5 \\
GJ 205 (M1.5V)   & AWSoM & & $1.1\times10^{6}$ & $1.5\times10^{5}$ & ZDI & $1.4\times 10^{-14}$  & 3 \\
TOI~700 (M2.0V)   & AWSoM & & $1.1\times10^{6}$ & $1.5\times10^{5}$ & ZDI\tnote{$\aleph$} & $1.0\times 10^{-14}$  & 6 \\
GJ 436 (M2.5)    & AWSoM & & $1.1\times10^{5}$ & $1.5\times10^{5}$ & ZDI & $1.1\times10^{-15}$ & 7 \\
---              & AWSoM & & $1.1\times10^{6}$ & $1.5\times10^{5}$ & ZDI & $2.5\times10^{-14}$ & 7 \\
EV Lac (M3.5V)   & AWSoM & & $1.1\times10^{6}$ & $1.5\times10^{5}$ & ZDI & $1.1\times 10^{-15}$ & 3 \\
---              & AWSoM & & $7.7\times10^{5}$ & $1.5\times10^{7}$ & ZDI & $3.0\times 10^{-14}$ & 8\\
AD Leo  (M3.5)   & AWSoM & & $7.7\times10^{5}$ & $1.5\times10^{5}$ & ZDI\tnote{*} & $4.0\times10^{-13}$ & 9\\
Barnard's Star (M4.0V)   & AWSoM & & $1.1\times10^{6}$ & $6.0\times10^{5}$ & ZDI\tnote{*,$\aleph$} & $1.5\times10^{-15}$ & 10 \\
YZ CMi (M4.5V)   & AWSoM & & $1.1\times10^{6}$ & $1.5\times10^{5}$ & ZDI & $4.1\times10^{-13}$ & 3\\
Prox Cen (M5.5V) & AWSoM & & $1.1\times10^{6}$ & $6.0\times10^{5}$ & ZDI\tnote{$\aleph$} & $1.5\times10^{-14}$ & 11 \\
---              & AWSoM & & $1.1\times10^{6}$ & $6.0\times10^{5}$ & Dynamo\tnote{$\sharp$} \,(min) & $6.0\times10^{-15}$ & 12\\
---              & AWSoM & & $1.1\times10^{6}$ & $6.0\times10^{5}$ & Dynamo\tnote{$\sharp$} \,(max) & $1.8\times10^{-14}$ & 12\\
--- & AWSoM & & $0.5\times10^{6}$ & $1.5\times10^{5}$ & ZDI & $5.0\times10^{-15}$ & 4 \\
GJ 1245 B (M6V)  & AWSoM & & $1.1\times10^{6}$ & $1.5\times10^{5}$ & ZDI & $3.6\times 10^{-15}$  & 3\\
Trappist-1 (M8V) & AWSoM & & $7.7\times10^{5}$ & $6.0\times10^{5}$ & ZDI\tnote{$\aleph$} & $3.0\times10^{-14}$ & 13 \\ 
--- & AWSoM & & $1.1\times10^{6}$ & ---$^{**}$ & Solar$^\beth$ & $2.0\times10^{-15}$ & 14\\ 
\hline
\end{tabular}
\end{scriptsize}
\begin{tablenotes}
\footnotesize
\item[$\dagger$] Typical parameters for 3D polytropic solar wind models are $\gamma = 1.05$, $n_{0} = 1.0\times10^8$~cm$^{-3}$ and $T_{0} = 2\times10^{6}$~K \citepads{2017ApJ...835..220C}.
\item[$\dagger$] The solar-validated AWSoM parameters correspond to $(S_{\rm A}/B)_{\odot} = 1.1\times10^{6}$~W m$^{-2}$ T$^{-1}$ and $(L_{\perp}\sqrt{B})_{\odot} = 1.5\times10^{5}$~m~$\sqrt{\rm T}$ (van der Holst et al.~\citeyearads{2014ApJ...782...81V}, \citeyearads{2022ApJ...925..146V}).  
\item[$\sharp$] These models employed a Dynamo-generated magnetic field distributions as boundary condition from the MagIC code \citepads{2016ApJ...833L..28Y}   
\item[*] Different ZDI maps were employed. The listed $\dot M$ value correspond to the average of all models.
\item[$\aleph$] The ZDI map used was the one from another star of similar spectral type.
\item[$\beth$] The magnetic map used was a scaled solar synoptic map.
\item[$**$] The $(L_{\perp}\sqrt{B})_{\bigstar}$ value was not specified in the publication.
\item[$\S$] \textbf{References:} 
---1: \citetads{2014MNRAS.438.1162V}
---2: \citetads{2011MNRAS.412..351V}
---3: \citetads{2023MNRAS.524.5060C}
---4: \citetads{2021MNRAS.504.1511K}
---5: \citetads{2022ApJ...928..147A}
---6: \citetads{2020ApJ...897..101C}
---7: \citetads{2023A&A...678A.152V}
---8: \citetads{2014ApJ...790...57C}
---9: \citetads{2026A&A...707A.397S}
---10: \citetads{2019ApJ...875L..12A}
---11: \citetads{2016ApJ...833L...4G}
---12: \citetads{2020ApJ...902L...9A}
---13: \citetads{2017ApJ...843L..33G}
---14: \citetads{2018PNAS..115..260D}
\end{tablenotes} 
\end{threeparttable}
\end{table*}

\subsection{Challenges and Prospects in Stellar Wind modeling} \label{subsec:sw_challenges}

As was discussed in the preceding sections, the development of realistic global 3D MHD models for the solar wind has enabled a broad range of applications in the stellar regime. Despite methodological differences between numerical schemes, the predicted mass-loss rates for well-observed systems often fall within similar ranges. For example, both \textsc{BATS-R-US} and \textsc{WindPredict} recover $\dot{M}\sim \mathrm{few}\times10^{-12}~M_\odot~\mathrm{yr^{-1}}$ for $\tau$~Boo, while \textsc{AWSoM} and \textsc{WindPredict} yield $\dot{M}\sim10^{-12}~M_\odot~\mathrm{yr^{-1}}$ for HD~189733. However, this apparent convergence should not be interpreted as robust physical agreement. Instead, it reflects compensating assumptions across models, because the dominant source of uncertainty in all current 3D wind simulations arises from the poorly constrained inner boundary conditions---this holds true even for the Sun, whose far-sided photospheric fields are not routinely observed \citep[e.g.,][]{2012LRSP....9....6M}. Given how critical these issues are, we provide a brief discussion in the most pressing aspects below:

\smallskip

\noindent \textbf{- Magnetic boundary conditions:} In stellar wind simulations, the inner magnetic boundary is typically prescribed using ZDI maps of the photospheric field. This choice is convenient, but it is also intrinsically limited by both the method and the quality of the underlying observations. ZDI is sensitive mainly to large-scale structures and to the non-cancelled component of the surface field, implying that it recovers only a small fraction of the total photospheric magnetic flux. Depending on stellar activity, the reconstructed flux may represent merely $10\%-0.1\%$ of the actual value \citepads{2020A&A...635A.142K}. 

In addition, the spatial resolution of the resulting map depends not only on stellar parameters such as projected rotational velocity, but also on the instrument resolving power, the phase sampling, and the signal-to-noise ratio of the data \citepads{2016LNP...914..177K}. These factors also enter the iterative ZDI inversion, for which no community-wide convergence criterion exists, thereby affecting the maximum field strength retained in the final ZDI reconstruction (see \citeads{2015A&A...582A..38A} for a proposed methodology). An additional limitation arises from the unknown or only approximately known inclination of the rotation axis, which controls the fraction of the stellar surface accessible by the image reconstruction procedure. In practice, a fiducial inclination of $60^{\circ}$ is often adopted even when direct constraints are available, as in the case of AU Mic (see the discussion in \citeads{2020ApJ...902...43K} and a fiducial ZDI reconstruction in \citeads{2021MNRAS.502..188K}). 

The ZDI maps also generally give constraints on the full magnetic vector. Nevertheless, most wind models rely only on the radial component of the field, as for the Sun it is generally good enough to model the 3D corona. In the context of other stars, many ZDI maps present strong toroidal field component, which inclusion as boundary conditions require more advanced initial magnetic extrapolations. Such efforts were started in e.g. \citetads{2013MNRAS.431..528J} and \citetads{2022MNRAS.512.4556S}. Ultimately, more generic extrapolations such as non-linear force-free extrapolation developed for solar applications (e.g. \citeads{2018SSRv..214...99Y} and references therein) would be good avenues to take into account the whole 3D magnetic field constrained by ZDI in the global stellar wind models.

From a practical standpoint, the lack of openly accessible ZDI maps as machine-readable data products also hinders intercomparison between different modeling groups, since identical magnetic inputs are not readily available for shared benchmarking. Thus, although ZDI has proved robust in many applications, further progress is required before such reconstructions can be regarded as true stellar analogues of solar magnetograms, which are the standard inputs for both steady-state and time-dependent solar coronal and wind simulations \citep[e.g.,][]{2017SSRv..210..249W, 2018LRSP...15....4G}.

\smallskip

\noindent \textbf{- Thermodynamic boundary conditions:} A second major obstacle concerns the specification of realistic coronal boundary conditions for cool stars. In the solar case, coronal temperatures, densities, and heating profiles are usually constrained by an extensive set of in~situ and remote-sensing measurements. For other stars, in contrast, these quantities must be inferred indirectly. In principle, X-ray spectroscopy can constrain coronal temperature and density, but this is only feasible for a relatively small set of bright nearby targets observed with facilities such as \textit{XMM-Newton} and \textit{Chandra}. The situation is even more restrictive for M dwarfs, whose optical spectra are dominated by molecular bands and whose faintness in X-rays limits the usefulness of standard emission-measure-distribution analyses.

As noted in Section~\ref{subsec:coolstarwinds}, polytropic corona and wind models require prescribed base values for the plasma temperature and density. In \textsc{AWSoM}, these quantities are chosen to match conditions near the top of the solar chromosphere \citepads{2014ApJ...782...81V} and are further tuned against EUV imagery and measured solar wind conditions at 1~au \citepads{2023SpWea..2103262J}. Such calibration is not possible for most stars, which effectively turns the base thermodynamic parameters into free variables. In that regard, \citet{2020A&A...635A.170A} proposed a parametrization of these free variables as a function of stellar mass, radius, rotation, and ultimately metallicity to tune stellar wind models to match all the available observational constraint to date. 
Nevertheless, such parametrizations only give a somewhat broad range of acceptable parameters. Together with the limited spatial and magnetic resolution of ZDI maps, this produces substantial uncertainty in the predicted stellar wind properties. 

The consequences are clearly visible in the emission measure distributions (EMDs) produced by global models. The EMD, which is central to the interpretation of optically thin plasma emission, can be reconstructed from X-ray and ultraviolet data and then confronted with model predictions. As an example, \citetads{2020A&A...636A..49C} compared the Sun with the young solar analog $\varepsilon$ Eridani and found that, whereas the solar corona peaks near 1~MK, $\varepsilon$ Eridani reaches a substantially hotter maximum above 3 MK, with a significant amount of plasma above 10 MK according to EUVE observations. By contrast, MHD models driven by ZDI maps of this star generate an un-balanced EMD, with not enough material at these high temperatures and an over-abundance of low-temperature plasma in the coronal steady-state solution \citepads{2016A&A...588A..28A}. This situation has been attributed to the limited ZDI resolution and the way this affects the resulting coronal structure, given that the highest temperature components are expected to emerge from active regions (see Section~\ref{subsec:quiescent}) missed by the ZDI technique. Attempts to alleviate these issues via the implementation of high-resolution dynamo-generated magnetic boundary conditions have only been partially successful. Recent studies have managed to reproduce the hot X-ray emitting component of active coronae (i.e. $\log(T) \geq 6.0$) but the solutions fail systematically in the EUV domain, where the predicted EMDs can exceed the observations by several orders of magnitude (\citeads{2024ApJ...971..153X}, \citeads{2025ApJ...995...83C}). 

The diagnostic value of EMDs for improving global 3D MHD models is fundamental, especially in the stellar regime where spatially resolved constraints are unavailable. It also motivates future ``Sun-as-a-star'' validation efforts based on disk-integrated EMDs rather than resolved solar images, since this framework more closely resembles the observational situation for other stars.

\smallskip

\noindent \textbf{- Assumptions underlying quasi-steady coronal heating and wind acceleration:} In addition to the base thermodynamic parameters, 3D MHD wind models generally require an explicit prescription for coronal heating and the driving mechanism of the wind (see also Section~\ref{subsubsec:stellarwind_models}). In \textsc{AWSoM}, for example, these processes are represented through Alfv\'en-wave turbulence and its dissipation, with a relatively small number of free parameters, including the Alfv\'en-wave correlation length and the Poynting flux (see Tables~\ref{tab:FstarWinds} to \ref{tab:MstarWinds}). The most recent version of the model, informed by Parker Solar Probe measurements, introduces four additional input parameters (see \citeads{2022ApJ...925..146V}). As in the case of the base temperature and density, the corresponding wave-related quantities are not directly measurable for other stars, although some constraints may eventually be extracted from far-ultraviolet spectral diagnostics \citepads{2023ApJ...950..124B}. Because these parameters control the heating rate, the acceleration profile, and the mass flux of the outflow, even modest changes can alter the predicted $\dot{M}$ by orders of magnitude, despite identical magnetic boundary conditions. Hence, in the absence of empirical constraints, current simulations cannot yet be tuned reliably to the physical properties of individual stars.

The uncertainty is further amplified by the possibility that the dominant coronal and wind-driving mechanisms in other stars differ substantially from the solar paradigm \citepads{2023hxga.book..132D}. In highly active stars, especially M dwarfs, the coronae may be in a near-continuous flaring state, with light curves often appearing as a superposition of flares with little or no quiescent emission (e.g.~\citeads{2007A&A...471..645C}, \citeads{2010ApJ...723.1558H}). In such cases, the assumption that Alfv\'en waves provide the principal source of coronal heating and wind acceleration becomes questionable, and additional energy input from transient phenomena may need to be incorporated.

\smallskip

\noindent \textbf{- Observational constraints of stellar winds:} Finally, the comparison between model predictions and observations is far more indirect for stellar winds than for the solar wind. In the heliosphere, model validation can be performed against individual observables such as density, velocity, and magnetic-field strength, which allows different physical components of the solution to be tested separately and modified when necessary. For other stars, however, available wind constraints are extremely sparse (Section~\ref{sec:SWinds}). In most cases, only mass-loss estimates are obtainable, generally through the product of density and speed, with no direct information on the separate density, velocity, or magnetic-field structure of the flow. 

Moreover, while solar wind models can be confronted with measurements taken at multiple heliocentric distances, most stellar wind constraints are inferred near the astrosphere--ISM interface, for example from the hydrogen wall method (Section~\ref{sec:Ly-a_Astrospheres}). Such diagnostics are affected by large uncertainties and degeneracies, and they probe only the far outer astrosphere. As a consequence, they provide little to no direct leverage on the physical processes that determine the wind characteristics. In their current form, therefore, stellar wind observations impose only limited constraints on theoretical models of outflows from cool stars.

\section{Coronal Mass Ejections (CMEs)}\label{CMEs}
The study of stellar CMEs is primarily rooted in our detailed understanding of solar eruptive phenomena. Accordingly, we first provide a brief review of CME observations in the solar system in both the remote-sensing and in-situ regimes, together with their impacts on planetary environments (Section~\ref{CME:Overview}). We then move onto describing the observational techniques used to detect and characterize stellar transients (Section~\ref{CME:Obs}), and how long-term solar monitoring has enabled empirical characterization of their occurrence rates and connection to stellar activity cycles (Section~\ref{CME:Empirical}). Finally, we discuss numerical models that have been developed to simulate CME initiation and evolution from the low corona into interplanetary space, capturing their interaction with the ambient wind (Section~\ref{CME:Models}). Building on the solar foundation, we review current approaches to extending CME physics to stellar regimes, with particular emphasis on the assumptions involved, the limitations imposed by the lack of direct observations, and the implications for stellar space weather and exoplanetary environments.

\subsection{Observing and Understanding CMEs: The Case of the Sun}\label{CME:Overview}

It is no surprise that the first observations of CMEs---and stellar eruptions in general---as well as the theories based on them came from the Sun. Given the wider range of data collection methods available in the solar system, CMEs and their related phenomena were initially observed at different stages of their evolution in various environments and only later recognized as manifestations of the same underlying process. For example, in-situ detections of CMEs in interplanetary space began in the 1960s--1970s with near-Earth spacecraft via the IMP satellite series as well as interplanetary missions such as Mariner, Pioneer, Voyager, and Helios \citep[e.g.,][]{steinolfson1975, smith1976}. On the other hand, the first remote-sensing observations of CMEs through white-light coronagraphs (such as those onboard OSO-7 and Skylab) took place in the early 1970s \citep[e.g.,][]{tousey1973, howard1982}; however, establishing a direct connection between in-situ interplanetary structures and these solar eruptions was not straightforward at first, as their solar counterparts were initially sought primarily in flares \citep{gosling1993}. It is now well known that CMEs and flares are, in fact, different expressions of the same eruption mechanism, typically described by the CSHKP scenario (see Section~\ref{sec:Rad_solarflare_mag}). In this framework, magnetic reconnection in the current sheet powers the flare energy release below the erupting structure, while simultaneously enabling the upward ejection of plasma and magnetic fields that form the CME. This close relationship is reflected in the strong statistical association between CME and flare occurrence, which increases steeply with flare class and reaches values of ${\sim}$90\% for X-class flares \citep[e.g.,][]{yashiro2005}. In this section, we will first summarize how CMEs are observed on the Sun and throughout the heliosphere (Section~\ref{CME:Over:Obs}), and then briefly review evidence of CME impact on solar system planets and other bodies (Section~\ref{CME:Over:Impact}).

\subsubsection{CME Observations through the Solar Corona and Heliosphere} \label{CME:Over:Obs}

\begin{figure*}[!t]
\centering
\includegraphics[width=0.99\textwidth]{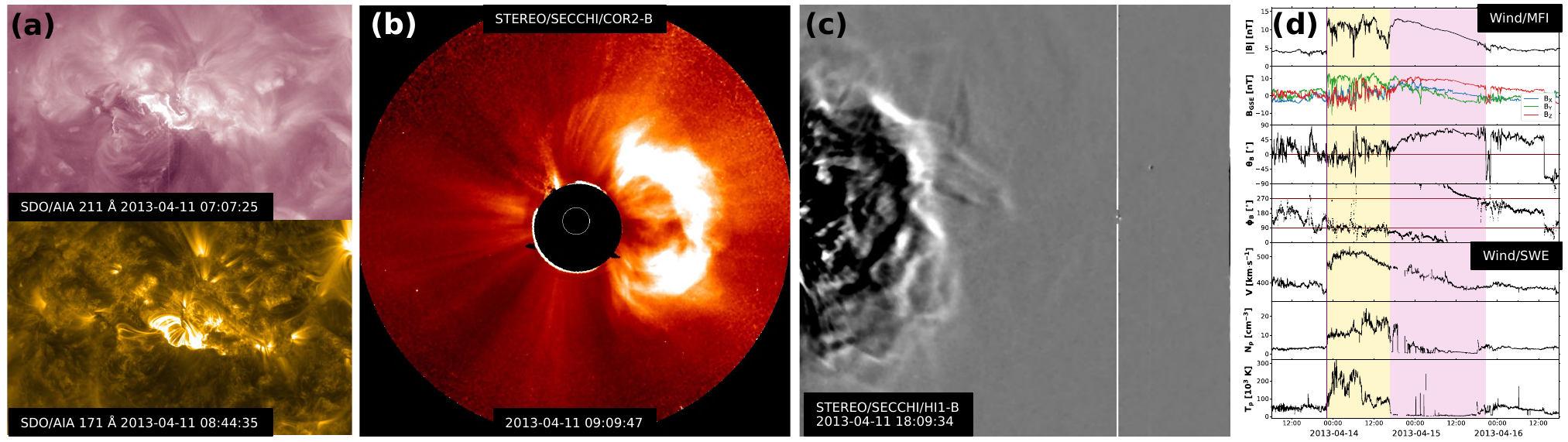}
\caption{Examples of different observational methods to detect CMEs from the Sun, illustrated using the April 11, 2013 event. (a) SDO/AIA EUV imagery of the eruption employing the 211~{\AA} (top, showing flare ribbons) and 171~{\AA} (bottom, showing post-eruptive arcades) passbands. (b) White-light coronagraph observations of the CME at ${\sim}10$\,$R_{\odot}$ as seen by STEREO/SECCHI/COR2-B. (c) White-light observations of the CME in the heliosphere as seen by STEREO/SECCHI/HI1-B. (d) The CME observed in situ by the Wind spacecraft at the Sun--Earth Lagrange L1 point. From top to bottom, the parameters shown are magnetic field magnitude, Cartesian components, and field angles by Wind/MFI as well as solar wind speed, proton density, and proton temperature from Wind/SWE. The purple vertical line marks the arrival of the CME-driven shock, while the yellow and pink shaded regions indicate the passage of the sheath region and magnetic ejecta, respectively.} \label{fig:solarcme}
\end{figure*}

The plethora of methods and signatures to observe and track solar CMEs has been reviewed extensively by \citet{webb2012} and include, as mentioned earlier, both remote-sensing and in-situ techniques (some of which are shown in Figure~\ref{fig:solarcme}). When monitoring the solar disk (usually in EUV and/or X-ray wavelengths), indications that a CME has erupted \citep[also known as low-coronal signatures; e.g.,][see also Figure~\ref{fig:solarcme}(a)]{hudson2001} include the appearance of coronal dimmings, post-eruptive arcades, flare ribbons, and EUV waves, as well as the disappearance of filament/prominence material and X-ray sigmoids. These proxies are generally understood to reflect distinct processes or outcomes within the standard CME–flare eruption scenario. Dimmings trace regions of expansion and evacuation of plasma along field lines that open up during an eruption \citep[e.g.,][]{jin2022, veronig2025}, arcades represent the manifestation of magnetic reconnection in the wake of the eruption \citep[e.g.,][]{tripathi2004, gopalswamy2017}, ribbons correspond to the footpoints of newly reconnected flux tubes in the overlying arcade \citep[e.g.,][]{masson2009, kazachenko2017}, waves are interpreted as low-coronal tracers of large-scale shock fronts \citep[e.g.,][]{zhukov2004, veronig2010}, rising filaments correspond to the destabilization of cool, dense plasma embedded in the CME \citep[e.g.,][]{parenti2014, mierla2022}, and sigmoids are considered signatures of the presence of a pre-eruptive flux rope configuration \citep[e.g.,][]{green2007, james2017}.

Coronagraph observations of CMEs \citep[e.g.,][see also Figure~\ref{fig:solarcme}(b)]{howard2023} are widely used to track their early evolution in the corona. In white-light imagery, CMEs appear as bright enhancements against the fainter ambient corona and are often accompanied by a diffuse outer envelope, which is commonly interpreted as the shock driven ahead of the eruption. The CME ejecta itself can assume a myriad of morphologies \citep[e.g.,][]{vourlidas2013} due to both the intrinsic nature and structure of the eruption \citep[e.g.,][]{vourlidas2017} as well as projection effects \citep[e.g.,][]{cremades2004, howard2017}. Generally, a CME that displays in coronagraph imagery the ``classic'' three-part structure \citep{illing1985}---consisting of a bright front, a dark cavity, and a bright core---is interpreted as possessing a well-defined flux rope configuration. White-light observations of CMEs and of the Sun's environment at large are not limited to the corona; heliospheric imagers extend these measurements into the inner heliosphere, enabling continuous tracking of CME evolution over large heliocentric distances \citep[e.g.,][see also Figure~\ref{fig:solarcme}(c)]{harrison2018}. However, these observations are subject to more pronounced projection effects than coronagraph data, owing to the large elongation angles involved, which introduce substantial line-of-sight ambiguities in the interpretation of CME kinematics and morphology \citep[e.g.,][]{barnes2019, barnes2020}. CMEs impacting a spacecraft can be identified via a number of signatures in in-situ measurements of magnetic field, plasma, and particles \citep[e.g.,][see also Figure~\ref{fig:solarcme}(d); note that any given event may display only a sub-set of signatures]{zurbuchen2006}. Events that display an enhanced magnetic field magnitude, smoothly rotating magnetic field vectors, a decreasing speed profile, as well as depressed temperature and plasma beta are usually classified as magnetic clouds \citep{burlaga1981} and are understood as a clear indicator of an underlying flux rope configuration \citep[e.g.,][]{kilpua2017}.

More indirect observables that can be leveraged to track and analyze CMEs in the solar corona and interplanetary space include radio type II bursts that provide information on shock speed and electron acceleration \citep[e.g.,][]{gopalswamy2001, morosan2020}, interplanetary scintillation data for 3D density and velocity reconstructions due to radio waves propagating through a fluctuating solar wind density \citep[e.g.,][]{lynch2002, fallows2023}, Faraday rotation measurements that remotely probe CME magnetic fields along the line of sight \citep[e.g.,][]{jensen2010, howard2016}, spectroscopic Doppler diagnostics of outflow speed and heating \citep[e.g.,][]{raymond2002, giordano2013}, SEP information to evaluate particle acceleration at CME-driven shocks as well as CME internal topology \citep[e.g.,][]{malandraki2002, dresing2025}, and cosmic-ray modulation such as Forbush decreases \citep[e.g.,][]{cane2000, dumbovic2020}. Together, the ensemble of direct and indirect observational methods described above has shaped our current understanding of CME structure and evolution, revealing both event-to-event variability and the factors that govern their propagation. It is now well established that CMEs do not all erupt in the same way \citep[e.g.,][]{green2018, patsourakos2020}, reflecting differences in their initial magnetic and plasma configurations. Their evolution is strongly influenced by the surrounding corona and solar wind, with interactions within the structured heliosphere---including CME--CME encounters \citep[e.g.,][]{lugaz2017}---producing deflections, rotations, deformations, and reconfigurations of their magnetic fields \citep[e.g.,][]{manchester2017, luhmann2020}. Moreover, the geoeffectiveness or impact of a CME at a given planetary or interplanetary body depends not only on its intrinsic properties but also on the characteristics of the impacted environment, as will be discussed in the next section.

\subsubsection{CME Impact on Solar System Bodies} \label{CME:Over:Impact}

The effects of CMEs on planets and other solar system bodies are not determined solely by the intrinsic properties of the CME, but are rather critically governed by the interaction between the CME and the environment of the impacted body. As a result, a single CME event can elicit profoundly different impacts at distinct objects, reflecting the diversity of their local magnetospheric and plasma conditions \citep[e.g.,][]{mostl2015, palmerio2022}. Overall, solar wind--planetary environment interactions can be broadly classified into three categories. The first involves bodies with an intrinsic magnetosphere, in which the planet generates a global magnetic field via dynamo processes (e.g., Earth). The second comprises induced magnetospheres, where the planet lacks a significant intrinsic magnetic field and the interaction is governed primarily by the solar wind and ionospheric currents (e.g., Venus). The third category includes hybrid magnetospheres, which exhibit characteristics of both intrinsic and induced systems, often due to localized crustal magnetic fields (e.g., Mars). CME-driven space weather effects have been extensively investigated across the solar system, with documented impacts at Mercury \citep[e.g.,][]{slavin2019, sun2020}, Venus \citep[e.g.,][]{luhmann2007, collinson2015}, Earth \citep[e.g.,][]{pulkkinen2007, zhang2007}, Mars \citep[e.g.,][]{jakosky2015, lee2018}, Jupiter \citep[e.g.,][]{prange2004, dunn2016}, Saturn \citep[e.g.,][]{palmerio2021, cecconi2022}, Uranus \citep[e.g.,][]{lamy2012, lamy2018}, and even comets \citep[e.g.,][]{kuchar2008, edberg2016}.

At planets with intrinsic magnetospheres, such as Earth and Jupiter, CMEs can couple directly to the local space environment and trigger magnetic disturbances---however, the magnitude and character of these interactions vary substantially across the solar system: smaller, inner planets such as Mercury and Earth can experience stronger magnetospheric responses, while the giant outer planets such as Jupiter and Saturn are comparatively less sensitive to individual CME events due to their prominent internal plasma dynamics. For bodies lacking strong intrinsic magnetic fields, such as Venus and Mars, CMEs interact primarily with the upper atmosphere and ionosphere, driving phenomena such as increased ionospheric currents, atmospheric escape, and diffuse and/or patchy auroras \citep[e.g.,][]{2014Icar..233..342G, 2018NatAs...2..802D}. Comets, with their extended plasma and neutral envelopes, offer further opportunities to observe dramatic CME-induced interactions, such as tail disconnection and subsequent regrowth \citep[e.g.,][]{shaik2026}. The study of CME-driven space weather across the solar system is complicated by the fact that there is no single, uniform set of instruments monitoring all planets and small bodies at all times. Consequently, the effects of CMEs have been analyzed using a variety of techniques, including in-situ plasma and magnetic field measurements, as well as remote-sensing observations, particularly of auroral emissions in XUV wavelengths and of kilometric radiation in radio \citep[e.g.,][]{lilensten2014}. These observational strategies provide an important framework for studying CME effects in exoplanetary systems, where in-situ measurements are infeasible. In particular, remote-sensing observations throughout the solar system demonstrate the diagnostic power of auroral monitoring to infer the properties of stellar wind interactions and magnetospheric dynamics, suggesting that exoplanet aurorae may serve as observable indicators of CME-driven space weather in distant planetary systems.

\subsection{Observations of stellar CMEs: Methodologies and Results}\label{CME:Obs}

The Kepler and TESS missions have greatly improved our understanding of the stellar flaring activity (see Section~\ref{sec:Rad_Stellar_Flares}). Observations show that young stars, cool M-dwarfs, and close-binary stars tend to produce more frequent and energetic flares, suggesting that active stars may also experience more frequent and/or very large CMEs/eruptions\footnote{Here we intentionally distinguish between the terms ``eruption" and ``CME." When referring to a ``CME," we mean the entire structure including the core, cavity, and front, whereas the term “eruption” is often used specifically for the core component, such as filament or prominence eruptions. In stellar observations, detections are more commonly associated with the core, which differs from the CME front in velocity. For this reason, maintaining this distinction is particularly important in the present section.}. 
However, until recently, it has been unclear whether active stars indeed experience them. 
Many active stars have different environments from those of the Sun \citep[see][in this collection]{Chapter1}, such as magnetic field strength, magnetic structure, rotation rate, and stellar wind conditions. 
These differences could either prevent CMEs from taking place in the first place, or modify their appearance in observational data (see Section~\ref{sec:CME-Suppression}), therefore, the presence of large flares alone may not guarantee that stellar CMEs can be detected in the same way as solar CMEs (see review in Section~\ref{CME:Overview}). 
This makes direct observations necessary to confirm whether stellar CMEs occur at all and whether they can be identified using solar-based techniques. Here, we briefly review previous efforts to detect stellar CMEs in Section~\ref{CME:Obs:method}, their properties in Section~\ref{CME:Obs:property}, and frequencies in Section~\ref{CME:Obs:freq}.

\subsubsection{Methodologies and Detections of Stellar CMEs/Eruptions}\label{CME:Obs:method}


Observing stellar flares is relatively straightforward due to their intrinsically high intensities, whereas, detecting stellar CMEs is much more difficult and uncommon due to their faint nature. 
In the solar case, CMEs are observed through coronagraphs (see Section~\ref{CME:Over:Obs}), which block the million-times brighter emission of the solar photosphere in order to make the faint CME plasma visible by measuring the Thomson scattering of photospheric light on coronal electrons.
Unlike the Sun, distant stars cannot be spatially resolved with present-day instruments, which makes it nearly impossible to directly capture the release and propagation of CMEs. 
Despite these challenges, researchers have sought to apply solar knowledge to search for indirect or direct evidence of stellar CMEs, as reviewed in Section~\ref{CME:Overview}.
Table~\ref{tab:methods-CME} summarizes the methods that have been explored to detect stellar CMEs/eruptions.
Among these, the most successful approaches so far in terms of numbers are detecting Doppler shifts caused by CME-related plasma motions and identifying coronal dimming signatures associated with coronal mass depletion. 
Recent reviews on the challenges and methods for the detection of stellar CMEs can be found in \cite{2022SerAJ.205....1L}, \cite{2024Univ...10..313V}, and \cite{veronig2025}.

\begin{table}[!t]
\centering
\begin{tabular}{lllllll}
\hline 
Methods & Wavelength & Sun & Stars &\textbf{Number$^\ddagger$} & \textbf{Ref.} \\
\hline
\hline
Doppler Shift & Optical/UV/X-ray & Yes & Yes & $\ge$100$^\dagger$ & [1] \\
\hline
Coronal Dimming & X-ray/EUV/FUV & Yes & Yes & 23 & [2]  \\
\hline
Absorption Dimming & X-ray & ? & Yes & 7 & [3]  \\
\hline
Type IV Radio Burst & Radio & Yes & Yes & 2 & [4]  \\
\hline
Type II Radio Burst & Radio & Yes & Yes & 2 & [5]  \\
\hline
Moving Radio Source & Radio & ? & ?$^{(\rm A)}$ & ? & [6]  \\
\hline
Pre-flare Dips & Optical/UV & Yes & ?$^{(\rm A)}$ & ? & [7]  \\
\hline
CME-related SPI & Radio/UV/Optical & Yes & ?$^{(\rm B)}$ & ? & [8]  \\
\hline
%
\end{tabular}
\caption{List of successful or tested stellar CME detection methods.  
$^{(\rm A)}$Reported but not well-justified.
$^{(\rm B)}$Attempted but not detected.
$^\ddagger$Limited to events reported before October 31, 2025. $^\dagger$Snapshot blueshifts have been many reported by \cite{2019A&A...623A..49V}. 
References: [1] e.g.,
\cite{1990A&A...238..249H};
\cite{1994A&A...285..489G};
\cite{1999MNRAS.305...45M};
\cite{2004A&A...420.1079F};
\cite{2016A&A...590A..11V};
\cite{2018PASJ...70...62H};
\cite{2019NatAs...3..742A};
\cite{2020PASJ..tmp..253M};
\cite{2022NatAs...6..241N};
\cite{2023ApJ...948....9I};
\cite{2024ApJ...961...23N}; \cite{2024ApJ...961..189N}; and many snapshot data from \cite{2019A&A...623A..49V}
[2] \cite{2021NatAs...5..697V} ; \cite{2022ApJ...936..170L}; \cite{2024ApJ...961...23N}, [3] \cite{2017ApJ...850..191M}; \cite{2019ApJ...877..105M}, 
[4] \cite{2020ApJ...905...23Z}; \cite{2024A&A...686A..51M}, 
[5] \cite{2018ApJ...862..113C,2018ApJ...856...39C}; \cite{2025A&A...703A.198K, 2025Natur.647..603C}
[6] \cite{1999A&A...344.1014L}; \cite{1999ApJ...517L..43L}; \cite{2002ApJ...572..487R}; \cite{2024ApJ...965...86G}, 
[7] \cite{1982ApJ...252L..39G}; \cite{2014MNRAS.443..898L}
[8] \cite{2020NatAs...4..577V}
}
\label{tab:methods-CME}
\end{table}

\begin{figure*}[!t]
\centering
\vspace{-0.36cm}
\includegraphics[width=0.99\textwidth]{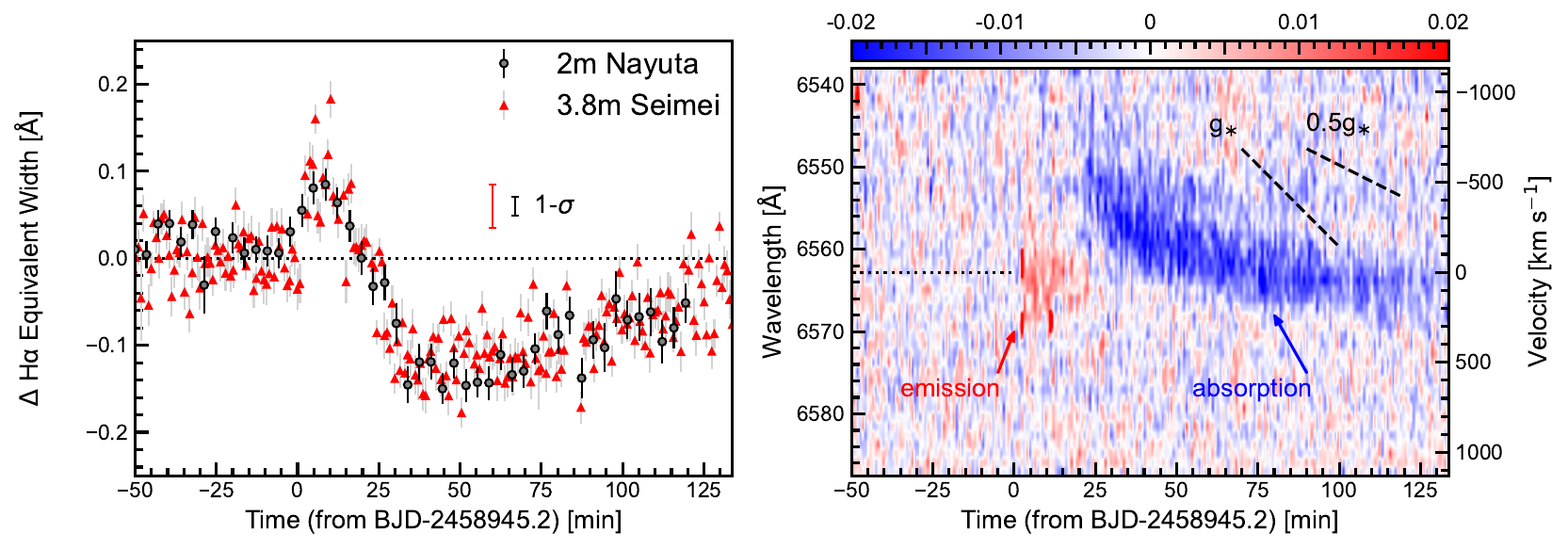}\vspace{-0.15cm}
\caption{A filament eruption on a young G dwarf, EK Draconis. The figure is adapted from \cite{2022NatAs...6..241N}. Left panel: light curve of H$\alpha$ equivalent width observed by ground-based telescopes with different colors. The dotted line represents the background level. Right panel: dynamic spectrum of the H$\alpha$ line where red represents emission and blue represents absorption compared to the pre-flare template spectrum.}\label{Fig:4-2-1}
\end{figure*}

\begin{figure}[!th]
\centering
\includegraphics[width=0.5\textwidth]{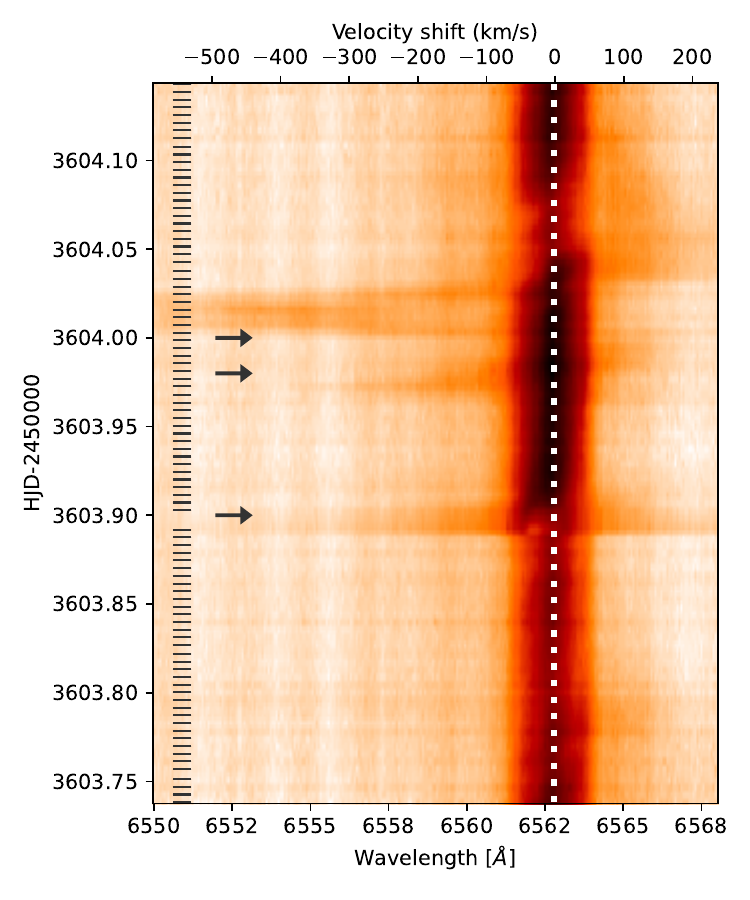}
\caption{Eruptions from an M-dwarf V374 Pegasi. The blue-wing enhancement in H$\alpha$ line are marked with arrows. Marks on the left show the time of observations \citep{2019A&A...623A..49V}.}
\label{fig:v374peg}
\end{figure}

Spectral-line Doppler shifts provide a direct diagnostic of plasma flows. The key difficulty lies in separating such shifts from other flare-related processes, such as upflows from chromospheric evaporation and downflows from chromospheric condensation and post-flare loop (coronal rain). Nevertheless, this method is still considered one of the most promising ways to trace plasma flows associated with stellar CMEs. 
Numerous reports have documented blueshifted chromospheric lines during stellar flares, primarily in the optical range \citep[e.g.,][]{1990A&A...238..249H,1994A&A...285..489G,2016A&A...590A..11V,2018PASJ...70...62H,2019A&A...623A..49V,2020PASJ..tmp..253M,2022NatAs...6..241N,2023ApJ...948....9I,2024ApJ...961...23N,2024ApJ...961..189N,2024MNRAS.tmp.1385L}, but also in the ultraviolet \citep{2007PASP..119...67H,2014MNRAS.443..898L} and X-rays \citep{2019NatAs...3..742A,2022ApJ...933...92C,2024ApJ...969L..12I}. Most of these observations target M–K dwarfs, but more recently similar signals have been found in G dwarfs \citep{2022NatAs...6..241N,2024ApJ...961...23N,2024MNRAS.tmp.1385L} and in RS CVn-type binaries \citep{2023ApJ...948....9I}.
In many studies, 
blueshifted chromospheric lines are interpreted as signatures of prominence or filament eruptions, while some authors \citep[e.g.,][]{2019A&A...623A..49V} suggest that redshifts may trace material falling back.
Figure~\ref{Fig:4-2-1} shows an example showing unambiguous detection of eruptive phenomena: a blueshifted absorption spectrum at H$\alpha$ line $6562.8$~{\AA} observed in a young G dwarf, EK Dra, providing clear evidence of a stellar filament eruption \citep{2022NatAs...6..241N}. 
Comparisons with Sun-as-a-star analyses \citep{2022NatAs...6..241N,2022ApJ...939...98O,2024A&A...682A..46P} and a simple hydrodynamic model \citep{2024ApJ...963...50I} of solar filament eruptions suggest that the G-dwarf event is likely analogous to solar ones.
In contrast to G-dwarf flares, all reported blueshift events in M dwarfs exhibit blueshifted wing enhancement rather than absorption, which have made it difficult to separate bright flaring emission, possibly cool plasma due to chromospheric evaporation, from eruptive contributions. 
Figure \ref{fig:v374peg} shows a prominent example of H$\alpha$ blueshift events from the M dwarf, V374 Peg.
The main drawback of the Doppler-shift method, in addition to uncertainties in interpretation, is its short-lived nature, which makes it difficult to determine the ultimate fate of the ejected material. Although interpretation remains somewhat uncertain, this method is most accessible observationally and offers one of the few direct ways to estimate the mass and especially the velocity of stellar eruptions, which is examined in more detail in Sections~\ref{CME:Obs:property}, \ref{CME:Obs:freq}, and \ref{CME:Empirical}.

\begin{figure*}[ht]
\centering 
\includegraphics[width=0.75\textwidth]{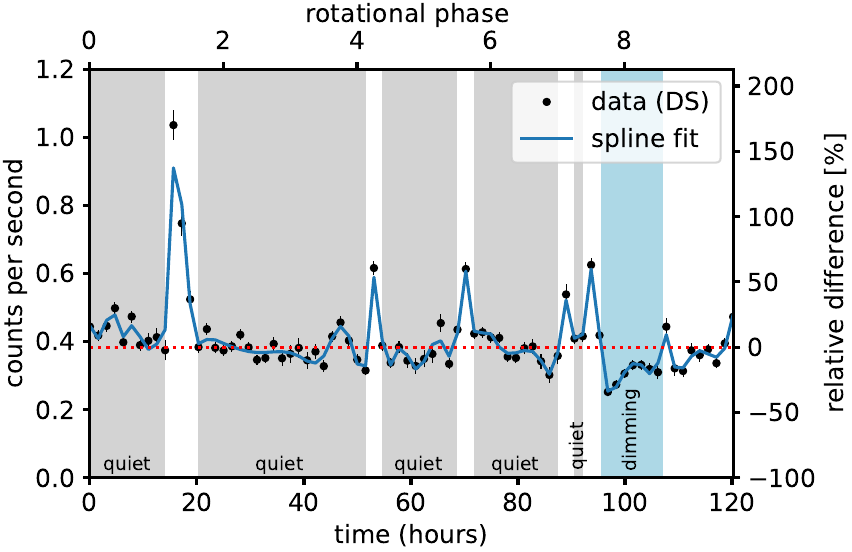} 
\vspace{-0.15cm}
\caption{Post-flare coronal dimming 
on an active K dwarf, AB Doradus. The data show an 
EUVE DS light curve in the EUV waveband. Grey-shaded areas indicate 
quiescent periods while the blue-shaded area shows the detected post-flare dimming. 
From \cite{2021NatAs...5..697V}.} \label{Fig:4-2-2}
\end{figure*}

\begin{figure*}
\centering
\includegraphics[width=0.8\textwidth]{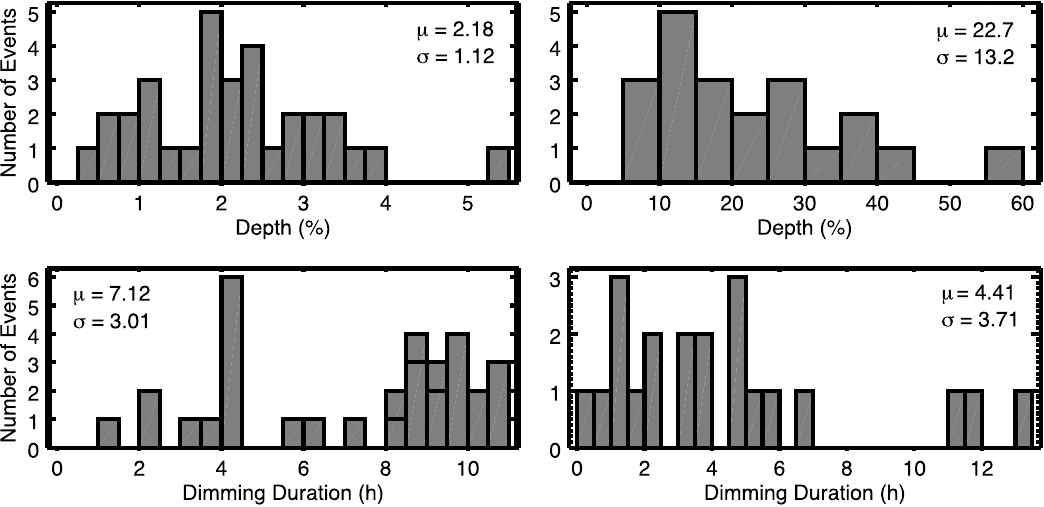} 
\caption{Distributions of dimming depth (relative to the pre-event level) and duration of post-flare coronal dimmings on the Sun (left) and late-type stars (right). The solar dimmings are derived from full-Sun SDO/EVE 15–25~nm light curves, the stellar dimmings from XMM-Newton, Chandra and EUVE. 
Adapted from \cite{2021NatAs...5..697V}.} \label{Fig:dimming2}
\end{figure*}

Reports of stellar coronal dimmings have become more frequent in recent years (see \cite{veronig2025} for a comprehensive review). On the Sun, coronal dimmings are frequently observed in association with the initiation and early evolution of CMEs. They appear as a sudden decrease of the coronal EUV (100-920 {\AA}) and soft X-ray ($<$100 {\AA}) emission, due to the density depletion and mass evacuation by the erupting structure. They are regularly observed in solar EUV and soft X-ray images, but have also been detected in full-Sun EUV observations \citep{2016ApJ...830...20M,2016SoPh..291.1761H}. \cite{2021NatAs...5..697V} found that about 80\% of CMEs associated with large solar flares reveal a significant dimming in Sun-as-a-star EUV observations, which makes them an excellent candidate for stellar CME detections. 
The first stellar coronal dimming were presented by \cite{2021NatAs...5..697V}, who analyzed archival observations of cool stars and reported post-flare dimming in X-ray and EUV bands (see Figure \ref{Fig:4-2-2}). They found 21 post-flare dimmings on 13 different stars (out of about 100 suitable data sets),
though it is important to note that most detections have so far been limited to M/K dwarfs. Comparison of the stellar dimmings with full-Sun solar dimmings showed that the stellar dimmings are about an order of magnitude stronger than the solar ones, with values of several 10\% of decrease with respect to the pre-flare emission (see Figure~\ref{Fig:dimming2}). The dimming duration is in both cases similar, ranging from 1 hour to $>$10 hours.
Subsequently, \cite{2022ApJ...936..170L} identified possible post-flare FUV dimming in the Fe~XII~1349 {\AA} (1.6 MK) and Fe~XXI~1354 {\AA} {(11 MK)} lines on the K-type star $\epsilon$ Eridani using Hubble observations.
More recently, \cite{2024ApJ...961...23N} reported a potential X-ray dimming associated with a massive prominence eruption detected in H$\alpha$ on the young G dwarf EK Draconis. 
These findings suggest that post-flare coronal dimmings are viable means for detecting stellar CMEs. 
Nevertheless, several uncertainties remain: interpretations depend sensitively on how the quiescent flux level is defined, which is itself variable over time in magnetically active stars.
The ESCAPE mission proposed to NASA will be able to improve the detection of coronal dimmings from other flare stars by providing high sensitivity spectroscopy in the EUV.

\begin{figure}
    \centering
    \includegraphics[width=0.7\linewidth]{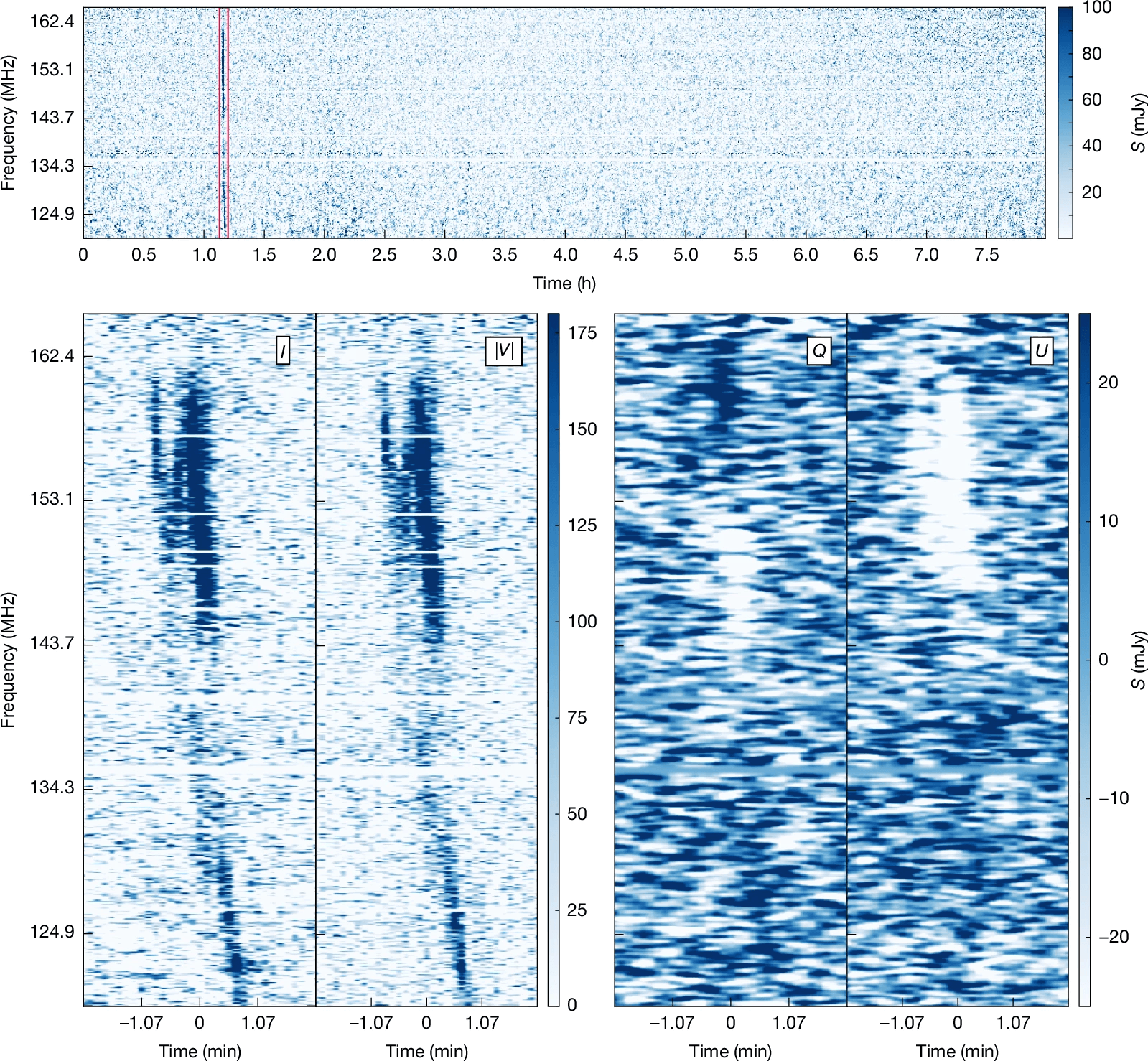}
    \caption{First detection of a Type II burst as a tracer of a stellar CME. 
    The top plot shows intensity dynamic spectrum for the entire 8\,h observation with the burst bracketed by the two red lines. 
The burst is shown in the bottom panels, with Stokes $I$, absolute $V$, $Q$ and $U$ shown from left to right, respectively. Note that the colour scale of Stokes $I$ and $V$ is different from that of Stokes $Q$ and $U$. Adopted from \cite{2025Natur.647..603C}.}
    \label{fig:typeiiburst}
\end{figure}
In addition to these strong pieces of evidence for flare-related eruptions, several further indications have been reported. In particular, the detection of radio bursts in the MHz band is considered a powerful diagnostic: Type II radio bursts probe shock waves driven ahead of a CME, while Type IV bursts trace high-energy particles trapped within a CME flux rope or flare loops. Despite intensive monitoring of flare stars \citep{2017PhDT.........8V,2018ApJ...862..113C,2018ApJ...856...39C}, Type II burst associated with a stellar coronal mass ejection were undetected for a long time, raising concerns whether shock waves fail to develop in stellar CMEs \citep{2019ApJ...873....1M}, or whether CMEs themselves are absent.
The recent detection of a Type II burst on StKM1-1262 \citep{2025Natur.647..603C}, followed by a second detection by \cite{2025A&A...703A.198K}, establish the first observational limits on the impact of CMEs on exoplanets. 
The Type II burst on StKM1-1262 lasted approximately 2 minutes, sweeping in frequency from 166 to 120~MHz. The emission exhibited identical frequency, time, and polarisation properties to fundamental plasma emission from a solar Type II burst. It was highly circularly polarised (average $\approx 90\%$) with a  median drift rate of $\approx -0.62\pm 0.22$~MHz\,s$^{-1}$. With this the derived shock velocity  at 144~MHz was $2400\pm600$~\kms. This speed is extremely fast: on the Sun, CMEs ${\geq}2400$~\kms{} are accompanied by a Type II burst in $\geq95\%$ of cases \citep{2008AnGeo..26.3033G,2008ApJ...674..560G,2023A&A...675A.102K}.
\cite{2025A&A...703A.198K} used the LOFAR Two Metre Sky Survey (LoTSS) to search all known stars within 100 parsecs  and accumulated a total of 107 years of stellar data. This survey  yielded the detection of two drifting stellar radio bursts:
the previously published burst from the M dwarf StKM1-1262 and  
a new 13-minute burst from the M dwarf LP 215-56.
The fitted luminosity distribution for the stellar Type II bursts was found to be consistent with the cumulative luminosity distribution derived from decametric 
solar Type II observations 
($\alpha=-0.81\pm0.06$).
The search also identified 19 additional circularly polarised stellar radio bursts that did not exhibit a time--frequency drift. 
Recent observations have also revealed possible Type IV radio bursts from nearby M dwarfs. For instance, \cite{2020ApJ...905...23Z} reported a candidate (possibly moving) Type IV burst from Proxima Centauri using ASKAP in Australia, which may point to the existence of a stellar CME flux rope. More recently, \cite{2024A&A...686A..51M} presented a potential Type IV burst from the active M dwarf AD Leo based on GMRT data from India. Although current detections remain rare, future observations in the MHz band with facilities such as SKA, GMRT, and LOFAR are expected to provide decisive progress.

Some other methods for stellar CME detection are also proposed as summarized in Table \ref{tab:methods-CME}. One is X-ray absorption dimming during stellar flares, considered a possible CME signature. It is inferred from spectral fits showing an increase in hydrogen column density $N_{\rm H}$, consistent with plasma obscuring the flare emission, likely due to filament eruptions \citep{1999A&A...350..900F,2017ApJ...850..191M}. Additional candidates are summarized by \cite{2019ApJ...877..105M}, though such evidence remains model- and fit-dependent and, without direct solar counterparts, can only be regarded as tentative CME signatures. Several additional CME detection techniques have been proposed—such as moving radio sources \citep{2024ApJ...965...86G}, pre-flare dips \citep{1982ApJ...252L..39G,2014MNRAS.443..898L}, and star–planet interaction signatures \citep[e.g.,][]{2020NatAs...4..577V}—which we do not discuss in depth here. Please refer to comprehensive discussions by \cite{2017IAUS..328..243O,2022SerAJ.205....1L,2022arXiv221105506N,2024Univ...10..313V,veronig2025} and the recent KISS workshop report \citep{kiss2025}.

Multi-method or multi-wavelength detections are required to further constrain the global picture of stellar CMEs and establish their occurrence conclusively. \cite{2024ApJ...961...23N} reported potential X-ray coronal dimming associated with prominence eruptions on a solar-type star. Furthermore, \cite{namekata2025natas} detected multi-temperature Doppler shifts in H$\alpha$ (${\sim}10^{4}$~K) and the FUV (${\sim}10^{4.8}$~K) from stellar flares on the same star. 
These findings indicate that a single CME event may exhibit multi-wavelength characteristics similar to those observed on the Sun.
Conversely, \cite{Notsu+2025_ApJ} reported no H$\alpha$ Doppler shift associated with the coronal dimming event reported by \cite{2021NatAs...5..697V}. In the future, increasing the number of such coordinated multi-wavelength observations will be essential {for robustly constraining CME phenomena as well as for future estimates of stellar CME occurrence rates (see Section \ref{CME:Obs:property} and \ref{CME:Obs:freq})}.



\subsubsection{Velocity, Mass and Kinetic Energy of Stellar CMEs/Eruptions}\label{CME:Obs:property}

\begin{figure*}[!t]
\centering
\vspace{-0.36cm}
\includegraphics[width=0.99\textwidth]{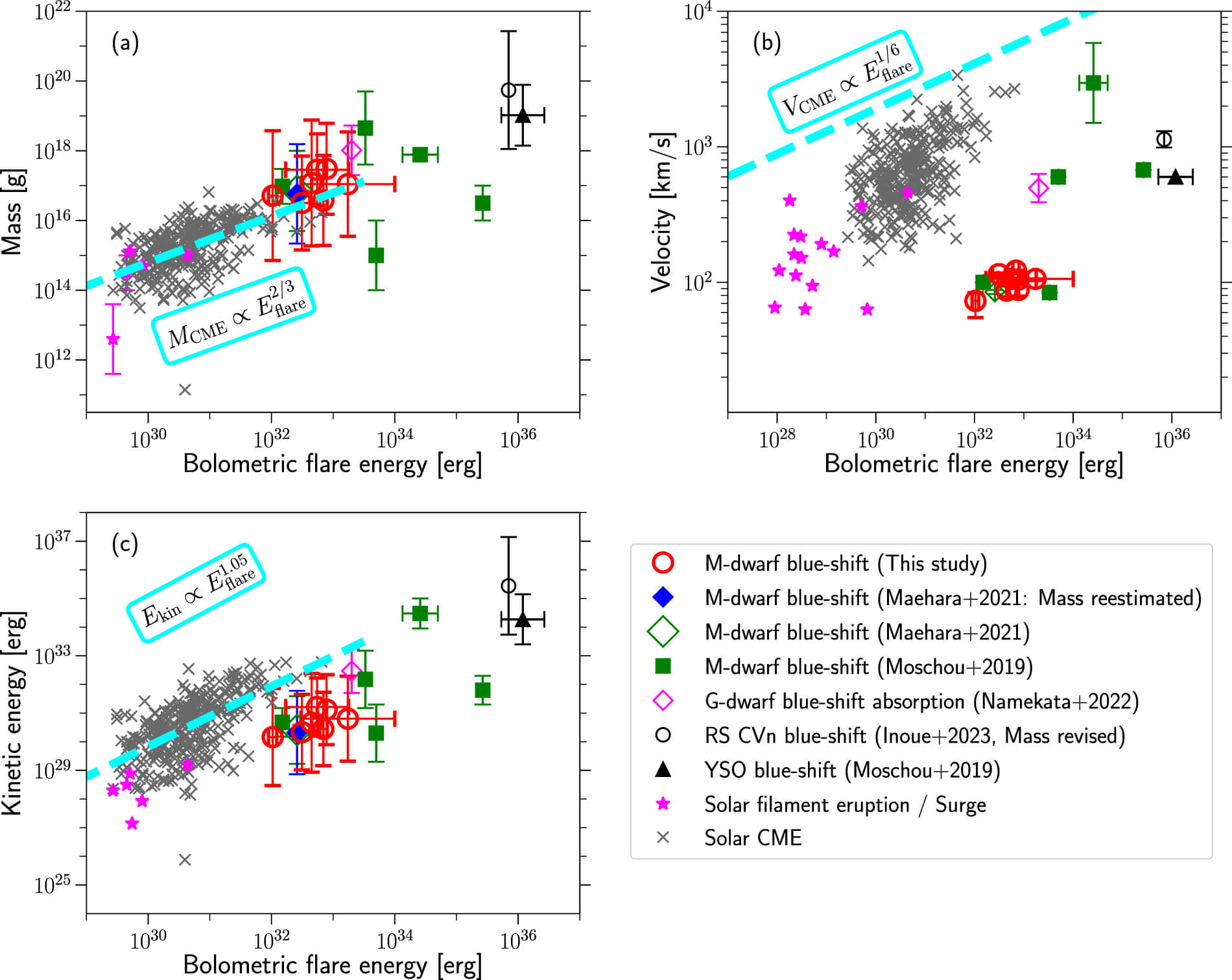}\vspace{-0.15cm}
\caption{Mass, velocity, and kinetic energy of solar eruptions/CMEs and stellar eruptions observed with Doppler method. The figure is taken from \cite{2024ApJ...961..189N}.}\label{Fig:4-2-3}
\end{figure*}

The velocities and masses of stellar CME candidates have been estimated mainly from the Doppler-shift method, with additional constraints sometimes obtained from X-ray absorption dimming. Figure~\ref{Fig:4-2-3} compiles values of velocity, mass, and kinetic energy derived from optical Doppler measurements as a function of flare energy (including only events with reported flare energies). 
An original summary was done by \cite{2019ApJ...877..105M} (see, e.g., \citealt{2024ApJ...961..189N}, for the latest statistics).

For M/G dwarfs, most reported velocities are on the order of a few hundred km~s$^{-1}$, although higher velocities have occasionally been observed \citep[see e.g.,][]{2019A&A...623A..49V,2024ApJ...961..189N,2024ApJ...961...23N}.
These speeds are comparable to, and sometimes exceed, those of solar filament or prominence eruptions (typically a few hundred km~s$^{-1}$; \citealt{2003ApJ...586..562G}), suggesting that fast events also occur on other stars. A common debate is whether such velocities, being comparable to or lower than the stellar escape velocity, imply that the eruptions fail to evolve into CMEs.
However, it should be noted that solar CMEs are often accelerated/heated during their ejection, and chromospheric lines such as the Balmer series may trace only the initial, relatively slow and cool phase of the eruption. 
Indeed, even for the Sun, the mean CME speed measured in white-light coronagraph observations ($\sim$500~km~s$^{-1}$) is slightly below the surface escape speed (617~km~s$^{-1}$; \citealt{2009EM&P..104..295G}).
Moreover, by the end of the observations, the ejected material may already have reached higher altitudes, where the local escape velocity is lower.
Taken together, these effects suggest that the actual fraction of successful CMEs could be significantly higher than what is inferred from the observed line-of-sight velocities alone.
Importantly, a few stellar events have exhibited extreme velocities of several thousand km~s$^{-1}$ \citep{2023ApJ...948....9I,2024ApJ...969L..12I}, reaching up to $\sim$5800~km~s$^{-1}$ \citep{1990A&A...238..249H}, which would almost certainly exceed stellar gravity. This suggests that at least some extreme events are likely to be associated with successful CMEs.

Mass estimates provide an additional perspective. In many studies, the masses of eruptions have been estimated \cite[][and more]{2013ApJ...764..170D,2019ApJ...877..105M,2019A&A...623A..49V,2022NatAs...6..241N,Notsu+2025_ApJ}.
For Doppler-shift events, the masses are usually estimated from excess line emission or absorption by assuming parameters such as the density, optical thickness and geometry/size. 
For absorption dimming events, they are inferred from the excess X-ray absorbing column density, \(N_{\rm H}\), by assuming an obscuring eruptive structure with a characteristic size, geometry, and expansion.
Depending on the flare magnitude, some events are found to eject masses comparable to those of solar CMEs, while eruptions associated with very energetic flares are estimated to have masses exceeding those of solar CMEs.
The gigantic eruptions have masses several orders of magnitude larger, yet they appear to follow the solar flare energy–CME mass scaling relation \cite[e.g.,][see Figure \ref{Fig:4-2-3}(a) for recent summary from \citealt{2024ApJ...961..189N}]{2012ApJ...760....9A,2013ApJ...764..170D}, regardless of CMEs and filament/prominence eruptions. 
This suggests that physics established for the Sun across a limited energy range ($\lesssim 10^{32}$ erg) may be extended to much larger energy scales in other stars, pointing to a shared underlying mechanism with stellar flares \cite[e.g.,][]{2016ApJ...833L...8T,2023ApJ...943..143K}.

Kinetic energy estimates are, in general, known to be lower than values extrapolated from the solar flare–CME relationship \cite[e.g.,][]{2019ApJ...877..105M}. 
Interpretation of this trend requires caution: as already mentioned above, in most stellar cases, the derived values correspond to the filament/prominence components, rather than the leading edge of a CME. 
Given that solar CMEs display a significant velocity contrast between the core and front \citep{2003ApJ...586..562G}, it is natural that stellar CME kinetic energies and velocities appear underestimated \citep[e.g.,][]{2022NatAs...6..241N}. 
With this background in mind, other possibility is also suggested that stellar eruptions from active stars are genuinely slowed by being suppressed by strong overlying magnetic fields \citep[e.g.,][]{2018ApJ...862...93A}. 
Future observational and theoretical studies will be essential to distinguish between these scenarios.

At present, while the velocities and masses of eruptive filaments/prominences (CME cores) are becoming better constrained, the speeds of CME fronts and their subsequent propagation and evolution remain poorly established observationally. 
Among the methods described in Section 4.2.1, the detection of coronal dimming and type-II bursts, if modeled properly, are expected to play a crucial role in estimating CME velocities and masses. 
Solar studies have demonstrated that there exist distinct correlations between dimming and CME parameters \citep{Dissauer2019}, and that CME mass can be inferred from the depth of dimming, and velocity from its slope \citep{2016ApJ...830...20M}. 
By appropriately modeling stellar coronae, it would become possible to empirically estimate CME masses and velocities. 
In addition, assuming reasonable coronal/interplanetary density profile, type-II bursts will enable us to derive CME velocities from the frequency drift. In the future, combining these methods with detailed modeling will provide a more comprehensive picture of stellar CMEs.

\subsubsection{Occurrence Rate of Stellar CMEs/Eruptions}\label{CME:Obs:freq}

Regarding occurrence rates, current observations do not yet provide strong constraints. Nevertheless, ground-based optical spectroscopic surveys have enabled intensive monitoring of individual stars and yielded empirical estimates of the frequency of eruption signatures. For example, \cite{2024ApJ...961..189N} and \cite{2025ApJ...979...93K} reported 7 and 3 eruption candidates from 41 and 27 H$\alpha$ flares (${\sim}10^{31}$--$10^{34}$ erg), respectively, on the nearby M dwarfs EV Lacertae, YZ Canis Minoris, and AD Leonis. In addition, \cite{2025ApJ...993...80N} detected 4 eruptions from 15 flares (${\gtrsim}10^{33}$ erg) in focused campaigns on the young solar-type stars EK Draconis and V889 Herculi and \cite{2024MNRAS.tmp.1385L} also reported 1 eruption from 5 events on EK Draconis. 
These results provide some indication of the occurrence rate of CMEs/eruptions (i.e., the fraction of flares accompanied by eruptions). However, these numbers are likely lower limits, as only line-of-sight Doppler shifts can be detected and other observational limitations apply. Other than the Doppler-shift methods, the absence of type-II burst detections within restricted observing windows reported by \cite{2018ApJ...856...39C,2018ApJ...862..113C} may suggest that CME-driven shocks are not frequent. Future intensive monitoring, particularly at wavelengths beyond the optical, will be essential to better constrain the occurrence rates of stellar CMEs.

\subsection{The Empirical Solar--Stellar CME Connection}\label{CME:Empirical}




\subsubsection{Mass and Energy Losses Associated to Stellar CMEs}\label{sec:CME_M-Kdot}

One of the first seminal papers exploiting the solar--stellar CME connection is the study performed by \citetads{2013ApJ...764..170D}, which used empirical solar flare--CME data to assess how CMEs would have contributed to mass and energy loss from magnetically-active, flare-dominated stars. The study began with a compilation of solar soft X-ray flares observed between 1996 and 2007 \citepads{2009IAUS..257..233Y} deriving power laws between flare X-ray fluence ($E$) and both the mass ejected by CMEs, $m_{c}(E)$, and their kinetic energy, $E_{ke}$. They obtained empirical relations in the form $m_{c}(E)=\mu E^{\beta}$ and $E_{ke}=\eta E^{\gamma}$ and parameterized the CME-to-flare association fraction as a function of fluence, finding that the association fraction reached unity for flares above approximately $3.5\times10^{29}$~erg. These semi-empirical relations are often used to interpret and diagnose observations of stellar CME candidates (see Figure \ref{Fig:4-2-3} in Section \ref{CME:Obs:property}).

To estimate stellar CME mass and kinetic energy loss rates the authors adopted the common flare frequency distribution $dn/dE=kE^{-\alpha}$ (see Section~\ref{sec:Rad_Stellar_Flares}) and used the observed stellar X-ray luminosity $L_{X}$ as the integrated flare power that set the normalization $k$. They combined the fitted solar scalings for $m_{c}(E)$, $E_{ke}(E)$ and the CME-to-flare association fraction, $f(E)$, with the flare frequency law and integrated over an energy interval $[E_{\mathrm{min}},E_{\mathrm{max}}]$ to obtain analytic expressions for the total CME mass loss rate $\dot{M}_{c}$ and the kinetic energy loss rate $\dot{E}_{ke}$. It is important to note that their formalism explicitly showed how the results depended on the flare power law index $\alpha$ and on the integration limits.

\begin{figure*}[!t]
\centering
\includegraphics[trim=0.0cm 3.25cm 0.0cm 3.25cm, clip=true, width=0.99\textwidth]{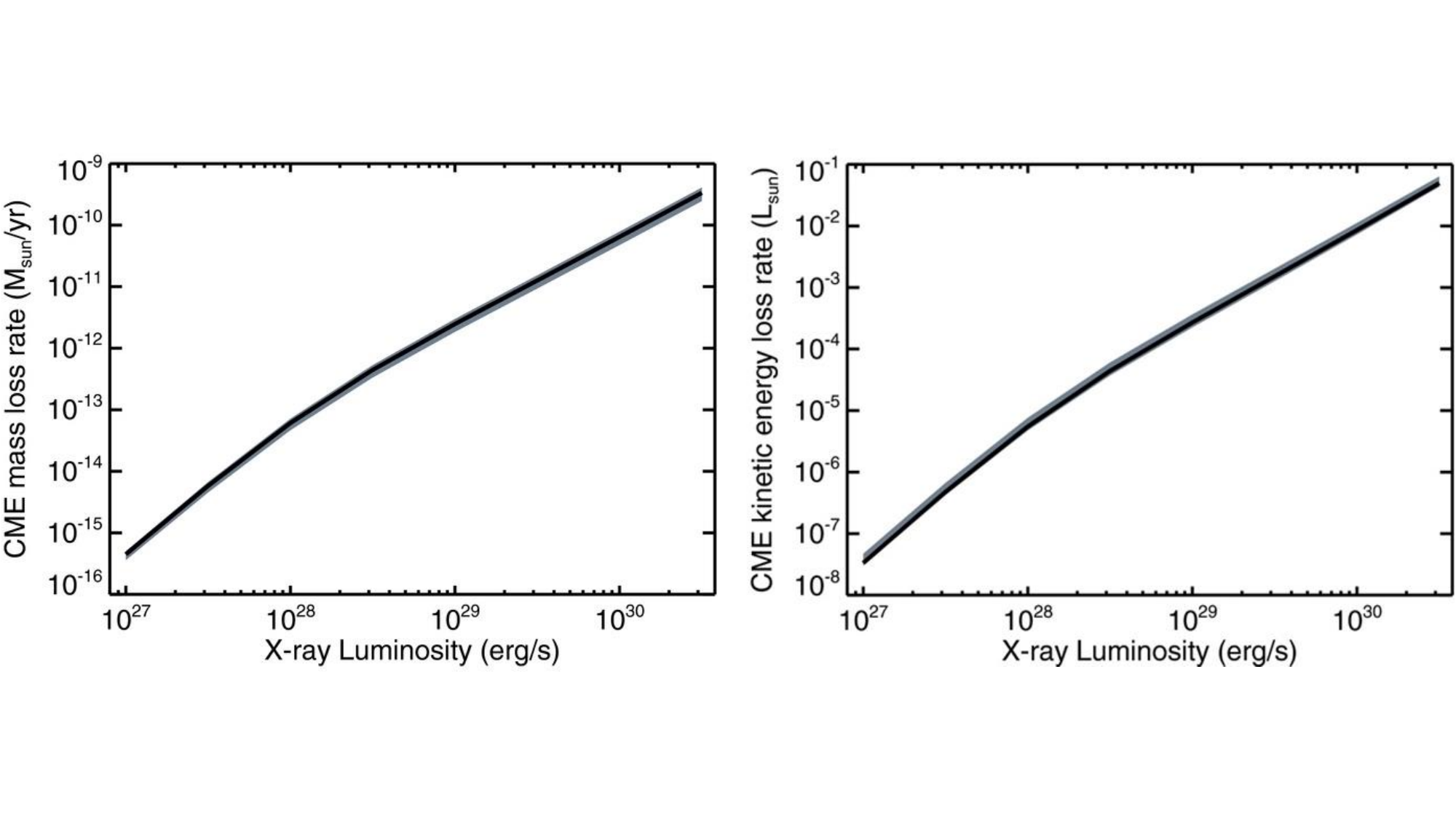}\vspace{-0.25cm}
\caption{Predicted stellar CME mass (left) and kinetic energy (right) loss rates vs. broadband X-ray luminosity based on the solar flare-CME extrapolation. The black solid curve represents the power-law index $\alpha = 2.25$. The gray shaded areas represent the range in the loss rates corresponding to the power-law index range $1.5 \leq \alpha \leq 3.0$. \textbf{Taken from:} \citetads{2013ApJ...764..170D}.}\label{Fig:Drake2013_CME_Mdot-Kdot}
\vspace{-0.4cm}
\end{figure*}

Applying this framework to a solar-like star at X-ray saturation (see Section~\ref{subsec:quiescent}) and adopting plausible integration limits yields very large values. For a saturated X-ray luminosity of order $10^{30}$~erg~s$^{-1}$ in the GOES $1-8$~\AA~band and for typical flare frequency indices $\alpha$ around 2 to 2.5, the associated CME mass loss rates reach $\dot{M}_{c}\sim 5\times10^{-10}~M_{\odot}~\mathrm{yr}^{-1}$ and kinetic energy loss rates that could approach $\dot{E}_{ke}\sim 0.1~L_{\odot}$. These results are only weakly sensitive to the precise value of $\alpha$ within the observationally supported range, and changing the lower integration limit had little effect because most of the integrated mass and energy in these extrapolations were weighted toward the highest energies. With these results the study generated simple power law scalings of $\dot{M}_{c}$ and $\dot{E}_{ke}$ against $L_{X}$ so that estimates could be obtained for stars of different activity levels (Figure~\ref{Fig:Drake2013_CME_Mdot-Kdot}).

\citetads{2013ApJ...764..170D} examined the plausibility of such large CME losses in the context of energetic constraints. They argued that a CME kinetic energy budget approaching a sizable fraction of the stellar bolometric luminosity was implausible because it implied magnetic energy extraction and dissipation rates far larger than the canonical X-ray saturation level of $L_{X}/L_{\mathrm{bol}}\sim 10^{-3}$. To try to reconcile the extrapolated solar scalings with energetic limits two possibilities were proposed. Either the empirical relations between flare fluence and CME mass and velocity flattened at the highest flare energies so that CME masses and speeds did not continue to grow as simple power laws, or the CME-to-flare association fraction decreased for the most energetic flares because overlying magnetic confinement prevented ejection (as sometime observed in highly energetic solar events; e.g., \citeads{2015ApJ...804L..28S}, \citeads{2017ApJ...844L..27T}). Both possibilities would have reduced the integrated mass and energy requirements. This study also discussed observational and theoretical hints of such limits, including evidence for cutoffs in solar energetic particle fluences (e.g., \citeads{2007ApJ...663L..45H}, \citeads{2022LRSP...19....2C}) and practical upper bounds on active region free energy, which could be related to a flattening of the scalings.

The principal implication of the analysis was that if active stellar coronae were truly dominated by solar-like flares and if solar flare-CME relationships held at much higher energies and frequencies, then CMEs could have dominated stellar mass loss at high activity levels and would have demanded a ``hidden'' energy budget far larger than indicated by X-rays alone. Under a more conservative assumption that magnetic processes consumed at most about one percent of the bolometric luminosity, the estimated CME mass loss would be of the order $5\times10^{-11}~M_{\odot}~\mathrm{yr}^{-1}$, which would still have made CMEs a major contributor to stellar mass loss in active stars. This estimate however is slightly at odds with the required radio transparency of the stellar wind (Sections~\ref{radiobrems}--\ref{sec:Wind_Radio3}) as well as our current knowledge on mass loss rates in cool stars from different methods (Section~\ref{sec:Wind_Overview}). This implies that solar empirical relations could not be extrapolated indefinitely. This work also highlighted the need of observational efforts to detect stellar CMEs and further theoretical work on confinement and energetic upper limits to determine where and why the solar scalings must break down.


Later, \citetads{2015ApJ...809...79O} developed a concise, physically-motivated framework that extended and complemented the empirical extrapolations explored by \citetads{2013ApJ...764..170D}, by translating multi-wavelength flare measurements into bolometric flare energies and by relating those energies directly to cumulative mass loss. They compiled observational flare energy partitions and derived bandpass conversion factors that allowed flare energies measured in different bands to be compared on a common bolometric scale. They then adopted an equipartition hypothesis between CME kinetic energy and bolometric flare energy together with the simplifying assumption that a characteristic CME speed scaled with the stellar escape velocity. From these ingredients, they obtained analytic expressions that linked the integrated CME mass loss to the total flare rate and to the flare frequency power law index, and they applied this formalism to published flare frequency distributions for solar analogs, young stars and M dwarfs after performing the appropriate bandpass conversions. 

Their results broadly agreed with the large transient mass loss rates inferred by empirical solar extrapolations such as those discussed by \citetads{2013ApJ...764..170D}, while emphasizing that the integrated CME contribution was controlled more directly by the total flare occurrence rate and by multi-wavelength energy partitions than by single band proxies alone. In their analysis, they obtained CME $\dot{M}$ values around $10^{-11}~M_{\odot}~\mathrm{yr}^{-1}$ for active Sun-like stars and within $10^{-11}-10^{-16}~M_{\odot}~\mathrm{yr}^{-1}$ for low-mass stars depending on their activity level. This study therefore reinforced the primary conclusion that the naive extension of solar flare-CME relations implied very large, and probably unrealistic, mass and energy demands on active stars. This investigation also exposed where the largest uncertainties lay, namely in the assumed equipartition factor, the adopted flare energy partitions, the flare frequency index and the characteristic CME speed, and they thus identified concrete observational priorities for constraining CME occurrence and energetics on active stars. It is good to note here that recent observational efforts indicate observationally-constrained lower limits on the order $10^{-13}-10^{-12}~M_{\odot}~\mathrm{yr}^{-1}$ on CME mass loss rate for active young Sun-like stars \citepads{2025ApJ...993...80N}.

\subsubsection{Occurrence Rate of Stellar CMEs: Empirical Modeling}

\begin{figure*}[!t]
\centering
\includegraphics[trim=0.0cm 3.0cm 0.0cm 3.0cm, clip=true, width=0.99\textwidth]{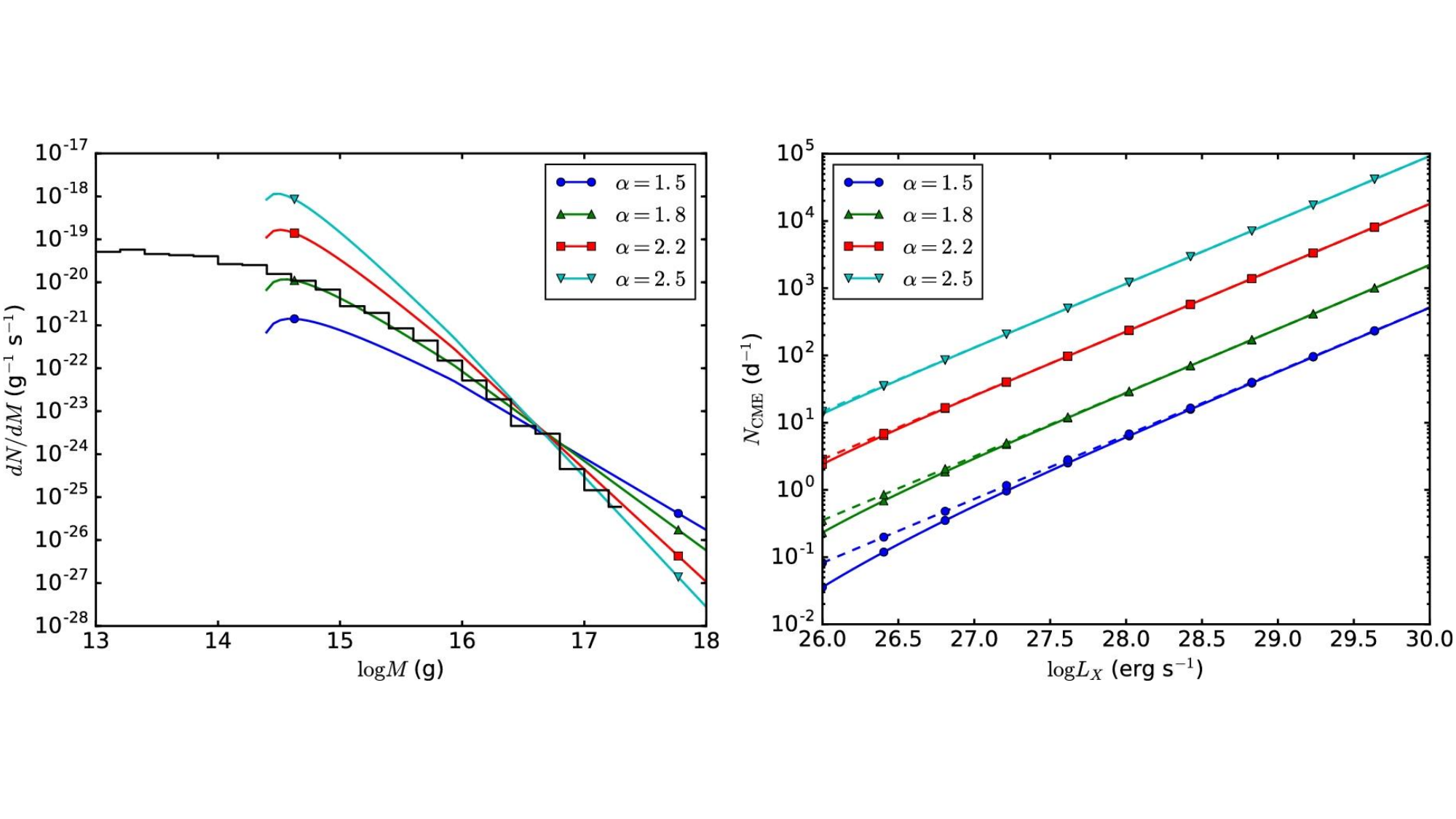}\vspace{-0.25cm}
\caption{\textit{Left:} Differential distribution of CMEs as a function of mass for a star with $\log L_{X}=26.5\ \mathrm{erg\ s^{-1}}$, representative of solar minimum conditions, shown for four values of $\alpha$ that span the commonly observed range (Section~\ref{sec:Rad_Stellar_Flares}). The black curve denotes the observed solar CME distribution from \citet{2012ApJ...760....9A}. \textit{Right:} CME occurrence rate as a function of $L_{X}$ and $\alpha$. Solid lines show rates obtained by integrating up to a specified maximum CME mass, whereas dashed lines give occurrence rates derived from the unbound cumulative distribution. \textbf{Taken from:} \citetads{2017MNRAS.472..876O}.}\label{Fig:Odert2017_CME_dM-NCME}
\vspace{-0.4cm}
\end{figure*}

Building on the studies mentioned in the previous Section, \citetads{2017MNRAS.472..876O} developed an empirical framework that directly translated observed stellar flare rates into predicted CME occurrence and mass loss. They formulated the differential CME distribution (Figure~\ref{Fig:Odert2017_CME_dM-NCME}, left) as an explicit chain of factors linking the stellar flare frequency distribution, band pass conversions between solar and stellar flare measurements, the empirical solar relation between flare energy and CME mass, and the probability that a flare produced a CME. They used this formulation to derive expressions for CME occurrence as a function of CME mass and to calculate the integrated CME mass and kinetic energy losses. A key new ingredient was an empirical estimate of the maximum CME mass a star with a given X-ray luminosity could generate (based on a maximum expected flare energy, $E_{\rm max}\approx10^{4.5}L_{X}$) which they expressed analytically as $\log(M_{\rm max}) \approx -3.955 + 0.7375\log(L_{X})$. This expression yields maximum CME masses in the range of about $10^{15}$ to $10^{18}$~g for stars with $L_{X}$ between $10^{26}$ and $10^{30}$~erg~s$^{-1}$. 

\citetads{2017MNRAS.472..876O} tested the model against solar CME statistics and then applied it across a broad stellar activity range to obtain estimates of CME rates versus $L_{X}$ and the flare power law index $\alpha$ (Figure~\ref{Fig:Odert2017_CME_dM-NCME}, right). Note that unlike the previous analysis of \citetads{2013ApJ...764..170D}, the results demonstrated a very strong sensitivity of the predicted rates to $\alpha$ (Section~\ref{sec:Rad_Stellar_Flares}) and to the adopted bandpass conversions. 
Another important result was the quantitative exploration of the flare to CME association probability as the dominant source of uncertainty in their model and the introduction of simple prescriptions for scaled associations that represented enhanced magnetic confinement in active coronae. These scaled prescriptions produced substantial reduction of CME occurrence and mass loss for very active stars and thereby provided a plausible resolution of the tension identified by \citetads{2013ApJ...764..170D} between naive solar extrapolations and observed total mass loss from Ly-$\alpha$ absorption (Section~\ref{sec:Ly-a_Astrospheres}). By confronting their model predictions with available observational constraints on stellar mass loss (Section~\ref{sec:Wind_Overview}) they also showed that extrapolations based on solar relations could exceed measured total mass loss by orders of magnitude for the most active stars and that accounting for a reduced eruptivity due to stronger confinement could diminish the predicted CME rates and mass loss substantially.

\subsubsection{Exoplanetary-Effective CMEs: Collision Rate Estimates}\label{sec:CME-Planet_Impact-Rates}

Taking advantage of the derived occurrence rate of CMEs, \citetads{2017MNRAS.472..876O} also estimated the collision rate of CMEs with exoplanets orbiting stars of different activity levels (given by their $L_{X}$). The impact frequency is expressed as
\begin{equation}
f_{\mathrm{imp}} = P_{1}P_{2}f_{\mathrm{CME}},
\label{eq:fimp}
\end{equation}
where $f_{\mathrm{CME}}$ denotes the intrinsic CME rate and $P_{1}$ and $P_{2}$ are the meridional and azimuthal probabilities, respectively. These angular probabilities are written as
\begin{equation}
P_{1} = \frac{\sin\left[(\Delta_{\rm CME} + \delta_{\mathrm{pl}})/2\right]}{\sin\Theta},
\qquad
P_{2} = \frac{\Delta_{\rm CME} + \delta_{\mathrm{pl}}}{2\pi},
\label{eq:P1P2}
\end{equation}
where $\Delta_{\rm CME}$ is the CME angular width, $\delta_{\mathrm{pl}}$ is the angular size of the planet as seen from the star and $\Theta$ represents the maximum stellar latitude of CME source locations. To determine the CME rate as a function of $\Delta_{\rm CME}$, an empirical solar relation  linking CME mass and apparent angular width, $\Delta_{\rm{app}}$, is used. This relation, proposed by \citetads{2005JGRA..11012S07G}, is expressed in the form $\log M = 12.6 + 1.3 \log(\Delta_{\rm{app}})$ with $M$ in grams and $\Delta_{\rm{app}}$ in degrees. The total impact rate on an orbiting exoplanet is then given by the integral
\begin{equation}
f_{\mathrm{imp}} = 
\int_{\Delta{0}}^{\Delta_{\rm{max}}}
P_{1}P_{2}\frac{dN}{dM}\frac{dM}{d\Delta_{\rm{app}}}\left|\frac{d\Delta_{\rm{app}}}{d\Delta_{\rm{true}}}\right|d\Delta_{\rm{true}},
\label{eq:full_integral}
\end{equation}
which includes the mass-dependent intrinsic CME distribution, $dN/dM$
(Figure~\ref{Fig:Odert2017_CME_dM-NCME}, left), the transformation to the apparent width $dM/d\Delta_{\rm app}$, and a correction factor $\left|d\Delta_{\rm{app}}/{d\Delta_{\rm{true}}}\right|$ that accounts for projection effects linking the true ($\Delta_{\rm{true}}$) and apparent angular widths of solar CMEs \citepads{2011RAA....11..237W}. For the integration limits, \citetads{2017MNRAS.472..876O} assumed $\Delta_0 = 20^{\circ}$ (associated with the lowest mass value in the solar flare-CME  distribution\footnote[2]{Taken as $2.2 \times 10^{14}$~g based on the compilation provided by \citetads{2009IAUS..257..233Y}.}) and $\Delta_{\rm max} = 180^{\circ}$, noting that the inferred impact rates were not too sensitive to the choice in the upper limit value. Similarly, for planets located in the classical temperature-based Habitable Zone (HZ, \citeads{1993Icar..101..108K}, \citeads{2014ApJ...787L..29K}) of main-sequence stars, the angular size of a planet is expected to be negligible compared to the average CME size, which allows to drop $\delta_{\rm pl}$ in Equation~\eqref{eq:P1P2} with minimal effect on the CME impact estimates.

\begin{wrapfigure}{r}{0.56\textwidth}
\vspace{-0.4cm}
\centering 
\includegraphics[trim=4.85cm 0.3cm 4.85cm 0.3cm, clip=true, width=0.55\textwidth]{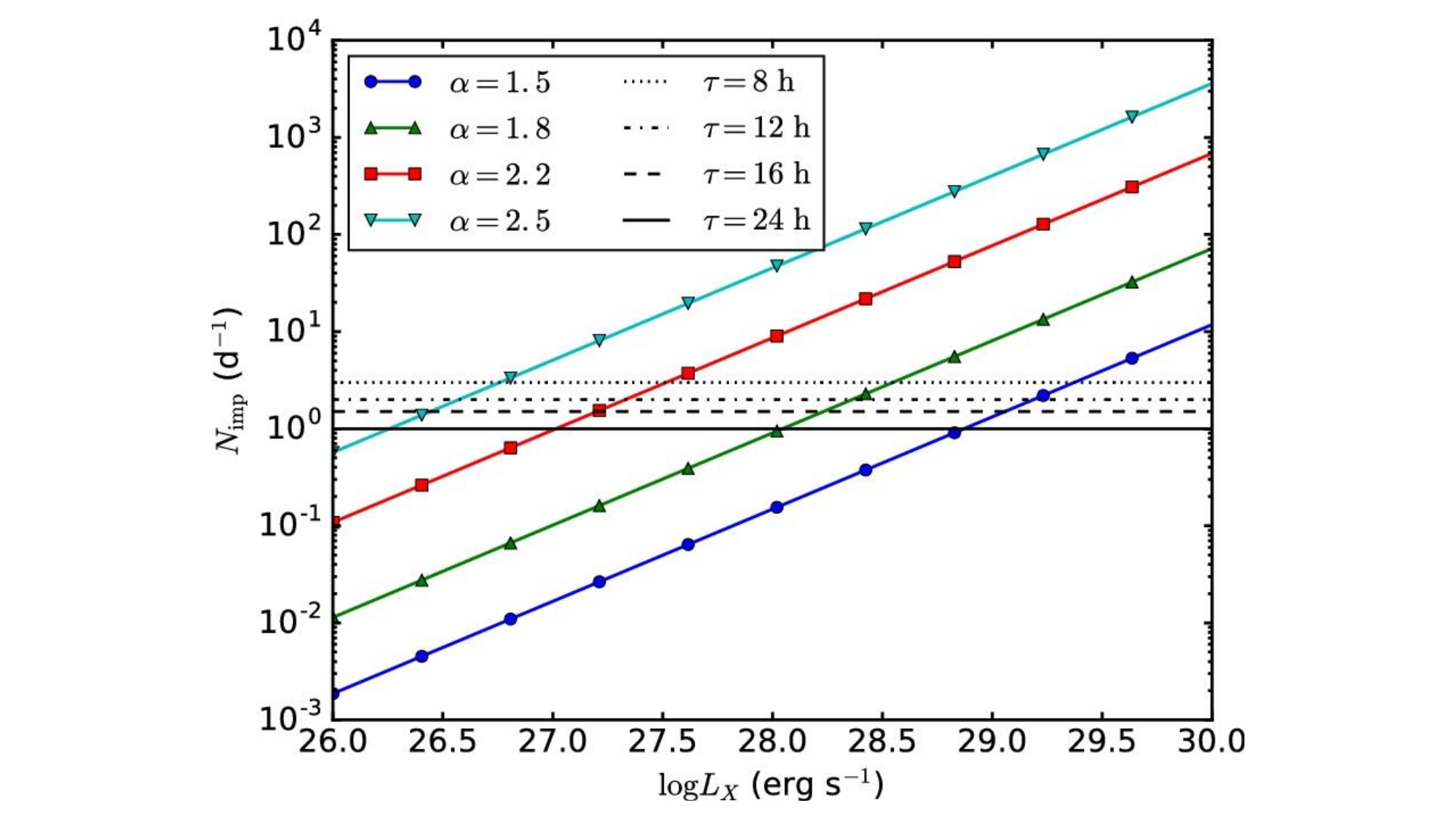}
\vspace{-0.2cm}
\caption{CME impact rates on a planet located in the classical HZ as a function of $L_{X}$ and $\alpha$. For different CME durations $\tau$, planets around stars with $N_{\mathrm{imp}}$ above the plotted lines would experience continuous exposure to CME plasma.
\textbf{Taken from:}~\citetads{2017MNRAS.472..876O}.}\label{Fig:Odert_CME-Planets}
\vspace{-0.3cm}
\end{wrapfigure}

Under the previous assumptions and a maximum CME source latitude of $90^{\circ}$, the resulting CME impact rates ($N_{\rm imp}$, Figure~\ref{Fig:Odert_CME-Planets}) show a very strong dependence on the flare power-law index (via the intrinsic CME distribution), while projection corrections and the adopted CME source latitude band have minimal effect (factors of a few compared to the orders of magnitude variance given by $\alpha$). For solar parameters ($2.7 \times 10^{26}$~erg~s$^{-1}$ $\leq L_{X} \leq 4.7 \times 10^{27}$~erg~s$^{-1}$, $\Theta = 30^{\circ}$) and a flare index of about $1.8$, the model produced CME impact numbers of order 10 to 100 per year, which are on the same order as front facing halo CMEs and geoeffective events observed on the Sun over a solar cycle (\citeads{2002JGRA..107.1340W}, \citeads{2007JGRA..112.6112G}). The authors also evaluated the condition for effectively continuous CME exposure by requiring $N_{\mathrm{imp}}\tau \ge 1$, where $\tau$ was a characteristic CME duration. Using plausible values for $\tau$ in the range of $8$ to $24$ hours (black horizontal lines in Figure~\ref{Fig:Odert_CME-Planets}), they found that planets in close orbits around very active stars could be effectively under permanent CME plasma exposure. The astrophysical consequences of such extreme space weather conditions could include enhanced ion pick up loss rates and rapid atmospheric erosion (see \citeads{2020JGRA..12527639G}), underscoring the potential importance of CME impacts for planetary atmosphere evolution and habitability.

Finally, \citetads{2017MNRAS.472..876O} noted some important caveats of their empirical model. The usage of solar empirical relations to map mass to apparent angular width and the use of geometric probabilities neglected many propagation details such as non-radial motion, deflection, CME shape and magnetic orientation (\citeads{2008ApJ...677.1378K}, \citeads{2016ApJ...826..195K}, \citeads{2024ApJ...962..193L}). They showed that ignoring the projection correction would have increased impact rates by factors of two to four for reasonable choices of the flare index. They also pointed out that uncertainties in the intrinsic CME distribution, in the association between flares and CMEs, and in the true distribution of CME source latitudes introduced orders of magnitude uncertainty in predicted planetary impact statistics. As such, the final conclusion was that a reliable assessment of exoplanet space weather required both improved intrinsic CME statistics on active stars and a better understanding of the propagation and magneto-hydrodynamical (MHD) properties of stellar CMEs. 

\subsection{Modeling Stellar CMEs}\label{CME:Models}

\subsubsection{From the Sun to the Stars: CME Modeling}

Modeling CMEs from the Sun is considerably more demanding than simulating the ambient solar wind, largely because these transients exhibit pronounced spatio-temporal complexity. At the core of most CME descriptions lies the magnetic flux rope (MFR), which is generally regarded as the principal eruptive structure and the agent that powers the event. Yet, because direct measurements of the coronal magnetic field remain extremely difficult, observational characterization of MFRs is typically inferred from EUV and/or soft X-ray imaging \citep[e.g.][]{2017SoPh..292...39P}, and the internal organization of erupting flux ropes is therefore still only partially constrained (see Section~\ref{CME:Overview} for general aspects of solar CMEs). Consequently, the majority of CME models are constructed to reproduce the flux rope geometry, its triggering, and its subsequent dynamical evolution. In broad terms, these approaches are usually grouped into data-inspired, data-constrained, and data-driven frameworks.

Data-inspired CME models are designed to capture, through idealized or analytic prescriptions, the canonical properties repeatedly identified in observational studies. Representative examples include the Gibson--Low (GL; \citeads{1998ApJ...493..460G}) and Titov--Démoulin (TD; \citeads{1999A&A...351..707T}) flux rope solutions, together with later numerical refinements of these formulations (e.g., \citeads{2014ApJ...790..163T}, \citeads{2023ApJ...955..126S}). The GL construction was motivated by the characteristic three-part density morphology of white-light CMEs, featuring a bright outer front, a dark cavity, and a bright inner core (see, e.g., Figure~\ref{fig:solarcme}(b)), whereas the TD model is able to reproduce the filament-channel-like structure aligned with the polarity inversion line in H$\alpha$ observations. To represent an actual eruption, such analytic flux rope configurations must be embedded within a realistic coronal atmosphere and stellar wind background (see Section~\ref{sec:Sun-to-Stars_Models}). They are also commonly employed in zero-beta MHD experiments to isolate the essential physics governing CME initiation and propagation (see \citeads{2023BAAS...55c.393T}). Nevertheless, whether a coherent flux rope exists prior to eruption remains an open question, which has motivated alternative data-inspired scenarios, including the breakout model \citepads{1999ApJ...510..485A, 2008ApJ...683.1192L} and the tether-cutting picture \citepads{1989ApJ...343..971V, 2001ApJ...552..833M}, both of which posit in-situ formation of the rope during the eruptive process.

Data-constrained flux rope models, in contrast, are anchored to measurements of the photospheric magnetic field, especially the vector field at a specified epoch. Within this framework, magnetofrictional techniques (e.g., \citeads{2004ApJ...612..519V}, \citeads{2009ApJ...703.1766S}) and nonlinear force-free field relaxation methods (e.g., \citeads{2014ApJ...783..102M}, \citeads{2023ApJ...956..119G}) are used to reconstruct magnetic topologies that are consistent with the structures inferred from EUV and X-ray data. As in the data-inspired case, these reconstructed configurations can then be supplied as initial conditions for global MHD calculations aimed at following CME onset and expansion.

Data-driven approaches currently represent the most advanced category of solar CME modeling, since in these schemes the flux rope is not imposed a priori but instead emerges and erupts self-consistently under the action of time-dependent photospheric forcing. Such forcing can be supplied either through flux-transport models (e.g., \citeads{2008ApJ...680L.165Y}), convective flux-collision simulations \citepads{2023ApJ...955..105R} or by assimilating successive vector magnetograms (e.g., \citeads{2016ApJ...828...62J}, \citeads{2024ApJ...975..206F}). In this way, the eruptive evolution is tied directly to the temporal development of the lower boundary conditions rather than to an externally prescribed flux rope configuration.

Each of these modeling classes has achieved varying degrees of success when confronted with observations. For example, data-inspired flux rope prescriptions have been widely adopted in global MHD studies intended for space-weather forecasting. Likewise, data-driven simulations have yielded encouraging results, including the production of M- and X-class flares under time-dependent boundary driving, although they still omit the particle beams that are believed to play a central role in such energetic events. Progress in solar CME modeling depends not only on improved numerical frameworks, such as model coupling across a wider range of spatial and temporal scales, but also on more complete observational coverage of the Sun, especially of the magnetic field, which remains indispensable for driving global MHD simulations. Equally important is the availability of more robust coronal magnetic field diagnostics, which are required to constrain the internal magnetic structure of the flux rope itself.

In the stellar context, existing studies have so far relied on the most idealized solar eruption prescriptions, namely the data-inspired class of models. In the remaining part of this section, we summarize recent works on stellar CMEs based on these numerical approaches.

\subsubsection{CME Confinement in Active Stars by Overlying Large-Scale Magnetic Fields}\label{sec:CME-Suppression}

\begin{figure*}[!t]
\centering
\vspace{-0.36cm}
\includegraphics[trim=0.0cm 4.5cm 0.0cm 4.5cm, clip=true, width=0.99\textwidth]{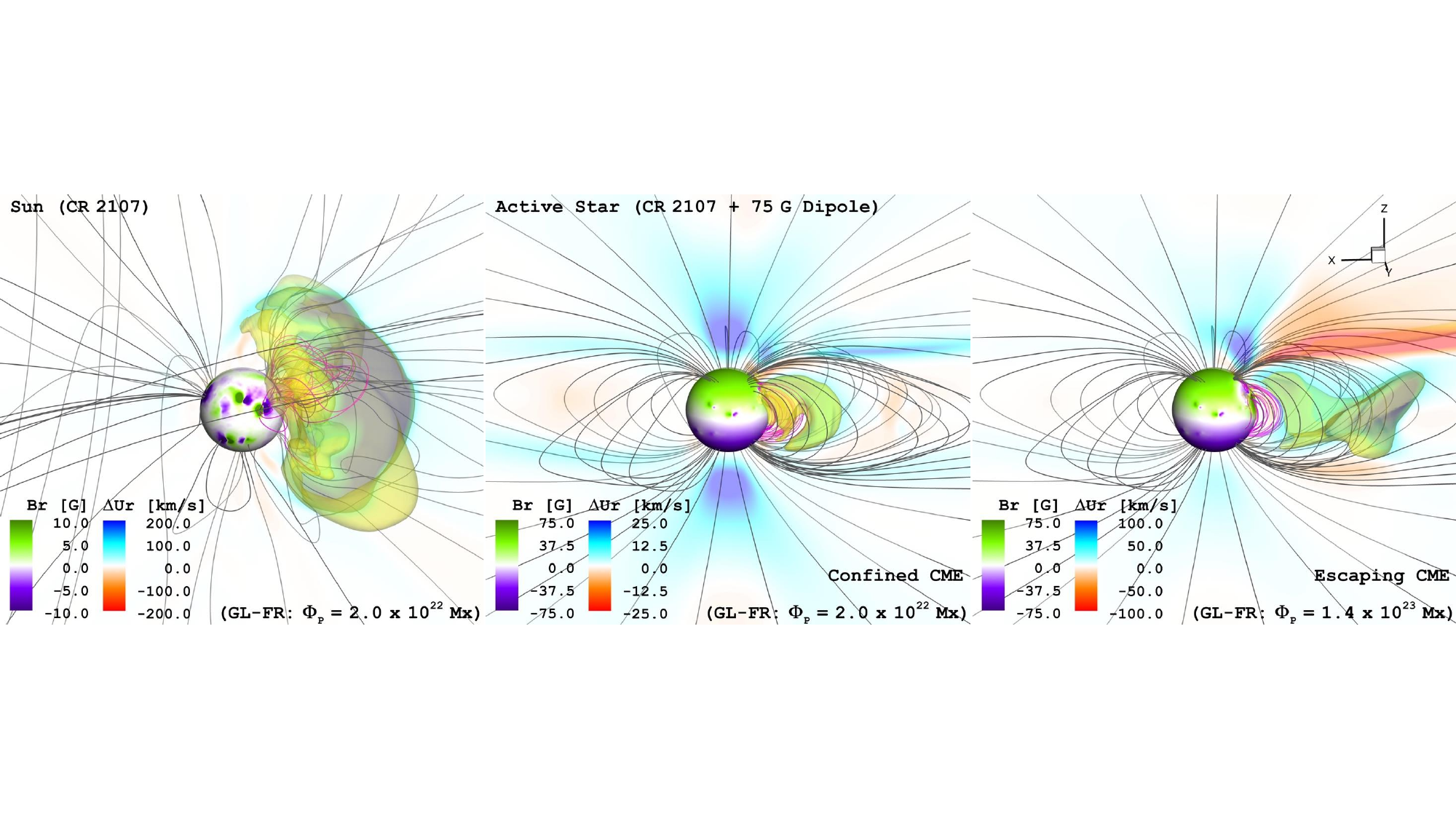}\vspace{-0.15cm}
\caption{Results from different 3D MHD simulations employed to investigate the suppression of CMEs in active stars. The stellar surface (central sphere) is colored by the magnetic field driving each model. A fiducial solar magnetic field (left) and a stellar counterpart with an enhanced $75$~G large-scale dipole configuration (middle and right) are compared. The coronal Doppler-shift radial velocities ($\Delta U_{\rm r}$), calculated with respect to the pre-CME conditions, are indicated by a secondary color scale. A translucent yellow iso-surface identifies the eruption in each case. While the same flux-rope eruption (with a magnetic flux variation equivalent to average values observed during $\sim$\,X5.0 GOES class flares) produces a relatively strong CME in the solar case ($M^{\rm CME} \sim 10^{17}$~g, $E_{\rm K}^{\rm CME}~\sim~10^{32}$~erg; left), it is totally confined in the stellar simulation (middle). The perturbed material follows the overlying field (gray lines), remaining bound to the lower regions of the corona. The right panel contains a third model with a stronger flux-rope eruption ($\sim$\,10 times the magnetic flux of the previous case) which is able to break free from the large-scale magnetic field confinement. \textbf{Adapted from:}~Alvarado-G\'omez~et~al.~(\citeyearads{2018ApJ...862...93A}, \citeyearads{2022AN....34310100A}).}\label{Fig:AG_Confined-CMEs}
\end{figure*}

Following up on the quandary identified by \citetads{2013ApJ...764..170D} and its possible mitigation via magnetic confinement of CMEs in active stars, \citetads{2018ApJ...862...93A} carried out the first investigation on this mechanism by performing a controlled suite of 3D MHD CME simulations. The study coupled the self-consistent Alfvén wave-driven solar wind model AWSoM (see also Section~\ref{sec:SW_Models}) with the data-constrained GL flux rope CME model, both incorporated as part of the SWMF. AWSoM evolves the mass, momentum, energy and magnetic induction conservation equations on a spherical grid while computing coronal heating and wind acceleration from Alfvén wave propagation, reflection, and turbulent dissipation. The GL MFR model prescribes an initially magneto-hydrostatic, twisted flux rope anchored to a mixed-polarity active region and triggers eruption by magnetic tension imbalance with the ambient field in the time-accurate MHD evolution. Both models have been used in several validation campaigns within the solar system context \citep[e.g.,][]{2014JGRA..119.5449M, 2017ApJ...834..173J}.

Some of the key assumptions are discussed below which help delimit the parameter space explored by \citetads{2018ApJ...862...93A}. The global surface field is constructed as the superposition of a simple large-scale dipole aligned with the rotation axis and a small-scale distribution taken from a solar GONG synoptic map (CR~2107). The dipole magnitude is fixed at $75$~G as a representative but computationally tractable strong large-scale field. Solar values for stellar mass, radius and rotation rate are assumed so that the setup can be compared directly to solar CME studies. The GL flux rope parameters are mostly those used in prior solar validations \citep{2017ApJ...834..173J}, with only the size and a magnetic-strength parameter varied across runs. The GL implementation does not add filament mass explicitly to the flux rope; the eruptive mass in the simulation therefore comes from the ambient coronal plasma that is perturbed and swept up during expansion. The authors generated a steady-state corona driven by the combined surface magnetic field and then insert GL flux ropes with varied size and magnetic strength to produce twelve eruption runs whose poloidal fluxes span roughly $10^{21}$--$10^{24}$~Mx. These modeling choices are intended to isolate the effect of an overlying large-scale field on eruption dynamics while acknowledging that real stars may differ in field complexity, filament loading, or fundamental stellar parameters. 

The numerical results show a strong confining influence of the $75$~G dipole on eruptions that, under solar conditions, would produce fast CMEs (see Figure~\ref{Fig:AG_Confined-CMEs}). For eruptions driven with poloidal fluxes comparable to those used in solar CME validations, the flux ropes are arrested within the low corona and remain magnetically bound; these confined CMEs roughly maintain their shape after an initial rise and never escape the large-scale field. Simulations with much larger poloidal fluxes produce escaping events, but all eruptions (whether confined or escaping) exhibit substantially reduced radial speeds compared with the speeds one would expect from events occurring under a nominal solar configuration (i.e. lacking a significant large-scale magnetic field). Escaping CMEs often fragment into fast polar components and slower equatorial components because of the dipolar wind and density structure, and a significant fraction of the erupted mass can remain confined (up to $60\%$). Importantly, by mapping simulated poloidal flux to observational reconnection flux and then to GOES soft X-ray flare class using published empirical solar scalings, the authors identify a suppression threshold for their configuration: CMEs with kinetic energies below approximately $3\times10^{32}$~erg are confined by the $75$~G dipole, which corresponds (under the aforemetioned solar flare-CME scalings) to flares up to roughly GOES X20 being suppressible and escape requiring far larger flare energies (the authors estimate escape for solar-like relations near GOES X70). The masses of the perturbed coronal material fall broadly within solar CME observational scalings, in contrast to the strong speed reduction (with the corresponding decrement in CME kinetic energy). Nevertheless, the largest simulated ``monster" CMEs reach masses and kinetic energies well above typical solar values.

The main implications of these results are twofold. First, a substantial large-scale field can reduce the flare-CME association and thus mitigate the enormous mass-loss and energetic demands that arise from naively extrapolating solar flare-CME relations to very active stars (Section~\ref{sec:CME_M-Kdot}). Second, the finding that escaping CMEs are slowed down (and therefore less energetic) and can fragment means that even when eruptions leave the corona, their interplanetary properties and potential impacts on exoplanet environments (Section~\ref{sec:CME-Planet_Impact-Rates}) will be altered by stellar magnetospheres and winds. The authors emphasize, however, that their quantitative suppression threshold applies strictly to the specific geometry and parameters assumed. As discussed in \citet{Chapter1} in this collection, similar
dipole field strengths and axisymmetric topologies have been reported on very young (~250-500 Myr old) Sun-like stars (e.g. \citeads{2018MNRAS.474.4956F}) as well as in M-dwarfs of different ages (see \citeads{2021A&ARv..29....1K}).
All in all, this study highlights a potentially important mechanism shaping stellar CME occurrence and impact, and calls for future work to explore varying large-scale field strengths and topologies, different stellar properties, and the role of filament mass loading to generalize these findings.

\subsubsection{A Large-Scale Carrington-Class Stellar CME}\label{sec:Lynch_SuperCME}

\begin{wrapfigure}{l}{0.54\textwidth}
\vspace{-0.3cm}
\centering 
\includegraphics[trim=0.0cm 0.0cm 0.0cm 0.0cm, clip=false, width=0.53\textwidth]{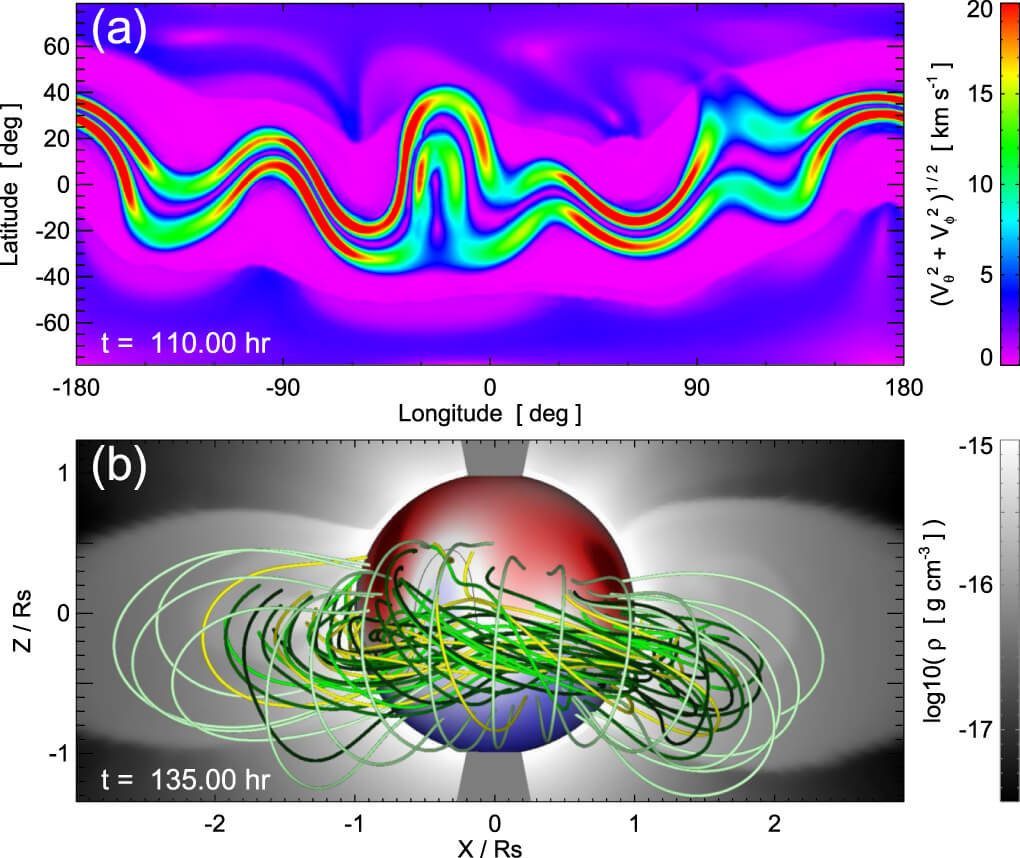}
\vspace{-0.2cm}
\caption{\textbf{(a)} Magnitude of the tangential velocity component, $(V_{\theta}^{2}+V_{\phi}^{2})^{1/2}$ (rainbow color bar), at the lower radial boundary at $t = 110$ hr, illustrating the shearing flow pattern $\lvert V_{\mathrm{shear}}\rvert$ that energizes the magnetic field along the global polarity inversion line provided by the ZDI map. \textbf{(b)} Illustrative magnetic field lines at $t = 135$~hr displaying the energized, globally-extended erupting magnetic configuration. The grayscale shows the plane-of-the-sky ($x-z$ plane) projection of the plasma density $(\log\rho)$. \textbf{Taken from:}~\citetads{2019ApJ...880...97L}.}\label{Fig:Lynch2019_SuperCME}
\vspace{-0.3cm}
\end{wrapfigure}

The CME suppression investigation of \citetads{2018ApJ...862...93A} explored the interplay between a stellar large-scale magnetic field with an eruption powered by a small-scale structure (i.e., a FR bearing active region). However, what if the source of the eruption would correspond to the stellar large-scale magnetic field itself? This was the approach followed by \citetads{2019ApJ...880...97L}, who carried out a 3D MHD simulation of a global-scale stellar CME from the young solar analog $\kappa^1$ Ceti. The study used a data-inspired approach in which the observed surface magnetic flux distribution from ZDI served as the lower boundary condition and the Adaptively Refined MHD Solver \citep[ARMS;][]{2008ApJ...680..740D} was employed to solve the ideal MHD equations on a spherical, adaptively refined grid. The background stellar wind was represented by an isothermal Parker solution with a uniform temperature of $2\times10^{6}$~K and the simulation was advanced through a relaxation phase, a quasi-static energy accumulation phase driven by prescribed tangential boundary flows that traced contours of the radial field, and a dynamic eruption phase in which reconnection naturally developed at the large-scale current sheet and launched a coherent flux rope and its associated CME (see Figure~\ref{Fig:Lynch2019_SuperCME}).

The main assumptions of the study included an MHD system that was treated as ideal so that reconnection occurred through numerical diffusion at grid scales rather than explicit resistivity. The driving flows preserved the radial flux distribution so that increases in global magnetic energy represented free magnetic energy introduced by shearing and twisting above the polarity inversion line. The stellar magnetogram did not resolve putative starspots or compact active regions, thus the experiment intentionally explored the most extreme global-scale eruption consistent with the measured background magnetic flux without assuming additional unresolved field. Because of this, the simulation intrinsically neglected the influence from a strong confining overlying field, as the largest possible magnetic field scale in the system (i.e., the star) was hosting the eruption. Furthermore, the isothermal assumption simplified thermodynamics and precluded explicit heating and cooling processes or a full internal energy evolution. Synthetic emission diagnostics were obtained with proxy measures that used field line current density and density squared quantities rather than a self-consistent temperature dependent~emissivity.

Under these simplifications, the principal results demonstrated that the observed background field of $\kappa^1$ Ceti could store and then release an amount of magnetic free energy sufficient to drive a Carrington-scale CME. The simulation released a magnetic energy of order $7\times10^{33}$ erg and produced a maximum increase in kinetic energy of order $3\times10^{33}$ erg, with a magnetic to kinetic conversion efficiency of ${\sim}$30--70\% during the impulsive phase depending on the metric used. The cumulative unsigned reconnection flux reached about $2.3\times10^{23}$ Mx and the inferred proxy X-ray flux corresponded to an X-class flare of roughly X58, comparable to estimates for the historic Carrington event. The erupting structure displayed the canonical three-part morphology in plane-of-the-sky cross sections and propagated with a strong leading shock. Shock compression ratios in the isothermal model reached values of order four or greater in places and the model predicted initial plasma frequencies in the few to a few tens of megahertz, implying Type II radio signatures in a frequency band that is challenging but in principle accessible to low frequency observations. Synthetic emission proxies exhibited pre-eruption dimming followed by post eruption brightening and decay, qualitatively like in observed solar and stellar flare light curve phenomenology.

The study had several important implications. It showed that even a background global field, without invoking unresolved extreme star spots, could produce a Carrington-scale powerful CME if large-scale shear accumulated around the main polarity inversion line. The presence of such global eruptions implies potentially severe space weather for exoplanets through enhanced energetic particle fluxes, shock driven acceleration, and dense CME sheaths capable of eroding atmospheres or altering atmospheric chemistry. Importantly, this scenario is not without solar precedent, as global-scale eruptions involving extended polarity inversion lines spanning at least half of the solar disk in longitude have been observed on the Sun \citep[e.g.,][]{2007ApJ...664L.131Z, 2021ApJ...914...39L}. The synthetic emission and shock diagnostics provided mock-up observational signatures and frequency bands that could be targeted by coordinated multi-observatory campaigns to attempt detecting these global stellar CMEs and their associated Type II radio bursts.

\subsubsection{CMEs in M-dwarf Stars: Cutting-Edge Simulations and Predictions}\label{sec:M-dwarf_CME-Sims}

The first 3D MHD simulations of CMEs evolving within the magnetic environments expected for moderately active M-dwarf stars were carried out by \citetads{2019ApJ...884L..13A}. Building directly on the CME suppression mechanism proposed in \citetads{2018ApJ...862...93A}, the authors examined how different large-scale magnetic fields suppress or modify CME evolution and investigated the various responses of the coronal structure to this process in low-mass main-sequence~stars.

Similar to \citetads{2018ApJ...862...93A}, the stellar corona and wind are modeled with AWSoM within the SWMF, where Alfvén wave turbulent dissipation heats the corona and drives the stellar wind. Unlike their initial investigation (using a modified solar magnetogram) the authors employed as boundary conditions surface magnetic fields distributions extracted from a global dynamo simulation of a fully convective star matched to Proxima Centauri \citepads{2016ApJ...833L..28Y}. The dynamo field is scaled to surface radial strengths between $\pm 600$ and $\pm 1400$ G, consistent with observations of moderately active M dwarfs (\citeads{2008A&A...489L..45R}, \citeads{2022A&A...662A..41R}).

CMEs are initiated using the TD flux rope \citepads{1999A&A...351..707T}, anchored to the strongest mixed-polarity region on the stellar surface. The inserted structure carries magnetic free energy $E_{\rm FR}^{B} \approx 6.6\times10^{34}\,$erg, slightly above upper solar CME values (\citeads{2009EM&P..104..295G}, \citeads{webb2012}, \citeads{2017ApJ...834...56T}), and an initial mass comparable to solar eruptions \citepads{2011LRSP....8....1C}.

\begin{figure*}[p!]
\centering
\vspace{-0.36cm}
\includegraphics[trim=0.0cm 0.0cm 0.0cm 0.0cm, clip=false, width=0.99\textwidth]{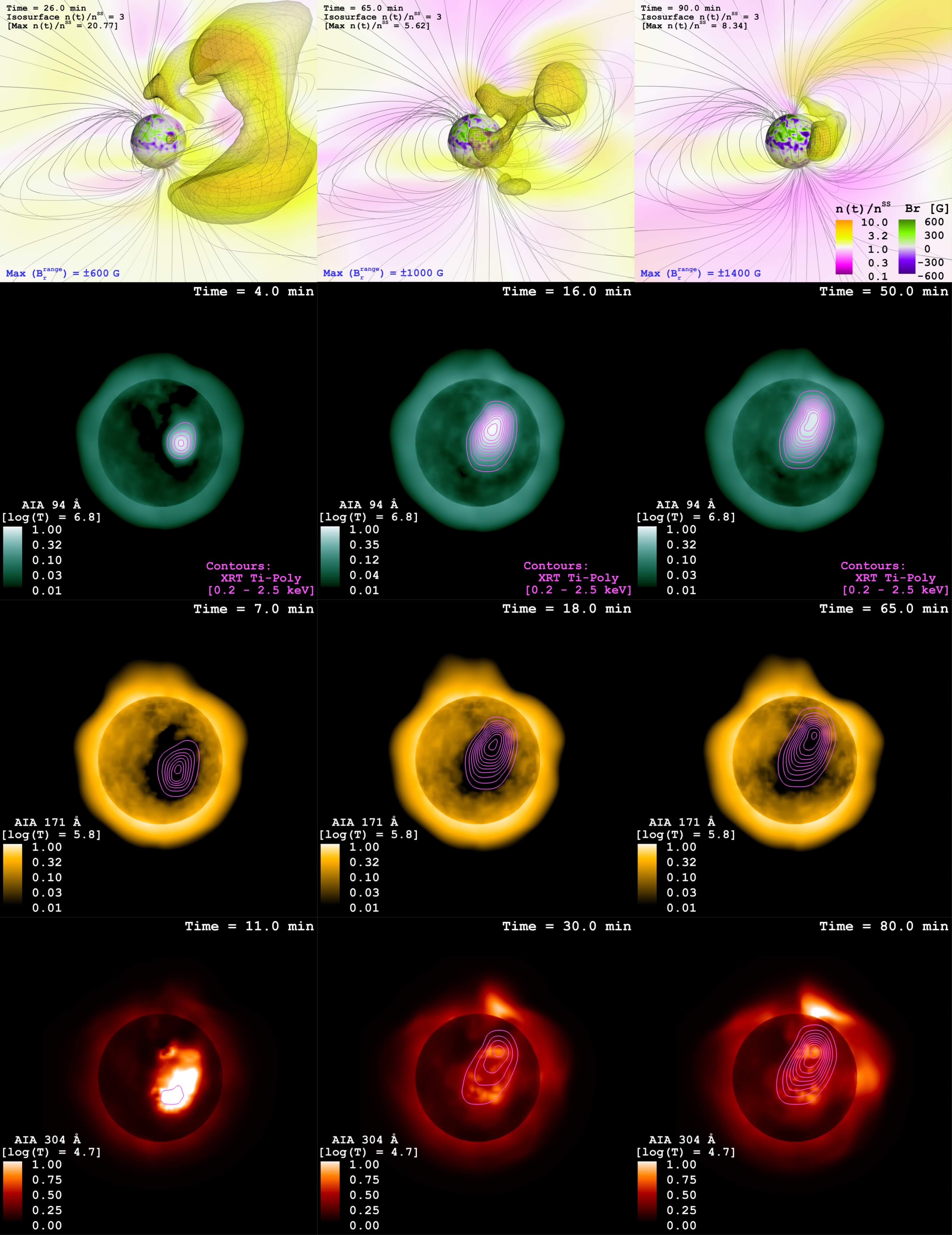}\vspace{-0.1cm}
\caption{\textit{Top:} Snapshots of a weakly (left), partially (center), and strongly confined CME (right) extracted at different times for each simulation. The color scales indicate the radial surface magnetic field (clipped at $\pm600$~G) and the coronal density contrast ($n(t)/n_{\mathrm{SS}}$). Representative large-scale magnetic field lines are shown in gray. \textit{Bottom:} Coronal signatures associated with the flux-rope eruptions in the top row, illustrated through simulated line-of-sight images of the stellar corona (arbitrary normalized units) synthesized in three AIA/SDO channels (second row: 94\,\AA; third row: 171\,\AA; fourth row: 304\,\AA) at selected times during the CME evolution. Magenta contours outline the corresponding X-ray emission (Ti-poly filter of XRT/Hinode; $0.2-2.5$~keV), plotted in $10$\% increments of the maximum mean X-ray intensity. The viewpoint places the eruption on the stellar disk with a field of view spanning five stellar radii. \textbf{Adapted from:}~\citetads{2019ApJ...884L..13A}.}\label{Fig:AG_M-Dwarf_CMEs}
\end{figure*}
\clearpage

Three confinement regimes arise (see Figure~\ref{Fig:AG_M-Dwarf_CMEs}). In the weak-field case ($B^{\rm max}_{\rm R}=\pm 600$~G), the flux rope collapses along overlying field lines, compressing and heating the ambient plasma by factors of $5$--$10$. This produces flare-like enhancements in EUV and X-ray emission, with increases exceeding an order of magnitude in the hottest channels. The Doppler shift signature evolves from red to blue as the perturbation first falls and then escapes, with characteristic doppler velocities of $90$--$150$ km\,s$^{-1}$. This behavior resembles small-scale solar responses (e.g., \citeads{2014ApJ...796...73H}, \citeads{2015Natur.523..437S}), although no explicit reconnection is included in the model.

Intermediate field strengths ($B^{\rm max}_{\rm R}=\pm 800$--$1000$~G) generate partial confinement: slower collapse and escape, broader and weaker flare-like light curves, and fragmentation of the eruption. Portions of the CME remain trapped and exhibit rising and falling motions with speeds up to ${\sim}100$~km\,s$^{-1}$, similar to solar coronal rain \citepads{2012ApJ...745..152A} and condensations \citepads{2018ApJ...864L...4L}. A persistent ${\sim}15\%$ deficit appears in mid-temperature integrated emission (e.g., the 171\AA~band), analogous to solar coronal dimming events (\citeads{2016SoPh..291.1761H}, \citeads{2016ApJ...830...20M}). However the authors caution on the interpretation of such signatures as unequivocal diagnostics of mass loss due to possible thermal effects in the corona, as well as their persistence even for cases of full CME confinement.

For the strongest fields ($B^{\rm max}_{\rm R}=\pm 1200$--$1400$~G), the CME is fully suppressed. The flux rope collapses and spreads into a large, slowly evolving cloud of in-falling material, producing a gradual X-ray brightening by factors of $2$--$3$ and predominantly redshifted velocities of tens of km~s$^{-1}$. As mentioned before, a transient coronal dimming signature in some EUV bands also appears for these confined eruptions. This could be the result of both, variations in the local thermodynamic conditions or removal of coronal material that will form a coronal rain cloud and appears prominently in cooler EUV channels. These results support the interpretation of coronal dimmings in different high-energy channels as indicators or CME activity on stars (see Section~\ref{CME:Obs:method}). Nevertheless they also bring a word of caution on their usage for direct assessments on CME escape with the corresponding consequences (i.e. associated mass and angular momentum losses, CME impacts on orbiting exoplanets).

Taken together, these results suggest that the magnetic environments of active M-dwarfs fundamentally reshape CME behavior relative to the solar case. The simulations show that CME suppression can occur over a relatively narrow range of field strengths, and that confined eruptions generate coronal signatures that may mimic flares or coronal rain rather than classical CME outflows. This implies that CME-driven mass loss and angular momentum evolution in active low-mass stars may be significantly lower than predicted by solar scalings, and that the space weather conditions impacting close-in exoplanets are likely governed more by magnetic confinement than by frequent high-mass ejections. The study also emphasizes that these predicted signatures (short-lived Doppler shifts, compression-driven brightenings, and cool in-falling clouds) are beyond the capabilities of current X-ray observatories, underscoring the need for next-generation missions with higher spectral resolution and larger effective area to detect and characterize stellar CME suppression.

\subsubsection{Radio Quiet CMEs: The Role of Stellar Magnetism}

Following with their investigation on stellar CMEs from a numerical perspective, \citetads{2020ApJ...895...47A} explored the conditions under which CMEs from active stars could generate observable type~II radio bursts. Motivated by the persistent lack of stellar type~II detections despite extensive radio campaigns (e.g.,~\citeads{2016ApJ...830...24C}, \citeads{2017PhDT.........8V}, Crosley \& Osten \citeyearads{2018ApJ...856...39C}, \citeyearads{2018ApJ...862..113C}) and by theoretical suggestions that strong stellar magnetic fields might render stellar CMEs radio-quiet (\citeads{2019ApJ...871..214V}, \citeads{2019ApJ...873....1M}), the study combined state-of-the-art three-dimensional MHD modeling of stellar coronae and winds with time-dependent CME simulations. Note that, as presented in Section~\ref{CME:Obs:method}, two recent papers reported the first detections on stellar Type II radio bursts in two M-dwarf stars (\citeads{2025Natur.647..603C}, \citeads{2025A&A...703A.198K}). Based on those new observational constraints and following the methodology described below, numerical investigations aimed at understanding the CME properties on those objects are currently underway. 

In their study, \citetads{2020ApJ...895...47A} analyzed how the ambient Alfvén speed distribution, set by the global magnetic geometry and wind structure, shaped the ability of eruptions to become super-Alfvénic and thus capable of driving shocks that produce type~II emission.  
\begin{figure*}[!t]
\centering
\vspace{-0.45cm}
\includegraphics[trim=0.5cm 0.0cm 0.0cm 0.8cm, clip=true, width=0.99\textwidth]{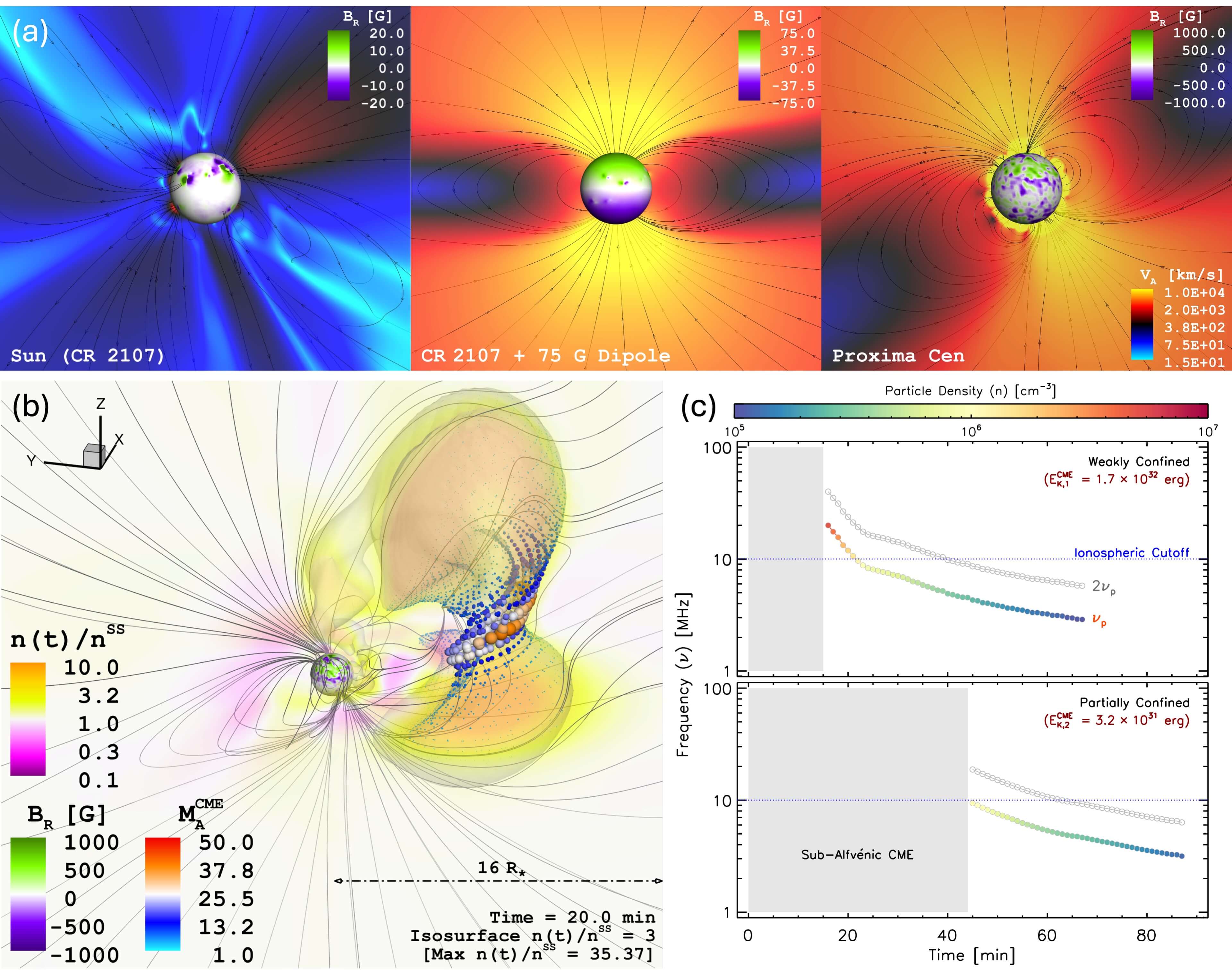}\vspace{-0.25cm}
\caption{(a) Meridional views of the Alfvén speed ($V_{A}$) from AWSoM steady-state models of the Sun (left), a young Sun-like star (middle), and the M-dwarf Proxima Cen (right). Stellar surfaces are colored by the radial field ($B_{R}$) driving the models and selected magnetic field lines are shown in black. A common $V_{A}$ color scale is used and the field of view is $12\,R_{\bigstar}$. (b) Snapshot of the weakly confined CME ($E_{\rm B,1}^{\rm FR} \simeq 4.1 \times 10^{35}$~erg, $E_{\rm K,1}^{\rm CME} \simeq 1.7 \times 10^{32}$~erg, $M_{1}^{\rm CME} \simeq 9.4 \times 10^{15}$~g) in the M-dwarf model. The stellar surface shows the radial field, while the density contrast $n(t)/n^{\rm SS} = 3.0$ highlights the CME front. Shock conditions from the Alfvénic Mach number ($M_{\rm A}^{\rm CME}$) are indicated by the color and size of the scattered spheres. The field of view extends to $32\,R_{\bigstar}$ with selected large-scale field lines in gray. (c) Time evolution of the fundamental and harmonic type II radio frequencies ($\nu_{p}$, $2\nu_{p}$) for the simulated M-dwarf CMEs. Colors indicate the mean plasma density weighted by $M_{\rm A}^{\rm CME}$. Gray intervals mark sub-Alfvénic phases.
\textbf{Adapted from:}~\citetads{2020ApJ...895...47A}.}\label{Fig:AG_CMEs_TypeII}
\end{figure*}
Employing the SWMF/AWSoM, steady-state solutions were obtained for three magnetic configurations: the Sun during an active phase (represented by the CR~2107), a young Sun-like star with an enhanced $75$~G dipolar component, and Proxima Cen using a surface field derived from the fully convective dynamo model of \citetads{2016ApJ...833L..28Y}. These steady-state models self-consistently yielded the coronal density, temperature, wind speed, and magnetic field distributions, from which the global Alfvén speed structure followed (see Figure~\ref{Fig:AG_CMEs_TypeII}).

The modeled Alfvén speed distributions revealed that the Sun exhibited the lowest values, consistent with observations (e.g., \citeads{2014A&A...564A..47Z}). In contrast, both the young solar analog and the M~dwarf displayed values up to an order of magnitude larger. In these stars, strong large-scale dipolar fields and low-density polar wind channels produced extended regions of high Alfvén speeds, while streamer-like structures allowed only localized minima. The simulations also recovered characteristic features known from the solar corona, including elevated $V_A$ over active regions and local minima produced by the superposition of large- and small-scale fields \citepads{2003A&A...400..329M}. Overall, the Alfvén speed distributions in the active stars were shown to greatly reduce the volume in which shocks could plausibly form.

To test whether CMEs in such environments could nonetheless become super-Alfvénic, the authors performed time-dependent CME simulations in the M-dwarf model by inserting unstable TD flux ropes at the inner boundary of their simulation. Two eruptions with different magnetic energies were initialized in order to probe regimes of weak and partial suppression by the large-scale stellar field. The flux ropes were anchored at the strongest mixed-polarity region of the surface field, and the simulations tracked their three-dimensional evolution and their interaction with the ambient stellar wind.

Both eruptions generated shock structures, but only after the CME fronts reached large heights above the star. Analysis of the CME radial velocities in the stellar-wind frame and the spatially averaged Alfvén speed showed that the transition to super-Alfvénic propagation occurred at several stellar radii, between approximately $10\,R_\bigstar$ and $20\,R_\bigstar$, depending on the event. This contrasted sharply with the solar case, where shocks typically form closer to the surface. The authors attributed this difference to the enhanced magnetic energy in the corona and the resulting suppression of CMEs, consistent with earlier predictions of magnetic confinement (\citeads{2016IAUS..320..196D}, \citeads{2018ApJ...862...93A}). The strong coronal fields kept the CME speeds relatively low while maintaining elevated Alfvén speeds, forcing the shock-formation region outward.

This shift in shock formation had major implications for the resulting type~II radio burst frequencies. The plasma frequency, and therefore the expected type~II emission frequency, is given by

\begin{equation}\label{eq:plasma_nu}
\nu_p = \left( \frac{n e^2}{\pi m_e} \right)^{1/2} \simeq 8980\,\sqrt{n}~\mbox{[Hz]},
\end{equation}

where $n$ is the ambient number density, $e$ is the electron charge, and $m_e$ is the electron mass. Because the density in the stellar wind decreases rapidly with radius, the shocks formed in extremely low-density regions, leading to predicted type~II frequencies near or below the terrestrial ionospheric cutoff (around $10$~MHz; \citeads{2018assi.book.....Y}). Under these conditions, the radio emission would be undetectable from the ground. The authors thus concluded that CMEs on active stars might frequently be radio-quiet not because they failed to produce shocks, but because the shocks formed too far from the~star.

A significant assumption in the modeling was that the coronal heating mechanism based on Alfvén waves used could be translated from the Sun to low-mass stars. While the authors try to adjust for the large difference in field strength and complexity in the model parameters, it is still an open question on how such processes take place on the Sun itself and within the strong field regime expected for M-dwarfs. This could modify the resulting corona and wind structure and therefore the associated CME shock formation heights. The results also depended on the large-scale magnetic field structure derived from the dynamo model of Proxima Cen, whose uncertainties could influence the resulting Alfvén-speed distributions and, consequently, the shock-formation sites. Despite these caveats, the study employed all relevant information on the Prox Cen system (i.e. surface field strength, stellar wind properties, flare energies) to constrain as much as possible the simulation and obtain realistic results. In this context, this investigation offered a consistent physical explanation for the scarcity of type~II detections in active stars. It demonstrated that the combination of strong magnetic confinement and elevated Alfvén speeds naturally pushed shock formation into the upper corona and wind, thereby shifting the emitted radio frequencies into an observationally inaccessible regime from the ground. 

\begin{figure*}[!t]
\centering
\vspace{-0.45cm}
\includegraphics[trim=0.0cm 0.0cm 0.0cm 0.0cm, clip=true, width=0.99\textwidth]{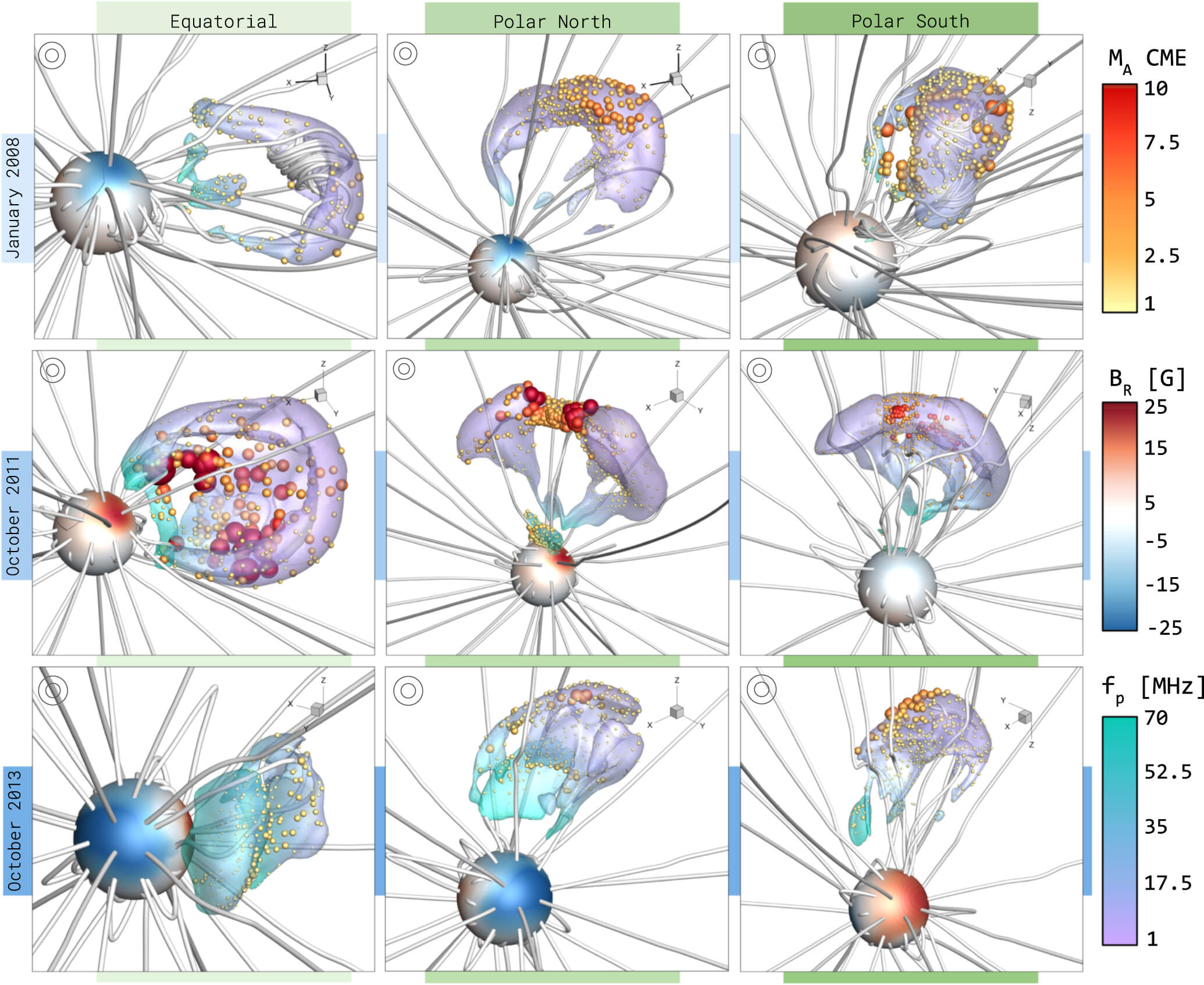}\vspace{-0.35cm}
\caption{Numerical models of equatorial (left) and polar (middle, right) CMEs in the stellar wind of $\epsilon$ Eri at different ZDI epochs shown 10 minutes after eruption. The CME is identified by a four-fold increase in density relative to the steady state wind. The magenta to turquoise contour traces the local plasma frequency on the CME surface. Symbol sizes indicate Alfvénic Mach numbers of $M_{A}=5$ and $M_{A}=10$. The stellar surface is color coded by the magnetic field strength in G with selected magnetic field lines shown in gray. \textbf{Taken from:}~\citetads{2022ApJ...924..115O}.}\label{Fig:OF2022_CMEs_TypeII}
\vspace{-0.3cm}
\end{figure*}

Later, \citetads{2022ApJ...924..115O} used the 3D MHD CME framework introduced in \citetads{2020ApJ...895...47A} on the nearby late-type K dwarf $\epsilon$ Eridani ($\epsilon$ Eri), applying the same numerical methodology to explore how stellar magnetic cycle variability could modulate CME-driven type II radio emission. The authors performed three dimensional MHD simulations using surface magnetic field maps derived from ZDI at three epochs (between 2008 and 2013) over the course of the magnetic cycle (Jeffers et al. \citeyearads{2014A&A...569A..79J}, \citeyearads{2017MNRAS.471L..96J}, \citeyearads{2022A&A...661A.152J}). For each epoch they initialized both polar and equatorial flux rope eruptions with solar-scaled CME masses and energies and followed their propagation through a self-consistent stellar wind. As in the earlier solar and M dwarf studies, shock formation was identified through the Alfvénic Mach number and synthetic pseudo radio spectra were constructed by converting the plasma density at the CME front into expected fundamental and harmonic plasma emission frequencies. The simulations showed that all modeled eruptions formed shocks and produced type II emission with fundamental frequencies reaching up to approximately 165 MHz, demonstrating that CME driven radio bursts should be physically viable around young solar analogs. In contrast to strongly magnetized M dwarfs, the relatively weak large scale field of $\epsilon$ Eri did not lead to systematic magnetic confinement of CMEs, in line with the confinement trends established by \citetads{2018ApJ...862...93A}.

The study found that the detectability of type II bursts was governed primarily by eruption latitude and magnetic geometry rather than the magnetic cycle phase itself. Polar CMEs remained above the ionospheric cutoff for only 10 to 20 minutes with occasional extension to about 30 minutes for harmonic emission, whereas equatorial CMEs propagated more slowly through denser plasma and could remain detectable for slightly longer intervals. Only one case in the 2013 epoch showed partial suppression of an equatorial eruption associated with locally stronger magnetic fields, providing limited evidence that magnetic variability could affect CME propagation. The authors further placed the simulated CME mass loss in the context of the stellar wind and found that CMEs could contribute between a few percent and up to roughly half of the total mass loss rate (see Section~\ref{sec:Wind_Overview}). Apart from the intrinsic limitation in the spatial resolution of the ZDI maps driving the simulations, additional caveats were linked to the uncertainties in shock location along the CME front and the assumed solar-like flux rope masses and energies. The results therefore provided optimistic upper limits on detectability. Overall, the work reinforced the conclusion of \citetads{2020ApJ...895...47A}  that stellar magnetic environments critically control both CME evolution and radio observability, while demonstrating that late-type stars such as $\epsilon$ Eri represent more favorable targets for low frequency CME detection than strongly magnetized M dwarfs.

\subsubsection{Stellar CMEs and Angular Momentum Loss}\label{sec:CMEs-AML}

\begin{figure*}[!t]
\centering
\vspace{-0.45cm}
\includegraphics[trim=0.0cm 0.0cm 0.0cm 0.0cm, clip=true, width=0.99\textwidth]{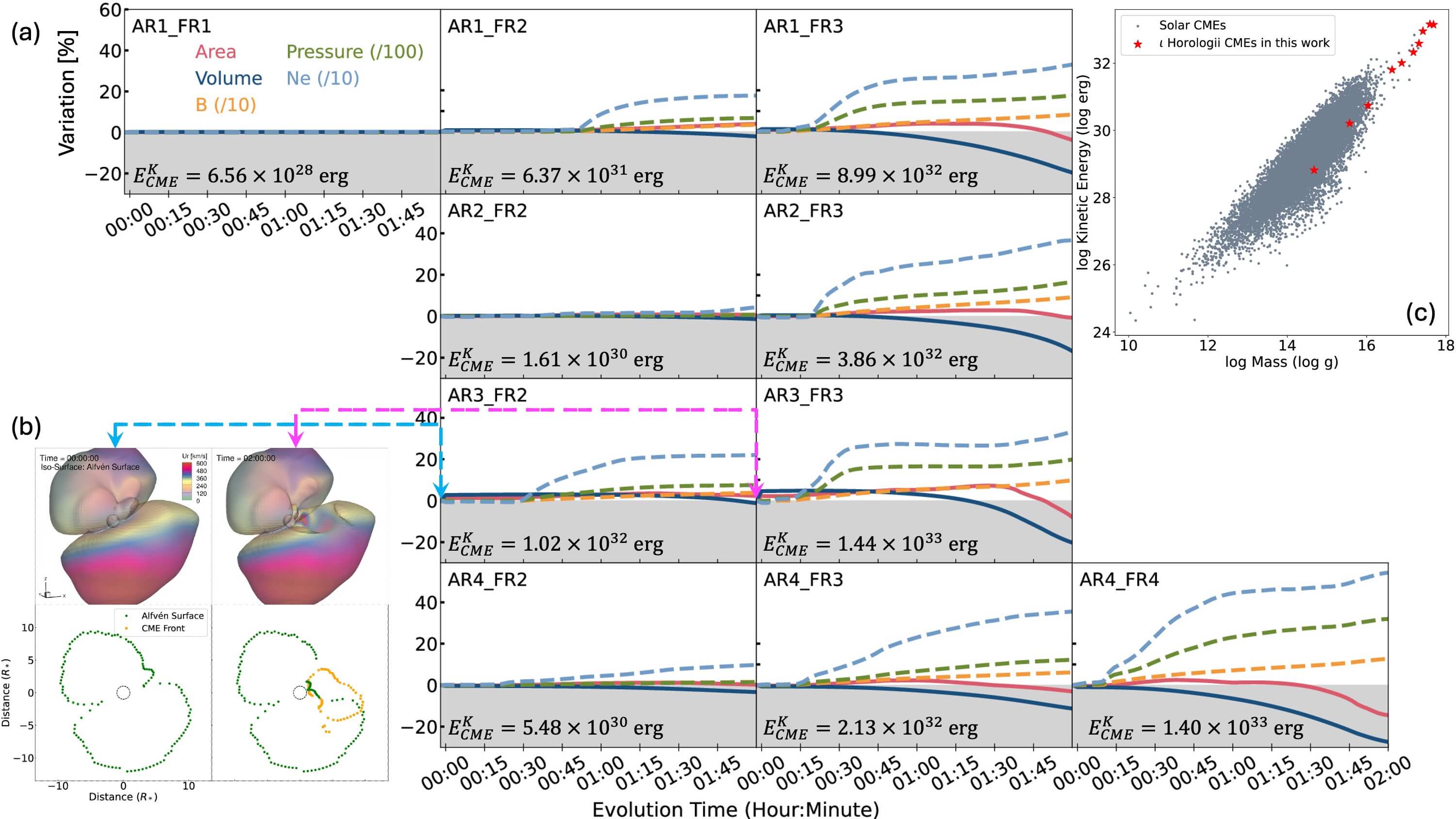}
\caption{Results from a 3D MHD model grid quantifying the instantaneous angular momentum loss associated with CME events. (a) Temporal evolution of the Alfv\'en surface area (solid red) and volume (solid dark blue), together with the variation of several physical parameters averaged over the instantaneous Alfv\'en surface relative to their steady-state values: magnetic field strength (dashed orange), pressure (dashed green), and electron density (dashed light blue). The magnetic field strength and density are divided by a factor of 10, and the pressure by a factor of 100, to fit within the plotting range. The gray shaded region marks intervals where the parameters are lower than their steady-state values. (b) Inset for the AR3\_FR2 case ($E_{\rm CME}^{\rm K} = 1.02 \times 10^{32}$~erg), showing the 3D morphology of the Alfv\'en surface (top) at two time steps over the CME evolution (cyan and magenta arrows). The color bar indicates the stellar wind speed in km~s$^{-1}$. The bottom panels show corresponding snapshots with the projection of the Alfv\'en surface (green) and the CME front (orange) on a plane aligned with the flux-rope main axis. The black circle denotes the stellar surface. (c) Parameter space explored by the stellar CME cases (red stars) in the mass-kinetic energy plane compared with solar events. The solar data are taken from the SOHO/LASCO CME catalog \citep{2009EM&P..104..295G}. \textbf{Adapted from:}~\citetads{2024ApJ...971..153X}.}\label{Fig:Xu2024_CMEs-AML}
\vspace{-0.3cm}
\end{figure*}

\citetads{2024ApJ...971..153X} conducted a detailed numerical study of stellar CMEs with the aim of quantifying their contribution to stellar angular momentum loss and assessing how eruptive transients modify the torque exerted by magnetized stellar winds. Similar to previous studies (see Section~\ref{sec:SW_Models}), the authors employed the SWMF/AWSoM to model the corona and wind of a young solar-type star inspired by the well-studied system $\iota$ Horologii. The large-scale magnetic field used as the inner boundary condition was generated by a global dynamo simulation \citepads{2021AGUFM.U44B..02G}, specifically tailored to reproduce the magnetic field geometry and cycle properties inferred from ZDI monitoring of this star (\citeads{2023MNRAS.524.5725A}, \citeads{2025A&A...704A..68A}). This approach ensured a physically consistent, divergence-free magnetic topology that captured small-scale features as well as the large-scale properties inferred from observations.

A steady-state stellar wind solution was first established, producing a coronal temperature and emission measure distribution broadly consistent with X-ray spectroscopic constraints \citepads{2019A&A...631A..45S}. CMEs were then initiated through the insertion of twisted magnetic flux ropes into selected active regions, following eruption-triggering techniques adapted from solar CME modeling (e.g., \citeads{torok2005}, \citeads{2014ApJ...790..163T}, \citeads{2018ApJ...856...75T}). By varying the flux rope parameters, \citetads{2024ApJ...971..153X} generated a suite of ten eruptive events with masses, velocities, and kinetic energies spanning more than an order of magnitude and extending beyond typical solar CME values (Figure~\ref{Fig:Xu2024_CMEs-AML}). This parameter range allowed the authors to explore how the strength and geometry of stellar magnetic fields modulate CME propagation and angular momentum transport.

The simulations tracked the full three-dimensional evolution of each eruption and quantified angular momentum loss by evaluating the transport of mass and magnetic stresses across the Alfv\'en surface, following classical formulations of stellar wind torques (\citeads{weber1967}, \citeads{1968MNRAS.138..359M}, \citeads{1970MNRAS.149..197M}). \citetads{2024ApJ...971..153X} found that CMEs induced strong, transient perturbations in the stellar angular momentum loss rate, including both positive and negative excursions relative to the steady-state wind. These variations emerged as a result of strong CME-induced changes of the geometrical properties at Alfv\'en surface including its area and volume, but also with modification of the ambient wind parameters such as density, magnetic field strength and total pressure (see Figure~\ref{Fig:Xu2024_CMEs-AML}). While predicting the absolute angular momentum change during individual eruptions proved difficult due to partial cancellations and geometric effects, the amplitude of the angular momentum loss rate variation exhibited strong correlations with CME mass, radial bulk velocity, and kinetic energy. The correlation with kinetic energy was the most robust, with Pearson correlation coefficients approaching unity within the simulated sample. These results supported the idea that energetic magnetic transients can temporarily dominate angular momentum transport in active stars, opening the path for future theoretical and numerical studies.

The authors highlighted several caveats in their approach. The correlations were derived from a limited number of simulated cases and a restricted velocity range, and their validity across a broader parameter space remained uncertain. The MHD treatment neglected kinetic-scale processes, and the simplified eruption mechanism could affect the detailed physical structure of the corona, including its associated Alfv\'en surface. Moreover, the correlations constrained only the amplitude of angular momentum loss variations rather than their absolute values, limiting their direct applicability to stellar spin-down models. Despite these limitations, \citetads{2024ApJ...971..153X} demonstrated that CMEs can introduce highly variable and potentially significant contributions to stellar angular momentum loss, motivating further investigation of their long-term role in stellar rotational evolution.

\subsubsection{Detectability of Stellar CMEs from Doppler Signatures in Coronal Emission Lines}\label{sec:Doppler-EUV_CMEs}

\begin{figure*}[!b]
\centering
\vspace{-0.45cm}
\includegraphics[trim=0.85cm 2.5cm 0.85cm 2.5cm, clip=true, width=0.99\textwidth]{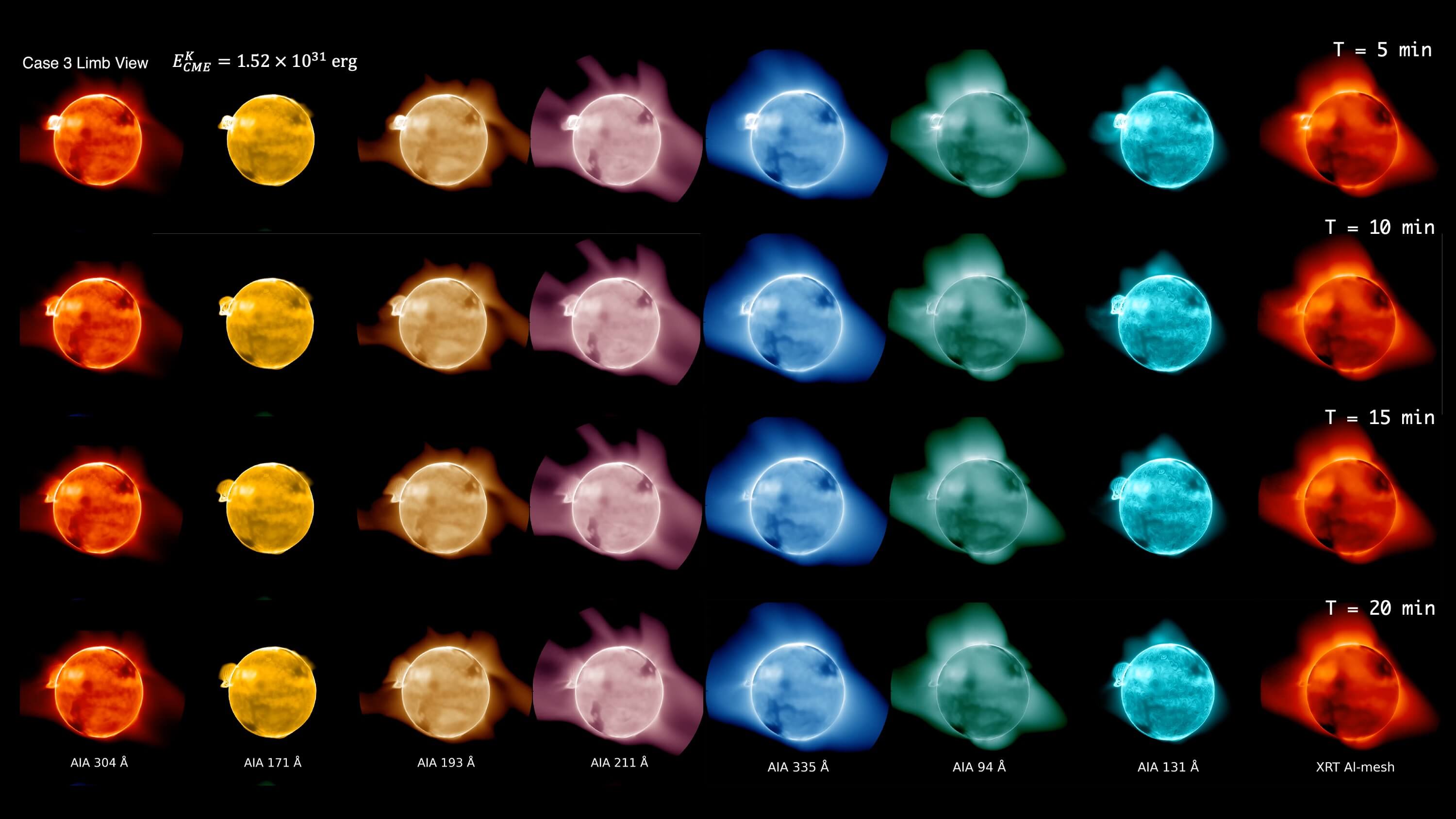}\vspace{-0.1cm}
\caption{Simulated line-of-sight images during a stellar CME synthesized in various EUV and X-ray wavelengths matching solar instrumentation (SDO/AIA, Hinode/XRT). The images are in arbitrary units and logarithmic scale. The kinetic energy event is $E_{\rm CME}^{\rm K} = 1.52 \times 10^{31}$~erg as indicated. Different rows contain snapshots extracted over the first $20$~min of evolution, matching the time-scale employed for the analysis of Doppler-shift asymmetries in the coronal emission lines (see Figure~\ref{Fig:Xu2025_CMEs-Doppler}). \textbf{Adapted from:}~\citetads{2025ApJ...985..219X}.}\label{Fig:Xu2025_CMEs-XUV}
\vspace{-0.3cm}
\end{figure*}

In a follow-up study, \citetads{2025ApJ...985..219X} extended the investigation on stellar CMEs by focusing on their spectroscopic detectability (via Doppler-shift asymmetries) in unresolved stellar observations. Using the same three-dimensional MHD set-up (see Section~\ref{sec:CMEs-AML}), the authors simulated representative eruptive and confined events, synthesizing disk-integrated line profiles for a set of optically thin extreme ultraviolet and X-ray emission lines with formation temperatures spanning $\log T \approx 5.8$ to $7.3$ (see~Figure~\ref{Fig:Xu2025_CMEs-XUV}). The goal was to determine whether CME-driven plasma motions could produce detectable Doppler signatures in stellar spectra and to identify the instrumental conditions under which such detections would be feasible.

\begin{figure*}[!t]
\centering
\vspace{-0.45cm}
\includegraphics[trim=0.0cm 0.0cm 0.0cm 0.0cm, clip=true, width=0.99\textwidth]{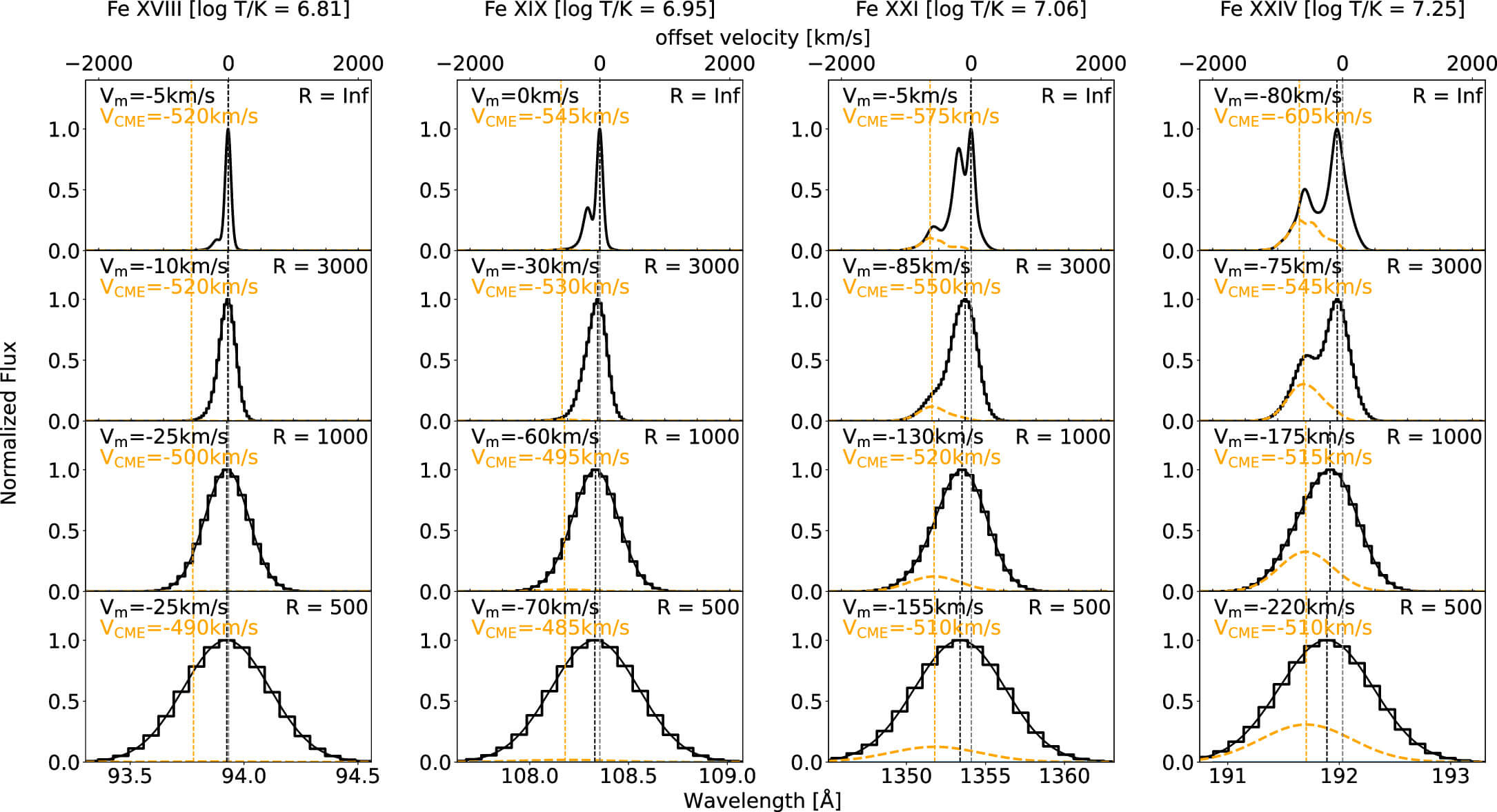}

\vspace{0.25cm}
\includegraphics[trim=0.0cm 0.0cm 0.0cm 0.0cm, clip=true, width=0.99\textwidth]{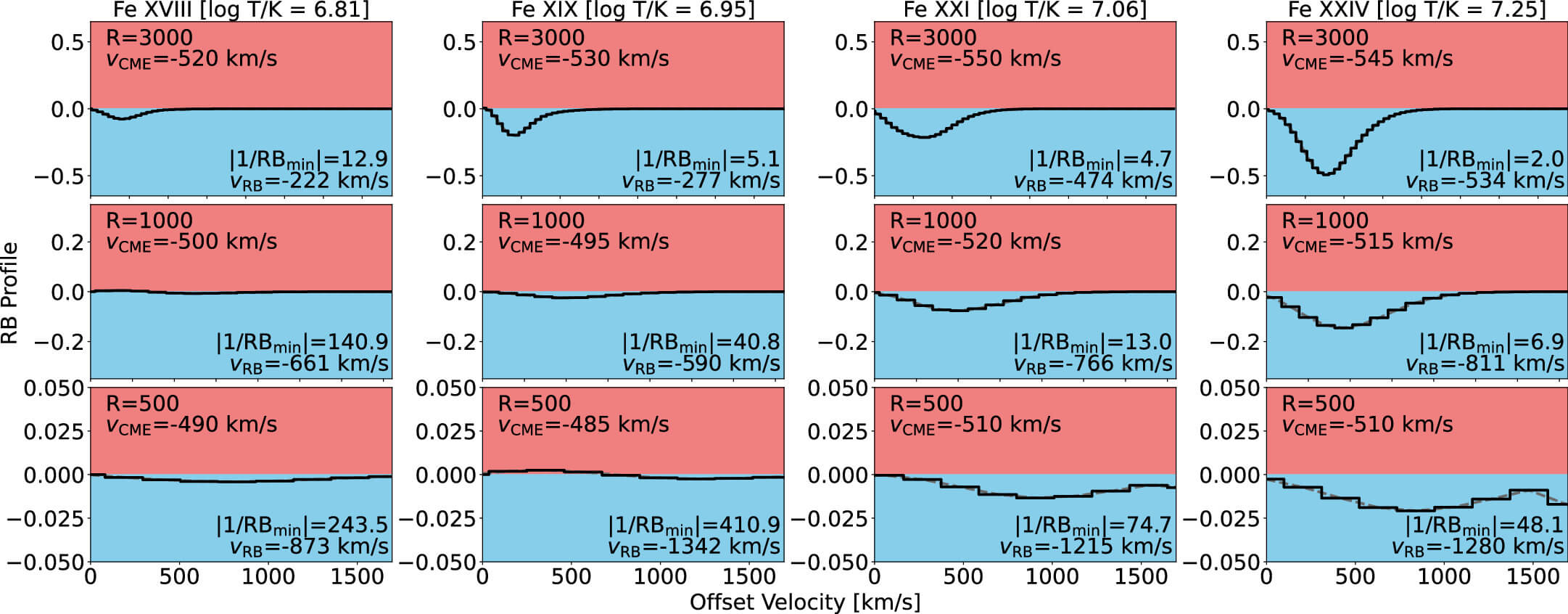}
\vspace{-0.1cm}
\caption{Synthetic line and RB profiles for the CME event of Figure~\ref{Fig:Xu2025_CMEs-XUV}. \textit{Top:} Synthetic line profiles integrated over the first 20~min. Black curves denote the modeled profiles and steps the sampled spectra. The gray dashed and black vertical lines mark the rest wavelength and line centroid, respectively, with the centroid shift labeled as $v_{\rm m}$. Each column shows one line, identified by ion and formation temperature, at different spectral resolutions $R$. The orange dashed curve (not always visible) denotes the escaping-mass component; its peak is marked by an orange dashed vertical line and its velocity by $v_{\rm CME}$. \textit{Bottom:} RB profiles for Fe~\textsc{xviii}~94~\AA, Fe~\textsc{xix}~108~\AA, Fe~\textsc{xxi}~1354~\AA, and Fe~\textsc{xxiv}~192~\AA\ at $R=3000$, 1000, and 500. Black steps show the sampled profiles and gray dotted curves the interpolated profiles. The velocity $v_{\rm CME}$ is adopted from the corresponding top panels, while $v_{\rm RB}$ marks the local minimum, $\rm RB_{\rm min}$. The minimum signal-to-noise ratio required to detect the asymmetries is $|3\sqrt{2}/\rm RB_{\rm min}|$. \textbf{Adapted from:}~\citetads{2025ApJ...985..219X}.}\label{Fig:Xu2025_CMEs-Doppler}
\vspace{-0.3cm}
\end{figure*}


The line synthesis incorporated thermal and nonthermal broadening motivated by solar observations \citepads{1998ApJ...505..957C}, and instrumental broadening parameterized by spectral resolution. Atomic data and contribution functions were taken from the CHIANTI database \citepads{2021ApJ...909...38D}, and the integration was performed over the full simulation domain excluding regions occulted by the stellar disk. To mimic realistic stellar observations, the line profiles were integrated over exposure times of order tens of minutes, comparable to or shorter than those of past EUV missions and consistent with expectations for future high-throughput spectrometers (\citeads{1997ApJS..113..131C}, \citeads{2014AdSpR..53.1003B}, \citeads{2022JATIS...8a4006F}).

\citetads{2025ApJ...985..219X} showed that eruptive CMEs produced clear asymmetries in the blue wings of high-temperature coronal lines, whereas confined eruptions primarily resulted in line broadening without pronounced asymmetries (see Figure~\ref{Fig:Xu2025_CMEs-Doppler}). The strength of the asymmetries depended on the spectral resolution, signal-to-noise ratio, and the match between the line formation temperature and the dominant temperature of the ejecta. For spectral resolutions below $R \sim 2000$, CME signatures manifested mainly as net blueshifts of the line centroids, while at higher resolutions secondary components associated with the erupting plasma became visible in some lines, most notably Fe~\textsc{xxiv}.

Two quantitative techniques were applied to characterize the asymmetries, namely single-Gaussian residual fitting and red-blue wing analysis (\citeads{2012ApJ...759..144T}, \citeads{2022ApJS..260...36Y}). The velocities derived from these methods were shown to include contributions from both CME material and accelerated stellar wind plasma, implying that spectroscopically inferred velocities could not always be directly interpreted as bulk CME speeds. The authors further demonstrated that increasing the spectral resolution beyond $R \sim 3000$ yielded diminishing returns, as intrinsic thermal and nonthermal broadening dominated the line widths.

A major limitation discussed in \citetads{2025ApJ...985..219X} concerned the determination of CME masses from spectroscopic data. The authors intentionally avoided estimating masses from the synthetic spectra, noting that commonly used methods based on coronal dimming or assumed radiative cooling times (e.g.~\citeads{2016ApJ...830...20M}, \citeads{2022ApJ...936..170L}, \citeads{2019NatAs...3..742A}) were not robust within their simulation framework given the limited temporal coverage of the model and the multi-thermal nature of the ejecta. They emphasized that the instrumental requirements derived in their study should be interpreted qualitatively rather than as strict design constraints, as more refined MHD simulations and a broader exploration of CME properties were required.

Overall, \citetads{2025ApJ...985..219X} demonstrated that stellar CMEs can imprint observable signatures in disk-integrated high-energy spectra under favorable conditions. By linking synthetic spectroscopic diagnostics to physically motivated MHD simulations, the study provided a critical step toward interpreting future EUV and X-ray observations of stellar eruptive activity and placed important constraints on the detectability of stellar CMEs beyond the solar system.

\section{Energetic Particles}\label{particles}
\begin{wrapfigure}{r}{0.5\textwidth}
\vspace{-0.3cm}
\centering 
\includegraphics[width=0.5\textwidth]{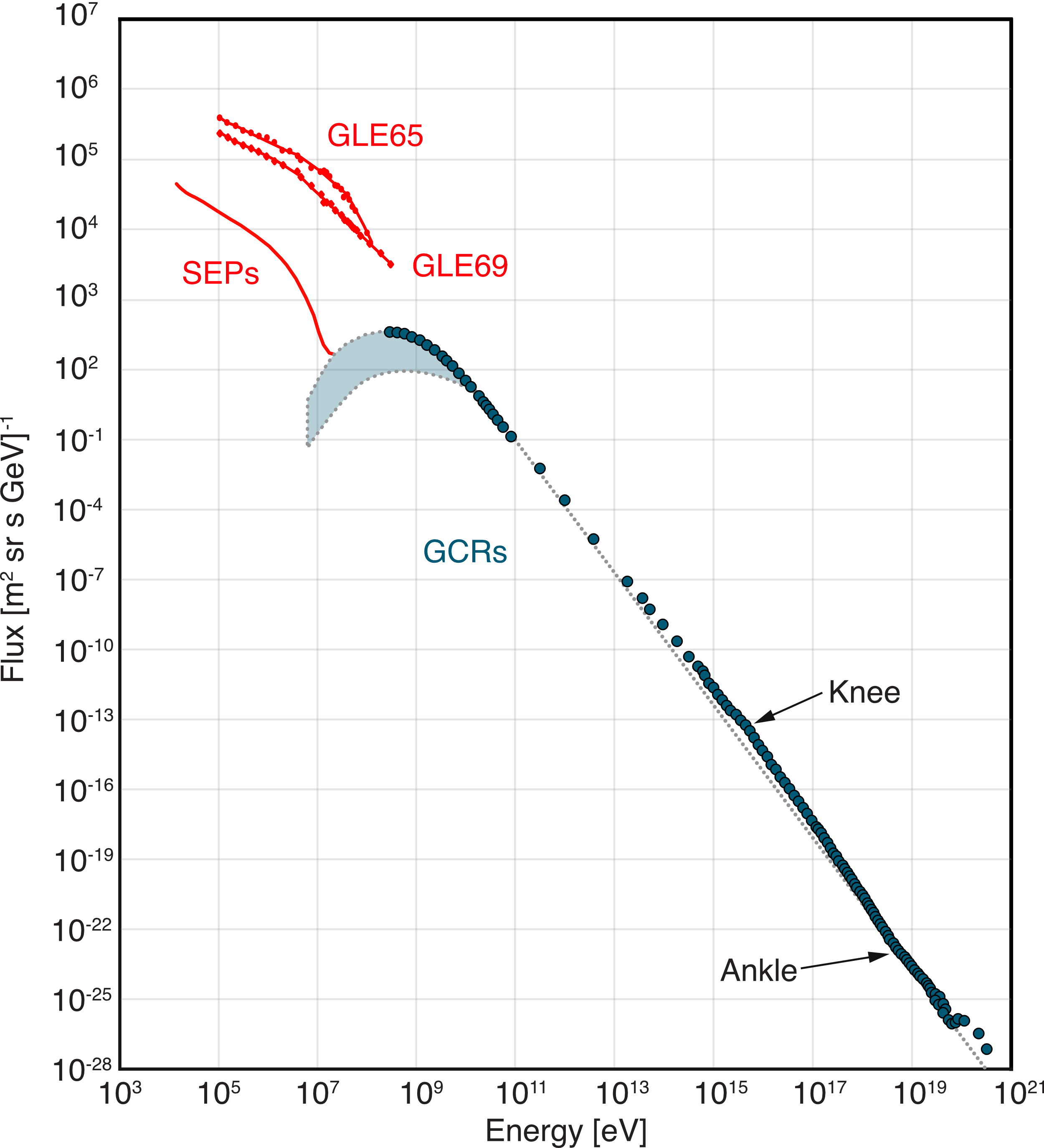}
\caption{Differential CR proton spectra at Earth. GCRs are displayed in petrol \citep[see][]{Swordy2001}, SEPs
in red. In addition the spectra of two strong GLEs are shown: GLE65 (dots) and GLE69 (diamonds).}\label{fig:CRspec}
\vspace{-0.3cm}
\end{wrapfigure}

From studies within the solar system, it is well established that high-energy particles strongly affect the atmospheres of planets such as Earth \citep[e.g.,][]{Banjac-etal-2019a, acp-23-6989-2023}, Venus \citep[e.g.,][]{Nordheim-etal-2015, Herbst-etal-2019b, Herbst-etal-2020a}, and Mars \citep[e.g.,][]{Guo-etal-2019, 2021A&ARv..29....8G, 2019JGRA..124.4556S}. Incoming particles induce atmospheric secondary-particle cascades that can lead to substantial atmospheric changes. Motivated by the potential influence of a star's radiation environment on planetary habitability \citep[see, e.g.,][]{Linsky14,Tabataba-Vakili-etal-2016}, possible alterations to upper-atmospheric chemistry that may obscure biosignatures \citep{Scheucher-etal-2018,Herbst-etal-2019c,Herbst_2024}, and the potential erosion of planetary atmospheres \citep{airapetian2016,Griessmeier-etal-2016}, the study of cosmic ray effects on exoplanetary atmospheres has become increasingly important in the context of exoplanets. Accordingly, this section provides a brief introduction to cosmic rays and their heliospheric transport, followed by a discussion of recent progress in modeling their transport around other stars and assessing their impact on exoplanetary radiation environments.

Cosmic rays, despite their name, are charged, sometimes highly energetic particles. Their origins are linked to key processes and dynamics involved in star formation, stellar evolution, supernova explosions, and the conditions of the interstellar medium within galaxies \cite[see, e.g.,][]{BTM20}. Based on their origin, cosmic rays can be primarily classified into two distinct categories: Galactic cosmic rays (GCRs, Section~\ref{sec:GCRs}) and solar/stellar energetic particles (SEPs and StEPs, respectively, Section~\ref{sec:StEPs})\footnote{Note that particles originating from the Jovian magnetosphere \cite[e.g.][and references therein]{StraussEA24} or accelerated at the heliospheric termination shock \cite[e.g.][]{GiacaloneEA22} are also classified as cosmic rays, but are not discussed here.}. These populations are also mostly distinct in terms of energy. As shown in Figure~\ref{fig:CRspec}, SEPs (red line) and their most energetic subclass, known as ground level enhancement (GLE) events (red circles), are restricted to the lower-energy regime, ranging from approximately $10^{6}$~eV to (typically) around $10^{9}$~eV \cite[see, e.g.,][]{Reames-1999} for protons. The high-energy differential energy spectrum (shown for protons) in general follows a power law with GCRs (in green) occupying a broad range of the high-energy regime reaching energies of up to 10$^{21}$~eV, with a drop-off in intensity below approximately $10^{10}$~eV. This decrease is due to the modulating influence of the heliosphere and will be discussed in more detail below.
 
\subsection{Galactic Cosmic Rays}\label{sec:GCRs}
\subsubsection{Galactic Cosmic Rays and their Heliospheric Transport}
\label{sec-gcr}

GCRs originate from far outside our solar system and are accelerated at the shock fronts of supernovae \citep[on average occurring once every 50 years within our galaxy;][]{2006Natur.439...45D}, supernova remnants \citep{2005ApJ...619..314B}, and pulsars \citep{j.asr.2007.05.051}. As such, they consist mainly of 98\% hadrons \citep[i.e., 87\% protons, 12\% helium nuclei and 1\% heavier elements up to iron, e.g.,][and references therein]{annurev.33.120183.001543,1998A&A...330..389W} and 2\% electrons and photons \citep{j.physletb.2008.07.018}. The relative galactic abundance of hydrogen and helium is much smaller compared to the solar abundances as both elements have high first ionization potentials (FIPs) and can therefore not be ionized. Consequently, they cannot be accelerated to GCR energies, and thus cannot be detected \citep[see, e.g.,][]{Mueller-etal-1991}. However, as GCRs interact with the ISM, elements like Sc, Ti, V, Cr, and Mr are much more abundant compared to the solar system composition \cite[for more detail, see][]{Wiedenbeck}. 

Once accelerated, GCRs must propagate through the ISM, encountering matter, magnetic fields, and radiation fields, all of which serve as targets for cosmic ray interactions, before reaching the solar system. GCR protons scatter within these magnetic fields and gradually diffuse away from their sources, while GCR electrons interact with the magnetic and radiation fields and generate synchrotron radiation and boost gamma rays with the inverse Compton effect \citep[][]{2021ASSL..462.....S}. GCRs with energies above 10$^{14}$~MeV have a Larmor radius of more than 5$\times10^{17}$~km when they arrive at Earth, and thus are expected to originate even beyond our galaxy. Observations suggest that GCRs impinge on the Solar System in a largely isotropic manner \cite[see][and references therein]{A06}. As such, the GCR population represents an ambient contribution to the radiation environment that the heliosphere, or any astrosphere for that matter, would encounter as it traverses its galactic orbit. Indeed, the GCR intensity is expected to change depending on the star's position in the galaxy \citep{amato,pez,arm}, and would, for example, be expected to be larger in regions of active star formation such as within the Galactic spiral arms \citep{Zucker22} as opposed to, e.g., the interspiral regions \citep{aviv,SchererEA06,j.asr.2007.05.051}. However, the intensities of GCRs at Earth, or any exoplanet (and therefore the resulting radiation environment), would be different from their local interstellar intensity outside the astrosphere, as GCRs are actively modulated within any star's astrosphere. The degree of this modulation, which essentially amounts to a reduction in the intensity of these particles, combined with changes in the energy dependence of their differential intensity spectra, depends strongly on the plasma properties of the astrosphere. This modulation arises through several physical mechanisms whose relative importance is itself highly sensitive to the specific characteristics of the astrosphere in question \citep[see, e.g.,][]{LightEA25}. As such, cosmic-ray modulation provides a direct link between the stellar properties of the host star and the exoplanetary radiation environment.

\begin{figure}[!t]
    \centering
    \includegraphics[width=0.7\linewidth]{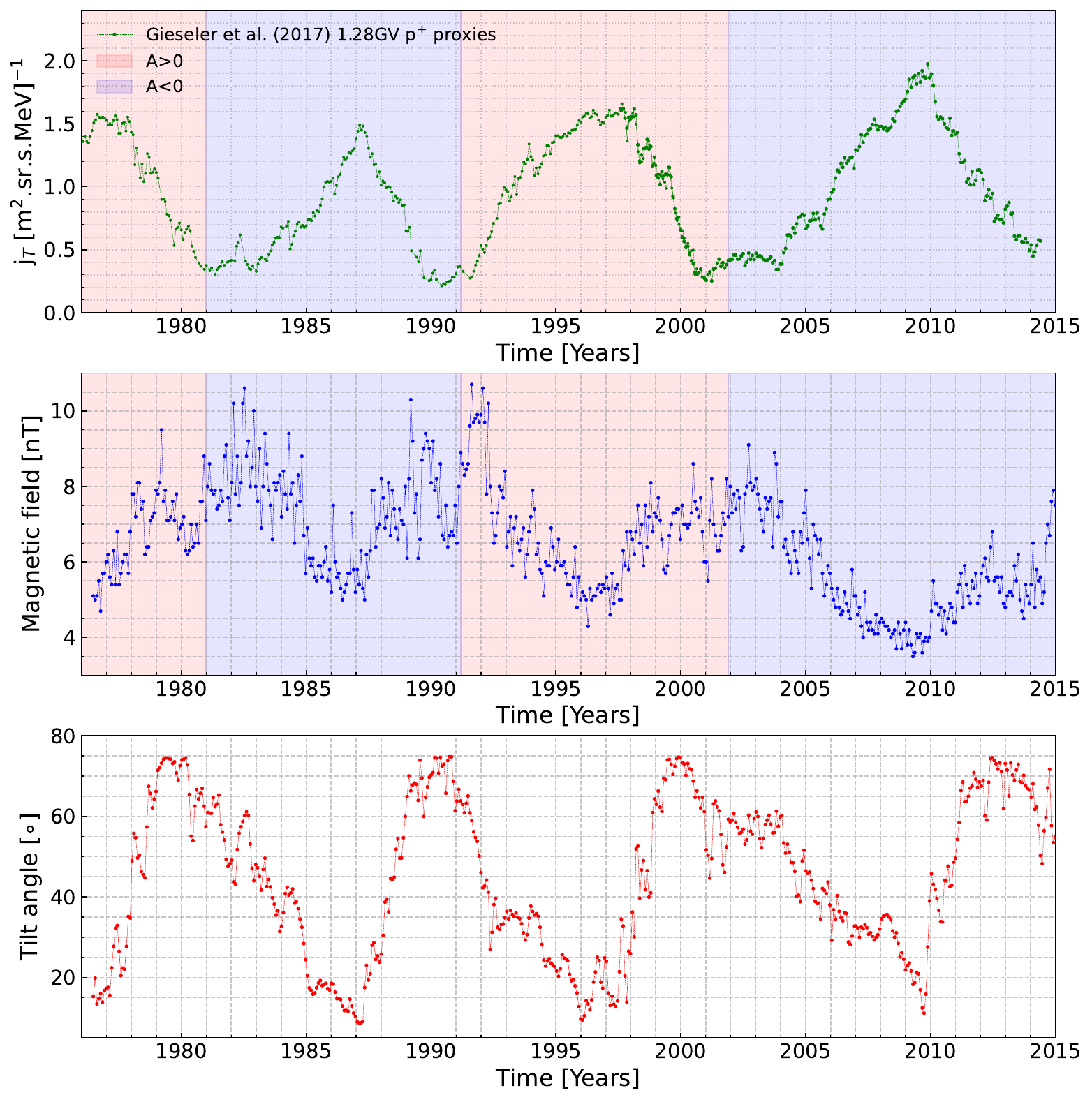}
    \caption{Top panel: Heliospheric $1.28$~GV GCR proton proxy intensities at 1~au reported by \citet{GieselerEA17} as function of time. Magnetic polarity reversals are indicated by shaded bands, after \citet{Pish}. Middle panel: Heliospheric magnetic field magnitude at Earth taken from OMNI data (\url{https://omniweb.gs.fc.nasa.gov/}, \citet{KingPap}). Bottom panel: Heliospheric tilt angles from the classical model results from the Wilcox Solar Observatory (\url{http://wso.stanford.edu/Tilts.html}, \citet{Hoeksema}).}
    \label{fig:heliomod}
\end{figure}

As an illustrative example of the effects of modulation in the heliosphere, the top panel of Figure~\ref{fig:heliomod} shows spacecraft observations at Earth of the differential intensities of GCR proton proxies at a rigidity of $1.28$~GV reported by \citet{GieselerEA17}, as a function of time. A clear alternating maximum/minimum pattern can be discerned with a periodicity of approximately $11$ years, corresponding to the solar cycle, and reflected in the temporal variations of the heliospheric magnetic field (HMF) magnitude at $1$~au and the heliospheric tilt angle, the angle between the Sun's magnetic and rotational axes, shown in the middle and bottom panels of Figure~\ref{fig:heliomod}, respectively. Peaks in GCR intensities correspond to periods of minimum solar activity, characterized (amongst other quantities) by smaller HMF magnitudes and tilt angles, while the smallest GCR intensities correspond with periods of high solar activity. As no two solar cycles are exactly the same, the details of the resulting GCR intensity profiles differ, reflecting the differences in heliospheric plasma parameters between the different cycles \cite[see, e.g.,][]{CliverEA13,MolotoEA18,EngelbrechtWolmarans20,Li2021,FuEA21}. This reflects the modulating influence of the heliosphere: during periods of lower solar activity, the inward transport of GCRs is governed (mostly) by a combination of diffusion perpendicular to the \cite[relatively tightly wound, see][]{frans} HMF and drift, while during solar maxima, turbulence levels in the HMF reduce drift effects, thereby rendering the inward transport of GCRs across the HMF less effective, resulting in the (roughly) anticorrelated relationship that can be seen between the HMF magnitude and GCR intensity \cite[see, e.g.,][]{Moloto20,MolotoEA23}. A slight lag of approximately a year can be discerned between the features of the temporal plots of the GCR intensities and HMF magnitudes \citep{Forbush58,UsoskinEA98}. This is a consequence of GCR transport throughout the heliosphere: temporal changes in heliospheric plasma conditions that affect GCR propagation require time to travel outward through the heliosphere, while GCRs themselves typically take about a year to propagate to Earth \cite[see, e.g.][]{StraussEA11}. An approximately $22$-year cyclic pattern of peaked and plateaued temporal profiles can also be seen in solar minimum GCR intensities corresponding to the 22-year Hale cycle of reversing HMF polarity, with the former corresponding to periods of negative HMF polarity, denoted by $A<0$ and blue shading in Figure~\ref{fig:heliomod}, and the latter to periods of positive HMF polarity (red shading, $A>0$)\footnote{By convention, the HMF in the northern hemisphere points away from the Sun and towards it in the southern hemisphere during $A>0$, and vice versa during $A<0$.}. The influence of HMF polarity on GCR intensities during solar minima arises because these particles experience drift effects due to gradients and curvatures of the HMF as well as due to the heliospheric current sheet: during $A>0$, positively charged particles drift into the heliosphere from the polar regions and out along the current sheet, with this pattern reversing during $A<0$ (not that negatively charged CRs, such as electrons and antiprotons, will drift in the opposite sense) \cite[see, e.g.,][]{JokipiiEA77,JokThom81,TroskieEA24}. Temporal variations on smaller timescales have also been observed, such as 27-day variations due to the presence of SIRs in the solar wind \cite[see, e.g.,][]{renata}, and diurnal variations due to small anisotropies in the GCR intensity arising from the interplay between drift and diffusion perpendicular and parallel to the HMF \cite[e.g.][]{pom}. As can be seen from this example, modeling the transport of GCRs within the heliosphere is a complex endeavor, requiring not only insight into the various transport mechanisms these particles encounter as they traverse the heliosphere, but also of the large- and small-scale heliospheric plasma quantities that influence those mechanisms.

The transport of GCRs is described by the \citet{Parker-1965} transport equation, which models the evolution of the omnidirectional GCR phase space density $f(\vec{r},p,t)$ as a function of position, momentum, and time\footnote{Related to the differential intensity in Figure~\ref{fig:heliomod} by $j_T=p^2 f$ \citep{Moraal13}.}, and given by
\begin{equation}
\frac{\partial f}{\partial t}
    =
      \underbrace{\nabla \cdot \left( \overleftrightarrow{K} \cdot \nabla f \right)}_{(1)}
    -  \underbrace{\vec{V}_\text{sw} \cdot \nabla f}_{(2)}
    +  \underbrace{\frac{1}{3} \left( \nabla \cdot \vec{V}_\text{sw} \right) 
            \frac{\partial f} {\partial \ln p}}_{(3)}\, .
\label{eq:TPE1}
\end{equation}
In the above, term (1) describes the diffusion of GCRs both parallel and perpendicular to the HMF due to interactions with the turbulent heliospheric/astrospheric magnetic field as well as drift effects due to gradients and curvatures in said field, alongside drifts along the heliospheric current sheet; term (2) describes the convection of GCRs with the solar/stellar wind, with speed $\vec{V}_\text{sw}$; and term (3) describes the adiabatic energy changes of these particles as they encounter the solar/stellar wind. As can be inferred from the discussion of Figure~\ref{fig:heliomod}, diffusion and drift are particularly important mechanisms. From first principles, the latter mechanism must be modeled either in 2D or 3D, while, given the inherent differences in charged particle diffusion parallel versus perpendicular to a guide field such as the HMF, the former also requires modeling in at least two spatial dimensions. These processes are described by the diffusion tensor, given in HMF-aligned coordinates as \citep[see, e.g.,][]{BurgerEA08}
\begin{equation}
\overleftrightarrow{K}=\overleftrightarrow{K}_{s}+\overleftrightarrow{K}_{a}=\left[\begin{array}{ccc}
\kappa_{\perp,3} & \kappa_{A} & 0\\
-\kappa_{A} & \kappa_{\perp,2} & 0\\
0 & 0 & \kappa_{\parallel}\end{array}\right]|, ,
\label{eq:K}
\end{equation}
which can be separated into symmetric and antisymmetric tensors (indicates by subscripts `s' and `a') which separately describe diffusion and drift. Both processes in turn are governed by diffusion coefficients $\kappa$, where subscripts denote diffusion parallel or perpendicular to the HMF, $\kappa_A$ being the drift coefficient. This is related to the drift velocity by \cite[e.g.][]{JokipiiEA77,MinnieEA07}
\begin{equation}
    \overrightarrow{v}_{d}=-\nabla\cdot\overleftrightarrow{K}_{a}=\nabla\times\kappa_{A}\hat{e}_B\, ,
    \label{eq:drift}
\end{equation}
where $\hat{e}_B$ denotes a unit vector in the local direction of the HMF. All coefficients in the diffusion tensor depend in some way on the turbulence embedded in the HMF, which directly influences parallel \cite[e.g.][]{TeufelSchlickeiser2002,Shalchi05} and perpendicular diffusion \citet[e.g.][]{MattEA03,Shalchi10,RuffoloEA12}, and acts to reduce drift effects \citet{MinnieEA07,TS12,EngelbrechtEA17,ElsEA24,WangEA26}.

In the heliosphere, the modulation of GCRs has been studied extensively since the beginning of the Space Age, using a continuously refined set of numerical modulation codes that solve the Parker equation with increasing levels of complexity \cite[see, e.g.,][and references therein]{Quenby84,frank,Kota13,EngelbrechtEA22}. This progress has been driven by the increasing availability of high-quality observations throughout the heliosphere not only of GCR intensities \cite[see][]{Bindi23}, but also of solar wind plasma parameters that influence the transport of these particles. These developments have also driven advances in our understanding of the transport mechanisms experienced by GCRs, including diffusion \cite[see, e.g.,][]{Shalchibook,Shalchi20} and drift \cite[e.g.][]{EngelbrechtEA17,MohloloEA22} as well as the broader evolution of the HMF and solar wind plasma properties, both on large scales \cite[e.g.][]{WiengartenEA16,PogorelovEA17,FraternaleEA23,UsmanovEA25} and on the (turbulent) small scales \cite[e.g.][]{BC13,EngelbrechtStrauss18,OughtonEngelbrecht21,AdhikariEA21}, which are again implemented in heliospheric modulation models \cite[for a review, see][]{EngelbrechtEA22}. Such direct observations are not yet available for exoplanet-hosting astrospheres, but the theoretical framework provided by heliospheric studies can provide a useful basis for extrapolation.

\subsubsection{Galactic Cosmic Ray Modulation in Astrospheres}

The mechanisms governing GCR transport in a given astrosphere depend strongly on its plasma characteristics, which in turn depend strongly on the type of star, which can often be very different from the Sun. As such, MHD simulations, motivated by extant observations of the star in question, can provide some insight into the behavior and evolution of large scale plasma quantities such as the stellar wind and astrospheric magnetic field (AMF) that influence GCR transport, as well as the large-scale structure of the astrosphere such as its size, and salient features such as the termination shock location, which influence the modulation volume of said astrosphere and thus indirectly the intensity of GCRs at exoplanets close to the star. At the time of writing, many such simulations have been reported for a broad range of stellar types, including G-type stars \cite[e.g.][]{2021ApJ...916...96A, EngelbrechtEA26}, O-type stars \cite[e.g.][]{Baalman22}, M-type stars \cite[e.g.][]{HerbstEA20, 2020MNRAS.494.1297M, RL23, Scherer-etal-2025}, red supergiants \cite[e.g.][]{MeyerEA21, MeyerEA24}, among others \cite[see, e.g., the review by][]{HerbstEA22}, often specifically for exoplanet-hosting astrospheres. Such simulations provide the basic input required by modulation models to calculate GCR differential intensities at the locations of exoplanets. Often, analytic formulae based on hydrodynamic theory are used to calculate, e.g., distances to astrospheric termination shocks \cite[e.g.][]{Parker-1963,Wilkin-2000} where direct MHD simulations, which are complex to perform and computationally expensive, are not available. However, these formulae are subject to limitations in that they do not adequately account for {\it magneto}hydrodynamic effects and therefore may overestimate distances compared to those obtained from MHD simulations \cite[see discussions in][]{Scherer-etal-2025,EngelbrechtEA26}. 

Initial GCR modulation studies in astrospheres other than our own were done using either the force field approximation \citep{Caballero-Lopez-Moraal-2004} or solving a 1D Parker transport equation \cite[see][]{EngelbrechtDiFelice20} numerically, motivated by the relative ease with which both approaches can provide a computationally-inexpensive estimate for GCR intensities, as well as by the uncertainties implicit to estimating GCR transport parameters (and the plasma conditions they depend on) in distant astrospheres. The 1D Parker transport equation is given by
\begin{equation}
    \frac{\partial{f}}{\partial{t}} = \underbrace{\frac{1}{r^2}\frac{\partial}{\partial{r}}(r^2\kappa_{rr}\frac{\partial{f}}{\partial{r}})}_{(1)}- \underbrace{\frac{1}{r^2}\frac{\partial}{\partial{r}}(r^2V_sf)}_{(2)} + \underbrace{\frac{1}{3r^2}\frac{\partial}{\partial{r}}(r^2V_s)\frac{1}{p^2}\frac{\partial}{\partial{p}}(p^3f)}_{(3)}\, ,
    \label{eq:TPE1D}
\end{equation}
assuming a radially constant solar wind speed, and where terms (1) through (3) correspond to the same physical processes discussed for Equation~(\ref{eq:TPE1}). The diffusion here is governed by a single radial diffusion coefficient $\kappa_{rr}$, which is a key limitation, as it does not allow for differences between diffusion parallel and perpendicular to the HMF. This is also a limitation of Equation~(\ref{eq:TPE1D}), since drift is an inherently higher-dimensional transport phenomenon. Technically, a radial component of the drift velocity could be incorporated, but this would mean that the drift velocity as modeled would not be divergence-free, which in the case of Equation~(\ref{eq:drift}) would be a strict requirement, rendering such an approach mathematically incorrect as well as unphysical. It should be noted that differences in the GCR intensities reported by various modulation studies reflect differences in the input parameters employed in the various models, ranging from different stellar wind speeds and magnetic fields, to differences in assumptions as to particle transport parameters such as $\kappa_{rr}$, to differences in the assumed local interstellar spectrum. As such, 1D studies can yield contradicting results as to GCR intensities at a particular exoplanet. An example of this are GCR intensities computed for the Proxima Centauri astrosphere, specifically at Proxima Centauri b: \citet{Sadovski-etal-2018} and \citet{MesquitaEA22} reported that GCR intensities are significantly lower than those at Earth, while \citet{HerbstEA20} computed intensities almost an order of magnitude higher. Similarly, for the TRAPPIST-1 system, \citet{strum} found that GCR intensities are negligible, while \citet{Herbst_2024} reported on intensities comparable to observations at Earth. As such, 1D studies of GCR modulation have been performed for an array of solar-type stars \citep{RL21a}, M-dwarf systems \citep{HerbstEA20,MesquitaEA21,MesquitaEA22,RL23}, young-Sun proxy stars \citep{RodgersEA21,RaesideEA25}, and even more exotic astrospheres such as those of Luminous Blue Variable stars \citep{LightEA22}.

As has been noted above, a higher-dimensional approach is needed when modeling the transport of GCRs, in order to appropriately model the various transport mechanisms known from heliospheric studies to play significant roles in this process, such as diffusion parallel and perpendicular to the magnetic field, and drifts. Even if model input parameters are open to debate, it is vital to model their influence on GCR transport in an appropriate, self-consistent manner. In this particular context, a physics-first approach, whereby the influence of plasma quantities on particle transport parameters such as diffusion coefficients is modeled from first principles, is most useful, as it allows for extrapolation beyond the heliospheric context. The first 3D study of the astrospheric transport of GCRs was done by \citet{EngelbrechtEA24} for the Proxima Centauri system, using astrospheric plasma parameters yielded by the CRONOS MHD code \citep{Kissmann-etal-2018} as inputs. These authors computed GCR intensities considerably larger than those observed at Earth at Proxima Centauri b. The additional insight gained from 3D modeling was the reason why this was so: the longer rotation period of Proxima Centauri resulted in an AMF that was highly underwound in comparison to the HMF. This means that modulation in this atmosphere was fundamentally different to the heliospheric case. In the heliosphere, the HMF direction is essentially transverse to the radial direction beyond approximately $10$~au, which means that GCRs primarily rely on perpendicular diffusion (combined with drift during solar minima) to reach Earth. For Proxima Centauri, the underwound AMF means that GCRs can now also depend on diffusion parallel to the AMF to reach Proxima Centauri b. As the parallel diffusion coefficient is considerably larger than the perpendicular diffusion coefficient\footnote{Approximately two to three orders of magnitude larger at $1$~au in the heliosphere, see \citet{Palmer82,MinnieEA07b,EngelbrechtEA22,ElsEA24}}, this results in larger GCR intensities due to easier access. Furthermore, \citet{EngelbrechtEA24} also reported that drift effects play a significant role in the transport of these particles, albeit a more muted one than observed in the heliosphere due to the underwound AMF. Such an interplay of transport mechanisms can, as noted above, only be modeled in 3D. The relationship between stellar rotation period and GCR intensities was further investigated in 3D by \citet{LightEA25} for several different stellar systems, where it was found that fast rotation of the host star leads to lower GCR intensities at exoplanetary locations, and vice versa. \citet{Scherer-etal-2025} directly compared GCR intensities calculated at LHS1140b using both a 1D and a 3D GCR modulation code, using the same plasma input parameters as yielded by the CRONOS MHD code. The 3D code yielded intensities essentially identical to the local interstellar spectrum assumed outside the astrosphere, implying no modulation at all. This is not entirely unexpected, as the astrosphere of LHS1140 is extremely small (with a termination shock at $3$~au or less). However, the 1D model yielded modulated spectra, with differential intensities at LHS1140b between one and two orders of magnitude smaller than the interstellar spectrum at the lowest GCR energies considered in that study, which would not be expected of such a small astrosphere. Recently, \citet{EH26} implemented some observationally-motivated stellar cycle-related changes in a 3D GCR modulation study within the astrosphere of Proxima Centauri. Although the computed cyclic variations were modest compared with those observed in the heliosphere, GCR proton and helium intensities remained well above those observed in the heliosphere throughout the stellar cycle, with drift effects remaining important even during stellar maximum. This latter finding is in contrast with what is observed in the heliosphere, where drift effects become essentially negligible during solar maximum \cite[see, e.g.,][]{Moloto20}, and highlights the importance of higher-dimensional GCR transport models, in order to take into account the nuances of particle transport within astrospheres often very different from our own.

While the larger-scale plasma quantities of an astrosphere, such as the stellar wind speed or AMF, can be computed using a MHD model, smaller-scale turbulence quantities such as magnetic variances and correlation scales are harder to pin down. This presents some difficulties, as scattering theories used to derive GCR diffusion coefficients require these quantities as direct inputs. The 3D modulation studies mentioned above employ simple power law scales, motivated from heliospheric observations, to model the spatial evolution of turbulence quantities. This, of course, is not ideal, as they may not be appropriate to the astrosphere in question. An alternative to this approach would be to employ turbulence transport models (TTMs), derived from the MHD equations and modeling the spatial evolution of various turbulence quantities, subject to different driving mechanisms. In the past decades, several TTMs of increasing refinement have been proposed \cite[see, e.g.,][]{BreechEA08,OughtonEA11,ZankEA12,ZankEA17,BishopEA25}, which yielded results in good agreement with heliospheric spacecraft observations, some of which have been successfully employed in heliospheric GCR modulation studies \cite[see, e.g.,][]{EngelbrechtBurger13,Engelbrecht24}. Although direct observations of turbulence in astrospheres have not yet been made, such models can provide insights as to the behavior of these quantities in the same way that MHD codes inform us as to large-scale plasma quantities. Such an approach was taken by \citet{EngelbrechtEA26}, who employed the \citet{OughtonEA11} TTM to study the evolution of turbulence in the astrosphere of $\kappa^1$ Ceti, a young Sun proxy, finding turbulence levels significantly higher than those encountered in the modern heliosphere. Furthermore, these model results are used as inputs for diffusion (and turbulence-reduced drift) coefficients, to compute GCR intensities at $1$~au. These intensities were found to be significantly lower than modern heliospheric observations. This would seem to be counterintuitive, as higher turbulence levels imply more effective perpendicular diffusion, but they also lead to a greater reduction in drift effects, which hampers the inward transport of GCRs in an astrosphere with a fast-rotating host star.

\subsection{Solar/Stellar Energetic Particles}\label{sec:StEPs}

\subsubsection{Solar Energetic Particles (SEPs)}\label{subsec:seps}

Due to their acceleration mechanisms, solar flares (direct current acceleration during the magnetic reconnection in flares), and CMEs (diffusive shock acceleration at CME-driven shock fronts) are efficient drivers of SEPs (see Figure~\ref{fig:CRspec}, red solid line). Depending on their parent solar source, these events have been traditionally divided into two different classes: impulsive and gradual events \citep[see, e.g.,][]{Reames-2013}. Impulsive events that originate in stochastic acceleration and magnetic reconnection processes \citep[][]{Aschwanden-2002}, where particles are accelerated up to a few GeV, are commonly associated with flares \citep[see, e.g.,][]{Klein-Posner-2005}. Although being accelerated to such high energies, most of the accelerated particles are not able to escape the confined magnetic field configurations and instead are accelerated back towards the solar surface. Deceleration thereby causes the characteristic X-ray, synchrotron, EUV, and in rare cases even white-light emissions that typically can be observed within minutes to hours. Nevertheless, particles that can escape the confined magnetic field configuration travel into the HMF along open magnetic field lines. Flare-associated SEP events, thereby, have rather small acceleration regions, leading to a narrow spread within about $40^{\circ}$ \citep[][]{Reames-1999}, and tend to last a few hours. Gradual events, on the other hand, tend to occur due to continued particle acceleration at CME-driven shocks \citep[see, e.g.,][]{Cane-Lario-2006, Kahler-Vourlidas-2013, Desai-Giacalone-2016}, where mainly protons and ions are accelerated. Due to different seed particle populations, varying elemental compositions and charge states are produced \citet{Tylka-etal-2005}. Because of the typically large spatial extent of CME-driven shocks, the spread of SEPs can be much broader, reaching about $180^{\circ}$ \citep[see][]{Cane-1996}, with a few events reported to exhibit particle spreads extending around the entire Sun \citep[e.g.,][]{2015ApJ...799...55G, 2025A&A...695A.127D}. In contrast to magnetic reconnection, particles being accelerated at moving CME shock fronts show a rather gradual event profile. These events, however, produce some of the most energetic events observed at Earth \citep[][]{Mewaldt-etal-2012, Kuehl-etal-2017}. However, as shown by \citet{Papaioannou-etal-2016}, gathered observations indicate that a much wider variety of SEP events exists, and the existence of yet another class of so-called hybrid events has to be assumed. These events show signatures of both flare and CME-driven shock accelerated particles leading to strong SEP events \citep[see, e.g.,][]{Kallenrode-2003}. According to \citet{Vainio-etal-2007}, such events show properties of impulsive events, although mimicking gradual events.

GLE events are rare, extreme SEP events in which relativistic particles---primarily protons---reach energies sufficient to produce secondary particle cascades in Earth's atmosphere, leading to a detectable increase in ground-based neutron monitor counts (see Figure~\ref{fig:CRspec}, red dots). These particles typically have energies exceeding $\sim$400~MeV. During their propagation through interplanetary space, they can penetrate both the terrestrial magnetosphere and atmosphere. In the atmosphere, they significantly affect upper and middle atmospheric chemistry \citep[see, e.g.,][]{Sinnhuber-etal-2012} through the production of secondary particles.
As a result, GLEs can be observed by ground-based neutron monitors (NMs) \citep[see][]{Simpson-2000}. They are rare events, occurring on average about once per year. As of today, 77 GLEs have been recorded\footnote{See, for example, \href{https://www.nmdb.eu/nest/}{https://www.nmdb.eu/nest/} or \href{https://gle.oulu.fi/}{https://gle.oulu.fi/}}, with the strongest recorded event occurring on February 23$^{\mathrm{rd}}$ 1956 (GLE05).

Extreme SEP events (ESPEs) have been inferred from cosmogenic radionuclide records. Cosmogenic radionuclides are produced by the interaction of extraterrestrial energetic particles (cosmic rays, mainly protons and some heavier ions) with atmospheric molecules. Unexpected increases in annually resolved $^{14}$C concentrations recorded in tree rings were observed around AD~775 \citep{2012Natur.486..240M} and AD~993 \citep{2013NatCo...4.1748M}, which can be explained by enhanced production due to extreme SEP events. Subsequent multi-radionuclide analyses using high-resolution $^{10}$Be and $^{36}$Cl ice core data \citep{2015NatCo...6.8611M} strengthened the ESPE interpretation and ruled out alternative explanations such as extra-solar $\gamma$-ray bursts \citep[e.g.,][]{HambaryanEA2013}. Note that among these isotopes, $^{36}$Cl is particularly sensitive to SEP-induced changes in production rates.

In general, such extreme solar events often occur during times of low/moderate solar activity. To date, six ESPEs have been confirmed to occur in all three cosmogenic radionuclide datasets: AD~774/775 \citep{2012Natur.486..240M, 2015NatCo...6.8611M}, AD~993/994 \citep{2013NatCo...4.1748M, 2015NatCo...6.8611M}, 660~BC \citep{PNAS.10.1073/pnas.1815725116}, 1279~CE, and 1052~CE \citep{2021NatGe..14...10B}, 5410~BCE \citep{GRL_10.1029/2021GL093419}, while two candidates \citep[i.e., 7176~BCE and 5259~BCE][]{2022NatCo..13.1196B} are awaiting confirmation in the $^{10}$Be and $^{36}$Cl records. Accordingly, AD~774/775, 7176~BCE, and 5259~BCE are currently considered the strongest ESPE candidates occurring during the Holocene. Attempts to identify additional solar storms \citep{NeocleousEA2019, UsoskinEA2021} have been limited by the quality of the available data (i.e., resolution, precision, and noise) as well as by an incomplete understanding of the terrestrial paleo–cosmic-ray proxy system. The energy spectra of ESPEs cannot be directly reconstructed from isotope data; however, parameters assuming a prescribed spectral shape can be constrained using multiple cosmogenic radionuclide records. In practice, the spectral profile of ESPE fluence is often assumed to resemble that of strong modern GLEs, which have been reconstructed with comparatively high precision \citep{2018JSWSC...8A...4R}. However, none of the known GLE events, including the famous Carrington event of 1856, can be found in the records \citep[e.g.,][]{Herbst_2016, 2021JA029351}.

\begin{wrapfigure}{R}{0.5\textwidth}
\vspace*{-0.3cm}
\centering
\includegraphics[width=0.5\textwidth]{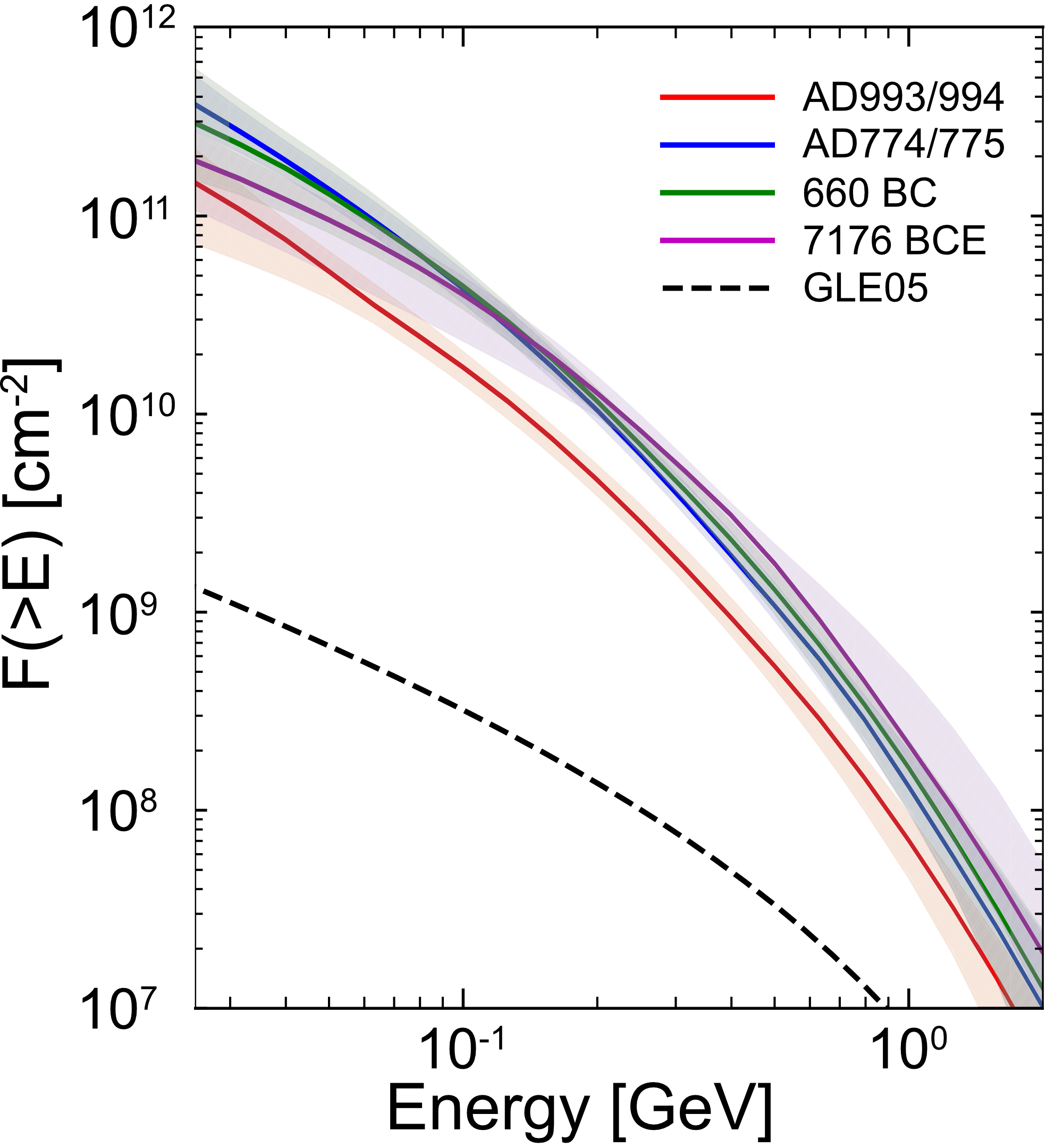}
\captionof{figure}{Most recent proton fluence reconstructions of AD 993/994, AD 774/775, 660 BC, and 7176 BCE (colored lines) in comparison with the strongest GLE directly observed on February 23, 1956 (i.e., GLE05, black dashed line). Figure after \cite{2023JGRA..12831186K}.}\label{fig:ESPE}
\vspace*{-0.3cm}
\end{wrapfigure}

The spectral shape of these ESPEs was reconstructed for the first time by \cite{2015NatCo...6.8611M} utilizing $^{36}$Cl data and selected strong GLE events to calculate the production rates and compare the modeled $^{36}$Cl/$^{10}$Be results to the observed values. Ever since, the methodology has been improved, first by \citet{Paleari2022} and, most recently, by \citet{2023JGRA..12831186K}. The latest fluence estimates of AD~774/775, AD~993/994, and 660~BCE yield softer ESPE spectra than the earlier estimates. As shown in Figure~\ref{fig:ESPE}, their reconstructed integral fluences (colored lines) likely were up to two orders of magnitude stronger than the strongest of the modern events (i.e., GLE05, black dashed line), which likely caused the observed spike-like increases in the atmospheric production of $^{14}$C, $^{10}$Be, and particularly $^{36}$Cl \citep[for more details, see][]{Usoskin+2023_SSRv}. However, most recently, \citet{2025A&A...702A.195K} suggested that these events not necessarily have to be connected to extreme solar flares (i.e., superflares with bolometric energies $>10^{33}$ erg). Instead, they could be produced by moderately strong flares comparable to those observed over the past few decades, provided that other coronal and interplanetary factors align favorably.

\subsubsection{Heliospheric Transport of Solar Energetic Particles}

After they are injected at the Sun, SEP propagation in interplanetary space is influenced by their movement along the HMF lines. In the absence of significant heliospheric disturbances, such as CMEs or interplanetary shocks, the HMF can often be effectively represented by a smooth large-scale structure such as the \citet{parker1958a} spiral model on which is superimposed a fluctuating component corresponding to magnetic turbulence. Particles exhibit adiabatic guiding-center motion along the HMF, which dictates their large-scale transport from the Sun into interplanetary space, while the presence of magnetic turbulence causes pitch-angle scattering. The latter is quantified by the pitch-angle diffusion coefficient, while cross-field transport is quantified by the perpendicular diffusion coefficient \citep{Strauss_2017, Laitinen_2019}. Because of the efficient focusing of SEPs close to the Sun, particle distributions tend to be highly anisotropic. As a direct consequence, Equation~(\ref{eq:TPE1}) cannot be applied to describe SEP transport within the HMF, and a more general transport approach is needed. 

The heliospheric evolution of the phase space density of a particle population is treated within a generalization of the Fokker--Planck equation. Adaption to a non-uniform, spatially-dependent HMF results in the focused transport equation \citep{Roelof-1969} given by 
\begin{equation}\label{eq:roeloff}
    \frac{\partial f}{\partial t} + \underbrace{v \mu \frac{\partial f}{\partial z}}_{(1)} + \underbrace{\frac{1-\mu^2}{2\xi} v \frac{\partial f}{\partial \mu}}_{(2)} - \underbrace{\frac{\partial}{\partial \mu}\left(D_{\mu \mu} \frac{\partial f}{\partial \mu}\right)}_{(3)} = Q(s, \mu, t)\, , 
\end{equation}
where $f$ represents the phase space density, $v$ the speed of the particles, $\mu$ the cosine of the pitch angle of the particles, $D_{\mu \mu}$ the pitch-angle diffusion coefficient, $s$ the spatial coordinate along the HMF, $\xi$ the focusing length given by $\xi = - \frac{B(s)}{\partial B / \partial s}$\footnote{In interplanetary space, focusing effects would reduce the pitch-angle of a SEP from $\sim 90^\circ$ to 0.7$^\circ$ at Earth, but would be significant only in the inner heliosphere \cite[see, e.g.,][and references therein]{frans}.}, and $Q$ the particle source. With that, Equation~(\ref{eq:roeloff}) describes the transport of charged energetic particles by means of (1) streaming along the mean HMF (field parallel propagation), (2) focusing due to the gradient in the HMF, which causes particles to experience a force along the HMF due to the conservation of their magnetic moment \cite[see also][]{BieberEA86}, and (3) pitch-angle scattering and diffusion. Note that Equation~(\ref{eq:roeloff}) does not contain the effects of perpendicular diffusion, modeled using a pitch angle dependent perpendicular diffusion coefficient \cite[see][and references therein]{Engelbrecht19}, which has been demonstrated to influence the transport of SEPs \cite[e.g.][]{Strauss_2017}. Drift effects are also not contained in Equation~(\ref{eq:roeloff}). However, it is debatable whether these will influence the transport of SEPs, given the high levels of turbulence in the very inner heliosphere \citep{jabus}.

Models like the 1D SEP-propagator model \citep{VanDenBerg2020}\footnote{\href{https://github.com/RDStrauss/SEP_propagator}{https://github.com/RDStrauss/SEP\_propagator}} based on a finite-difference scheme solve (Equation~(\ref{eq:roeloff})) by assuming a delta-like injection as well as specific functional forms of the HMF (i.e., the Parker spiral) and the pitch-angle diffusion coefficient \cite[for a comparative review on this and other existing models, see][]{Whitman}. Thereby, the quantity $D_{\mu\mu}$ is often parameterized as $D_{\mu\mu} = D_0 \left(|\mu|^{q-1} + H\right)\left(1-\mu\right)^2$\citep{2000ApJ...537.1073D, 2024ApJ...971..105L}, where the Kolmogorov inertial range spectral index is given by $q = 5/3$, $H = 0.05$ represents dynamical effects, and $D_0$ is the scattering amplitude \citep{VanDenBerg2020, 2024ApJ...971..105L}. Note, however, that $D_{\mu\mu}$ can also be modeled from first principles \cite[see, e.g.,][]{1994ApJ...420..294B,TeufelSchlickeiser2002,Shalchibook,vandenberg24}. Most recently, the model was successfully used to study the SEP transport of the relativistic widespread event of October 28, 2021 \citep{2026A&A...Lavasa}.

However, despite having been modeled extensively, the transport of SEPs within the heliosphere remains challenging, as key information on, for example, magnetic field turbulence \citep{Raouafi}, injection mechanisms \citep[e.g.,][]{Vlahos}, and ``seed'' particle populations \citep[see, e.g.,][and references therein]{WijsenEA} is either incomplete or not well constrained. In particular, the small-scale structures that govern particle scattering, as well as the properties of the initial suprathermal particle population, are difficult to quantify. This is mainly because direct measurements in the relevant coronal and heliospheric environments are not available, which introduces substantial uncertainties into transport modeling. In addition to these small-scale processes, SEP transport is also strongly modulated by the large-scale structure of the heliosphere. In particular, SIRs and transient structures such as CMEs can significantly alter particle propagation conditions at a given location. These large-scale structures can act both as barriers and as guiding channels for particle transport, by modifying the local magnetic connectivity, enhancing turbulence, and changing the diffusion conditions along field lines. As a result, SEP access to different heliolongitudes can be either suppressed or facilitated depending on the relative position of the observer with respect to these structures \citep[e.g.,][]{2021SpWea..1902654P, 2022ApJ...934...55L, 2025A&A...694A..64R}. Consequently, the observed SEP intensities and profiles at a given heliocentric distance can vary substantially even for nominally similar solar events, reflecting not only properties of the particle acceleration source but also the intervening heliospheric environment.

\subsubsection{Established (semi-)Empirical Statistical Solar Relations}

To decode the characteristics of SEPs and their heliospheric transport, over the past decades several empirical and/or semi-empirical statistical solar relations have been introduced and revisited.

\begin{wrapfigure}{r}{0.5\textwidth}
\vspace{-0.3cm}
\centering 
\includegraphics[width=0.5\textwidth]{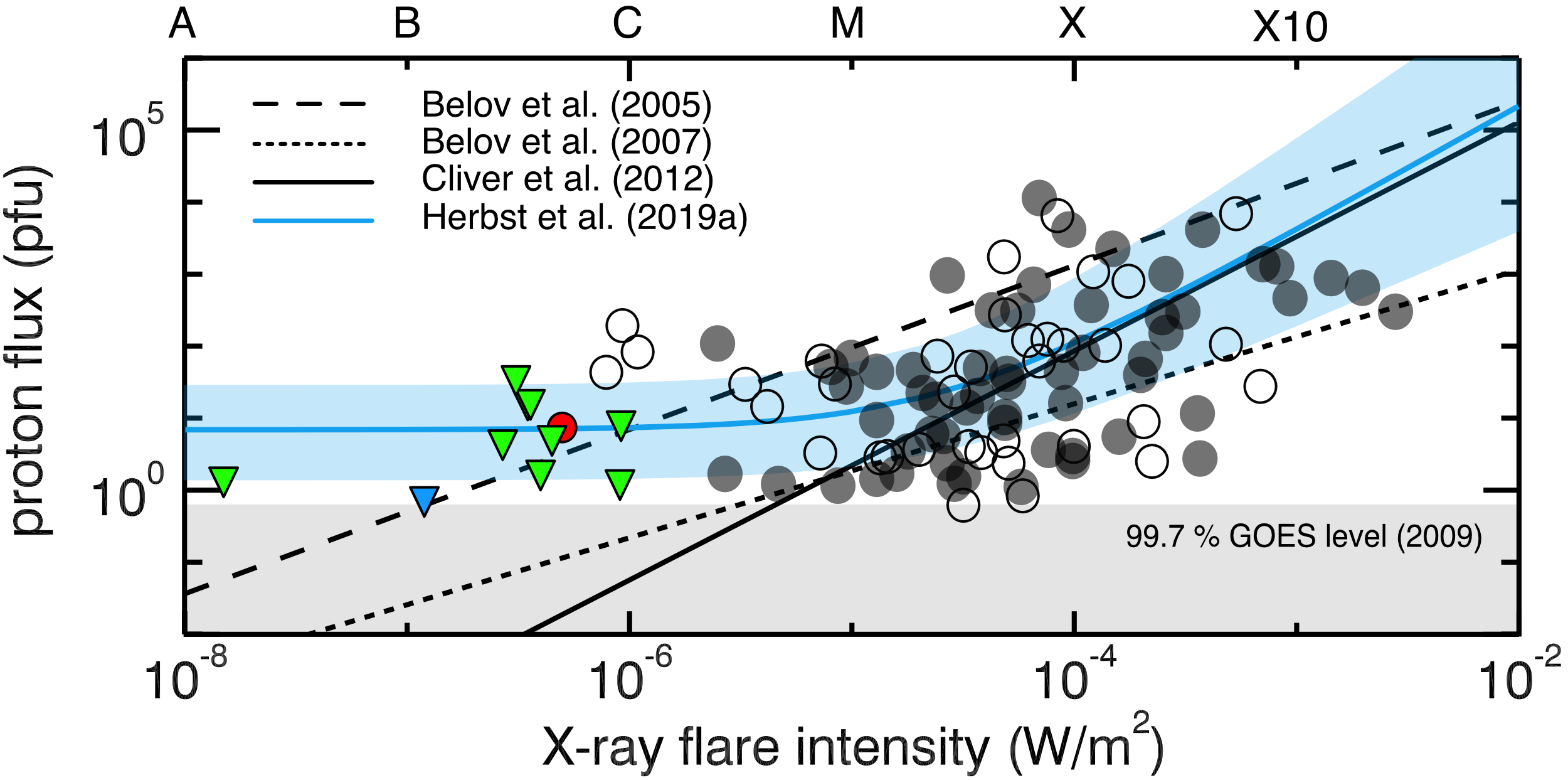}
\caption{Peak size distribution functions available in literature: \citet{Belov-etal-2005} as dashed line, \citet{Belov-etal-2007} as dotted line, \citet{Cliver-etal-2012} as solid black line, and \citet{2019A&A...621A..67H} given by the blue line including the estimated error band. Symbols represent solar observations (see text for details). Figure based on \cite{2019A&A...621A..67H}.}\label{fig:PSDs}
\vspace{-0.3cm}
\end{wrapfigure}

Besides the CME-driven correlations highlighted in Section~\ref{CME:Obs:property}, it has been demonstrated that X-ray and charged particle fluxes originating from the same solar event have an almost linear relation and that these so-called peak size distributions (PSDs) follow a power law \citep{hudson1978threshold, Belov-etal-2005, Cliver-etal-2012, Papaioannou-etal-2016, 2019A&A...621A..67H, 2018ApJ...864...48C}. Recent updates have been proposed \citet{2019A&A...621A..67H}. Figure~\ref{fig:PSDs} shows the different PSDs in comparison with solar observations based on \citet{Cliver-etal-2012} (black  circles, W20--W80), \citet{2017ApJ...843...31Y} (open circles, based on CaII), GOES events between 1975--2005 \citep[green triangles]{2019A&A...621A..67H}, SphinX data \citep[blue triangle][]{2017SoPh..292...77G}, and \citet{Cane-Lario-2006} (red circle, 25 MeV).

Furthermore, the soft X-ray flux is  closely correlated with the total energy released during solar flares. It is commonly accepted that the kinetic energy of the associated CMEs is directly proportional to the flare energy and that the mass of CMEs can be derived from the total mass contained within the gravitationally stratified active regions  \citep[e.g., ][see Section~\ref{CME:Obs:property} for further details]{2016ApJ...833L...8T}. 

Based on these assumptions, \citet{2023A&A...671A..66P} and \citet{2024A&A...690A..60P} provided the most recent PSD updates and the most recent updates on the solar proton fluxes and CME scaling laws, respectively.

\subsubsection{Deriving Stellar Energetic Particle Environments}

Stellar flare surveys from missions such as Kepler and TESS demonstrated that energetic flares are common among active stars (see Section~\ref{sec:flares}). Unfortunately, direct information on the energetic particle environment beyond the solar system is inaccessible, motivating the need for predictive stellar particle environment scaling/modeling. Yet, deriving the stellar particle environment is currently in its infancy.

First attempts to quantify the StEP environment explored incorporating known solar scaling relations and extreme-event methodologies developed in recent years \citep{2023A&A...671A..66P, 2024A&A...690A..60P} and enabled robust estimates of the worst-case stellar particle environments. As an extrapolation to the observed stellar flare energy regimes \citep[e.g.,][]{2017ApJ...843...31Y, 2019A&A...621A..67H} often results in much higher proton fluxes than observed in the solar system \citep[e.g.,][]{Scheucher_2020, EngelbrechtEA24, Herbst_2024}, their applicability is questionable because the soft X-rays--proton scaling relations should break down for events $>$X10 class \citep[e.g.,][]{2007ApJ...663L..45H, 2013ApJ...764..170D}. Nevertheless, recent reconstructions of the AD~774/775 event indicate that the associated flare was likely an X400$\pm$200 class flare \citep{2023A&A...671A..66P, 2024SoPh..299...39H}, consistent with flare energies observed in Sun-like stars directly comparable to the present Sun \citep[i.e., faculae-dominated sun-like stars,][]{2025A&A...697A..80H}, and potentially representing a solar worst-case scenario \citep{2025A&A...696A..63H}. Thus, although known scaling relations might not be applicable for all types of stars (i.e., K and M dwarfs), they might be a valuable first-order approximation for the G star regime. For more details on the extent of scaling relations, see \cite{2023A&A...671A..66P}.

\begin{figure}[!t]
    \centering
    \includegraphics[width=\textwidth]{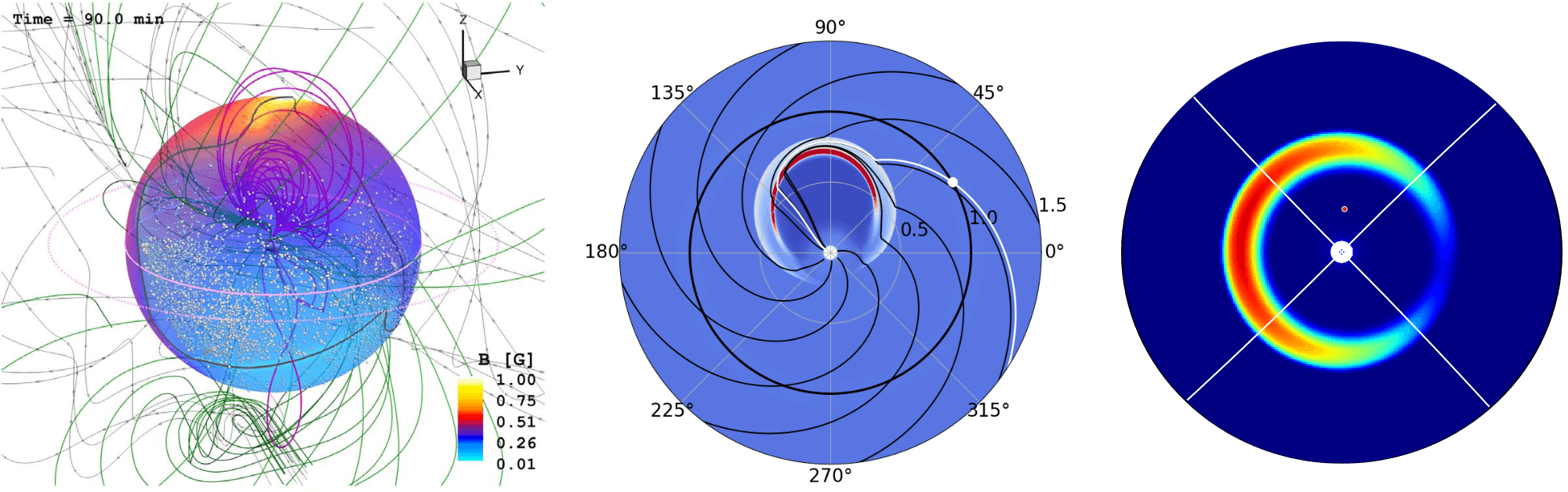}
    \caption{Model efforts to derive the StEP environment around cool stars: CME-driven model by \citet{Fraschetti_2022} (left panel) and \citet{10.1126/sciadv.abi9743} (middle panel), as well as the flare-driven model discussed in \citet{EngelbrechtEA24}.}
    \label{fig:StEPmodels}
\end{figure}

As of the time of writing, only a few attempts to numerically model the StEP transport have been reported in the literature. The first modeling effort was published by \citet{2019ApJ...874...21F}, where 3D MHD CME modeling outputs were fed into a test particle approach to calculate StEP fluxes within the TRAPPIST-1 system. Their results indicate proton fluxes around TRAPPIST-1 e that are up to six orders of magnitude higher than those observed at Earth. These findings align well with the extrapolated results based on the PSD functions \citep[][]{2019A&A...621A..67H, Herbst_2024}. \citet{Fraschetti_2022} further explored the StEP environment in the turbulent and CME-disrupted surroundings of the highly active M dwarf AU Mic (see the left panel of Figure~\ref{fig:StEPmodels}). Following the methodology of \citet{2019ApJ...874...21F}, they calculated the time-forward propagation of test particles for two distinct scenarios: one under quiescent interplanetary magnetic field conditions and the other when the CME front had reached regions beyond 100~$R_{\star}$. Their results suggest that, as is the case for the Sun, CME eruptions alter the structure of the large-scale magnetic field and primarily influence the StEP flux, surpassing the effects of magnetic turbulence. They also found that even for highly energetic CMEs, the StEP flux along the planetary orbit is dependent on the region of CME initiation, similar to the behavior observed within solar CMEs. In addition, \citet{10.1126/sciadv.abi9743} utilised the 2D Particle Acceleration and Transport in the Heliosphere \citep[iPATH;][]{2017JGRA..12210938H} model to simulate gradual StEP events of the Sun and Sun-like stars driven by fast CMEs with energies corresponding to stellar superflares (i.e., 10$^{33}$ to 10$^{35}$ erg, see middle panel of Figure~\ref{fig:StEPmodels}). Their study revealed that CME-driven shocks propagating in turbulent magnetic fields produce much stronger StEP events with spectra much harder than those observed on the Sun, and concluded that known solar relations cannot be applied to other G-stars.

As mentioned in Section~\ref{subsec:seps}, besides CME shock-front acceleration, SEPs are also accelerated during flares and can lead to an enhanced energetic particle environment around Earth. Based on astrospheric 3D MHD modeling \citep{HerbstEA20}, \citet{Engelbrecht24} solved a 2D version of the focused transport equation, where pitch angle-dependent diffusion perpendicular to the mean magnetic field \cite[see][and references therein]{Engelbrecht19} was included to derive the flare-induced StEP environment of our nearest neighbor Proxima Centauri (see the right panel of Figure~\ref{fig:StEPmodels}). Thereby, temporally resolved TESS flare observations \citep[e.g.,][]{vidaetal2019} were used to constrain the stellar injection profile. The authors found the astrospheric magnetic field to be underwound compared to the HMF, providing a highway for StEPs onto Prox Cen b. Thus, although the TESS observations displayed only moderate bolometric flare energies comparable to the Sun, within the observed 52 days three GLE05-type events (comparable to the strongest event directly observed at Earth via ground-based instrumentation) likely impacted the atmosphere of Prox Cen b, with a correspondingly significant effect on chemistry and climate.

\subsection{Future Prospects for Exoplanetary Sciences}

\textbf{Galactic Cosmic Ray Environment:} 
Further insight into stellar winds, as well as the large- and small-scale plasma properties of stellar astrospheres, will directly improve our understanding of energetic charged particle transport in plasma environments that often differ substantially from those encountered in the heliosphere, which has been comparatively well studied over the past decades \cite[see, e.g.,][and references therein]{Quenby84,frank,Moraal13,EngelbrechtEA22,RankinEA22}. Nevertheless, although the broader aspects of GCR modulation in the heliosphere are relatively well understood, many uncertainties remain, and these naturally carry over to astrospheric modulation studies. For example, the diffusion coefficients governing GCR transport in the heliosphere are still the subject of active research, both theoretically and through comparisons between transport model predictions, direct spacecraft observations \cite[see, e.g.,][]{BurgerEA00,2000ApJ...537.1073D,EngelbrechtEA22b,LangEA24}, and direct numerical test particle simulations \cite[e.g.][]{MaceEA00,MinnieEA07b,dalena,ElsEngelbrecht24,ElsEA24}. Numerous theoretical approaches have been developed to describe GCR diffusion coefficients, often yielding substantially different results even when based on identical turbulence input parameters \citep[see][]{EngelbrechtEA22}. A natural first step in an astrospheric context is therefore to employ transport approaches that have proven successful in heliospheric applications. However, as noted above, turbulence conditions in other astrospheres may differ substantially from those observed in the heliosphere. This was demonstrated, for example, by \citet{EngelbrechtEA26} for the astrosphere of the young solar proxy star $\kappa^1$ Ceti. Such differences may present significant challenges for particle transport modeling. For instance, under conditions of strong turbulence, expressions for the parallel diffusion coefficient derived from quasilinear theory \citep[e.g.,][]{Jok66}---an approach commonly employed in heliospheric modulation studies---may no longer be valid, requiring the development or application of more appropriate transport theories. Diffusion coefficients employed in phenomenological studies of GCR transport in the heliosphere should also be treated with extreme caution, as such coefficients are often chosen so that model results agree well with particular sets of spacecraft observations at Earth, and therefore are very likely not appropriate to use in an astrospheric context. Consequently, future astrospheric modulation studies will require greater care in selecting appropriate transport coefficients and turbulence models. In this regard, turbulence transport modeling may provide useful insights in terms of which theories would be most appropriate for use in a specific astrophysical context. The same argument can also be made in terms of other transport processes that these particles undergo, and, as such, astrospheric modeling endeavors would benefit from further observational and theoretical insights as to stellar winds.

The local interstellar spectrum is a key input for all cosmic ray modulation models, as it represents a fundamental boundary condition. Even in the heliospheric context it is relatively unknown, as it has only been directly measured by the Voyager spacecraft \citep{CummingsEA16} at relatively low energies. Typically, in heliospheric studies local interstellar spectra are modeled using GCR transport models such as GALPROP\footnote{\url{https://galprop.stanford.edu/}} \citep{Mos}. Many astrospheric modulation studies assume local interstellar spectra used in heliospheric studies. Given that the GCR intensity is expected to vary throughout the Galaxy \cite[e.g.][]{amato}, the question naturally arises as to whether that is an appropriate assumption to make. Another source of uncertainty is the nature of the ISM in the vicinity of any given star. Variations in parameters associated with the LISM (neutral hydrogen density, for example) have been demonstrated through MHD modeling to directly influence the large-scale characteristics of the heliosphere \cite[e.g.][]{ZankFrisch99,SchererEA08,MullerEA09}. This, at the very least, would influence the modulation volume of an astrosphere, which in turn would influence the intensity of GCRs at the location of an exoplanet within \cite[see, e.g.,][]{SchererEA08,fm13,Stepan}. For example, in the last approximately 10 million years, LISM conditions surrounding the heliosphere have changed considerably due to its motion through the Galaxy \cite[see, e.g.,][]{Wyman,Schullocalbubble,linsky,YoungbloodEA25,ZuckerEA25}. Many astrospheric modulation studies rely on MHD simulations performed under the assumption of LISM conditions similar to those encountered by the modern heliosphere, an assumption that may not be correct. Another factor that has hitherto remained unconsidered is that pertaining to the possible influence of stellar cyclic changes in astrospheric plasma parameters and their influence on GCR transport. In the heliosphere, solar cycle-related changes are well attested in observations and theoretically well-understood (see Figure~\ref{fig:heliomod} and its discussion). Cyclic behavior has been observed for several stars \cite[see, e.g.,][]{Macarena,Obridko-etal-2022,Jeffers-etal-2023}, including Proxima Centauri \cite[e.g.][]{WargelinEA17,KleinEA2021}, with evidence of magnetic polarity reversals \cite[see][]{Nigro,Becool}. An initial study of the influence of stellar cyclic behavior on GCR intensities in the astrosphere of Proxima Centauri by \citet{EH26} does indeed show that possible high-altitude atmospheric ionization profiles can vary by up to $28\%$ over a stellar cycle. Taking into account such cyclic temporal changes, particularly for stars with shorter (and stronger) cycles, may provide useful additional insights into the transport of GCRs in astrospheres, as well as a nuanced view as to temporal changes in atmospheric responses to these particles.

\textbf{Stellar Energetic Particle Environment:}
In terms of energetic particle environments within astrospheres, it is essential to emphasize that all the aforementioned StEP modeling efforts significantly depend on established solar relations due to the absence of critical information in the stellar case. While \citet{2019ApJ...874...21F}, \citet{Fraschetti_2022}, and \citet{Engelbrecht24} employed the PSDs \citep[see also][]{SteynEA20} to derive proton flux values, the findings of \citet{10.1126/sciadv.abi9743} are grounded in known solar CME relations, particularly the relationships $F_{SXR}\propto E_{CME}^{1.28}$ and $E_{CME} \propto M_{CME} \cdot V_{CME}^2$ \citep{Aschwanden_2017}, as well as the scaling relations between CME characteristics and peak proton fluxes discussed in \citet{2016ApJ...833L...8T} (e.g., Figure~\ref{Fig:4-2-3} in Section~\ref{CME:Obs:property}). Studies of the transport of such particles in astrospheric environments would undoubtedly benefit from additional insights as to the stellar winds of the relevant stars.

The increasing availability of solar data in recent years has addressed both observational and knowledge gaps, necessitating updates to these established solar relations. Recent revisions of known solar flare and CME relationships have been published by \citet{2023A&A...671A..66P} and \citet{2024A&A...690A..60P}, respectively, introducing new worst-case solar scenarios that align closely with the latest reconstructions of the most significant known events detected in cosmogenic radionuclide records \citep[e.g.,][]{2025A&A...696A..63H}. In addition, although it is often argued that these relations are not applicable for other cool stars (including Sun-like stars), comparing their extrapolated outputs with the StEP model efforts discussed above shows reasonable agreement. This should not be surprising, as the models fundamentally rely on these known solar relations. Thus, to address the limitations of the current StEP models, future numerical studies will require innovative approaches.

\section{Outlook and Perspectives}

In this review article, we have discussed the four main stellar forcing processes that shape exoplanetary environments, namely radiation, magnetized stellar winds, CMEs, and energetic particles. Together, these processes regulate the irradiation, plasma conditions, and particle environment experienced by planetary atmospheres, and they set the stage for atmospheric escape, magnetospheric compression, star--planet interactions, and the general space weather conditions relevant to habitability \citep[see also][]{2020IJAsB..19..136A}.

While the quiescent high-energy emission of cool stars is relatively well documented and a solid foundation for the the rotation-activity relationship exists, recent discoveries such as the ``rising NUV spectra'' observed during megaflares (shown in Figure~\ref{fig:NUV_HST_flare}), suggest that historical flare paradigms underestimate their high-energy emission, underscoring the vital importance of future comprehensive multi-wavelength observations. To accurately understand the physical mechanisms driving lower-atmospheric heating and to robustly validate theoretical radiative-hydrodynamic (RHD) models, the community must bridge the critical observational gap through simultaneous, continuous EUV and FUV spectroscopy. Recent and future missions capable of long-baseline, multi-band monitoring, such as SPARCS \citepads{2025JATIS..11d2212S}, ESCAPE \citep{France2022} and ExtREAM \citepads{2025SPIE13625E..0AD}, will provide the essential empirical constraints needed to resolve true flare energy budgets and accurately define the extreme radiation environments shaping exoplanetary atmospheres.

Similarly, although flare frequency (or energy) distributions have been widely reported, mostly in the optical and XUV regimes, a better understanding of the phenomena accompanying a flare of a given energy is still needed, at least in a statistical sense. How does the energetic particle flux scale with flare energies up to ``superflare'' levels? How do CMEs correlate with flares across the broad range of flare energies? We comment further on CMEs below. How universal are FFDs among stars with different activity levels or spectral types? Power-law slopes of FFDs are very likely to depend on the spectral band used for observations, for example optical versus XUV versus hard X-rays and gamma rays. In the XUV, there is strong evidence that at least part of the ``quiescent'' emission is simply the integrated emission from the large number of superimposed, unresolved flares at the low-energy end of an FFD. Values of the XUV FFD power-law slope $\alpha > 2$ strongly support this interpretation. However, the situation may be different in the optical and hard X-ray and gamma-ray regimes. The latter flare emissions are mostly related to the impulsive flare phase, in which energy is rapidly released, for example through accelerated particles. XUV emission, by contrast, is a consequence of cumulative heating, in some ways related to the impulsive phase, followed by gradual cooling. Other energy-loss channels are present as well, suggesting that linearity between XUV and optical flare output should not be assumed. Understanding how stellar output affects planetary atmospheres requires a more complete picture of the connections between flares, CMEs, and energetic particles, particularly from the observational side.

In terms of magnetized stellar winds, various methods for tackling their direct and indirect detections are rapidly developing. With a new generation of instruments, radio emission from stellar winds, including winds of main-sequence stars, are to be expected in the near future. Recently, stellar astrospheres were detected for the first time in X-rays, and this method may lead to additional detections in the near future. The ``hydrogen wall'' method continues to be the most fruitful way to assess the properties of magnetized outflows from cool stars. Stellar winds have also been indirectly detected from exoplanetary observations, and in the future this procedure may develop further, possibly using ground-based observations in the He\,{\sc i}~1083~nm line. At present, however, many of the available detection methods are potentially biased, including in target selection, or rely on simplified interpretations, as is evident from the disagreement between mass-loss rate estimates obtained with different techniques. However, the stellar sample with multiple estimates is still very small, and some upper limits are exceedingly high. Detailed case studies comparing the various methods will therefore be required. Combining theoretical modeling with observational evidence for stellar winds is also of paramount importance, as is investigating their links to stellar rotation and overall activity. Observational benchmarking can help provide a more solid physical basis for the boundary-condition problems that still remain in many modern MHD models of stellar winds. From a stellar population viewpoint, it is important to understand why some stars are very inactive for their rotation period. It is also interesting to study low-mass stars that have never decelerated despite their relatively old age. Could this mean that they have abnormally weak stellar winds, such that their angular momentum was never removed? Many questions in the field of stellar winds remain unanswered, but we will likely see substantial progress in the near future thanks to direct measurements of astrosphere sizes in the EUV (see~\citeads{2025SPIE13625E..0AD}).

On the other hand, as discussed in \citet[][part of this collection]{Chapter1}, the expanding pool of ZDI surveys together with the increasing emphasis on exoplanet habitability and star--planet interaction signatures, makes accurate stellar wind modeling more important than ever. 3D MHD simulations provide access to global quantities that cannot be inferred observationally, such as the morphology and extent of the Alfv{\'e}n surface, the wind ram and magnetic pressures at planetary orbits, and the location of transitions between sub- and super-Alfv{\'e}nic flows. These quantities are essential for evaluating atmospheric escape, magnetospheric compression, star--planet interaction-driven radio and chromospheric emission, and the effective inner edge of the habitable zone. Improving coronal boundary conditions and producing higher-fidelity ZDI maps therefore represents a key step toward building physically robust and predictive stellar wind models. Furthermore, the numerical simulations briefly reviewed here highlight both the strengths and limitations of the current modeling framework. They yield realistic, observation-anchored mass-loss rate predictions across the cool-star sequence, capture how magnetic topology shapes the global flow, and generate detailed Alfv{\'e}n surface geometries crucial for exoplanetary studies. However, differences between numerical implementations, uncertainties in thermodynamic and heating boundary conditions, and the intrinsic loss of small-scale magnetic structure in ZDI reconstructions introduce unavoidable model-to-model variability. Nevertheless, 3D MHD simulations remain the most sophisticated and physically consistent method for characterizing the winds of cool main-sequence stars. Crucially, their compatibility with radio-emission frameworks such as the ExPRES code enables direct predictions of SPI-induced radio signatures \citep{louis2019, 2026A&A...705A.149C}. Further discussion of radio-emission modeling is provided in \citet[][part of this collection]{Chapter3}.

In terms of magnetic transients such as stellar CMEs, recent dedicated studies are showing that stellar eruptions are now becoming observationally accessible rather than being discussed only through extrapolations from the Sun. Future missions such as ESCAPE will be able to provide measurements of CME-associated coronal dimmings at the $\geq 3\sigma$ level, even for moderate dimming depths of a few percent, as on the Sun, for all stars with an X-ray flux $\gtrsim 2 \times 10^{-12}$~erg~s$^{-1}$~cm$^{-2}$ and regardless of ISM attenuation \citep{Mason+2025_ApJ}. Similar CME-driven coronal dimming yields are expected for ExTREAM, with the added benefit of having up to $R\sim 6500$ of spectral resolution across the EUV range, sufficient to detect moderate CME events via the Doppler-shift method (\citeads{2025SPIE13625E..0AD}; see also Sect.~\ref{sec:Doppler-EUV_CMEs}). 

A crucial aspect of stellar CME detections is to connect these individual signatures in a coherent multi-wavelength framework, and to examine whether observed filament and prominence eruptions can develop into successful, escaping CMEs. Future coordinated multiwavelength observations in optical, UV, X-ray, EUV, and low-frequency radio wavelengths, together with realistic coronal and stellar wind models constrained by magnetic field measurements, will be essential for estimating CME masses, velocities, kinetic energies, and occurrence rates beyond the current poor statistics. These constraints are necessary for testing whether solar flare--CME scaling relations hold, or break down, for highly active stars, and for assessing how strong large-scale magnetic fields may suppress CMEs. Similarly, such observational constraints are fundamental for future modeling studies of these transients, permitting a better understanding of the physical properties of the eruptions, the conditions over which they develop and evolve, as well as the possible implication for CME--exoplanet interactions \citep[see also][for a discussion on bridging the solar--stellar CME connection]{2023BAAS...55c.254L}.

Finally, understanding astrospheric radiation environments requires detailed particle transport modeling, whether these particles are GCRs or StEPs, in plasma environments that can differ greatly from those in the heliosphere. Within the past decade, astrospheric GCR modulation models have progressed from extremely preliminary 1D approaches to full 3D modeling, reproducing the evolution of such modeling efforts in the heliosphere over a much shorter timescale. This rapid development has been informed by our increasing understanding of stellar winds and other plasma parameters governing GCR transport, an understanding built on observations and MHD modeling. However, astrospheric GCR modulation models have not yet reached the level of sophistication achieved by heliospheric models. For example, little attention has so far been given to time-dependent modulation effects arising from stellar cyclic changes in astrospheric plasma parameters, or to shorter timescales associated with stellar CMEs, to name just two examples. These factors are known to influence GCR modulation in the heliosphere, and they could reasonably be expected to play some role in the transport of these particles in other astrospheres. 

Another consideration is the evolution of astrospheric magnetic field turbulence in GCR modulation efforts. Although no direct observations of such turbulence exist to date, it is in principle possible to model these turbulence parameters using one of the growing number of well-tested turbulence transport models, at least in the heliosphere. Such information feeds directly into the calculation of GCR diffusion coefficients, which play a key role in particle transport and have proved essential in more advanced heliospheric GCR transport studies. Future prospects for particle transport modeling in astrospheres will certainly incorporate both temporal effects and an increasingly detailed treatment of the microphysics of charged particle diffusion due to scattering by turbulent astrospheric magnetic field fluctuations. These considerations are equally applicable to StEP transport modeling efforts. Recent model efforts, heavily relying on known solar scaling relations for various model input assumptions, have provided useful first-order estimates of the StEP fluxes around other stars. As new observations will become available, statistical relationships between stellar flares, CMEs, and the resulting particle environments will improve and path the way for new numerical approaches that will allow for more robust predictions of the stellar particle regimes.

Taken together, these developments are steadily transforming our understanding of stellar space weather from a largely solar-inspired framework into a predictive astrophysical discipline in its own right, capable of assessing the long-term evolution and habitability of planetary environments across the Galaxy. Continued progress in observations, theoretical modeling, and numerical simulations will be essential for developing a unified picture of stellar space weather, ultimately linking stellar magnetic activity to the evolution of planetary atmospheres and habitability.

\section{Acknowledgements}

The authors gratefully acknowledge all the contributions from the participants of the International Space Science Institute (ISSI) Workshop on Stellar Magnetism and its Impact on (exo)Planets, held in Bern, Switzerland, June 2 - 6, 2025, and convened by R.~Fares, S.~J{\"a}rvinen, E.~Isik, H.~Korhonen, C.~Mandrini, D.~Nandi, and K.~Vida  (\href{https://workshops.issibern.ch/stellar-magnetism/}{https://workshops.issibern.ch/stellar-magnetism/}). In particular, we thank Allison Youngblood, Alison Farish, and Mark Cheung for their support and feedback during the preparation of this review. We also thank the ISSI and its staff for hosting and supporting the workshop.

\renewcommand{\thefootnote}{\fnsymbol{footnote}}

\section{Funding}


EP acknowledges NASA's Heliophysics Guest Investigators‐Open (HGI-O; Grant 80NSSC23K0447) and Living With a Star (LWS; Grant 80NSSC24K1108) programmes.
KH acknowledges support from the Research Council of Norway through the Centres of Excellence funding scheme, project number 332523 (PHAB). 
AS and JC acknowledge support from the European Research Council (ERC) under grant agreement 101125367 (ERC-CoG ExoMagnets).\footnotemark[2]
KV acknowledges support provided by the Hungarian National Research, Development and Innovation Office grant \'Elvonal KKP-143986. 
KGK acknowledges the support by the European Research Council (ERC) under grant agreement 101123041 (ERC-CoG EASE).\footnotemark[2]
KP acknowledges support from the European Research Council (ERC) under grant agreement 101170037 (ERC-CoG Evaporator).\footnotemark[2]

\footnotetext[2]{Views and opinions expressed are however those of the author(s) only and do not necessarily reflect those of the European Union or the European Research Council Executive Agency. Neither the European Union nor the granting authority can be held responsible for them.}

\section*{Conflicts of Interest}
The authors have no relevant financial or non-financial interests to disclose, and no competing interests to declare that are relevant to the content of this article.

\bibliographystyle{aasjournal}
\bibliography{Ch2-biblio}

\end{document}